\documentclass[12pt,preprint]{aastex}
\usepackage{epsfig}
\usepackage{natbib}                
\shorttitle{Search for Hierarchical Triples}
\shortauthors{Gies et al.}

\newcommand{\noprint}[1]{}
\newcommand{\figsetstart}{{\bf Fig. Set} }
\newcommand{\figsetend}{}
\newcommand{\figsetgrpstart}{}
\newcommand{\figsetgrpend}{}
\newcommand{\figsetnum}[1]{{\bf #1.}}
\newcommand{\figsettitle}[1]{ {\bf #1} }
\newcommand{\figsetgrpnum}[1]{\noprint{#1}}
\newcommand{\figsetgrptitle}[1]{\noprint{#1}}
\newcommand{\figsetplot}[1]{\noprint{#1}}
\newcommand{\figsetgrpnote}[1]{\noprint{#1}}

\begin{document}


\title{A Search for Hierarchical Triples using Kepler Eclipse Timing}

\author{D. R. Gies, S. J. Williams,  
 R. A. Matson, Z. Guo, S. M. Thomas}
\affil{Center for High Angular Resolution Astronomy and
 Department of Physics and Astronomy,
 Georgia State University, P. O. Box 4106, Atlanta, GA 30302-4106, USA} 

\email{gies@chara.gsu.edu, swilliams@chara.gsu.edu, rmatson@chara.gsu.edu,
 guo@chara.gsu.edu, thomas@chara.gsu.edu}

\author{J. A. Orosz}
\affil{Department of Astronomy, San Diego State University, 
 San Diego, CA 92182-1221, USA} 
\email{orosz@sciences.sdsu.edu}

\and 

\author{G. J. Peters}
\affil{Space Sciences Center and Department of Physics and Astronomy, 
 University of Southern California, Los Angeles, CA 90089-1341, USA}
\email{gjpeters@mucen.usc.edu}

\slugcomment{AJ, in press}



\begin{abstract}

We present the first results of a {\it Kepler} survey of 
41 eclipsing binaries that we undertook to search for 
third star companions.  Such tertiaries will periodically 
alter the eclipse timings through light travel time and 
dynamical effects.  We discuss the prevalence of starspots
and pulsation among these binaries and how these phenomena 
influence the eclipse times.  There is no evidence of 
short period companions ($P < 700$~d) among this sample, 
but we do find evidence for long 
term timing variations in 14 targets ($34\%$). 
We argue that this finding is consistent with the presence 
of tertiary companions among a significant fraction of 
the targets, especially if many  
have orbits measured in decades.   This result supports 
the idea that the formation of close binaries involves the 
deposition of angular momentum into the orbital motion 
of a third star.

\end{abstract}




\section{Introduction}                              

Star formation requires very efficient processes to remove 
angular momentum from protostars in order to avoid faster 
than critical rotation.  This may be accomplished by 
magnetic winds among lower mass stars \citep{2005ApJ...632L.135M},
but the fact that binary stars are common among the more
massive stars 
\citep{1998AJ....115..821M,2007A&A...474...77K}
suggests that much of their natal angular momentum is deposited 
into orbital motion 
\citep{2002MNRAS.332..155L,2007ARA&A..45..481Z}.
Models of massive star formation 
\citep{2009Sci...323..754K,2010ApJ...708.1585K}
show that binary and often multiple stars with
orbital dimensions measured in AU can form 
through disk fragmentation processes.  In order to shrink 
such orbits to periods of days, interactions with a third 
star may be required to carry away angular momentum 
(for example, through Kozai cycles with tidal friction;
\citealt{2001ApJ...562.1012E}).
There is now substantial 
evidence that many close binaries have distant tertiary companions
\citep{2006AJ....131.2986P,2006A&A...450..681T,2010ApJS..190....1R}.

One of the best methods to detect tertiary stars orbiting 
close, eclipsing binaries is to search for periodic variations 
in the eclipse times caused by the light travel delay associated
with orbital motion, the so-called light travel time effect or LITE
\citep{1959AJ.....64..149I,2004ASPC..318..233M,2005ASPC..335..103P}.
If the third 
star's orbital period is short ($<1$ year), then additional, dynamical 
perturbations of the inner orbit can occur that will also create 
changes in the eclipse times 
\citep{2011A&A...528A..53B}.
The triple star system IU~Aur may represent an example where such 
dynamical perturbations influence the eclipse timings
\citep{2003A&A...403..675O}.
Eclipse timing observations have led to the identification of many 
candidate binaries with tertiary companions 
\citep{2010MNRAS.405.1930L,2010KPCB...26..269Z}
and even the detection of planets around a binary 
\citep{2009AJ....137.3181L}.
However, caution is required in the interpretation of trends 
in the eclipse times since other long-term, secular processes 
can also affect the angular momentum of the orbit 
\citep{2002AJ....123..450Z,2006AJ....132.2260H,2007MNRAS.378..757P}.

The NASA {\it Kepler} spacecraft offers us an unprecedented opportunity
to search for tertiary companions of eclipsing binaries thanks to its
extraordinary photometric precision and long time span of uninterrupted 
observations 
\citep{2011AJ....141...83P}.
Early results from {\it Kepler} have already led to the discovery of stellar 
\citep{2011AJ....142..160S}
and planetary companions 
\citep{2011Sci...333.1602D,2012Natur.481..475W}
of binary stars. 
Here we present a first examination of the eclipse timing variations 
in 41 eclipsing binaries that were identified prior to the launch 
of {\it Kepler}.  These systems are characterized by short periods, 
deep eclipses, and primary stars more massive than the Sun, 
parameters that may favor the detection of tertiary stars.  
A more complete examination of eclipse timing variations among 
a large subsample of binaries in the {\it Kepler} field of view will 
appear shortly (J.\ Orosz et al., in preparation).  We describe 
the measurements in \S2, outline the different processes that
cause timing variations in \S3, and discuss our results in \S4. 


\section{Eclipse Timing Measurements}               

We began this project in a Cycle 1 Guest Observer program on 20 targets, and 
we enlarged the sample to 40 and 41 systems in Cycles 2 and 3, respectively.  
The targets were selected from the 
All Sky Automated Survey Kepler Field of View study 
\citep{2009AcA....59...33P}, 
the HATNET survey 
\citep{2004AJ....128.1761H}, 
Vulcan survey 
\citep{2001PASP..113..439B,2007AAS...210.0307M}, 
and early {\it Kepler} results 
\citep{2011AJ....141...83P}.
The binaries were chosen from semi-detached and fully detached systems 
with deep eclipses ($>0.2$ mag).  The final sample consists of 41 binaries 
with orbital periods of 0.6 to 6.1 d and with primary star effective temperatures
in the range 5200 to 11000~K according to the {\it Kepler Input Catalog}
\citep{2009yCat.5133....0K}. 

We obtained all the long cadence, light curve data available through 
Quarter 9 (2009.3 -- 2011.5).   
We used the Simple Aperture Photometry product that was processed with minimal 
assumptions about the long term flux variations.  However, we found that there 
were significant drifts in flux level within and between the data quarters.  
These trends were removed in each quarter by binning the data into six parts 
and fitting a cubic spline through the means of the upper $50\%$ of each sample. 
Then the full set was divided by the spline fit and the data from each quarter 
combined.  This method effectively flattened out all the long term trends 
(on timescales larger than 30 d) with the exception of some fast drifts that
are occasionally seen at the start of a quarter.  The final product is a 
list of barycentric Julian date, normalized flux, and its uncertainty. 
There are some systematic differences between quarters (discussed further in 
the Appendix and noted by the letter Q in the final column of Table~1), 
the most egregious being those for KID~04678873, a star that has 
a nearby companion that blends by different amounts each quarter.  
Removing blending problems is important for models of the light curve, 
but the changing flux normalization has little influence on the eclipse timings
we present here. 

We measured the instance of mid-eclipse by fitting a model template to the observations 
around the eclipse times.  The template was constructed by binning all the data
in orbital phase according to an adopted period and trial epoch of mid-eclipse 
(usually from the work of 
\citealt{2011AJ....142..160S}), 
and then forming the mean phase and flux for each bin.  We fit a parabola to 
the lowest $20\%$ of the eclipse template data to find the actual phase of 
minimum, and this was used to re-center the template and adjust the epoch of
minimum light.   This served to produce a linear ephemeris of predicted 
eclipse times for the entire duration of the observations, and fits were 
made to each eclipse where there were at least three photometric measurements 
in each of the eclipse itself and in both adjoining out-of-eclipse sections
(of length similar to the full eclipse duration). 
Each eclipse was then fit in orbital phase space using the template by 
a non-linear, least-squares solution based upon four parameters,
the relative flux level and slope outside of eclipse, the eclipse depth, 
and the time of mid-eclipse.  The uncertainties in the eclipse times were estimated 
using the actual scatter of the observations from the fit of the template. 

We made such template fits for both the primary and secondary eclipses, 
and we found improved periods by setting the slope of the observed minus 
calculated ($O-C$) times to zero.   There were many cases where the 
periods derived from the primary and secondary eclipses were significantly
different, and we simply set the adopted period $P_a$ to 
be the average of these two periods.  The results are summarized in Table~1
that lists the {\it Kepler} identification number KID 
(appended with a P or S for the primary or secondary eclipses),  
the average timing error (the internal error $I$), 
the standard deviation of the $O-C$ times (the external error $E$), 
the adopted period $P_a$, 
the epoch $T$ of the mid-eclipse that defines the zero-point for the $O-C$ residuals, 
and the period $P$ that yields a zero slope in $O-C$ diagram.  The final 
columns give the formal value of $\dot{P}/P$ derived from a weighted, quadratic fit
of the $O-C$ trend (a measure of curvature) and symbolic remarks about the 
character of the $O-C$ diagram (discussed in \S3 and the Appendix). 
Numbers in parentheses give the uncertainty in the last digit quoted. 

\placetable{tab1}      

We measured over 27000 eclipse times in total, and these are collected in 
Table~2 (given in full in the electronic version of the paper). 
The columns give the KID number, the time of the eclipse from the 
adopted linear ephemeris $T_E$, the eclipse type (1 for the primary and 
2 for the secondary eclipse), the $O-C$ measurement, and its uncertainty.  

\placetable{tab2}      


\section{Characteristics of the $O-C$ Variability}  

The internal errors $I$ associated with the timing measurements 
are very small (1 to 318 s) compared to the sampling time 
of the long cadence data (1765 s) thanks to the extreme 
precision of the {\it Kepler} observations.  There are 
a few surprising cases where the external error $E$ is 
less than the internal error $I$, and these correspond to 
systems containing pulsating stars, where the fast varying flux 
causes an increase in the estimate of the internal scatter. 
We find that $E>I$ for most of the systems, indicating that 
there is some real variation present in the eclipse times. 
However, it is very important to place any apparent 
$O-C$ variations in the context of the kinds of light curve
variability observed outside of eclipse.   We found that 
it was very useful to display the entire set of photometric 
measurements in a diagram showing differences from the 
mean light curve.  We constructed such a diagram by 
placing these differences in a gray scale image as a 
function of orbital phase (on the adopted linear ephemeris) 
and of orbital cycle number (from the first recorded eclipse).  
These diagrams are presented
in Figure Set 1 (given in full in the electronic version). 
Each figure shows the mean light curve in the lower panel 
(extended in orbital phase to aid the sense of phase continuity)
and presents the differences in a gray scale image in the upper panel. 
The gray intensity varies from the lowest point below 
the average (black) to the highest peak above the average (white)
over a range in normalized flux given in each caption.  
Gaps in the time series are 
indicated by a uniform, mid-range gray intensity.   
The corresponding $O-C$ diagrams are given Figure Set 2 
(again given in full in the electronic version) for 
each binary in the sample.  These display the measurements 
for the primary and secondary eclipses as $+$ and $\times$ 
symbols, respectively.  

\placefigure{fig1}     

\placefigure{fig2}     

There are a number of features in these diagrams that are 
useful for the interpretation of the eclipse timing variations. 
Those binaries with slowly changing eclipse times are immediately 
detected in the gray scale diagrams in Figure Set 1 by the appearance
of alternating regions of bright and dark intensity (over those parts 
of the eclipsing light curve where the absolute value of the 
time derivative is large).  We assigned to the category of 
candidate third body systems those cases where the deviations 
in the secondary eclipse track those of the primary star's changes 
and where $|\dot{P}/P|$ is significantly larger than its error. 
These 16 systems are discussed on a case by case basis in the Appendix 
and are noted by letter T (candidate tertiary) in the last column 
of Table~1.  

The system with the largest systematic variation is KID 9402652
(Fig.\ Set 2.26).  Like the other candidates, the variation observed 
here has not yet completed one cycle over the two year duration of 
the {\it Kepler} observations.  We made preliminary LITE fits
for both the primary and secondary eclipses (shown as solid lines
in Fig.\ Set 2.26), and we find a semiampitude of 
$119 \pm 25$ s, an eccentricity $e=0.58 \pm 0.04$, 
a longitude of periastron $\omega = 290 \pm 3$ deg, 
an epoch of periastron of BY $2009.1 \pm 0.1$, and a 
period of $3.1 \pm 0.3$ y for the third body reflex orbit. 
For an assumed total mass of $2.7 M_\odot$, this yields 
a third star mass product of $M_3 \sin i = 0.32 \pm 0.08 ~M_\odot$.
If correct, then the {\it Kepler} data have covered only 
two thirds of one orbit. 

We found one system, KID~4544587 (Fig.\ Set 2.6), where the deviations
in the secondary timings were a mirror image of those for the primary. 
This is an eccentric system ($e=0.31$; \citealt{2011AJ....142..160S})
where the secondary eclipse is well offset from phase 0.5 (Fig.\ Set 1.6). 
We think the simplest explanation is that we are detecting apsidal 
motion due to tidal effects, and we used the method of
\citet{1992AJ....104.2213L}
to fit a solution for the advance of perihelion.  This fit yields 
parameters $\omega = 313\fdg3$ (at $T=$ BJD 2,455,262.7977), 
$\dot{\omega} = 0.0001107~(4)$ radians per sidereal period, 
and an apsidal period $U = 340.3 \pm 1.3$~y.   It is possible 
that some of this motion might be caused by a third body 
(see eq.\ 60, 61, and 63 in \citealt{2011A&A...528A..53B}), 
but this is not a necessary component.  We found, for example, 
that using structure constants $k_2$ from \citet{1992A&AS...96..255C}
and estimates of the fractional radii from \citet{2011AJ....142..160S},
the predicted apsidal period ranges from 180~y (for synchronous 
rotation at periastron) to 590~y (for rotation synchronous with 
the mean orbital motion).  Consequently, we simply assumed that 
the eclipse timing variations are due only to the tidal apsidal 
advance (marked by A in the last column of Table~1), and this 
system was not counted among the candidate third body group. 
Less pronounced timing variations in two other binaries 
(KID 04851217 and 08196180) are probably related to apsidal motion. 

There is ample evidence that many of the systems (containing stars 
with $T_{\rm eff} < 6500$~K) experience starspot activity.  This is 
observed as flux variations outside eclipse that generally move 
with respect to the orbital period.  We find examples where the spot rotation 
is faster than the orbit so that the spots are seen progressively 
earlier with each orbital cycle (KID 5444392, Fig.\ Set 1.13),
slower than the orbit (KID 8552540, Fig.\ Set 1.21), and where 
both trends are visible (KID 09899416, Fig.\ Set 1.30).  
These apparent spot variations do influence the eclipse times
\citep{2002A&A...387..969K}.
For example, we see that the times when dark spot patterns
cross the secondary eclipse in KID 5444392 (Fig.\ Set 1.13)
correspond to extrema of the secondary's $O-C$ timings 
(Fig.\ Set 2.13).  We detected such spot activity in 25 
systems, and these are indicated by the letter S in the last 
column of Table~1. 

We also found 23 cases where there were fast flux variations 
that are probably due to pulsations \citep{2011A&A...534A.125U}.
These appear in the gray scale diagrams with closely spaced 
brightness variations.   In many of these cases, the brightness
oscillations form coherent patterns in the gray scale 
diagrams, indicating that the pulsation periods have a 
harmonic or near-harmonic relationship to the orbital period
(for example, KID 3440230, Fig.\ Set 1.5), as was found 
for the remarkable binary HD~187091 (KID 8112039 = KOI-54) 
by \citet{2011ApJS..197....4W}. 

In order to search for possible evidence of periodic signals 
in the $O-C$ residuals, we calculated the power spectra of 
the $O-C$ measurements for each target (for $P>10$~d).  
In general, no significant periodicities were found with the exceptions of 
those with starspot activity (where the periodicities corresponded to the 
intervals between spot crossing at times of eclipse) and two pulsator cases. 
We found periods of 40.4~d and 26.6~d for KID 8553788 and 9592855,
respectively, and these can be seen to be the intervals between 
successive large pulsation peaks crossing the eclipses in these 
near-resonant cases (see Fig.\ Set 1.22, 1.27).  The binaries 
displaying pulsation are indicated by the letter P in the final 
column of Table~1, and their actual pulsation amplitudes are 
probably large since they were detected in long cadence data 
that averages over 29 minutes of flux variability. 

\citet{2011AJ....142..160S} and \citet{2012Natur.481..475W} 
found a few cases of {\it Kepler}
observations of eclipsing binaries where extra eclipses were 
seen from transits of tertiary stars. 
We made a rudimentary search for such transits by comparing the 
difference fluxes from the mean eclipse curve with a smoothed 
version of the same and by identifying any observations where 
three consecutive measurements were significantly lower than 
expected (minimum more than three standard deviations below). 
No events were identified through this scheme.  It is certainly
possible that smaller amplitude transits were missed, but we 
would have found any transits with amplitudes as large as those 
detected by \citet{2011AJ....142..160S} (see their Fig.~3). 


\section{Discussion}                                

We cannot at this stage claim that the period variations of 
all the third body candidates are actually due to orbital motion. 
Such a statement must await a longer time span of observations 
that shows true periodic variability in the $O-C$ residuals 
for both eclipses.   In the meantime, we caution that 
long-term secular variations associated with mass exchange, 
systemic mass loss, tidal dissipation, and magnetic cycles 
may be present in some systems 
\citep{2006A&A...450..681T,2007MNRAS.378..757P,2010KPCB...26..269Z}.  
However, there are some situations where the evidence does
indeed point towards the third body explanation.  
For example, the system with the largest $|\dot{P}/P|$, 
KID 04848423 (Fig.\ Set 2.11), shows systematic differences 
between the primary and secondary eclipse curves that 
may result from the dynamical effects of the third
star \citep{2011A&A...528A..53B}. 
Another target, KID 02708156 = UZ~Lyr, has a very long observational
history of eclipse timings, and the {\it Kepler} estimate of 
$\dot{P}/P$ is consistent with a periodic $O-C$ variation 
but is inconsistent with a secular variation.  Thus, 
we think it is appropriate to consider all the long-term 
variable systems as candidates for a tertiary companion. 

It is also important to review the selection effects associated
with this study.  First, any system with a third body will show
a periodic $O-C$ variation that will have both locally linear 
and curved sections.  Since our selection of candidates is based
upon only detecting curvature in the $O-C$ diagram, we will 
miss any systems that were observed in the locally linear 
part of their cycle (compensated for by our choice of period
that keeps the $O-C$ curve flat over the duration of the 
observations).  Second, the current observational window covers
only about two years, so the results are relatively insensitive
to motion in orbits with much longer periods, because the changes
will be minor over this time span.    

We show examples of the predicted variations in Figure~3 that
shows the semiamplitude as a function of outer orbital period 
for several assumed tertiary masses.   The solid lines show 
the LITE semiamplitude, 
$${m_3\over m_{123}} {{a_2 \sin i}\over{c}}(1 - e_2^2)$$
where $m_3/ m_{123}$ is the fractional mass of the third star
compared to the total mass of the system,
$a_2$ is the semimajor axis, $i$ is the inclination and 
$e_2$ is the eccentricity of the outer orbit, and $c$ is the speed of light 
\citep{2011A&A...528A..53B}. 
For the purpose of this figure, we assumed an inner binary mass
of $3 M_\odot$, $\sin i = \pi /4$, and a circular outer orbit ($e_2=0$).  
The dashed lines show the semiamplitude of the dynamical terms, 
$${15\over 8}{m_3\over m_{123}}{P_1\over P_2} (1 - e_2^2)^{-3/2} 
  {P_1\over{2\pi}}(1 - e_1^2)^{1/2}$$
where $P_1$ and $P_2$ are the periods of the inner and outer systems
\citep{2011A&A...528A..53B}. 
Again for illustration, we assumed representative values 
$e_1 = e_2 = 0$ and $P_1 = 1.5$~d.  The plus sign marks the 
preliminary results for KID 09402652.  We see that with timing results
accurate to a few tens of seconds, we should have been able to 
detect LITE variations for companions as small as $0.2 M_\odot$
with periods $>200$~d, and we might have found dynamical variations 
for such stars with periods $<200$~d.  The lack of detected systems
with $P<1000$~d probably means that such tertiary systems are rare. 

\placefigure{fig3}     

The lack of short period companions is consistent with other 
results on tertiary companions of spectroscopic binaries
\citep{2006A&A...450..681T} 
and on those previously found through LITE methods
\citep{2010KPCB...26..269Z}. 
\citet{2006A&A...450..681T} made an adaptive optics survey 
of nearby, spectroscopic binaries consisting of solar-type stars. 
They found that the frequency of tertiaries was 
$63\%$ for the whole sample and rose to $96\%$ for 
systems where the close binary period was less than 3 d. 
Based on their results, we would expect that most of the 
eclipsing binaries in our sample have tertiary companions. 
However, most of these tertiaries have long orbital periods 
(with a mean value of 32 y in the sample of 
\citealt{2010KPCB...26..269Z}), 
and only $11\%$ \citep{2010KPCB...26..269Z} to 
$15\%$ \citep{2006A&A...450..681T} of these triples 
have outer periods of less than 10~y.  
Thus, if we assume that all of the eclipsing binaries in 
our sample are triple, then we would have expected to 
find only 4 to 6 systems in the period range we can detect, 
much smaller than the 14 candidates we present here.   
We suspect that the discrepancy 
may result from the lack of detection of lower mass tertiaries
at lower periods in the earlier work and/or our likely inclusion 
of candidate systems whose variability actually has an origin 
unrelated to a tertiary. 

Our results appear to be consistent with the general occurrence
of tertiary companions to close binaries, provided that most 
of these have periods longer than a few years.  This agrees 
with theoretical results that dynamically stable, hierarchical 
systems have a large ratio of outer to inner period
\citep{1999AJ....117..621H,2001MNRAS.321..398M}
and with observational studies that show that low values
of the ratio are rare (for example, the smallest ratio was 
$P_2/P_1 \approx 1000$ in the survey of \citealt{2006A&A...450..681T}). 
We are currently involved in a moderate resolution spectroscopic 
study of all the eclipsing binaries in the sample presented here. 
We will make complete light and radial velocity 
curve studies of the targets, and this may aid detection 
of tertiaries.  Once allowance is made for the flux any nearby 
stars within the {\it Kepler} point spread function, 
the light curve analysis will include a potential third-light
component that acts to dilute (weaken) the eclipse depths. 
A tertiary might also be detected spectroscopically through
the flux dilution of the spectral lines (making the 
lines of the primary and secondary appear weaker than expected)
and/or through detection of the tertiary's relatively stationary spectral lines. 
Such results will provide additional constraints on the flux 
of any tertiaries as well as accurate masses and other parameters
for the stars in the close binaries. 


\acknowledgments

We are grateful for the support of Martin Still and the staff of 
the {\it Kepler} Guest Observer Office.  
We also thank Natalie Batalha, for sharing her VULCAN results 
on eclipsing binaries in advance of publication, and 
the referee Pavel Mayer, whose comments were particularly helpful.  
{\it Kepler} was competitively selected as the tenth Discovery mission. 
Funding for this mission is provided by NASA's Science Mission Directorate. 
This study was supported by NASA grants NNX10AC39G, NNX11AB70G, and NNX12AC81G.
The data presented in this paper were obtained from the 
Multimission Archive at the Space Telescope Science Institute (MAST). 
STScI is operated by the Association of Universities for Research 
in Astronomy, Inc., under NASA contract NAS5-26555. Support for MAST 
for non-HST data is provided by the NASA Office of Space Science via 
grant NAG5-7584 and by other grants and contracts.


Facilities: \facility{Kepler}


\appendix 

\section{Notes on Individual Stars}

\noindent{\sl
KID 02305372.} A similar period was found in both the 
HATNET \citep{2004AJ....128.1761H}
and ASAS surveys \citep{2009AcA....59...33P}. 
Both the primary and secondary $O-C$ measurements show a significant 
positive parabolic trend.  However, there is a slowly varying 
trend in the light curve residuals that probably results from 
starspot activity and leads to differences in the primary and 
secondary $O-C$ measurements. 

\noindent{\sl
KID 02708156.} This star, UZ~Lyr, has many eclipse measurements going 
back to 1920 that are listed by \citet{2001aocd.book.....K}.
There are oscillations in the historic $O-C$ measurements that have an amplitude 
of $\approx 900$~s, so the parabolic trends in the {\it Kepler} data
are probably related to changes on decadal time scales.  There are 
faster ($\approx 100$~d) variations in the secondary $O-C$ measurements
that are due to starspots.  
\citet{2001AJ....121.2723V} note the presence 
of H$\alpha$ emission in this Algol-type system. 

\noindent{\sl
KID 03241619.} A similar period was found in the HATNET survey. 
The light curve is strongly modulated by migrating starspots 
that clearly influence the $O-C$ timings for both eclipses. 

\noindent{\sl
KID 03327980.} The VULCAN survey \citep{2007AAS...210.0307M}
determined a similar period and 
ephemeris.  Beyond quarter to quarter systematic differences, 
there is no evidence of significant variability in the $O-C$ 
measurements. 

\noindent{\sl
KID 03440230.} The star was identified in the VULCAN survey with 
twice the actual orbital period.  Both primary and secondary 
$O-C$ timings indicate a negative periodic trend, but starspot
activity appears to influence the results for the secondary. 

\noindent{\sl
KID 04544587.} Similar periods were found for this eccentric system
in both the VULCAN and ASAS surveys.  The $O-C$ measurements have opposite 
trends for the primary and secondary as expected for apsidal motion. 
Both the primary and secondary $O-C$ timings are affected by resonant 
pulsations described by 
\citet{Hambleton2011}\footnote{http://kepler.nasa.gov/Science/ForScientists/keplerconference/sessions/}.

\noindent{\sl
KID 04574310.} Similar periods were determined in the HATNET and ASAS
surveys.  The light curve is dominated by starspot activity that 
especially influences the secondary eclipse times.  

\noindent{\sl
KID 04660997.} This star V1130~Cyg has the shortest period in our sample. 
The first published period from \citet{1966RA......7..217M}
of 0.562561247 (48) d is 
about $4\sigma$ longer than we and \citet{2004AcA....54..207K} find.
The $O-C$ measurements show fluctuations related to starspot activity on both stars.  

\noindent{\sl
KID 04665989.} A similar period was found in the ASAS survey. 
There are slight systematic variations in the light curves between 
quarters, and the gray scale representation of the light curve 
suggests that pulsation is present.  

\noindent{\sl
KID 04678873.} The {\it Kepler} period is similar to that found
in the HATNET and ASAS surveys.  The light curve shows evidence of 
pulsation that introduces scatter into the $O-C$ timings of the 
eclipses.  This target has a close visual companion that
is located $5\farcs1$ north and that is 0.9 mag fainter in the UCAC3
catalog \citep{2010AJ....139.2184Z}, 
and the influence of blending varies with each quarter. 

\noindent{\sl
KID 04848423.} There are only two quarters available currently, 
but this target shows the largest period changes of any in our 
sample.  The period appears to be increasing, yet the period 
from {\it Kepler} is slightly lower than that found in the 
HATNET and ASAS surveys (where the presumed period was set at 
twice the actual value).  There appear to be small but significant
differences in the $O-C$ timings of the primary and secondary that 
hint that dynamical affects from a third body are present. 
Some modest starspot activity is also indicated in the gray scale 
diagram of the light curve. 

\noindent{\sl
KID 04851217.} This star, HDE~225524, shows fast variability 
related to pulsation.  The $O-C$ measurements show a modest 
sign reversal over the course of the {\it Kepler} observations
that are opposite for the primary and secondary eclipses. 
The light curve (Fig.\ Set 1.12) shows that the secondary eclipse
occurs early (near phase 0.48), consistent with a non-zero 
eccentricity.  This is a candidate apsidal motion system. 

\noindent{\sl
KID 05444392.} A similar period was estimated in the 
HATNET and ASAS surveys.  The $O-C$ variations are closely
related to starspot changes evident in the gray scale 
depiction of the light curve. 

\noindent{\sl
KID 05513861.} The ASAS catalog reports a similar period. 
The $O-C$ curves for both the primary and secondary eclipses
show a large, positive curvature that requires a cubic
polynomial for an acceptable fit.  These variations may 
be caused by motion about a third body.   There is also 
evidence of rapid flux variability related to pulsation. 

\noindent{\sl
KID 05621294.} A similar period was established by the VULCAN survey. 
The light curve shows rapid variability related to pulsation of
the primary star (plus some modest starspot activity).  
The pulsations influence the $O-C$ measurements, 
but there also appears to be a negative parabolic trend in the 
$O-C$ timings of both components (with systematic differences 
evident at both extremes of the observing window).   This is
suggestive of changes related to a third body.

\noindent{\sl
KID 05738698.} Similar periods were found by HATNET and ASAS 
(half the actual period for ASAS).  This is a hint of longer 
pulsation periods ($\approx 2P$) in the gray scale diagram of the 
light curve.  
   
\noindent{\sl
KID 06206751.} The {\it Kepler} period agrees with earlier results 
from HATNET and ASAS.  There is evidence of both pulsation and
starspot activity in the gray scale light curve.  The $O-C$ timings 
suggest a low amplitude and negative curvature for both components
that may be indicative of a third body. 

\noindent{\sl
KID 07368103.} The VULCAN survey found a period equal to twice the 
actual one.  There is clear evidence of fast pulsation in the light 
curve as well as low-level starspot activity.  

\noindent{\sl
KID 08196180.} This is an eccentric system with narrow eclipses. 
The VULCAN survey found a similar period.  The light curve 
shows evidence of both starspot and pulsational modulation. 
The periods derived from the $O-C$ diagram are significantly 
different for the primary and secondary, and this may imply 
a very long term variation due to apsidal motion 
(probably consistent with the small radii, $R/a$, indicated by
the narrow eclipses). 

\noindent{\sl
KID 08262223.} The {\it Kepler} and VULCAN period results agree. 
The light curve is modulated by pulsation in near resonance with 
the orbit.  The wander in the $O-C$ values probably results from
the pulsational variations.  

\noindent{\sl
KID 08552540.} This eclipsing binary, V2277~Cyg, was discovered by
\citet{2001IBVS.5060....1D}, 
and the periods from TrES \citep{2008AJ....135..850D}, 
ASAS, and {\it Kepler}
all agree.  The light curve is modulated by starspot activity in 
both stars, and the apparent $O-C$ variations track the starspot 
evolution. 

\noindent{\sl
KID 08553788.} There is good agreement among the periods from ASAS, VULCAN, 
and {\it Kepler}.  The gray scale depiction of the light curve 
shows that there is near resonant pulsation in the primary, 
and there is probably starspot activity in both stars. 
The $O-C$ timings are influenced by both pulsation and starspots, 
but there is also a marked negative curvature in the both sets 
of $O-C$ measurements.  We tentatively suggest that the latter 
is due to a third body. 

\noindent{\sl
KID 08823397.} The period from VULCAN agrees with the {\it Kepler} result.
Apart from quarter to quarter systematic differences, the light curve
and $O-C$ trends look stable. 

\noindent{\sl
KID 09159301.} VULCAN estimated a period twice the actual one. 
This star displays rapid pulsations (which may form a near resonant
beat pattern in the more recent data) and starspot activity in 
the light curve.  The $O-C$ timings of the primary show a positive
curvature that we tentatively assume is related to third body effects. 

\noindent{\sl
KID 09357275.} The VULCAN and {\it Kepler} periods agree. 
The light curve is shaped by starspot activity that is readily
seen in the $O-C$ measurements for secondary eclipse. 
There are also quarter to quarter systematic differences in 
the secondary eclipse depth.   

\noindent{\sl
KID 09402652.} V2281~Cyg was discovered as an eclipsing binary by
\citet{2001IBVS.5060....1D}. 
The orbital period estimates from ASAS, 
WASP \citep{Payne2012}\footnote{http://wasp.paynescape.com}, 
and {\it Kepler} all agree.  
Both the primary and secondary $O-C$ values display a 
large amplitude and negative curvature trend.  The curve is 
not well matched with a parabola, but it can be reproduced 
as the light travel time effect of motion about a third star (\S 3). 

\noindent{\sl
KID 09592855.} All the estimates of period from ASAS, VULCAN, 
and {\it Kepler} are in agreement.  The light curve shows 
rapid variations presumably due to pulsation.  
 
\noindent{\sl
KID 09602595.} The first period determination for V995~Cyg was 
made by \citet{1963IBVS...32....1S}, 
and the star has been well observed 
since.  The {\it Kepler} period is close to the estimate of 3.556509~d 
from \citet{2004AcA....54..207K}.
The light curve is influenced by starspots
that result in an especially large $O-C$ variation for the secondary
eclipse.  The presence of negative curvature in the primary eclipse 
$O-C$ and of large $O-C$ residuals in the historical record
\citep{2001aocd.book.....K}
suggest that a third body may contribute to the observed variations. 

\noindent{\sl
KID 09851944.} The ASAS and {\it Kepler} periods are the same. 
This system displays near harmonic pulsational variability, possibly on 
both stars.  

\noindent{\sl
KID 09899416.} BR~Cyg has a long history of eclipse timings that
show variations as large as $\approx 1000$~s 
\citep{2001aocd.book.....K}.
The {\it Kepler} period agrees with the historical value from 
\citet{2004AcA....54..207K}. 
A multicolor study of the light curve was made by 
\citet{2005IBVS.5646....1T}.
The {\it Kepler} light curve shows the presence of starspot activity
that moves both ahead and behind the orbital period advance. 
Most of the eclipse timing variations are probably related to 
the starspot activity. 

\noindent{\sl
KID 10156064.} The period from VULCAN matches the {\it Kepler} result.
There is some evidence of starspot activity in the light curve diagram,
but the eclipse timings show no obvious variability. 

\noindent{\sl
KID 10191056.} The periods from ASAS, TrES, and {\it Kepler} are all 
consistent.  The periods from the primary and secondary differ 
by a small but significant amount.  The eclipses are narrow and 
thus the radii are relatively small, so we suspect that apsidal 
motion cannot be the explanation.  A third body dynamical perturbation
is a possible cause. 

\noindent{\sl
KID 10206340.} The periods from ASAS, 
\citet{2004AcA....54..207K}, 
and {\it Kepler} 
are in agreement for this system, V850~Cyg.  The light curve diagram 
reveals starspot activity associated with the primary, and there 
is evidence of pulsation that is best seen around orbital phase 0.25. 
\citet{2011A&A...534A.125U} suggest that this is a $\gamma$~Dor pulsator. 
The large excursions in the $O-C$ timings are associated with 
the starspot features in the light curve. 

\noindent{\sl
KID 10486425.} The periods from VULCAN, TrES, and {\it Kepler} are 
consistent with each other.  The light curve shows evidence of 
pulsation (probably related to the primary), and short-term trends
in the $O-C$ measurements are due to the net flux changes associated
with these pulsations.

\noindent{\sl
KID 10581918.} This system is WX~Dra, and the periods from 
\citet{2004AcA....54..207K}, 
ASAS, TrES, and {\it Kepler} all agree.  The light curve, gray scale 
diagram shows slowly evolving, starspot structures that affect the
$O-C$ timings.  

\noindent{\sl
KID 10619109.} The period estimates from {\it Kepler}, TrES, and VULCAN 
are consistent (although twice the period is reported for VULCAN). 
The light curve shows starspot activity and low amplitude pulsation. 
The primary and secondary periods are different, but this is probably
due to the larger influence of starspots at the beginning and ending
of the time series. 

\noindent{\sl
KID 10661783.} The derived periods from ASAS and {\it Kepler} are identical. 
Pulsations are prominent in the light curve of this totally eclipsing system. 

\noindent{\sl
KID 10686876.} The VULCAN, TrES, and {\it Kepler} periods are consistent. 
The light curve of this totally eclipsing binary shows starspot and 
pulsational activity.  Furthermore, both primary and secondary $O-C$ 
measurements display a negative curvature (the differences between the 
two sets are probably due to starspot activity).  This may result from 
the influence of a third body. 

\noindent{\sl
KID 10736223.} The period of V2290~Cyg was first determined by 
\citet{2001IBVS.5018....1G}
using observations made over a 73~y range, 
and the light curve was subsequently analyzed by 
\citet{2002AN....323..462P}.
The periods from Kreiner (2004) and {\it Kepler} are slightly 
less than that from 
\citet{2001IBVS.5018....1G}.
The {\it Kepler} light curve displays variations from starspots and 
pulsation.  Furthermore, both the primary and secondary $O-C$ data 
show a net positive curvature, implying a slightly increasing period. 
These facts suggest that the changes are related to a third body. 

\noindent{\sl
KID 10858720.} The periods for V753~Cyg are consistent among the estimates
from \citet{2004AcA....54..207K}, 
ASAS, and {\it Kepler}.  The light curve indicates
the primary is a pulsator, but the $O-C$ timings appear relatively constant. 
\citet{1985PASP...97.1178K} checked for H$\alpha$ emission from circumstellar gas 
but found none. 

\noindent{\sl
KID 12071006.} The period of V379~Cyg was first determined by 
\citet{Belyawsky1936}, 
but the system has had little attention since then.  This is the longest period 
system in the sample, and it has a very deep primary eclipse.  
The light curve shows evidence of pulsation and starspots. 
The primary $O-C$ timings show a slight negative curvature, but 
we suspect this is due to the characteristics of the starspots 
near the beginning of the observations.



\bibliography{apj-jour.bib,ms}


\begin{deluxetable}{rrrcllcl}
\tabletypesize{\scriptsize}
\tablewidth{0pt}
\tablenum{1}
\tablecaption{Eclipsing Binary Properties\label{tab1}}
\tablehead{
\colhead{}  &
\colhead{$I$}      &
\colhead{$E$}  &
\colhead{$P_a$}  &
\colhead{$T$}  &
\colhead{$P$}  &
\colhead{$\dot{P}/P$}  &
\colhead{}  \\
\colhead{KID}  &
\colhead{(s)}  &
\colhead{(s)}  &
\colhead{(d)}  &
\colhead{(BJD-2,400,000)}  &
\colhead{(d)}  &
\colhead{($10^{-6}$ y$^{-1}$)}  &
\colhead{Comment\tablenotemark{a}} }
\startdata
 2305372P &   5.1 &  60.7 & 1.4046774 & 55075.52509 (6)  & 1.404678238 (8)  & 9.45 (4)         & Q, S, T    \\
 2305372S &  20.4 &  96.6 &  \nodata  & 55587.5342 (6)   & 1.40467658 (5)   & 15.4 (2)         &  \nodata   \\
 2708156P &   2.8 &  39.8 & 1.8912670 & 55438.49951 (2)  & 1.89127025 (1)   & 2.29 (2)         & S, T       \\
 2708156S &  17.8 &  46.5 &  \nodata  & 55227.6235 (3)   & 1.89126366 (7)   & $-$1.9 (1)\phs   &  \nodata   \\
 3241619P &  14.3 &  34.2 & 1.7033444 & 55159.64932 (6)  & 1.70334416 (3)   & 2.20 (7)         & S          \\
 3241619S &  41.4 &  44.2 &  \nodata  & 55574.4141 (7)   & 1.70334564 (5)   & 2.7 (2)          &  \nodata   \\
 3327980P &   5.2 &   2.6 & 4.2310219 & 55411.36394 (7)  & 4.23102181 (9)   & 0.01 (9)         & Q          \\
 3327980S &   6.4 &   2.9 &  \nodata  & 55735.03527 (9)  & 4.2310220 (1)    & $-$0.1 (1)\phs   &  \nodata   \\
 3440230P &   4.6 &  48.5 & 2.8811205 & 55537.69744 (3)  & 2.88111953 (3)   & $-$8.23 (6)\phs  & S, T       \\
 3440230S &  27.2 &  86.8 &  \nodata  & 55057.9937 (2)   & 2.8811285 (2)    & $-$2.0 (3)\phs   &  \nodata   \\
 4544587P &   4.8 & 150.5 & 2.1891140 & 55341.60581 (4)  & 2.18909716 (3)   & 0.38 (5)         & A, P       \\
 4544587S &   3.8 & 149.8 &  \nodata  & 55358.35084 (3)  & 2.18913086 (2)   & $-$0.17 (3)\phs  &  \nodata   \\
 4574310P &   1.8 &   4.8 & 1.3062191 & 55235.49909 (2)  & 1.306218991 (4)  & $-$0.18 (1)\phs  & S          \\
 4574310S &   6.2 &  18.6 &  \nodata  & 55614.9559 (1)   & 1.30621925 (1)   & $-$0.12 (4)\phs  &  \nodata   \\
 4660997P &  16.9 &  43.8 & 0.5625604 & 55177.9496 (3)   & 0.562560545 (6)  & 1.15 (3)         & S          \\
 4660997S &  22.7 &  53.5 &  \nodata  & 55009.4630 (1)   & 0.562560116 (4)  & $-$0.94 (3)\phs  &  \nodata   \\
 4665989P &   1.6 &   1.4 & 2.2480675 & 55626.68385 (2)  & 2.248067537 (8)  & 0.02 (1)         & Q, P       \\
 4665989S &   2.3 &   2.2 &  \nodata  & 55540.13337 (3)  & 2.24806754 (1)   & 0.01 (2)         &  \nodata   \\
 4678873P &  24.4 &  47.0 & 1.8788771 & 55486.5223 (3)   & 1.8788767 (1)    & $-$0.1 (2)\phs   & Q, P       \\
 4678873S & 160.4 & 375.3 &  \nodata  & 55579.525 (2)    & 1.87887732 (5)   & 10.4 (1)         &  \nodata   \\
 4848423P &   3.5 &  11.1 & 3.0035189 & 55505.20125 (5)  & 3.0035202 (4)    & 34. (2)          & S, T       \\
 4848423S &   4.9 &  11.0 &  \nodata  & 55608.82124 (7)  & 3.0035184 (5)    & 43. (3)          &  \nodata   \\
 4851217P &  62.6 &  13.1 & 2.4702796 & 55487.4806 (6)   & 2.4702807 (4)    & 0.8 (7)          & A, P       \\
 4851217S &  45.5 &  11.2 &  \nodata  & 55093.4215 (7)   & 2.4702788 (3)    & 0.9 (5)          &  \nodata   \\
 5444392P &  11.4 &  29.3 & 1.5195281 & 55609.83190 (4)  & 1.51952822 (2)   & $-$0.68 (6)\phs  & S          \\
 5444392S &  15.5 &  44.3 &  \nodata  & 55569.5644 (1)   & 1.51952794 (4)   & $-$1.36 (9)\phs  &  \nodata   \\
 5513861P &   1.3 &  34.2 & 1.5101839 & 55500.17856 (2)  & 1.510184171 (5)  & 8.53 (1)         & P, T       \\
 5513861S &   1.1 &  33.7 &  \nodata  & 55111.306244 (4) & 1.510183743 (4)  & 7.83 (1)         &  \nodata   \\
 5621294P &   7.4 &  15.7 & 0.9389071 & 54989.2511 (1)   & 0.938906670 (9)  & $-$1.73 (5)\phs  & P, S, T    \\
 5621294S &  45.7 &  60.0 &  \nodata  & 55657.2851 (7)   & 0.93890760 (3)   & $-$5.6 (1)\phs   &  \nodata   \\
 5738698P &   4.8 &   1.6 & 4.8087740 & 55100.85490 (5)  & 4.8087739 (1)    & 0.02 (7)         & P, Q       \\
 5738698S &   4.5 &   2.5 &  \nodata  & 55189.81827 (5)  & 4.80877398 (9)   & 0.03 (7)         &  \nodata   \\
 6206751P &  13.0 &  25.5 & 1.2453439 & 55702.4624 (2)   & 1.24534410 (2)   & $-$1.34 (5)\phs  & P, S, T    \\
 6206751S &  28.6 &  58.2 &  \nodata  & 55510.0573 (1)   & 1.24534372 (5)   & $-$0.7 (1)\phs   &  \nodata   \\
 7368103P &  81.2 &  30.2 & 2.1825141 & 55445.506 (2)    & 2.1825156 (4)    & $-$0.6 (8)\phs   & P, S       \\
 7368103S & 317.8 & 470.6 &  \nodata  & 55084.300 (7)    & 2.1825116 (1)    & $-$6.8 (3)\phs   &  \nodata   \\
 8196180P &   5.1 &   8.4 & 3.6716598 & 55372.64452 (5)  & 3.67166118 (5)   & 0.00 (6)         & A, P, S    \\
 8196180S &  13.9 &  25.9 &  \nodata  & 55465.9158 (1)   & 3.6716584 (1)    & 0.3 (2)          &  \nodata   \\
 8262223P &   6.1 &   6.3 & 1.6130147 & 55430.90874 (8)  & 1.61301466 (2)   & $-$0.10 (5)\phs  & P, Q       \\
 8262223S &  15.9 &  16.4 &  \nodata  & 55694.63694 (7)  & 1.61301467 (5)   & $-$0.1 (1)\phs   &  \nodata   \\
 8552540P &   9.6 &  23.7 & 1.0619344 & 55471.26705 (3)  & 1.06193406 (1)   & 0.06 (4)         & S          \\
 8552540S &  13.8 &  28.7 &  \nodata  & 55610.9120 (2)   & 1.06193481 (1)   & $-$0.08 (5)\phs  &  \nodata   \\
 8553788P &   3.9 &  23.2 & 1.6061743 & 55046.54573 (7)  & 1.60617393 (1)   & $-$4.04 (3)\phs  & P, S, T    \\
 8553788S &  23.5 &  42.6 &  \nodata  & 55142.1194 (2)   & 1.60617473 (8)   & $-$3.7 (2)\phs   &  \nodata   \\
 8823397P &   1.0 &   1.0 & 1.5065037 & 55440.539331 (8) & 1.506503705 (3)  & 0.006 (7)        & Q          \\
 8823397S &   2.3 &   3.4 &  \nodata  & 55646.17738 (3)  & 1.506503679 (7)  & 0.04 (2)         &  \nodata   \\
 9159301P &  12.1 &  17.4 & 3.0447717 & 55726.6290 (1)   & 3.0447698 (1)    & 1.4 (1)          & P, S, T    \\
 9159301S & 128.0 & 125.9 &  \nodata  & 55645.9444 (9)   & 3.04477509 (3)   & 2.6 (1)          &  \nodata   \\
 9357275P &   1.7 &   1.7 & 1.5882981 & 55573.50073 (3)  & 1.588298146 (5)  & 0.07 (1)         & Q, S       \\
 9357275S &   9.4 &  15.9 &  \nodata  & 55378.9344 (1)   & 1.58829803 (3)   & 0.13 (9)         &  \nodata   \\
 9402652P &   0.9 &  53.9 & 1.0731136 & 55132.422059 (5) & 1.073113953 (2)  & $-$10.227 (6)\phs & T         \\
 9402652S &   1.0 &  53.5 &  \nodata  & 55135.10484 (1)  & 1.073113272 (2)  & $-$10.182 (6)\phs &  \nodata  \\
 9592855P &  10.9 &  20.8 & 1.2193248 & 55656.3029 (1)   & 1.21932475 (2)   & $-$1.65 (5)\phs  & P, Q       \\
 9592855S &  13.2 &  22.9 &  \nodata  & 55424.0215 (1)   & 1.21932480 (3)   & $-$0.46 (8)\phs  &  \nodata   \\
 9602595P &   2.0 &  51.0 & 3.5565240 & 55375.52022 (1)  & 3.55651727 (2)   & $-$3.32 (2)\phs  & S, T       \\
 9602595S &  13.1 & 103.2 &  \nodata  & 54993.1977 (2)   & 3.5565296 (2)    & $-$11.3 (2)\phs  &  \nodata   \\
 9851944P &  14.5 &  16.1 & 2.1639018 & 55345.3670 (2)   & 2.16390189 (8)   & $-$0.2 (1)\phs   & P          \\
 9851944S &  14.9 &  14.9 &  \nodata  & 55339.9571 (1)   & 2.16390178 (8)   & 0.1 (1)          &  \nodata   \\
 9899416P &   2.1 &  10.2 & 1.3325638 & 55248.88593 (2)  & 1.332564453 (5)  & 0.19 (1)         & Q, S       \\
 9899416S &   4.1 &  22.0 &  \nodata  & 55532.05579 (9)  & 1.332563116 (9)  & $-$0.13 (2)\phs  &  \nodata   \\
10156064P &   4.5 &   3.2 & 4.8559364 & 54988.44967 (5)  & 4.85593639 (9)   & 0.01 (8)         & Q, S       \\
10156064S &   4.6 &   3.2 &  \nodata  & 55413.34462 (6)  & 4.85593643 (9)   & 0.04 (8)         &  \nodata   \\
10191056P &   1.6 &   1.7 & 2.4274949 & 55120.10058 (2)  & 2.42749482 (1)   & $-$0.03 (2)\phs  & Q, T       \\
10191056S &   1.8 &   1.6 &  \nodata  & 55521.85228 (1)  & 2.42749498 (1)   & $-$0.03 (2)\phs  &  \nodata   \\
10206340P &  13.5 & 123.3 & 4.5643870 & 55691.8342 (2)   & 4.5643908 (2)    & 5.5 (2)          & P, S       \\
10206340S &  24.8 & 180.6 &  \nodata  & 55137.26033 (7)  & 4.5643817 (3)    & $-$3.4 (3)\phs   &  \nodata   \\
10486425P &  29.1 &  94.1 & 5.2748090 & 55091.0297 (7)   & 5.2748069 (4)    & 2.1 (4)          & P          \\
10486425S &  51.6 & 232.2 &  \nodata  & 55452.353 (1)    & 5.274813 (1)     & 0. (2)           &  \nodata   \\
10581918P &   5.0 &  21.5 & 1.8018650 & 55434.86895 (4)  & 1.80186287 (3)   & $-$0.45 (9)\phs  & Q, S       \\
10581918S &  31.4 &  90.6 &  \nodata  & 55396.1296 (2)   & 1.8018668 (2)    & $-$2.5 (5)\phs   &  \nodata   \\
10619109P &   8.5 &  33.0 & 2.0451630 & 55086.03015 (9)  & 2.04516121 (4)   & 1.5 (1)          & P, Q, S    \\
10619109S &  45.4 & 137.1 &  \nodata  & 55631.0696 (7)   & 2.04516339 (2)   & $-$0.14 (4)\phs  &  \nodata   \\
10661783P &  29.1 &   9.0 & 1.2313633 & 55102.7180 (4)   & 1.23136331 (5)   & 0.0 (2)          & P, Q       \\
10661783S &  55.6 &  21.1 &  \nodata  & 54974.0406 (3)   & 1.2313633 (1)    & 0.2 (3)          &  \nodata   \\
10686876P &   3.2 &  39.6 & 2.6184286 & 55111.05523 (3)  & 2.61842922 (2)   & $-$6.99 (4)\phs  & P, Q, S, T \\
10686876S &  14.8 &  53.4 &  \nodata  & 55591.5375 (1)   & 2.6184274 (1)    & $-$8.5 (2)\phs   &  \nodata   \\
10736223P &   4.0 &  14.9 & 1.1050922 & 55136.83544 (4)  & 1.105091980 (7)  & 2.58 (3)         & P, Q, S, T \\
10736223S &  24.6 &  28.4 &  \nodata  & 55145.1241 (3)   & 1.10509236 (4)   & 2.6 (2)          &  \nodata   \\
10858720P &   2.6 &   5.5 & 0.9523776 & 55502.47547 (3)  & 0.952377636 (4)  & $-$0.07 (2)\phs  & P, Q       \\
10858720S &   2.2 &   4.4 &  \nodata  & 55137.23868 (2)  & 0.952377592 (3)  & 0.04 (1)         &  \nodata   \\
12071006P &   9.3 &  16.2 & 6.0960140 & 55181.39070 (6)  & 6.096022 (2)     & $-$19. (3)\phs   & P, Q, S    \\
12071006S & 140.0 & 182.3 &  \nodata  & 55172.245 (1)    & 6.09600 (2)      & $-$63. (46)\phs  &  \nodata   
\enddata
\tablenotetext{a}{A = apsidal motion; P = pulsation; Q = systematic variations between quarters; 
S = starspots; T = candidate third body system.} 
\end{deluxetable}

\newpage

\begin{deluxetable}{rccrr}
\tablewidth{0pt}
\tablenum{2}
\tablecaption{$O-C$ Eclipse Timing Measurements\tablenotemark{a}\label{tab2}}
\tablehead{
\colhead{KID}  &
\colhead{$T_E$}  &
\colhead{Eclipse}  &
\colhead{$O-C$}  &
\colhead{$\sigma (O-C)$}  \\
\colhead{Number}  &
\colhead{(BJD-2,400,000)}  &
\colhead{Type}  &
\colhead{(s)}  &
\colhead{(s)}  }
\startdata
 2305372 &  54965.26211 & 2 & $-$3.1      &    39.9 \\
 2305372 &  54965.96025 & 1 &    26.4     &     1.6 \\
 2305372 &  54966.66678 & 2 &    25.4     &    45.8 \\
 2305372 &  54967.36493 & 1 &    26.8     &     1.6 \\
 2305372 &  54968.07146 & 2 &    6.4      &    42.9 \\
 2305372 &  54968.76960 & 1 &    27.5     &     2.1 \\
 2305372 &  54969.47614 & 2 &    69.2     &    43.3 \\
 2305372 &  54970.17428 & 1 &    26.2     &     2.3 \\
 2305372 &  54970.88082 & 2 &    0.9      &    36.8 \\
 2305372 &  54971.57896 & 1 &    26.5     &     1.2 \\
 2305372 &  54972.28549 & 2 &    17.5     &    43.4 \\
 2305372 &  54972.98364 & 1 &    27.9     &     1.1 \\
 2305372 &  54973.69017 & 2 &    10.7     &    40.2 \\
 2305372 &  54974.38831 & 1 &    30.0     &     1.7 \\
 2305372 &  54975.09485 & 2 &    90.6     &    51.1 \\
 2305372 &  54975.79299 & 1 &    27.5     &     2.1 \\
\enddata
\tablenotetext{a}{The full table is available in the electronic version.}
\end{deluxetable}



\clearpage

\figsetstart
\figsetnum{1}
\figsettitle{Light curve variations}

\figsetgrpstart
\figsetgrpnum{1.1}
\figsetgrptitle{g1}
\figsetplot{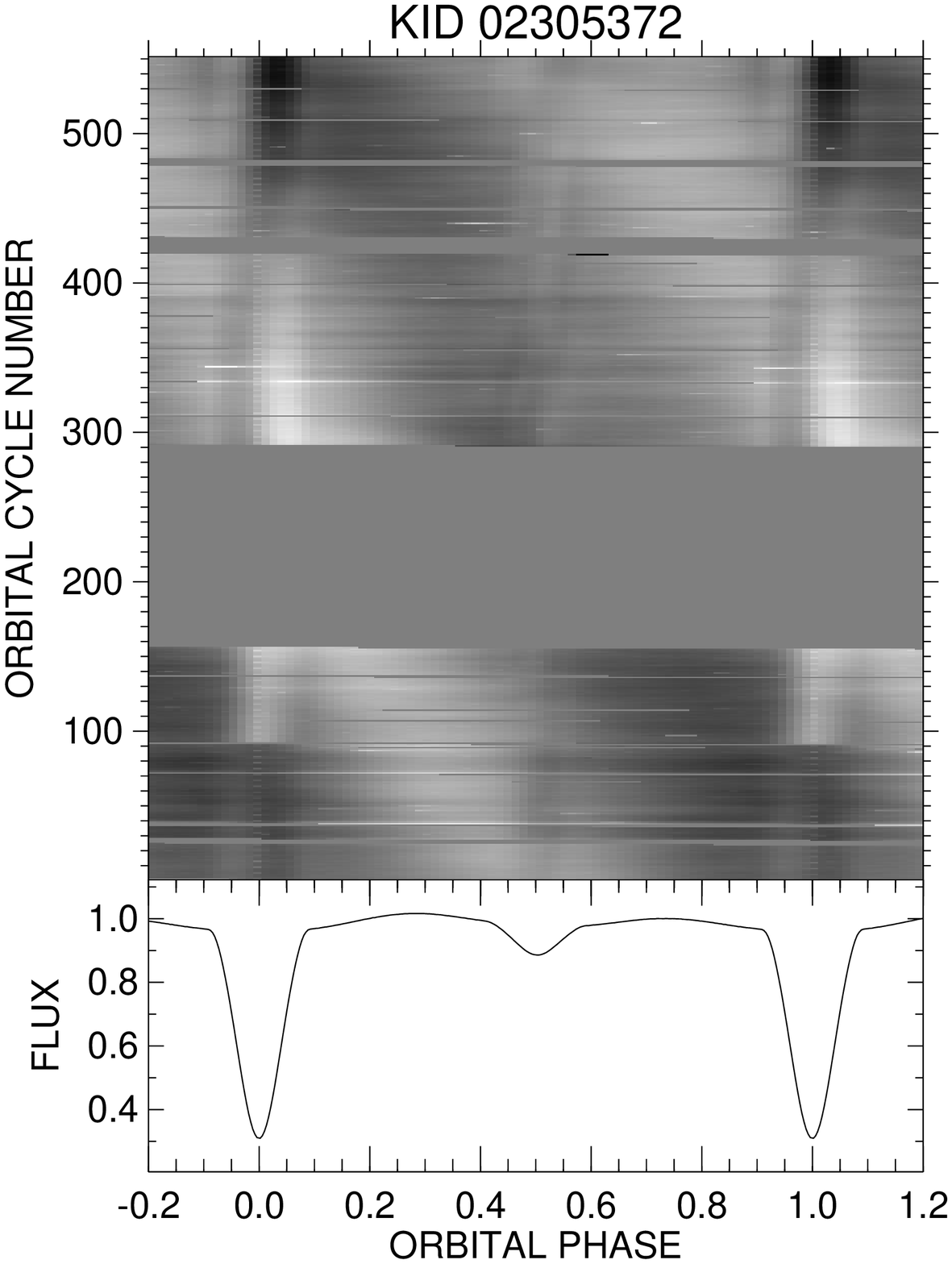}
\figsetgrpnote{The lower panel shows a mean, normalized light curve formed 
by binning in orbital phase.  The top panel shows the 
flux differences as a function of orbital phase and 
cycle number, represented as a gray scale diagram (range $\pm 3\%$). 
 }
\figsetgrpend

\figsetgrpstart
\figsetgrpnum{1.2}
\figsetgrptitle{g2}
\figsetplot{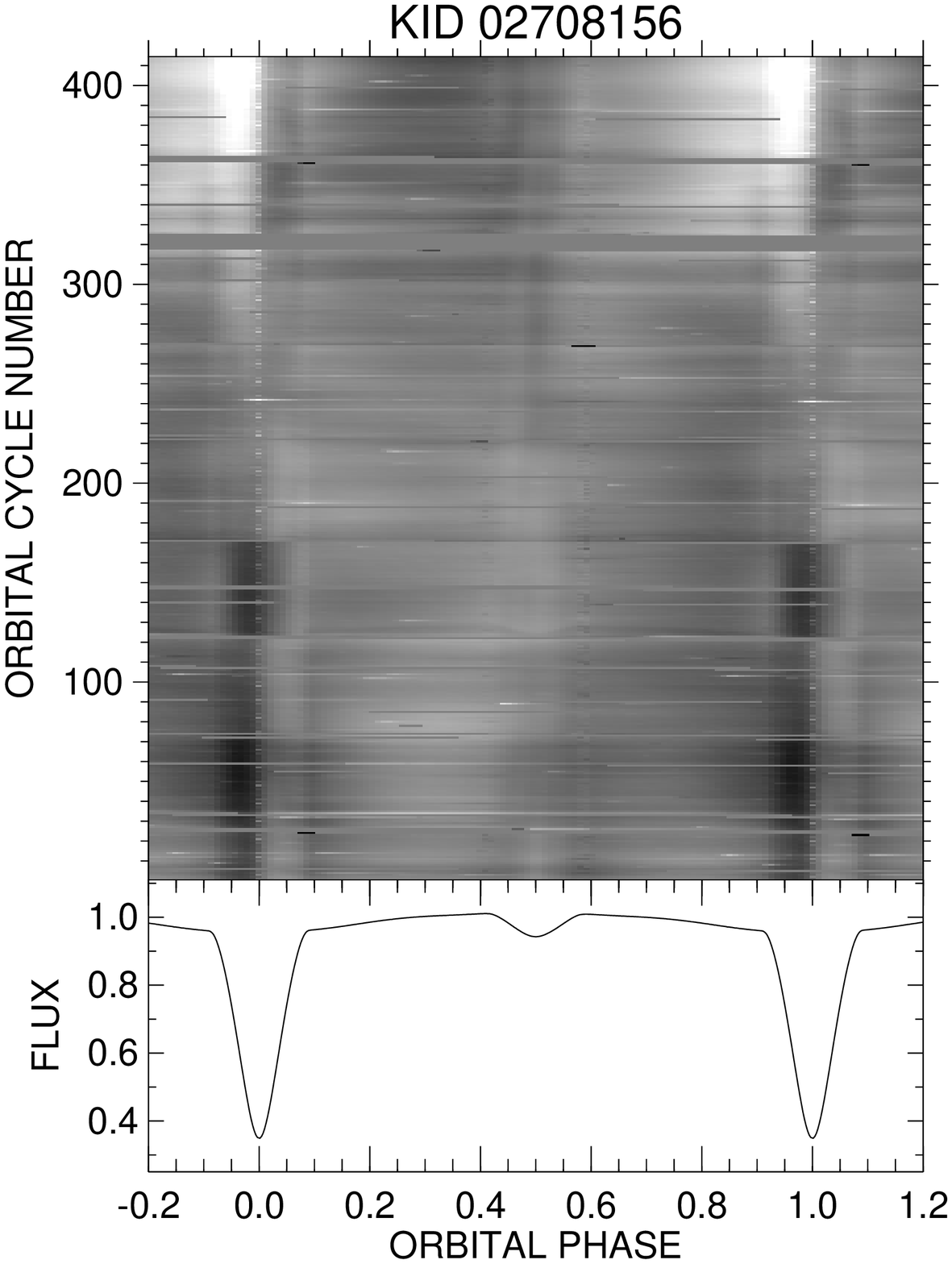}
\figsetgrpnote{The lower panel shows a mean, normalized light curve formed 
by binning in orbital phase.  The top panel shows the 
flux differences as a function of orbital phase and 
cycle number, represented as a gray scale diagram (range $\pm 1\%$). 
 }
\figsetgrpend

\figsetgrpstart
\figsetgrpnum{1.3}
\figsetgrptitle{g3}
\figsetplot{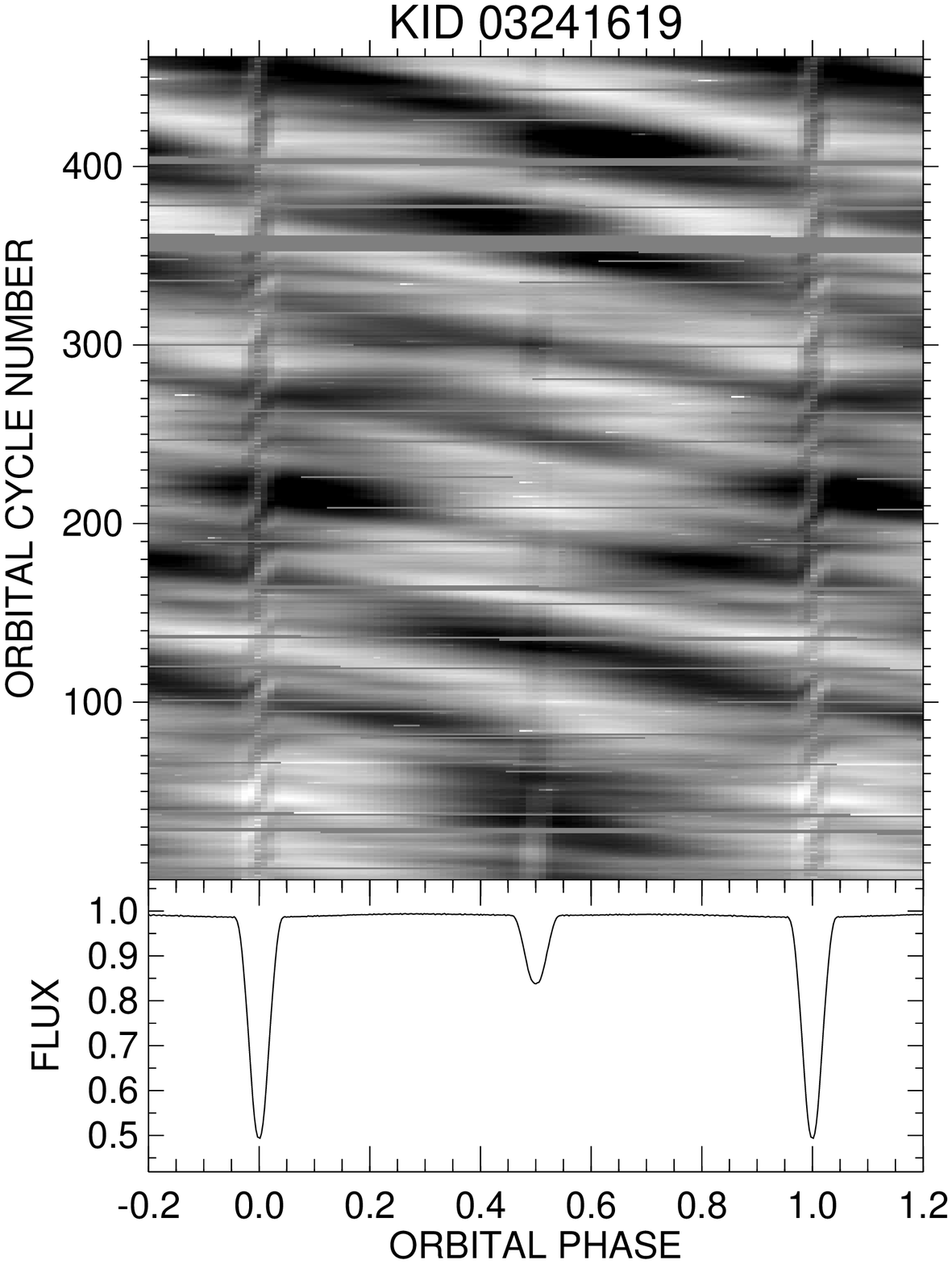}
\figsetgrpnote{The lower panel shows a mean, normalized light curve formed 
by binning in orbital phase.  ahe top panel shows the 
flux differences as a function of orbital phase and 
cycle number, represented as a gray scale diagram (range $\pm 3\%$). 
 }
\figsetgrpend

\figsetgrpstart
\figsetgrpnum{1.4}
\figsetgrptitle{g4}
\figsetplot{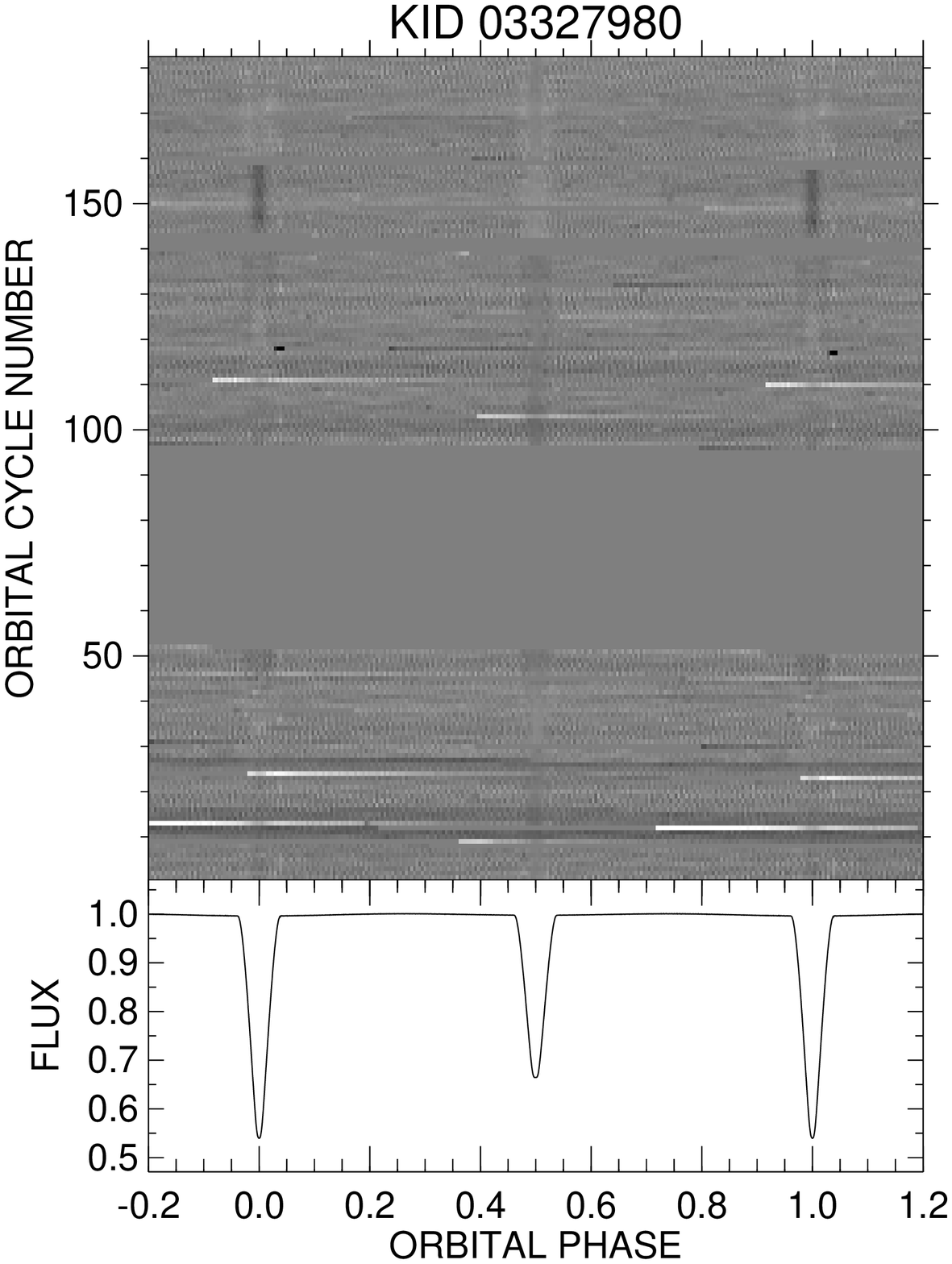}
\figsetgrpnote{The lower panel shows a mean, normalized light curve formed 
by binning in orbital phase.  ahe top panel shows the 
flux differences as a function of orbital phase and 
cycle number, represented as a gray scale diagram (range $\pm 0.5\%$). 
 }
\figsetgrpend

\figsetgrpstart
\figsetgrpnum{1.5}
\figsetgrptitle{g5}
\figsetplot{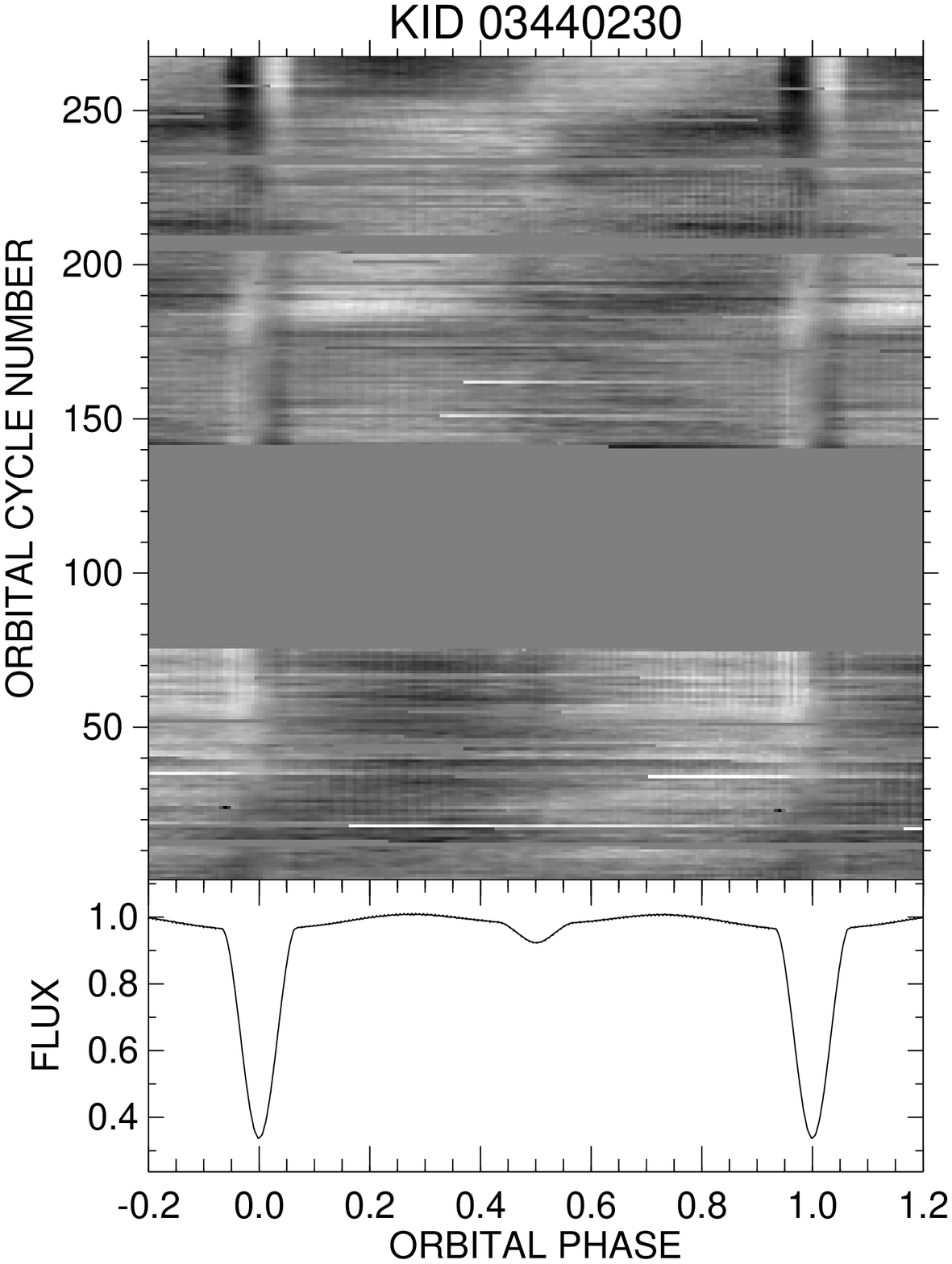}
\figsetgrpnote{The lower panel shows a mean, normalized light curve formed 
by binning in orbital phase.  The top panel shows the 
flux differences as a function of orbital phase and 
cycle number, represented as a gray scale diagram (range $\pm 1\%$). 
 }
\figsetgrpend

\figsetgrpstart
\figsetgrpnum{1.6}
\figsetgrptitle{g6}
\figsetplot{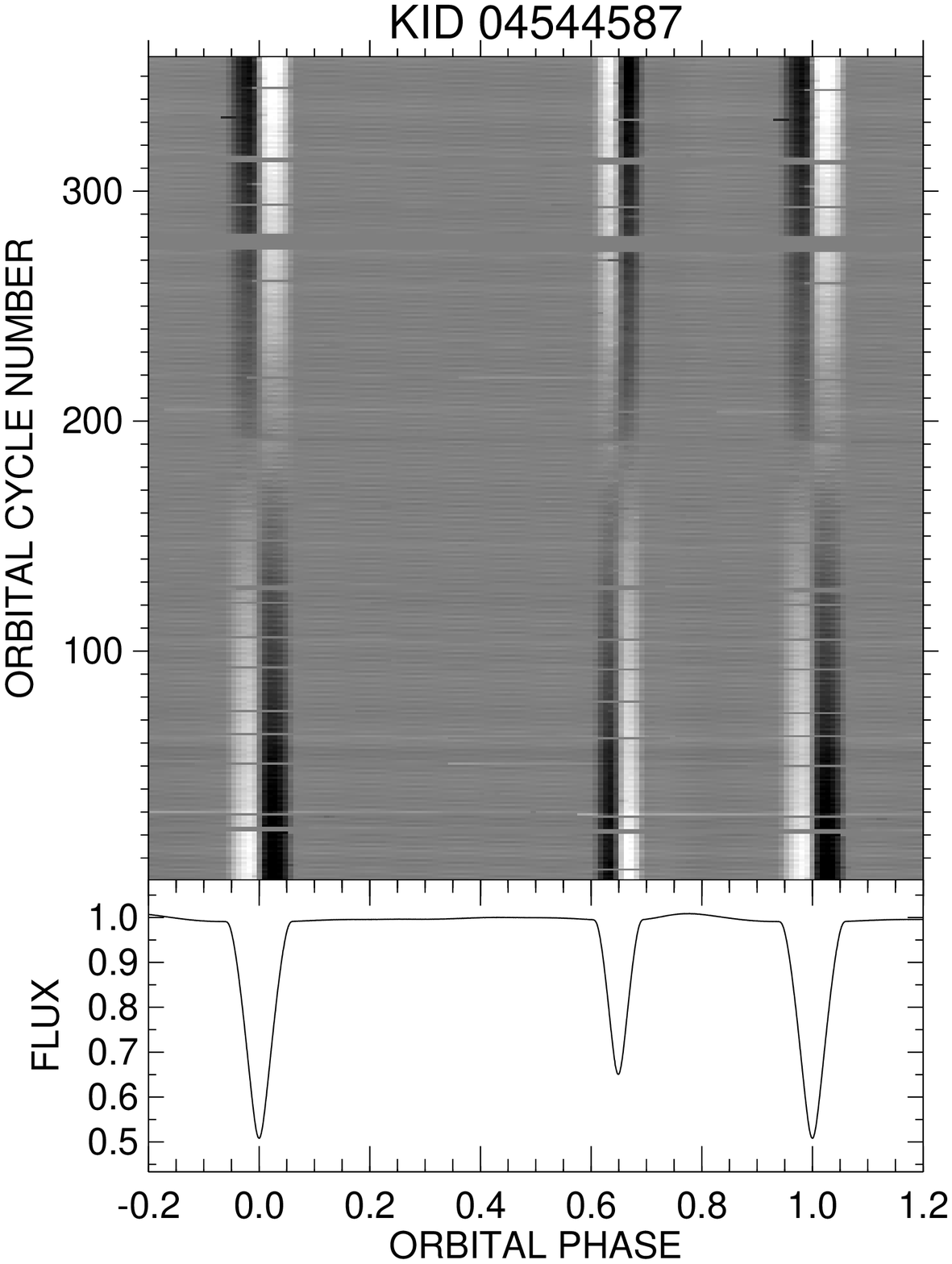}
\figsetgrpnote{The lower panel shows a mean, normalized light curve formed 
by binning in orbital phase.  The top panel shows the 
flux differences as a function of orbital phase and 
cycle number, represented as a gray scale diagram (range $\pm 1.5\%$). 
 }
\figsetgrpend

\figsetgrpstart
\figsetgrpnum{1.7}
\figsetgrptitle{g7}
\figsetplot{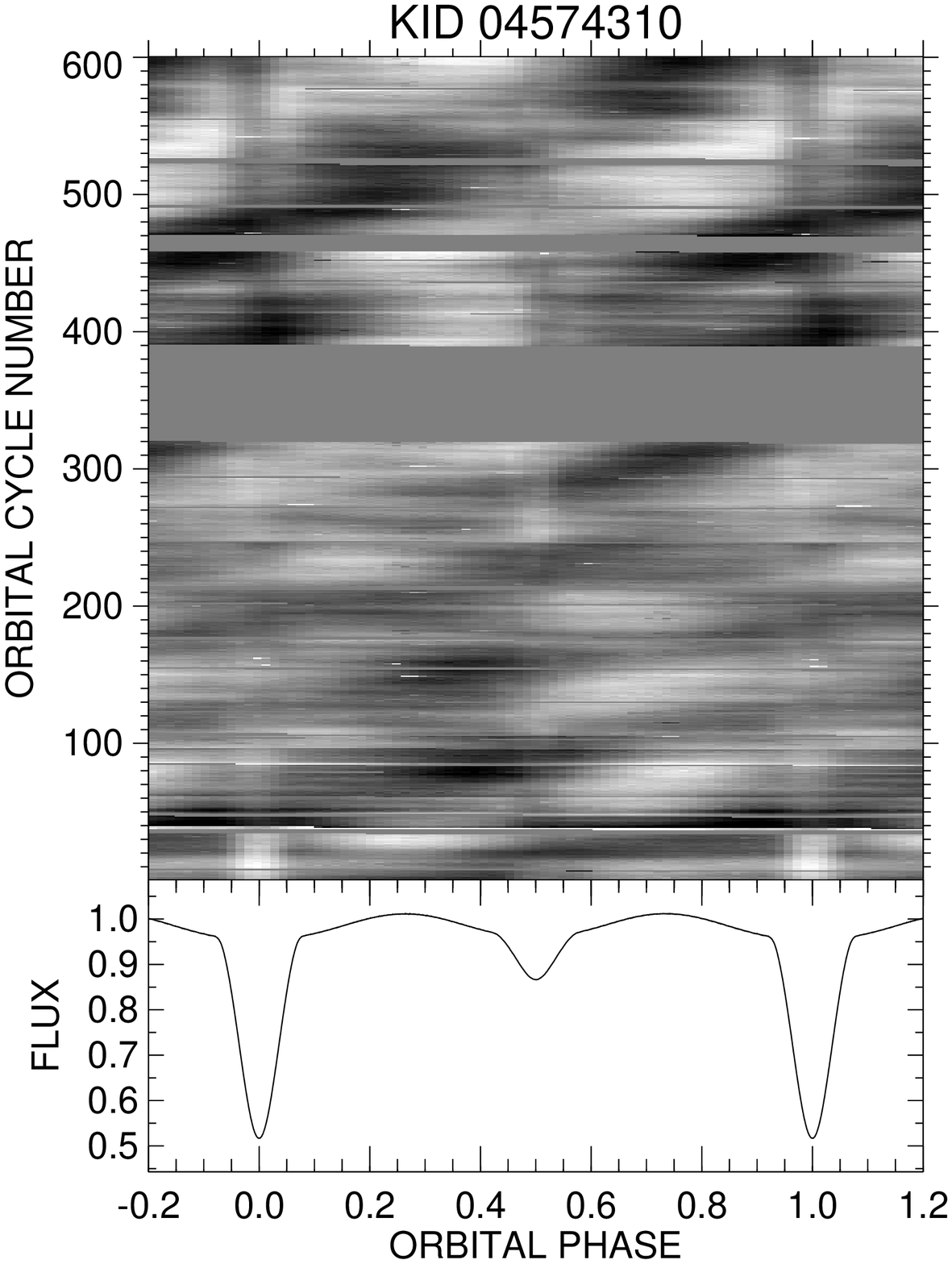}
\figsetgrpnote{The lower panel shows a mean, normalized light curve formed 
by binning in orbital phase.  The top panel shows the 
flux differences as a function of orbital phase and 
cycle number, represented as a gray scale diagram (range $\pm 0.5\%$). 
 }
\figsetgrpend

\figsetgrpstart
\figsetgrpnum{1.8}
\figsetgrptitle{g8}
\figsetplot{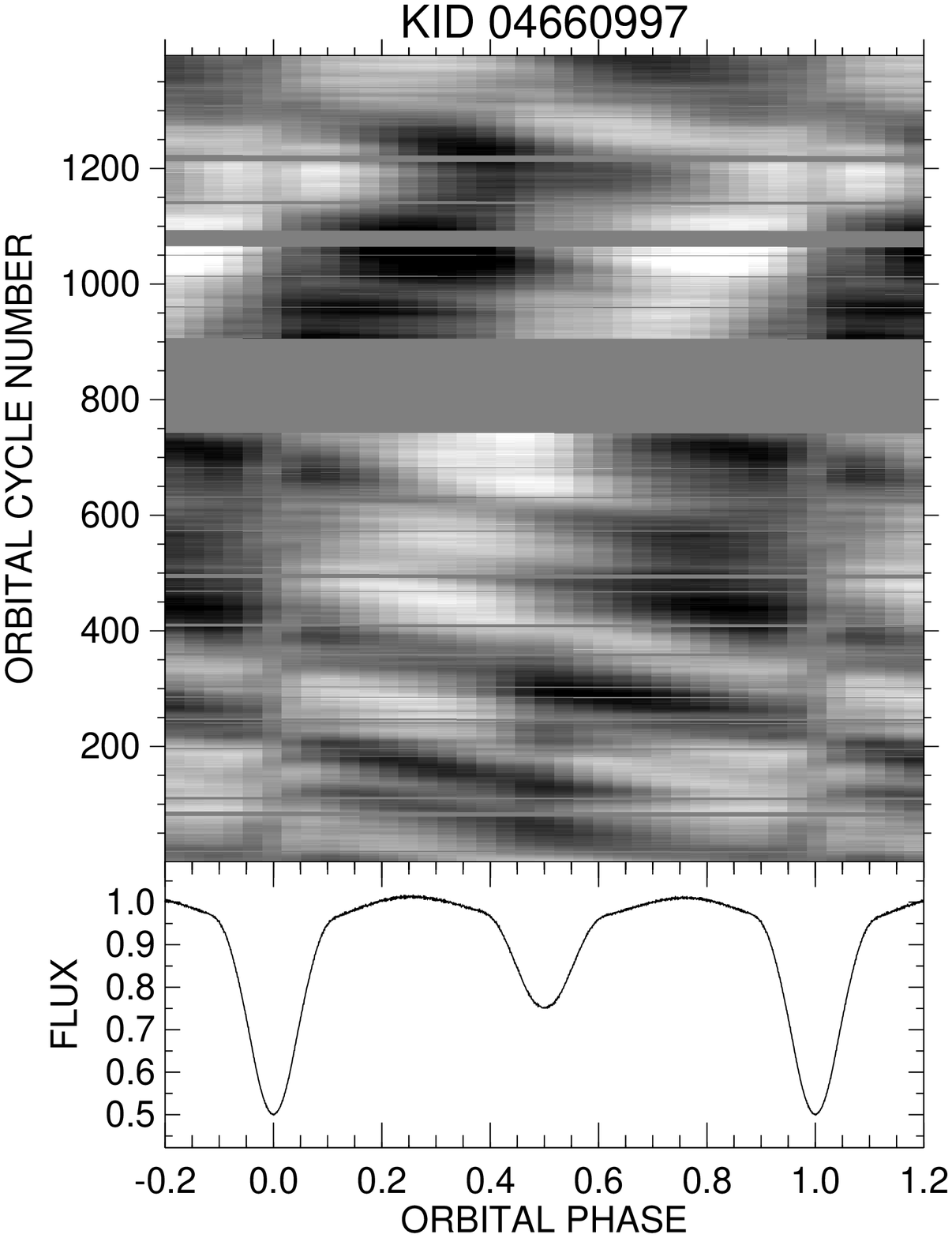}
\figsetgrpnote{The lower panel shows a mean, normalized light curve formed 
by binning in orbital phase.  The top panel shows the 
flux differences as a function of orbital phase and 
cycle number, represented as a gray scale diagram (range $\pm 4\%$). 
 }
\figsetgrpend

\figsetgrpstart
\figsetgrpnum{1.9}
\figsetgrptitle{g9}
\figsetplot{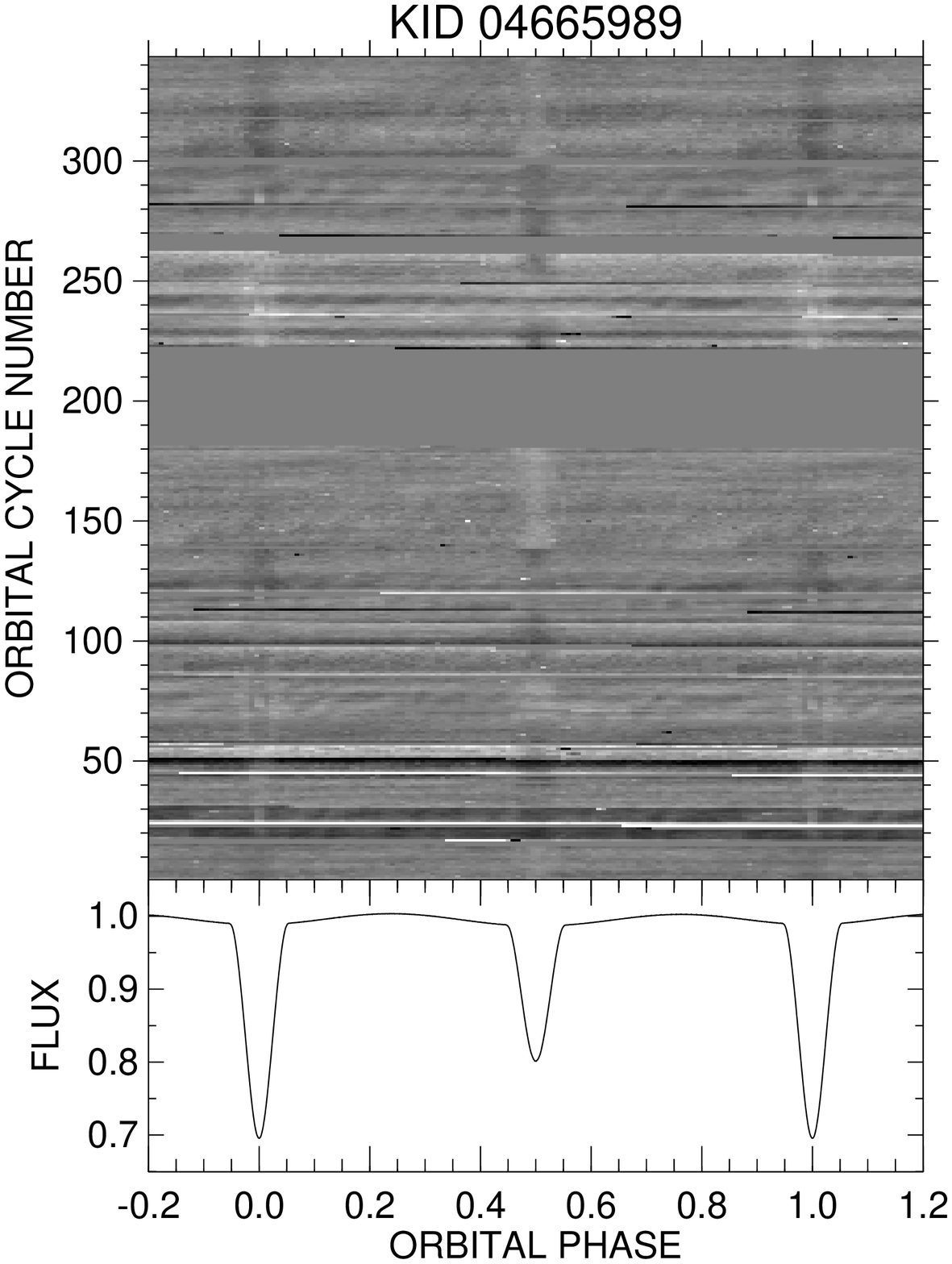}
\figsetgrpnote{The lower panel shows a mean, normalized light curve formed 
by binning in orbital phase.  The top panel shows the 
flux differences as a function of orbital phase and 
cycle number, represented as a gray scale diagram (range $\pm 0.2\%$). 
 }
\figsetgrpend

\figsetgrpstart
\figsetgrpnum{1.10}
\figsetgrptitle{g10}
\figsetplot{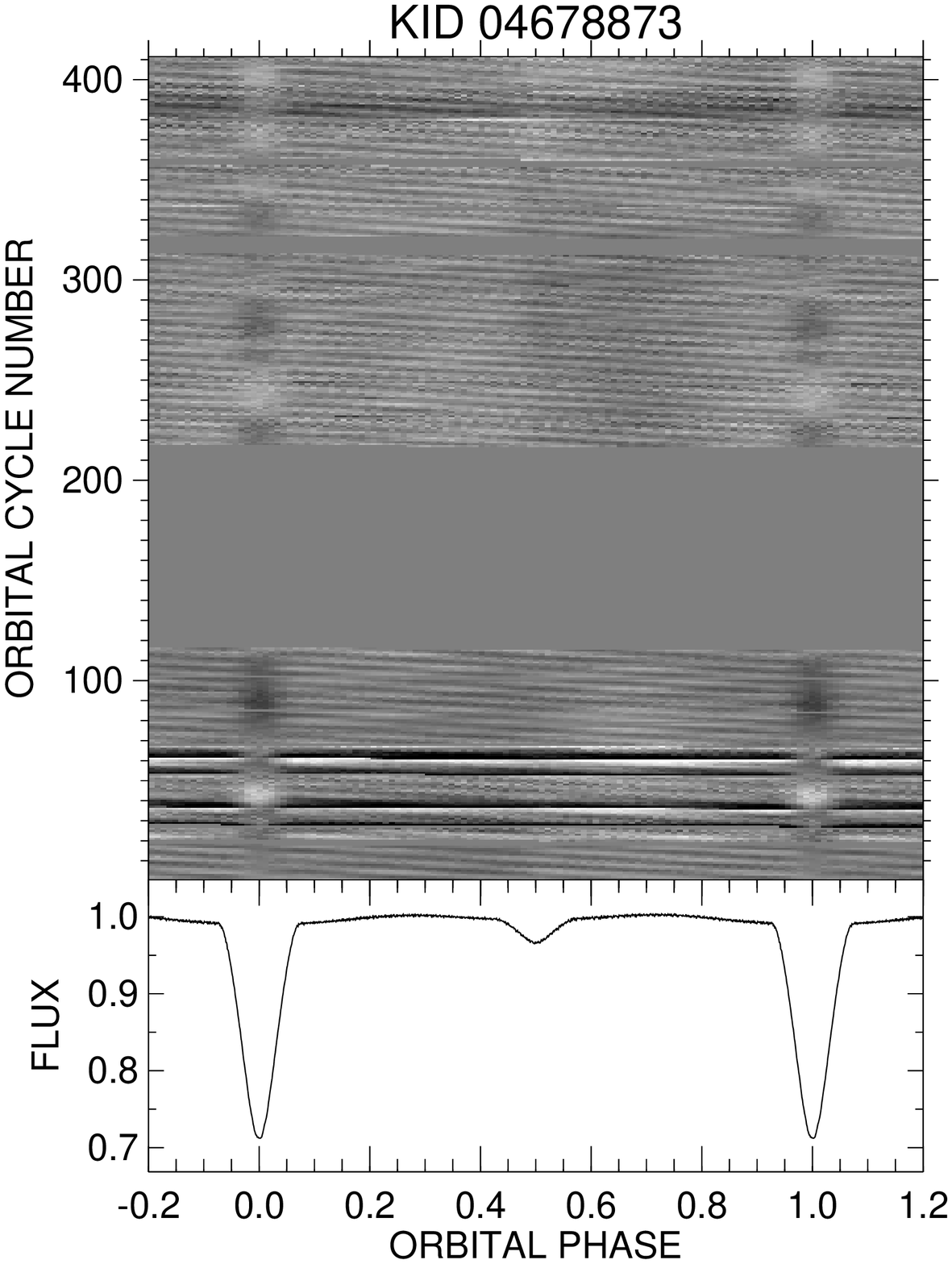}
\figsetgrpnote{The lower panel shows a mean, normalized light curve formed 
by binning in orbital phase.  The top panel shows the 
flux differences as a function of orbital phase and 
cycle number, represented as a gray scale diagram (range $\pm 2\%$). 
 }
\figsetgrpend

\figsetgrpstart
\figsetgrpnum{1.11}
\figsetgrptitle{g11}
\figsetplot{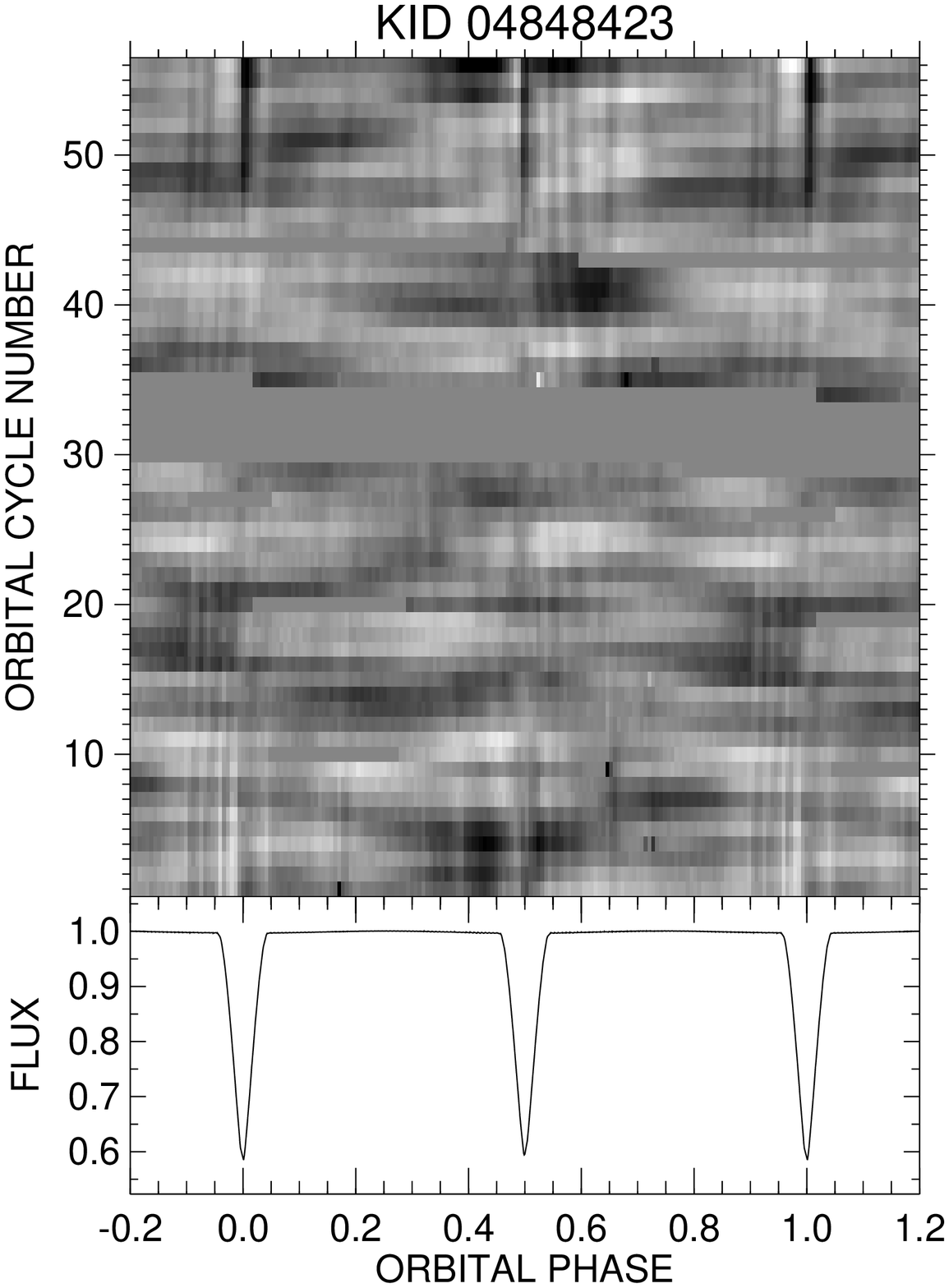}
\figsetgrpnote{The lower panel shows a mean, normalized light curve formed 
by binning in orbital phase.  The top panel shows the 
flux differences as a function of orbital phase and 
cycle number, represented as a gray scale diagram (range $\pm 0.2\%$). 
 }
\figsetgrpend

\figsetgrpstart
\figsetgrpnum{1.12}
\figsetgrptitle{g12}
\figsetplot{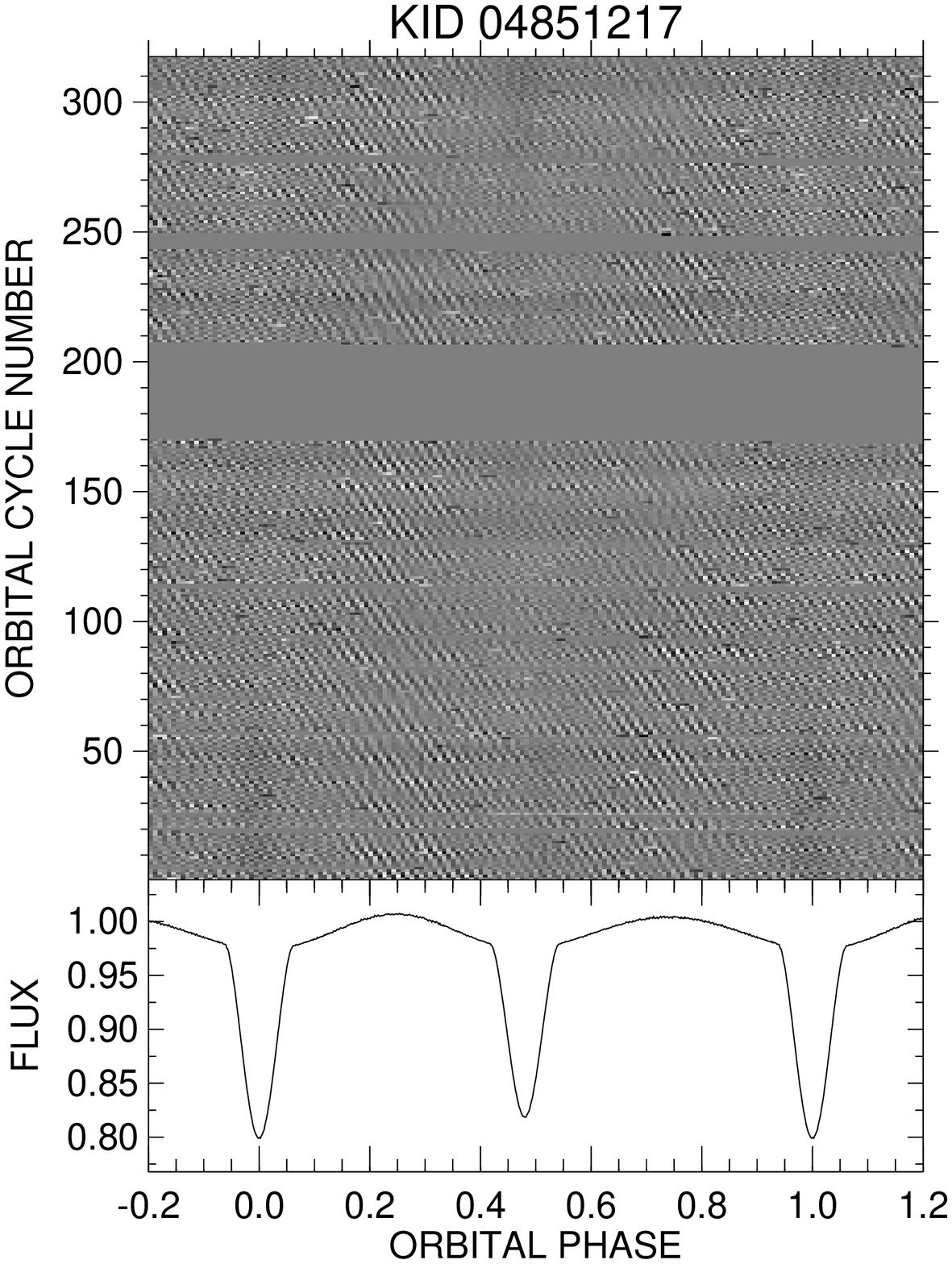}
\figsetgrpnote{The lower panel shows a mean, normalized light curve formed 
by binning in orbital phase.  The top panel shows the 
flux differences as a function of orbital phase and 
cycle number, represented as a gray scale diagram (range $\pm 1\%$). 
 }
\figsetgrpend

\figsetgrpstart
\figsetgrpnum{1.13}
\figsetgrptitle{g13}
\figsetplot{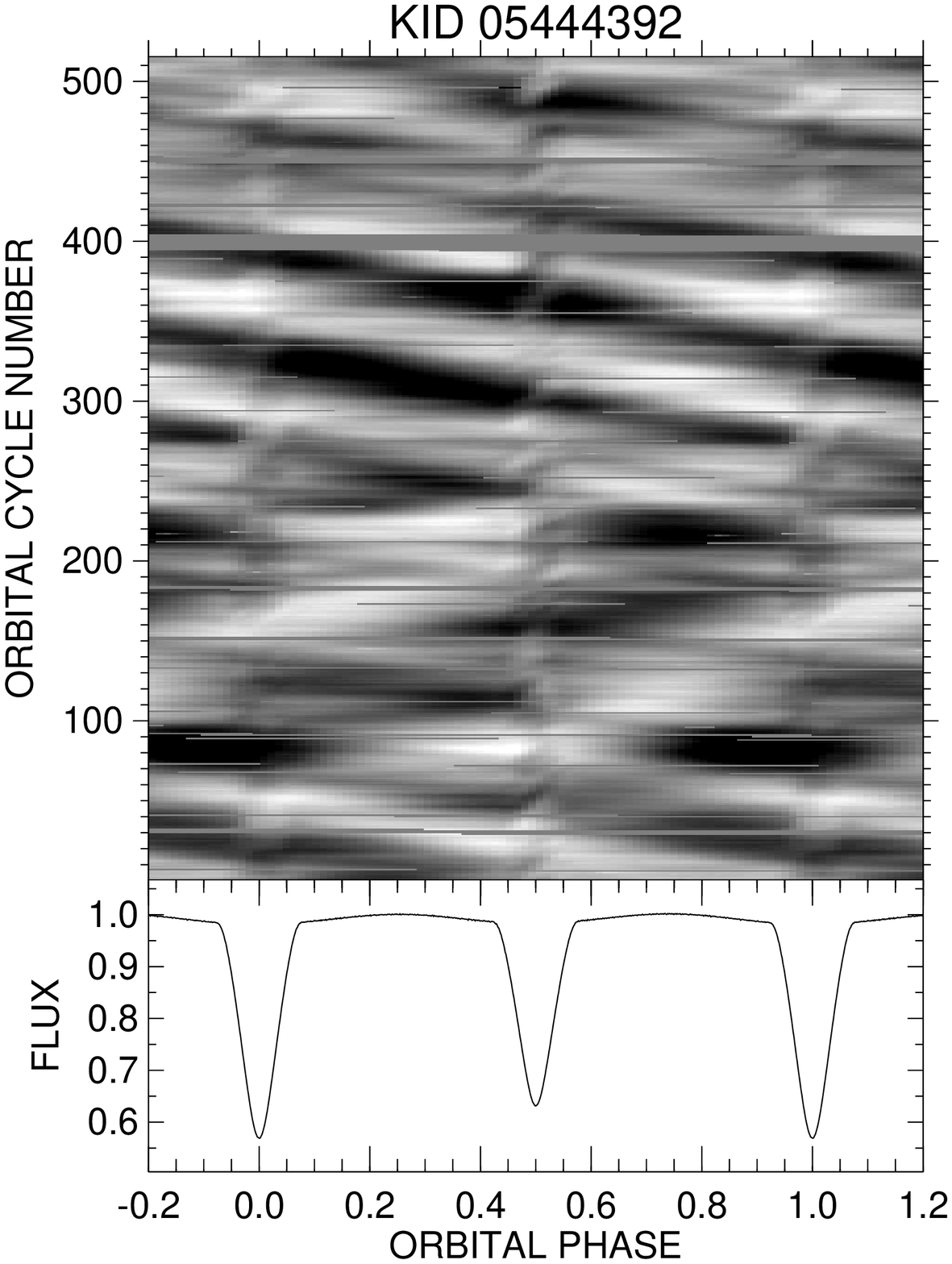}
\figsetgrpnote{The lower panel shows a mean, normalized light curve formed 
by binning in orbital phase.  The top panel shows the 
flux differences as a function of orbital phase and 
cycle number, represented as a gray scale diagram (range $\pm 2\%$). 
 }
\figsetgrpend

\figsetgrpstart
\figsetgrpnum{1.14}
\figsetgrptitle{g14}
\figsetplot{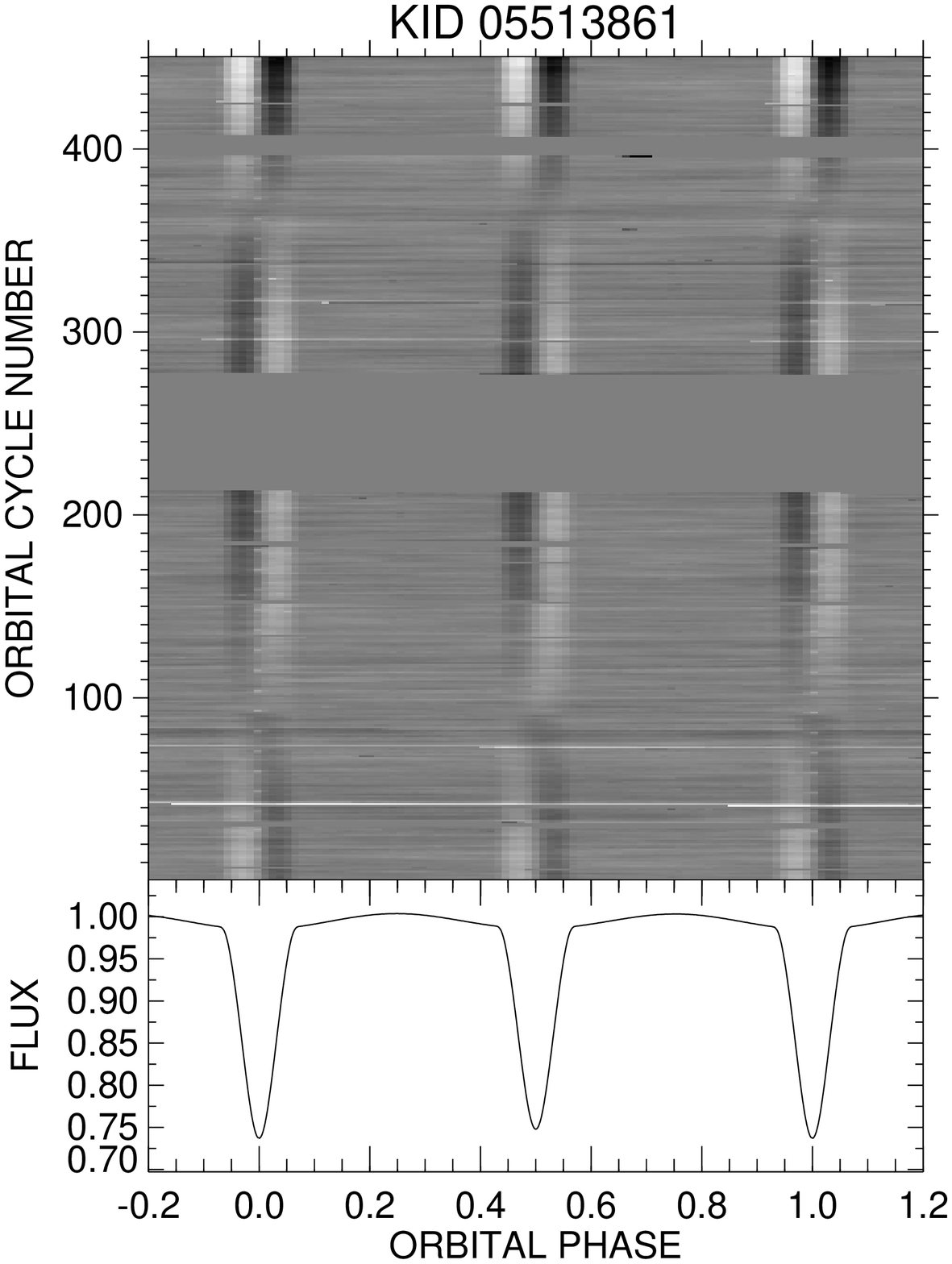}
\figsetgrpnote{The lower panel shows a mean, normalized light curve formed 
by binning in orbital phase.  The top panel shows the 
flux differences as a function of orbital phase and 
cycle number, represented as a gray scale diagram (range $\pm 0.5\%$). 
 }
\figsetgrpend

\figsetgrpstart
\figsetgrpnum{1.15}
\figsetgrptitle{g15}
\figsetplot{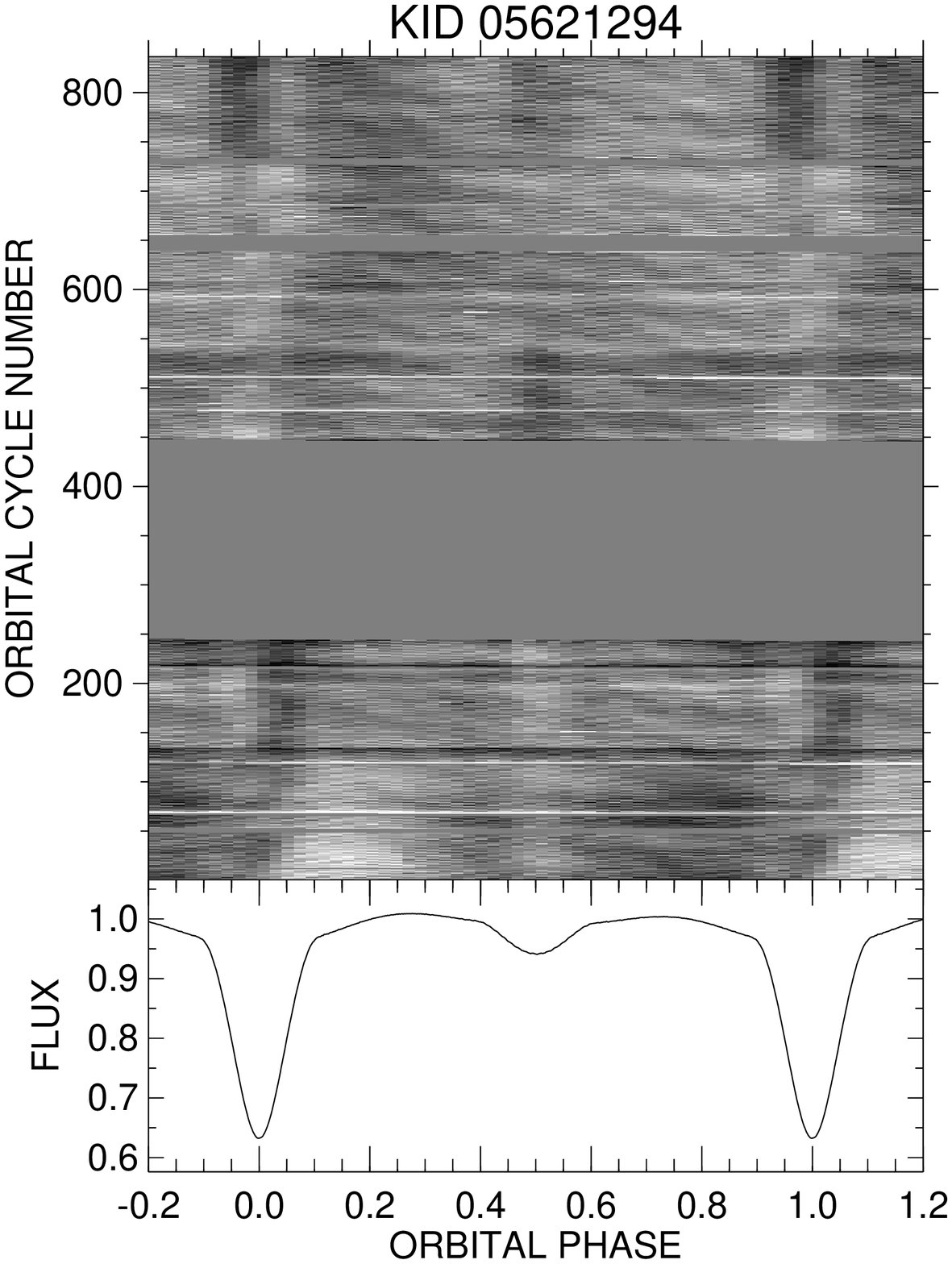}
\figsetgrpnote{The lower panel shows a mean, normalized light curve formed 
by binning in orbital phase.  The top panel shows the 
flux differences as a function of orbital phase and 
cycle number, represented as a gray scale diagram (range $\pm 0.5\%$). 
 }
\figsetgrpend

\figsetgrpstart
\figsetgrpnum{1.16}
\figsetgrptitle{g16}
\figsetplot{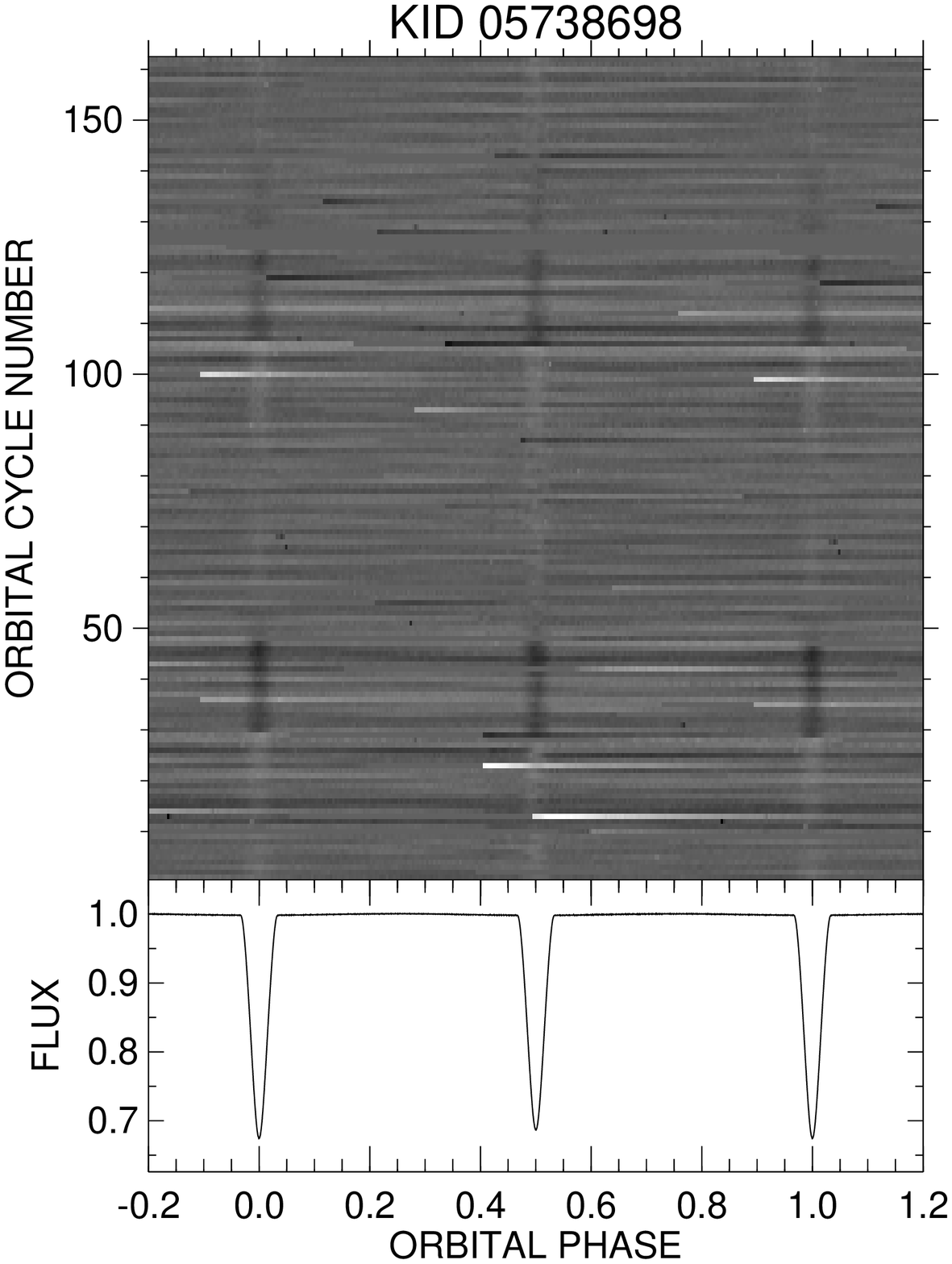}
\figsetgrpnote{The lower panel shows a mean, normalized light curve formed 
by binning in orbital phase.  ahe top panel shows the 4
flux differences as a function of orbital phase and 
cycle number, represented as a gray scale diagram (range $\pm 0.5\%$). 
 }
\figsetgrpend

\figsetgrpstart
\figsetgrpnum{1.17}
\figsetgrptitle{g17}
\figsetplot{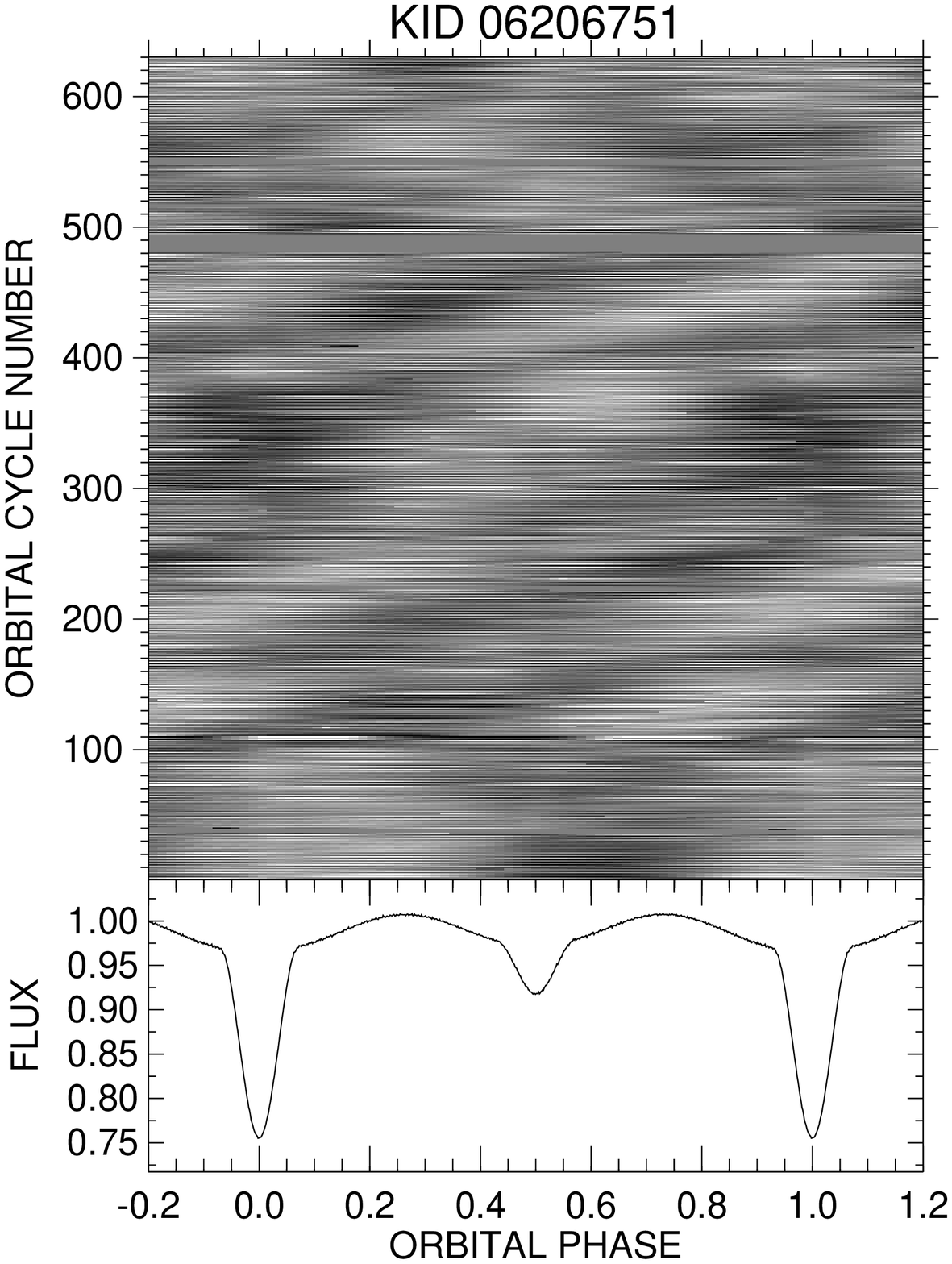}
\figsetgrpnote{The lower panel shows a mean, normalized light curve formed 
by binning in orbital phase.  ahe top panel shows the 4
flux differences as a function of orbital phase and 
cycle number, represented as a gray scale diagram (range $\pm 1.5\%$). 
 }
\figsetgrpend

\figsetgrpstart
\figsetgrpnum{1.18}
\figsetgrptitle{g18}
\figsetplot{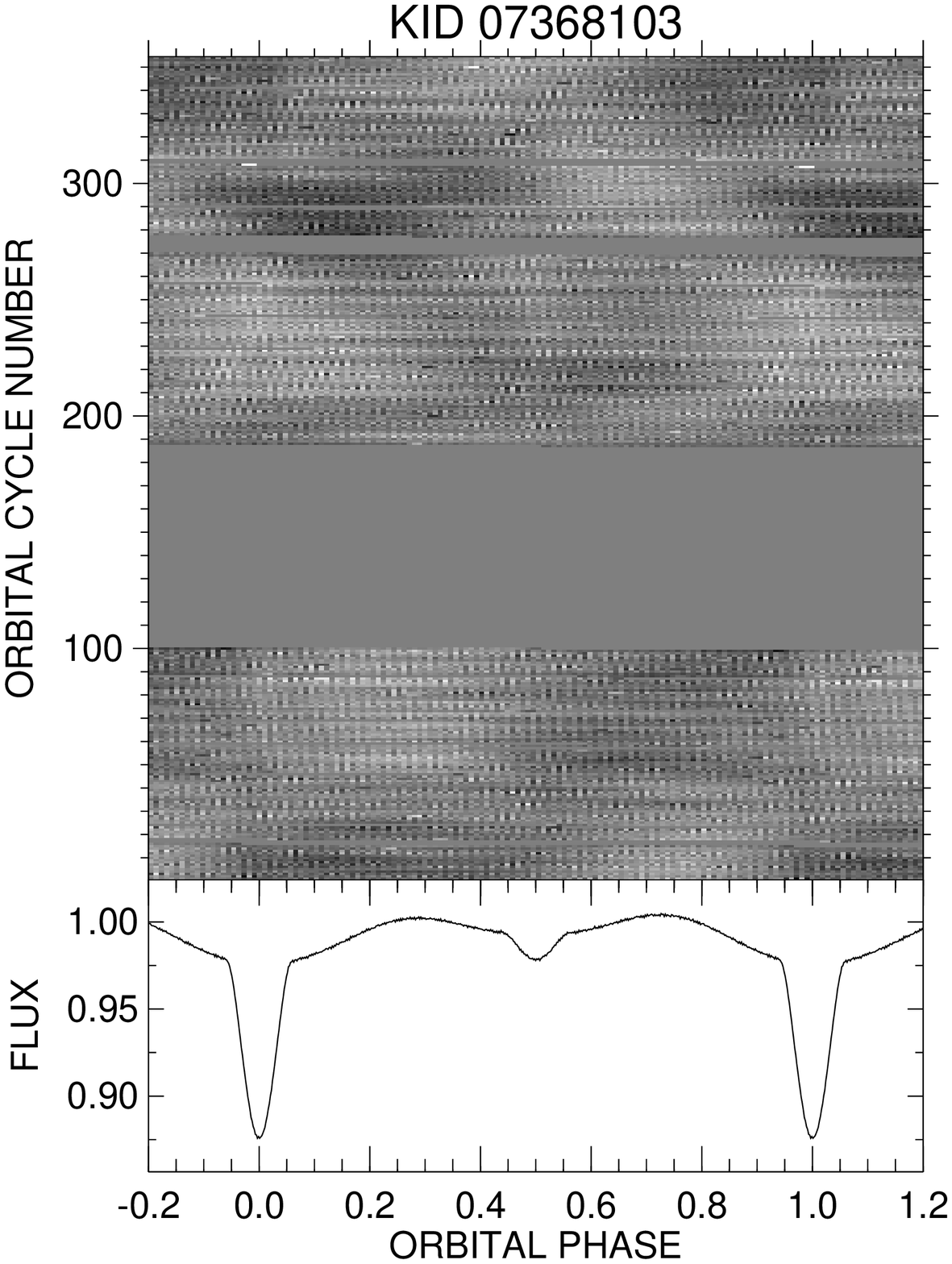}
\figsetgrpnote{The lower panel shows a mean, normalized light curve formed 
by binning in orbital phase.  The top panel shows the 
flux differences as a function of orbital phase and 
cycle number, represented as a gray scale diagram (range $\pm 1\%$). 
 }
\figsetgrpend

\figsetgrpstart
\figsetgrpnum{1.19}
\figsetgrptitle{g19}
\figsetplot{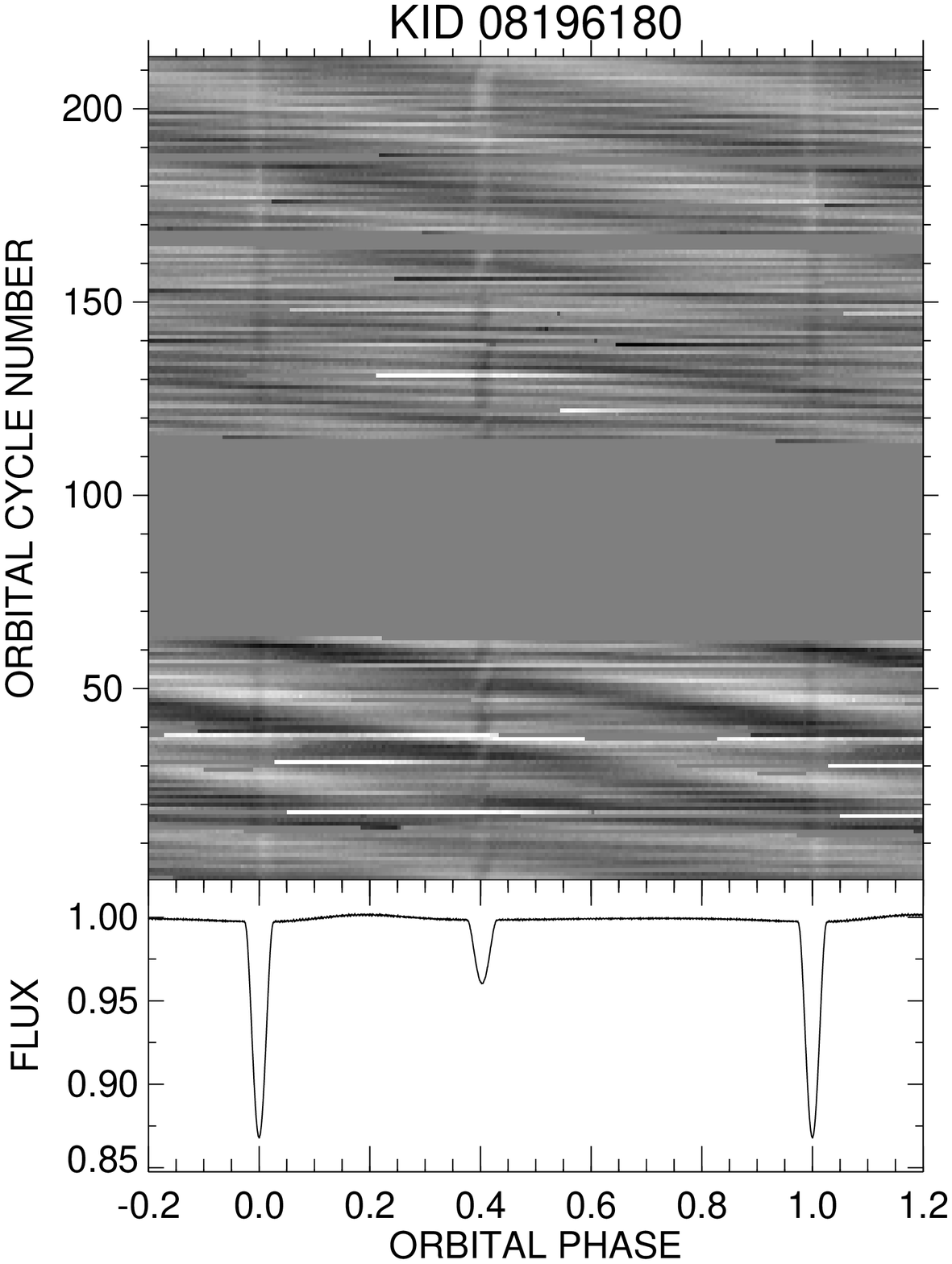}
\figsetgrpnote{The lower panel shows a mean, normalized light curve formed 
by binning in orbital phase.  The top panel shows the 
flux differences as a function of orbital phase and 
cycle number, represented as a gray scale diagram (range $\pm 0.5\%$). 
 }
\figsetgrpend

\figsetgrpstart
\figsetgrpnum{1.20}
\figsetgrptitle{g20}
\figsetplot{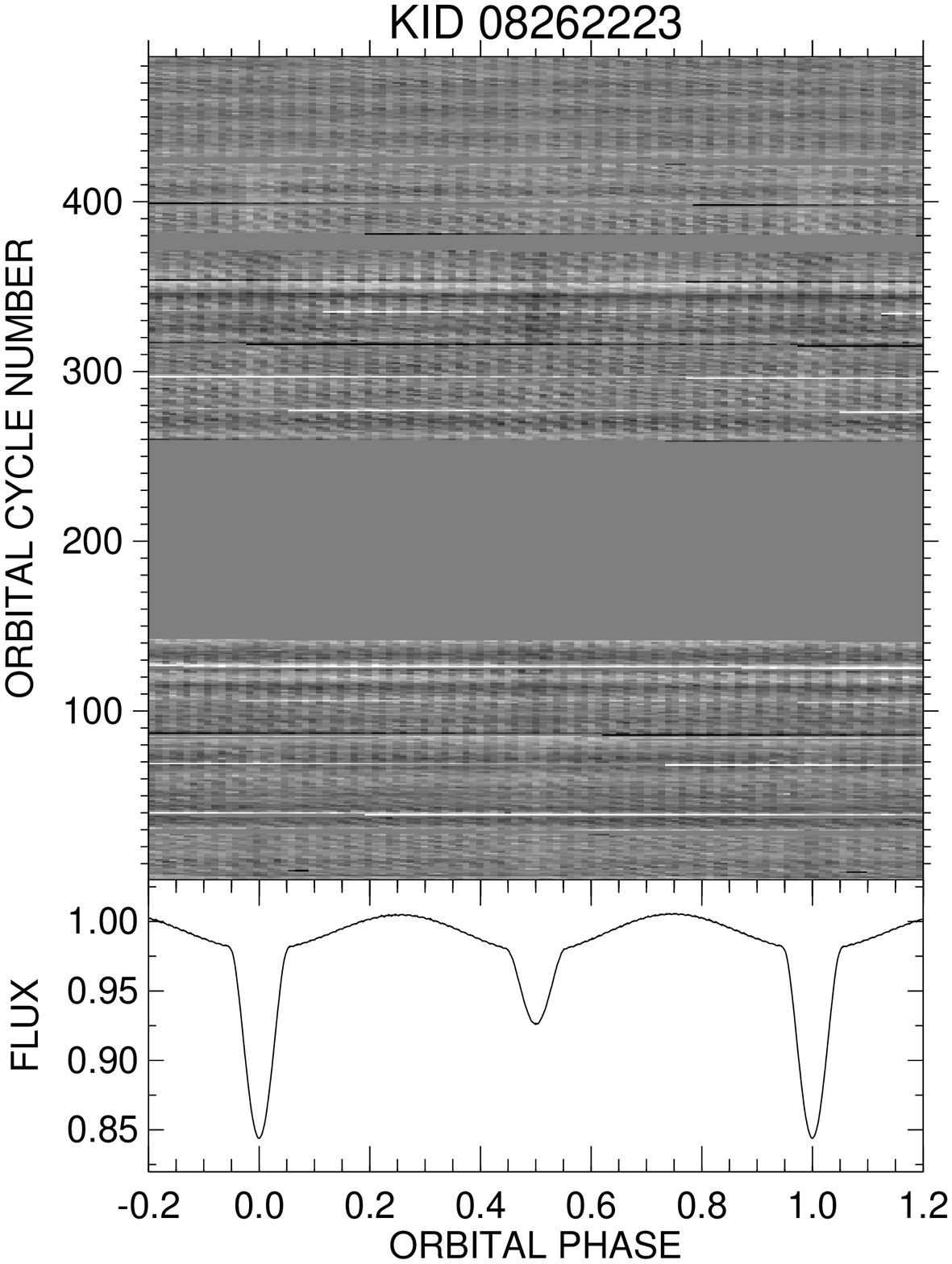}
\figsetgrpnote{The lower panel shows a mean, normalized light curve formed 
by binning in orbital phase.  The top panel shows the 
flux differences as a function of orbital phase and 
cycle number, represented as a gray scale diagram (range $\pm 0.3\%$). 
 }
\figsetgrpend

\figsetgrpstart
\figsetgrpnum{1.21}
\figsetgrptitle{g21}
\figsetplot{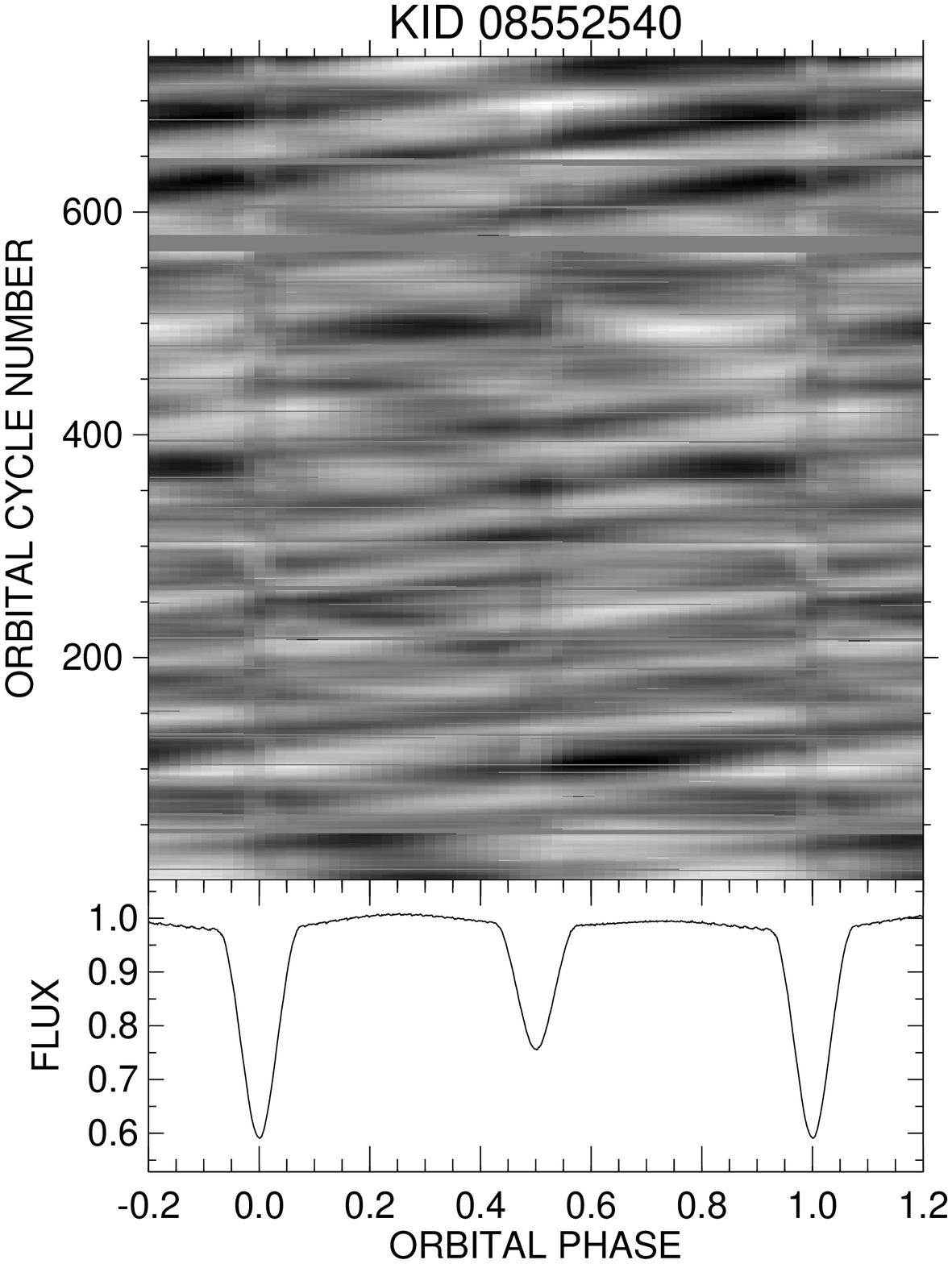}
\figsetgrpnote{The lower panel shows a mean, normalized light curve formed 
by binning in orbital phase.  The top panel shows the 
flux differences as a function of orbital phase and 
cycle number, represented as a gray scale diagram (range $\pm 3\%$). 
 }
\figsetgrpend

\figsetgrpstart
\figsetgrpnum{1.22}
\figsetgrptitle{g22}
\figsetplot{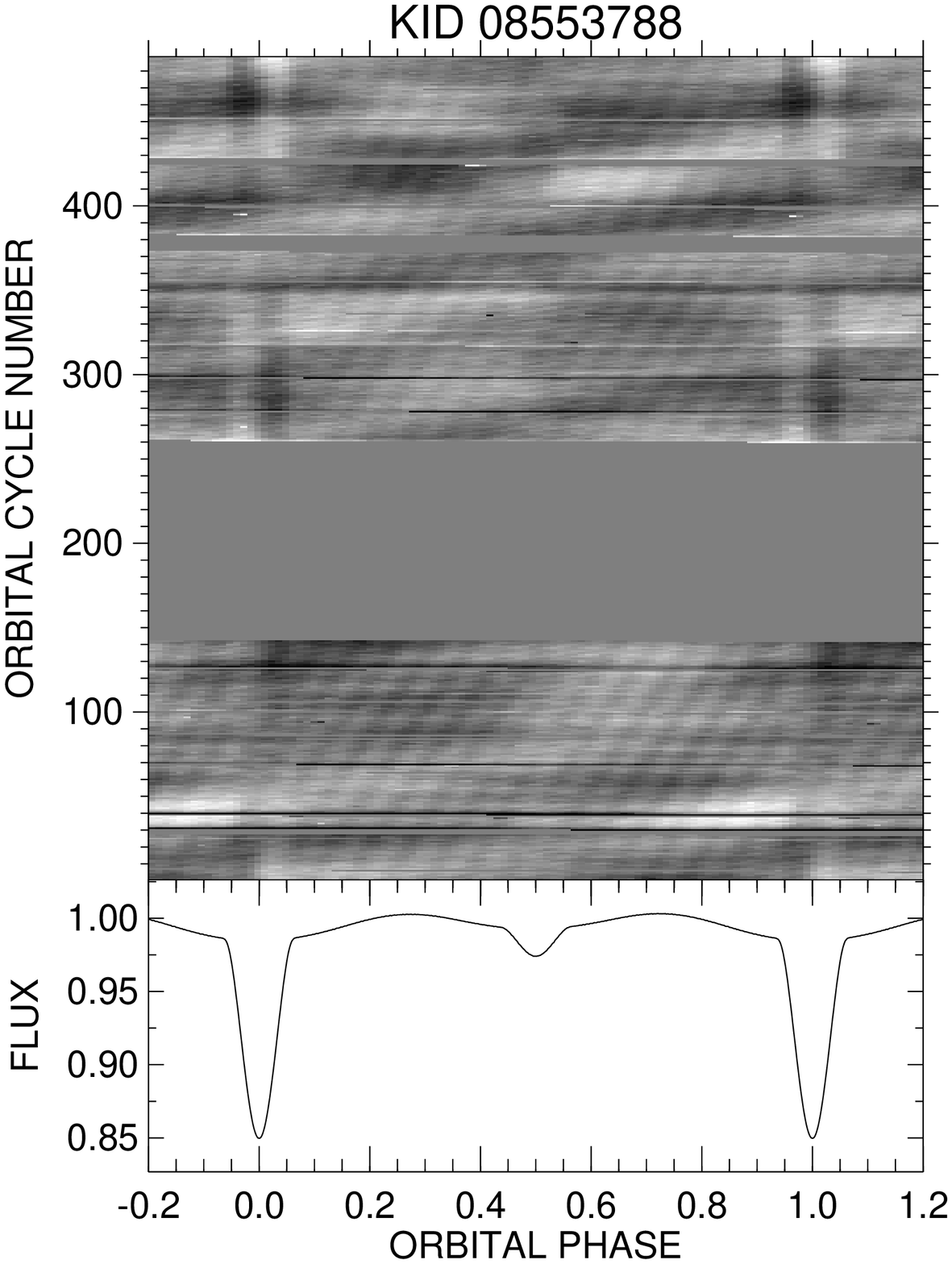}
\figsetgrpnote{The lower panel shows a mean, normalized light curve formed 
by binning in orbital phase.  The top panel shows the 
flux differences as a function of orbital phase and 
cycle number, represented as a gray scale diagram (range $\pm 0.3\%$). 
 }
\figsetgrpend

\figsetgrpstart
\figsetgrpnum{1.23}
\figsetgrptitle{g23}
\figsetplot{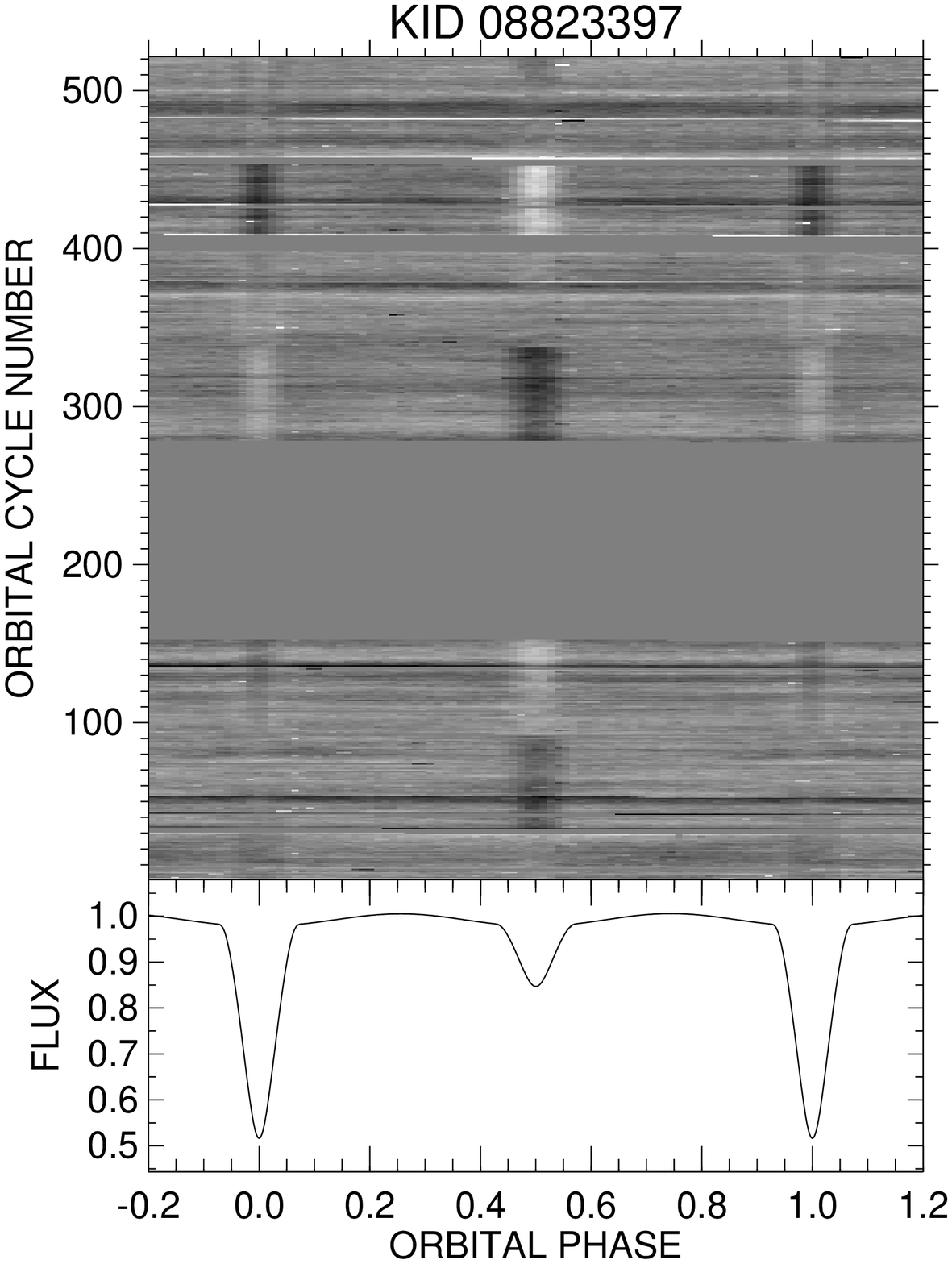}
\figsetgrpnote{The lower panel shows a mean, normalized light curve formed 
by binning in orbital phase.  The top panel shows the 
flux differences as a function of orbital phase and 
cycle number, represented as a gray scale diagram (range $\pm 0.2\%$). 
 }
\figsetgrpend

\figsetgrpstart
\figsetgrpnum{1.24}
\figsetgrptitle{g24}
\figsetplot{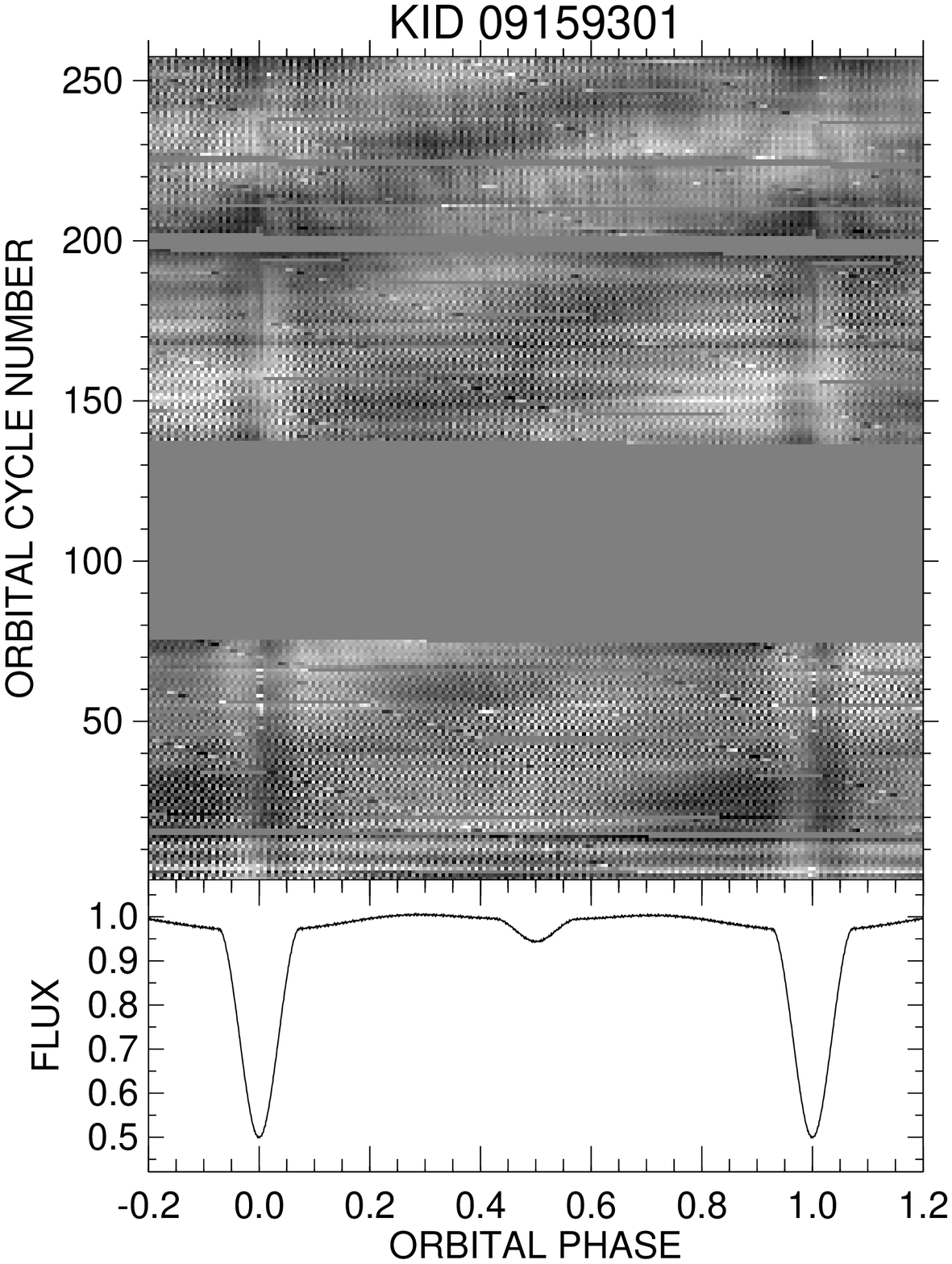}
\figsetgrpnote{The lower panel shows a mean, normalized light curve formed 
by binning in orbital phase.  The top panel shows the 
flux differences as a function of orbital phase and 
cycle number, represented as a gray scale diagram (range $\pm 0.4\%$). 
 }
\figsetgrpend

\figsetgrpstart
\figsetgrpnum{1.25}
\figsetgrptitle{g25}
\figsetplot{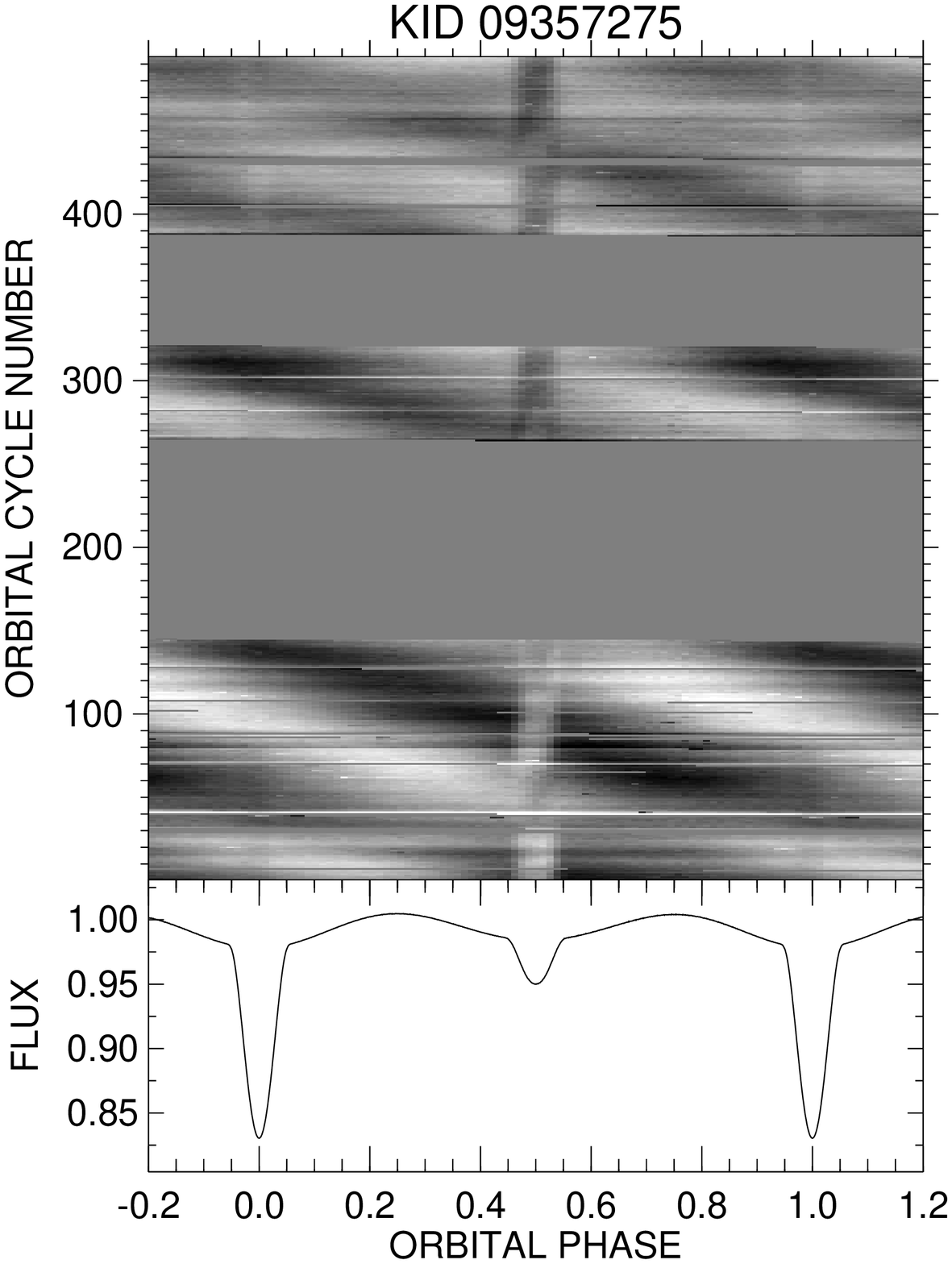}
\figsetgrpnote{The lower panel shows a mean, normalized light curve formed 
by binning in orbital phase.  The top panel shows the 
flux differences as a function of orbital phase and 
cycle number, represented as a gray scale diagram (range $\pm 0.2\%$). 
 }
\figsetgrpend

\figsetgrpstart
\figsetgrpnum{1.26}
\figsetgrptitle{g26}
\figsetplot{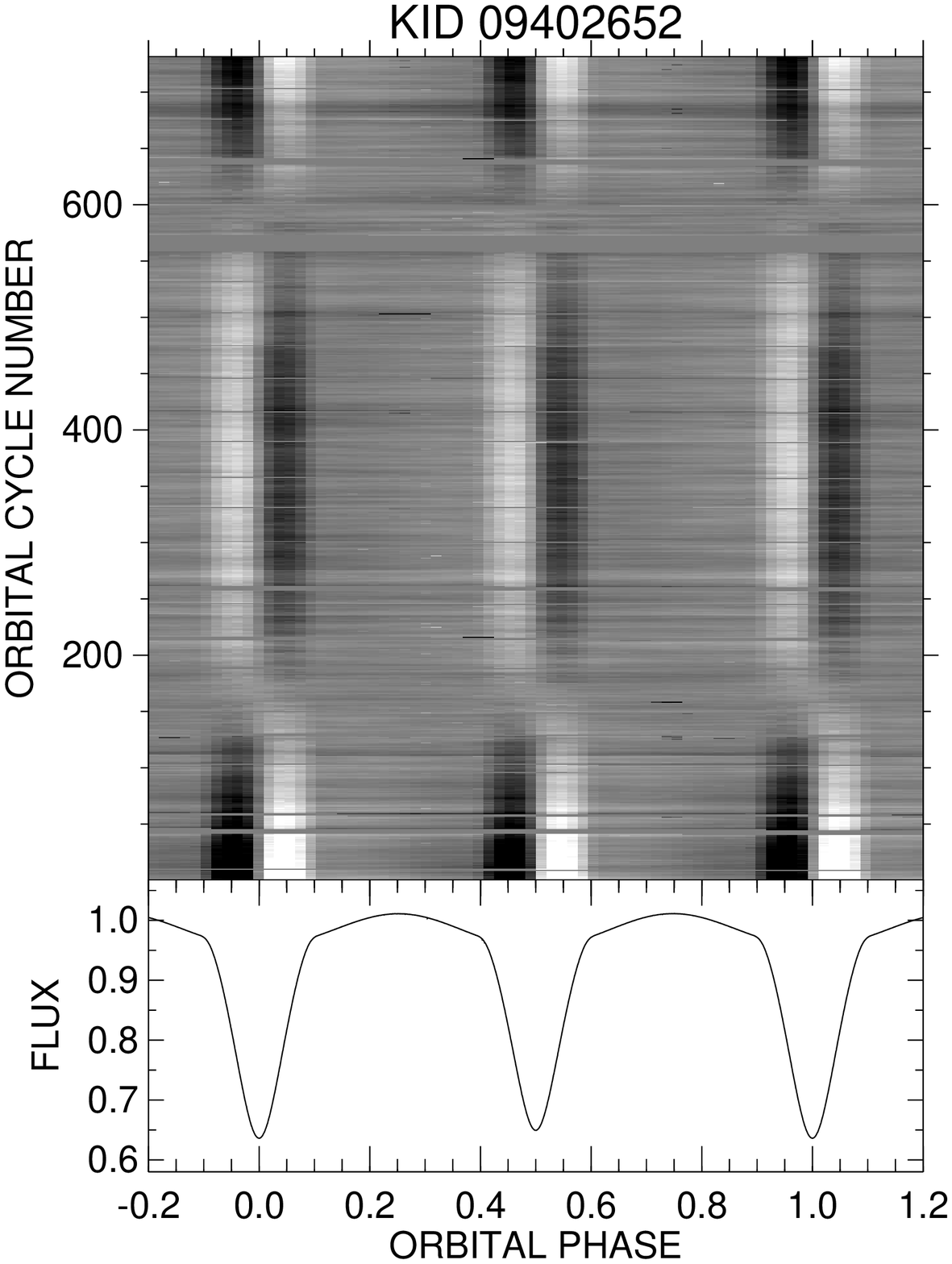}
\figsetgrpnote{The lower panel shows a mean, normalized light curve formed 
by binning in orbital phase.  The top panel shows the 
flux differences as a function of orbital phase and 
cycle number, represented as a gray scale diagram (range $\pm 0.5\%$). 
 }
\figsetgrpend

\figsetgrpstart
\figsetgrpnum{1.27}
\figsetgrptitle{g27}
\figsetplot{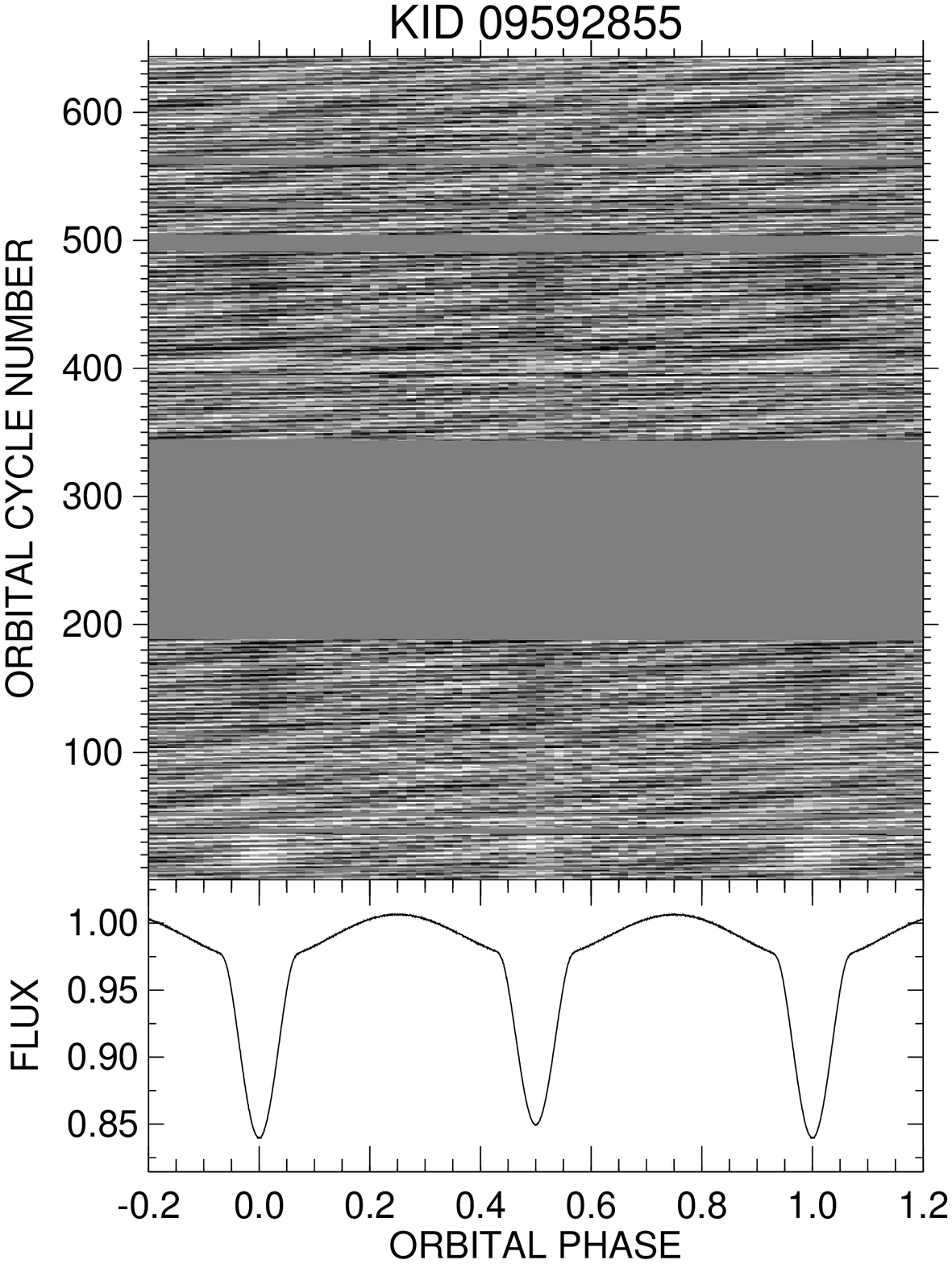}
\figsetgrpnote{The lower panel shows a mean, normalized light curve formed 
by binning in orbital phase.  The top panel shows the 
flux differences as a function of orbital phase and 
cycle number, represented as a gray scale diagram (range $\pm 0.3\%$). 
 }
\figsetgrpend

\figsetgrpstart
\figsetgrpnum{1.28}
\figsetgrptitle{g28}
\figsetplot{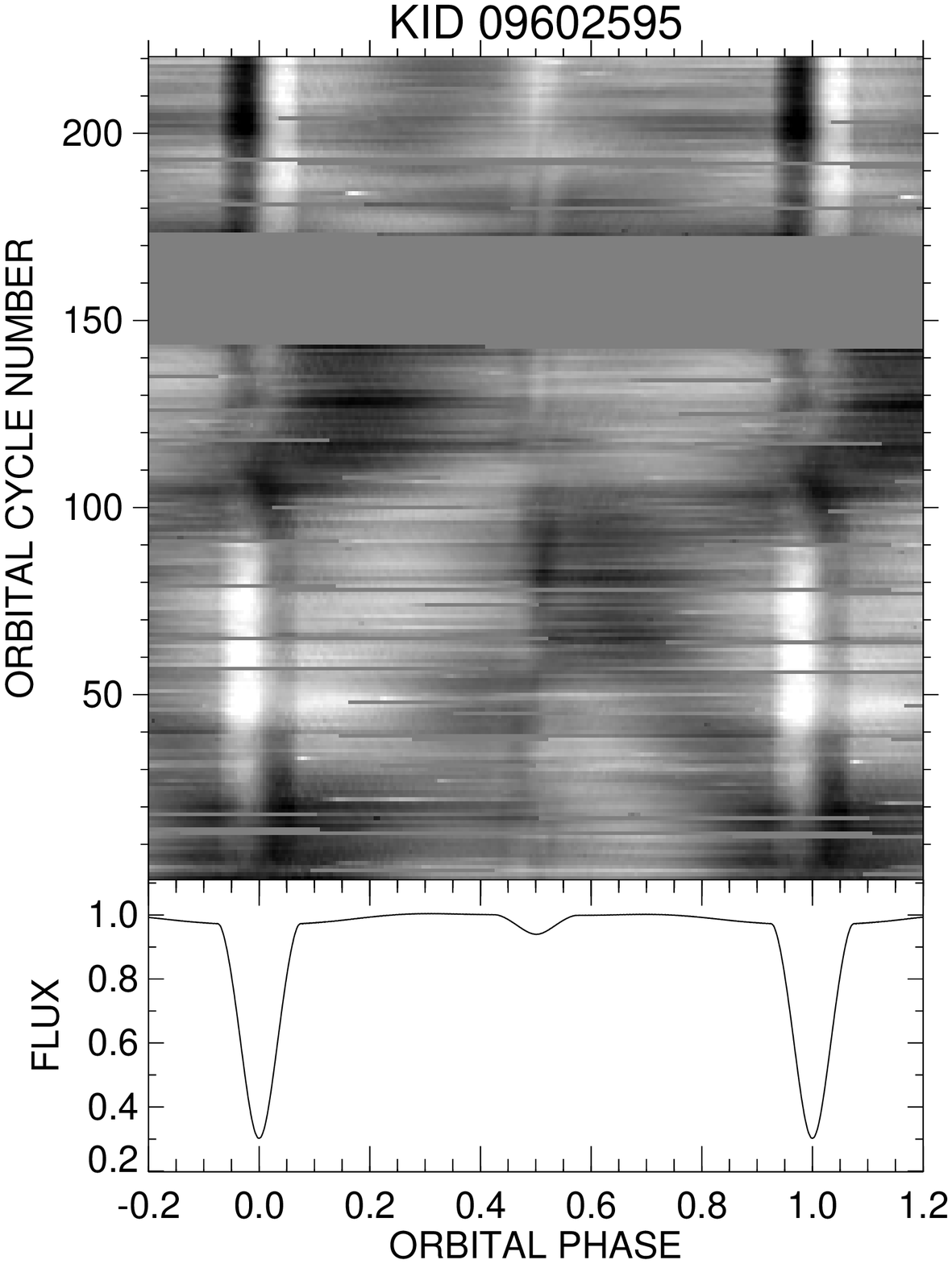}
\figsetgrpnote{The lower panel shows a mean, normalized light curve formed 
by binning in orbital phase.  The top panel shows the 
flux differences as a function of orbital phase and 
cycle number, represented as a gray scale diagram (range $\pm 0.5\%$). 
 }
\figsetgrpend

\figsetgrpstart
\figsetgrpnum{1.29}
\figsetgrptitle{g29}
\figsetplot{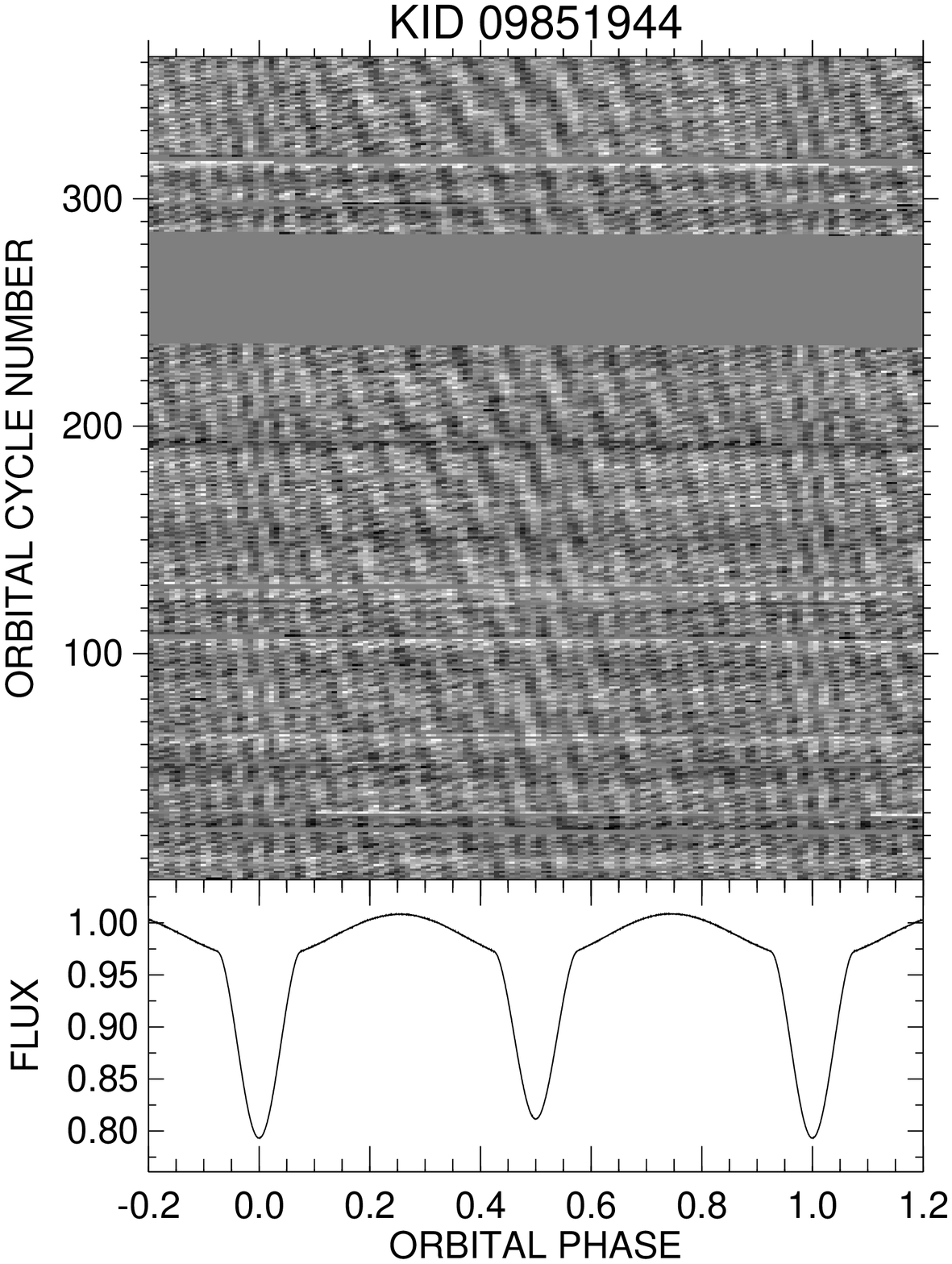}
\figsetgrpnote{The lower panel shows a mean, normalized light curve formed 
by binning in orbital phase.  The top panel shows the 
flux differences as a function of orbital phase and 
cycle number, represented as a gray scale diagram (range $\pm 0.3\%$). 
 }
\figsetgrpend

\figsetgrpstart
\figsetgrpnum{1.30}
\figsetgrptitle{g30}
\figsetplot{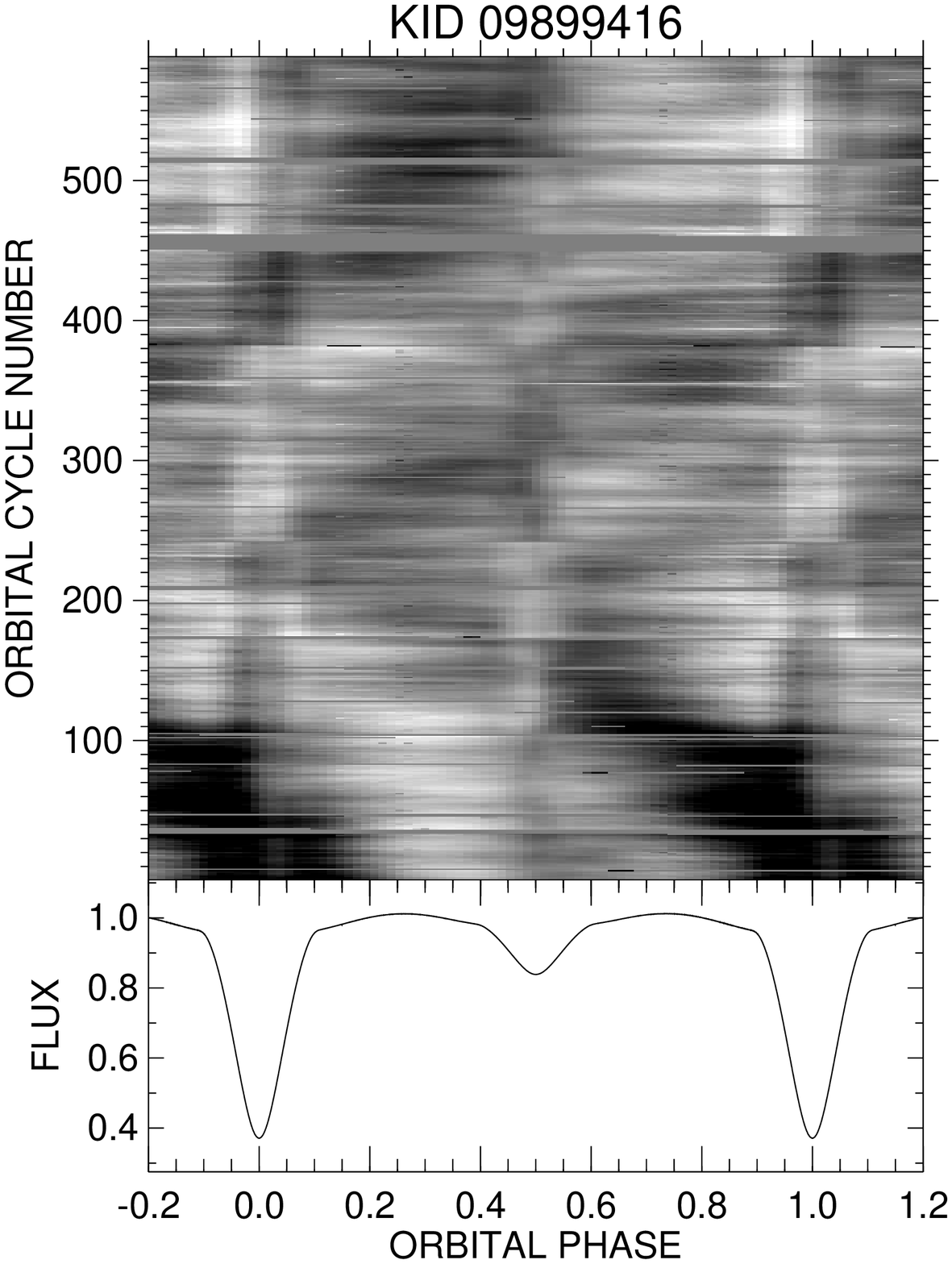}
\figsetgrpnote{The lower panel shows a mean, normalized light curve formed 
by binning in orbital phase.  The top panel shows the 
flux differences as a function of orbital phase and 
cycle number, represented as a gray scale diagram (range $\pm 0.4\%$). 
 }
\figsetgrpend

\figsetgrpstart
\figsetgrpnum{1.31}
\figsetgrptitle{g31}
\figsetplot{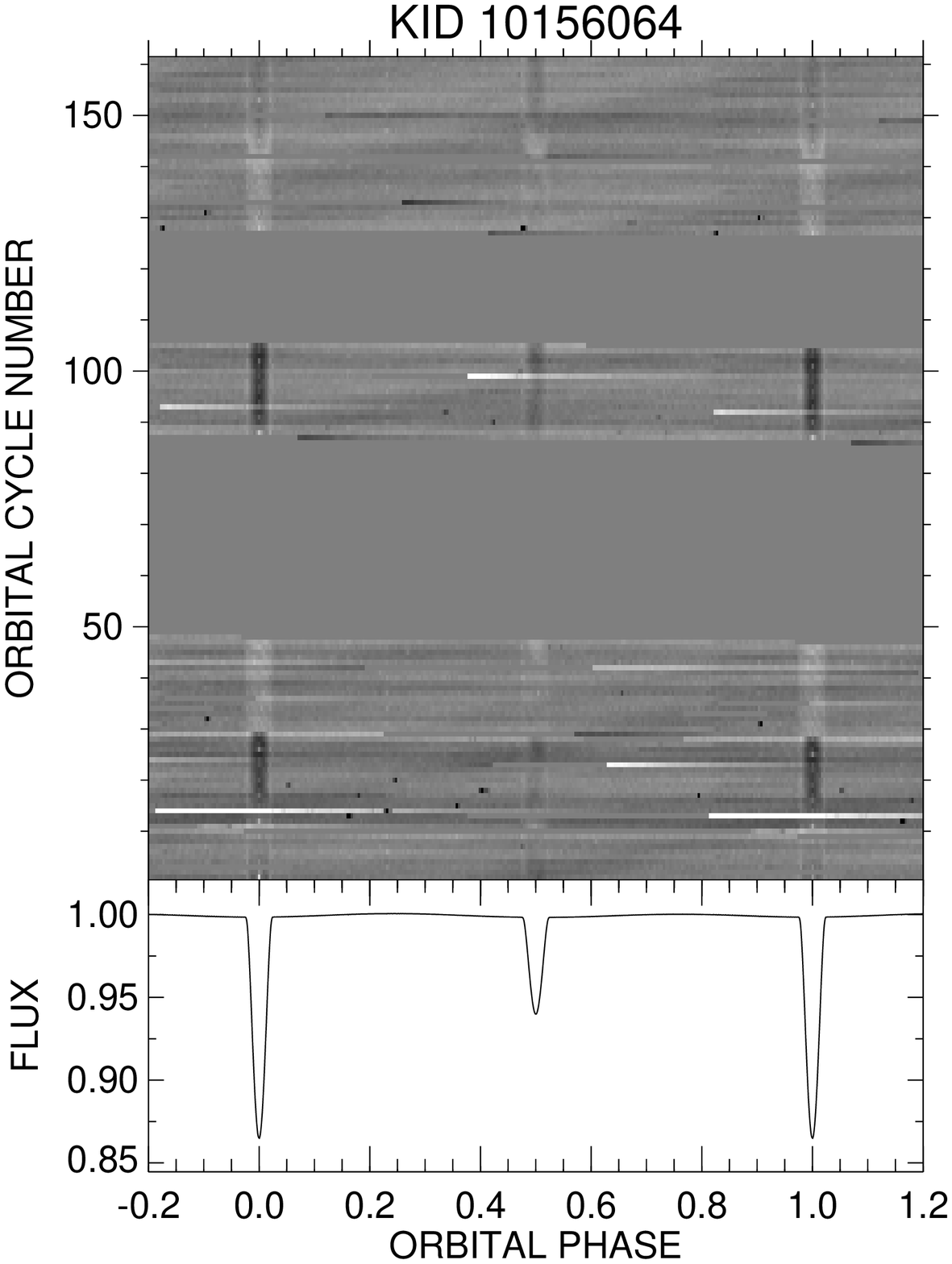}
\figsetgrpnote{The lower panel shows a mean, normalized light curve formed 
by binning in orbital phase.  The top panel shows the 
flux differences as a function of orbital phase and 
cycle number, represented as a gray scale diagram (range $\pm 0.1\%$). 
 }
\figsetgrpend

\figsetgrpstart
\figsetgrpnum{1.32}
\figsetgrptitle{g32}
\figsetplot{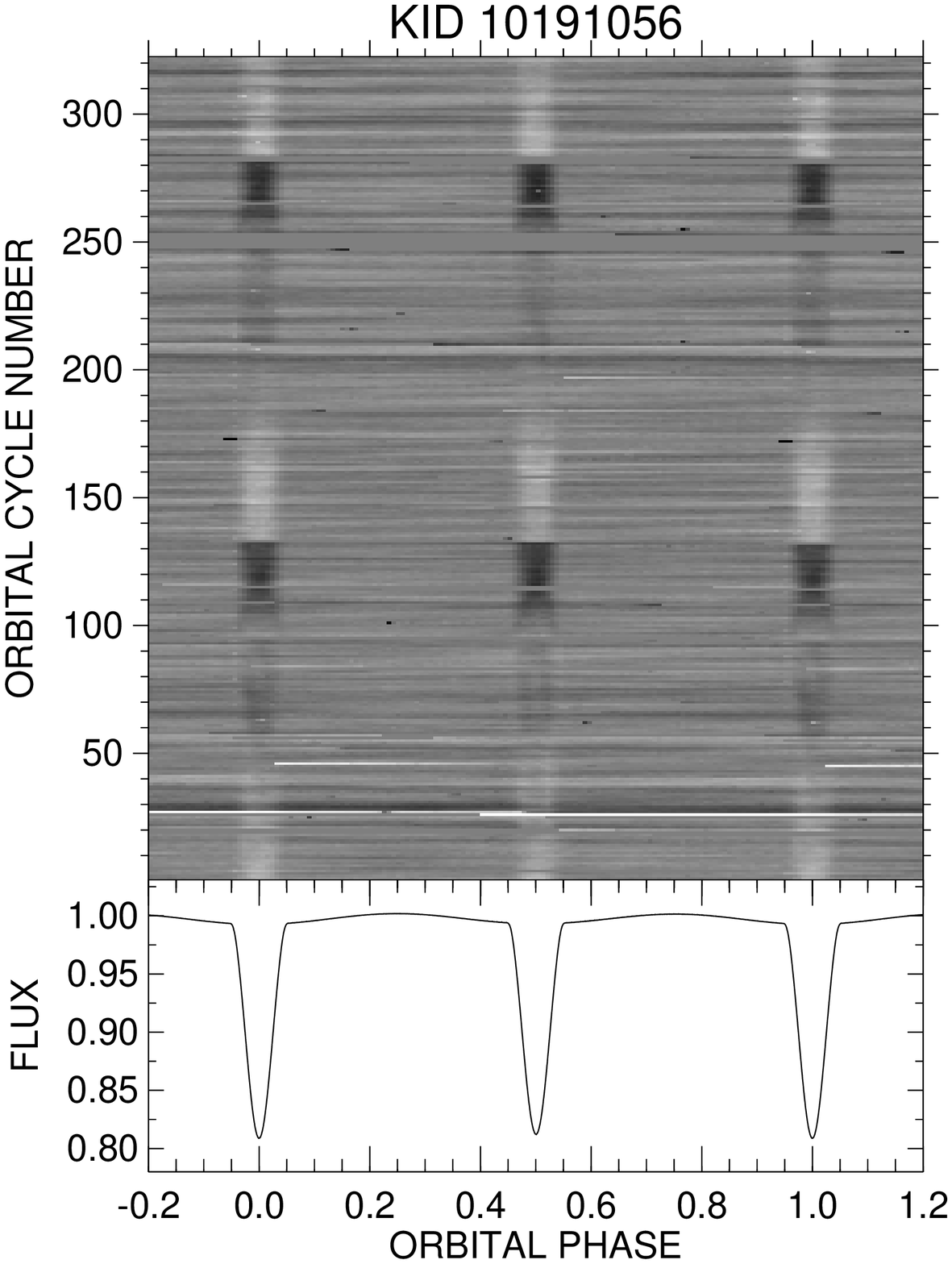}
\figsetgrpnote{The lower panel shows a mean, normalized light curve formed 
by binning in orbital phase.  The top panel shows the 
flux differences as a function of orbital phase and 
cycle number, represented as a gray scale diagram (range $\pm 0.2\%$). 
 }
\figsetgrpend

\figsetgrpstart
\figsetgrpnum{1.33}
\figsetgrptitle{g33}
\figsetplot{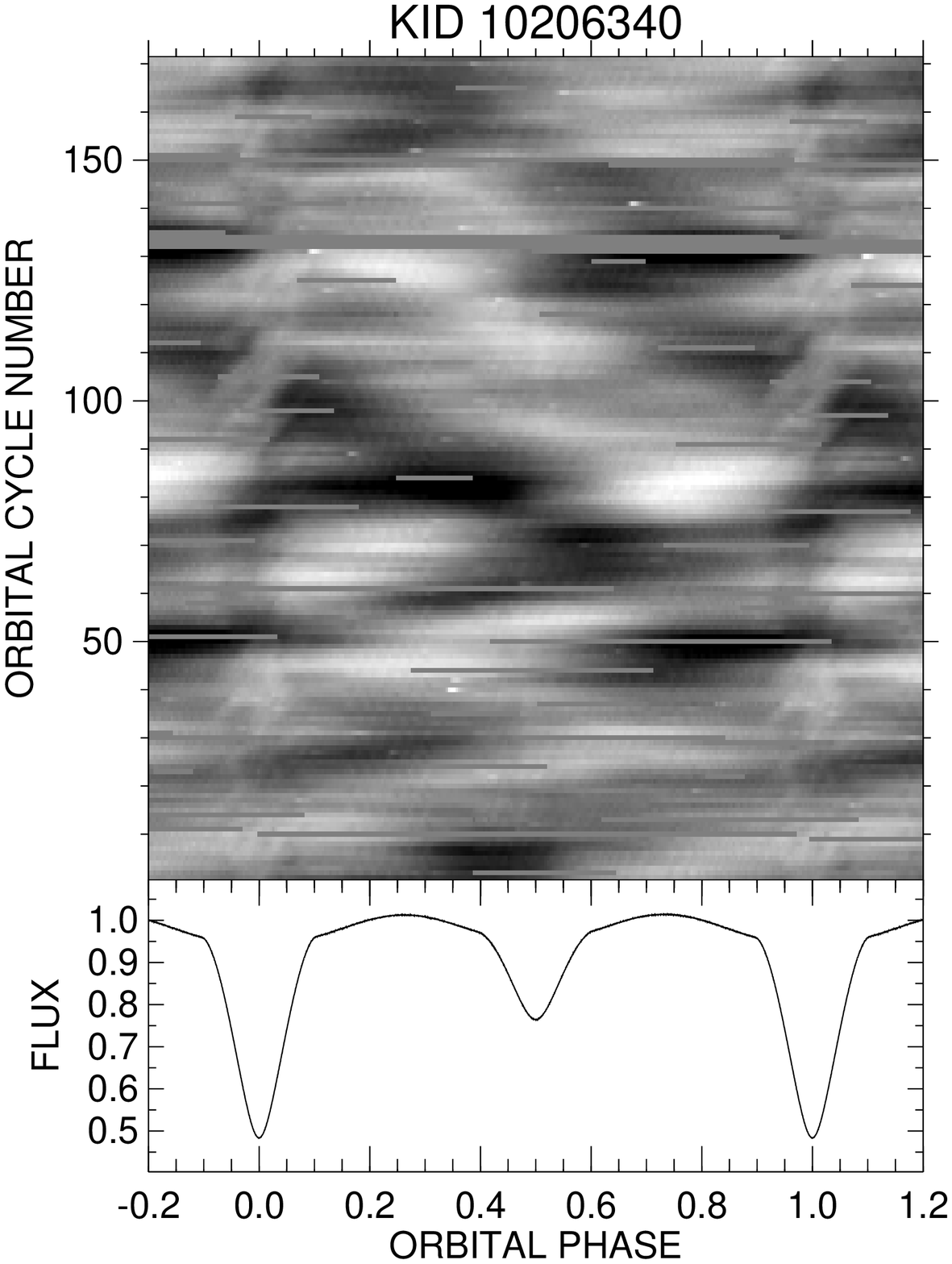}
\figsetgrpnote{The lower panel shows a mean, normalized light curve formed 
by binning in orbital phase.  The top panel shows the 
flux differences as a function of orbital phase and 
cycle number, represented as a gray scale diagram (range $\pm 2\%$). 
 }
\figsetgrpend

\figsetgrpstart
\figsetgrpnum{1.34}
\figsetgrptitle{g34}
\figsetplot{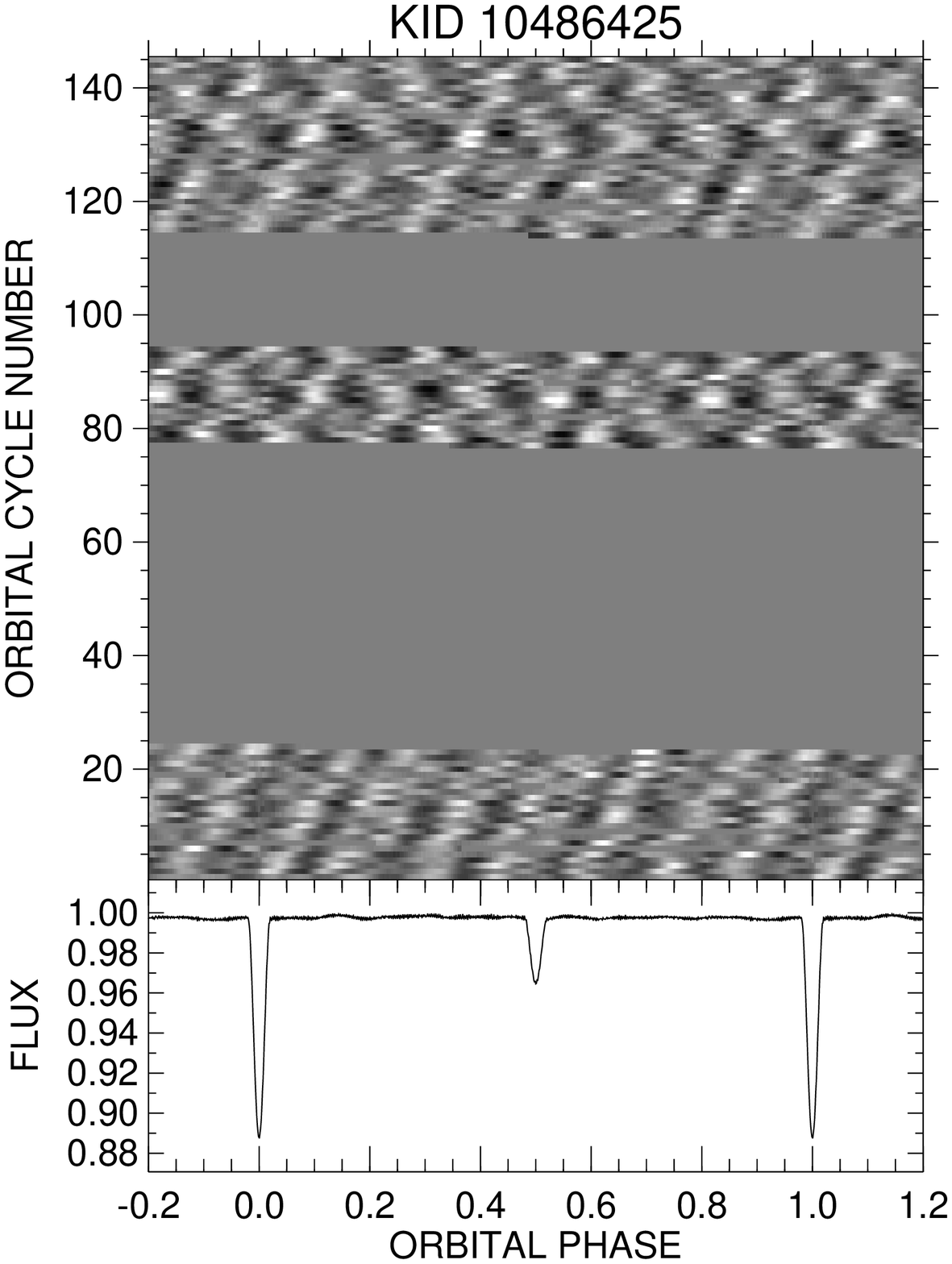}
\figsetgrpnote{The lower panel shows a mean, normalized light curve formed 
by binning in orbital phase.  The top panel shows the 
flux differences as a function of orbital phase and 
cycle number, represented as a gray scale diagram (range $\pm 1\%$). 
 }
\figsetgrpend

\figsetgrpstart
\figsetgrpnum{1.35}
\figsetgrptitle{g35}
\figsetplot{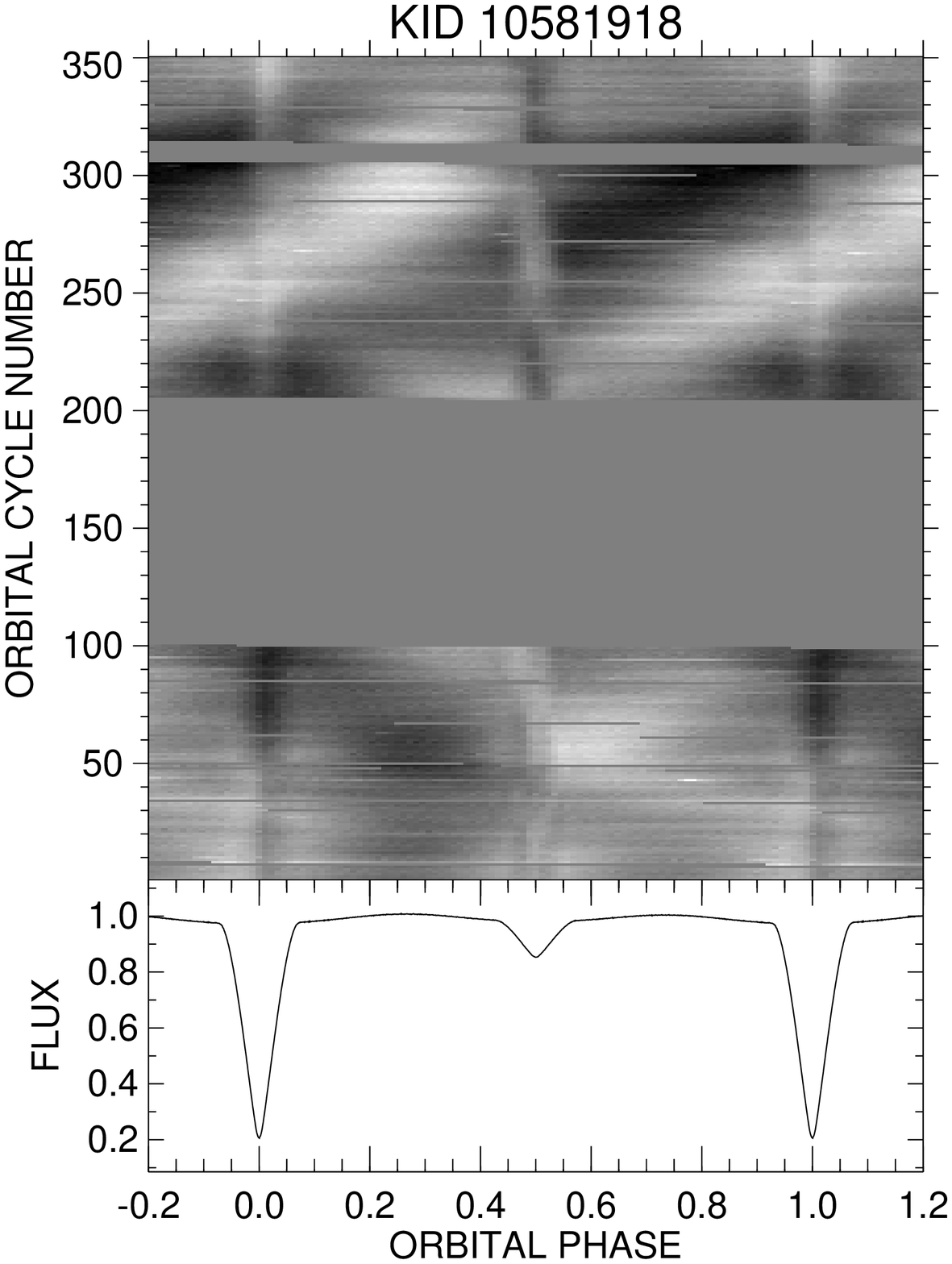}
\figsetgrpnote{The lower panel shows a mean, normalized light curve formed 
by binning in orbital phase.  The top panel shows the 
flux differences as a function of orbital phase and 
cycle number, represented as a gray scale diagram (range $\pm 2\%$). 
 }
\figsetgrpend

\figsetgrpstart
\figsetgrpnum{1.36}
\figsetgrptitle{g36}
\figsetplot{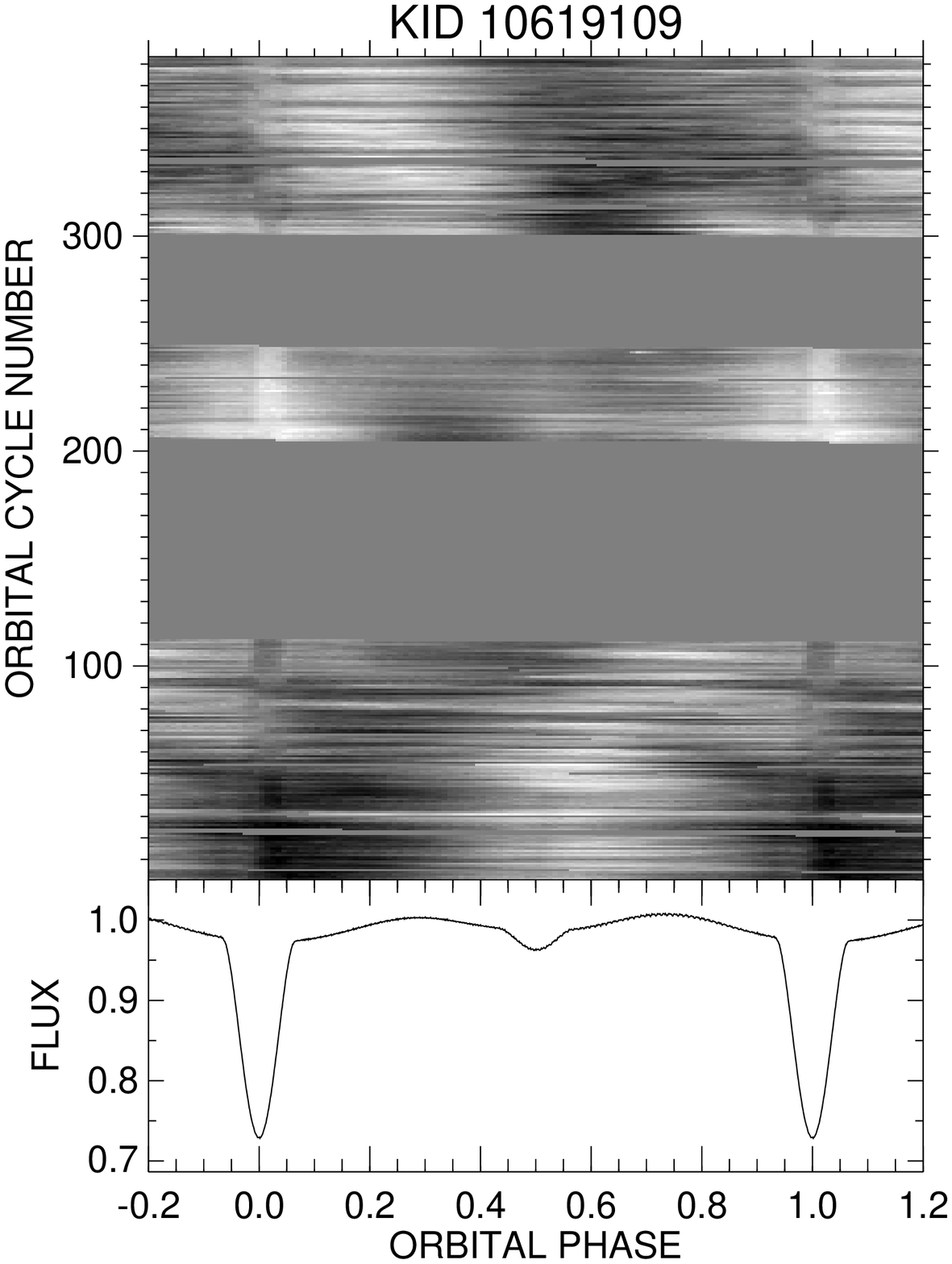}
\figsetgrpnote{The lower panel shows a mean, normalized light curve formed 
by binning in orbital phase.  The top panel shows the 
flux differences as a function of orbital phase and 
cycle number, represented as a gray scale diagram (range $\pm 1\%$). 
 }
\figsetgrpend

\figsetgrpstart
\figsetgrpnum{1.37}
\figsetgrptitle{g37}
\figsetplot{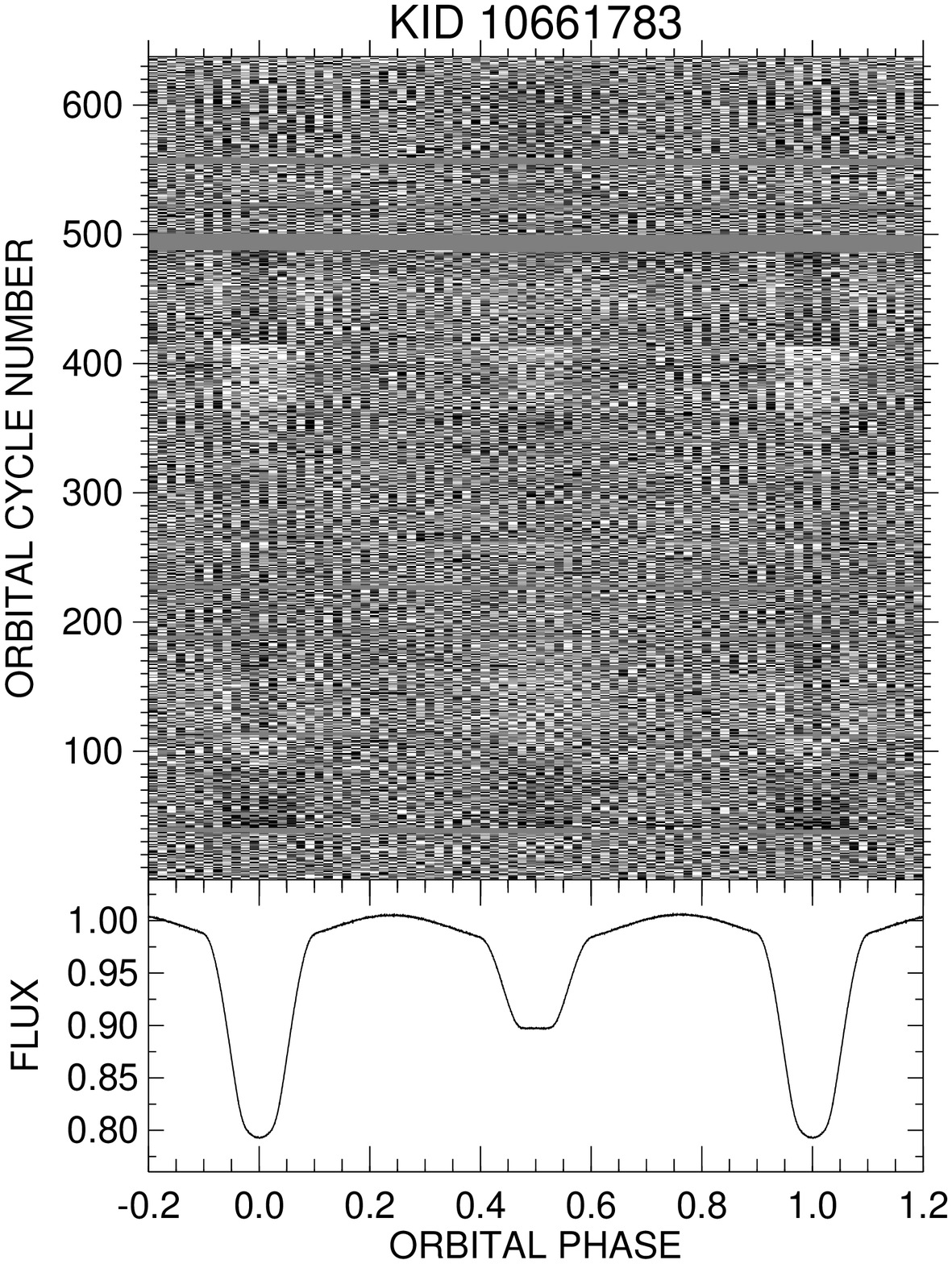}
\figsetgrpnote{The lower panel shows a mean, normalized light curve formed 
by binning in orbital phase.  The top panel shows the 
flux differences as a function of orbital phase and 
cycle number, represented as a gray scale diagram (range $\pm 0.2\%$). 
 }
\figsetgrpend

\figsetgrpstart
\figsetgrpnum{1.38}
\figsetgrptitle{g38}
\figsetplot{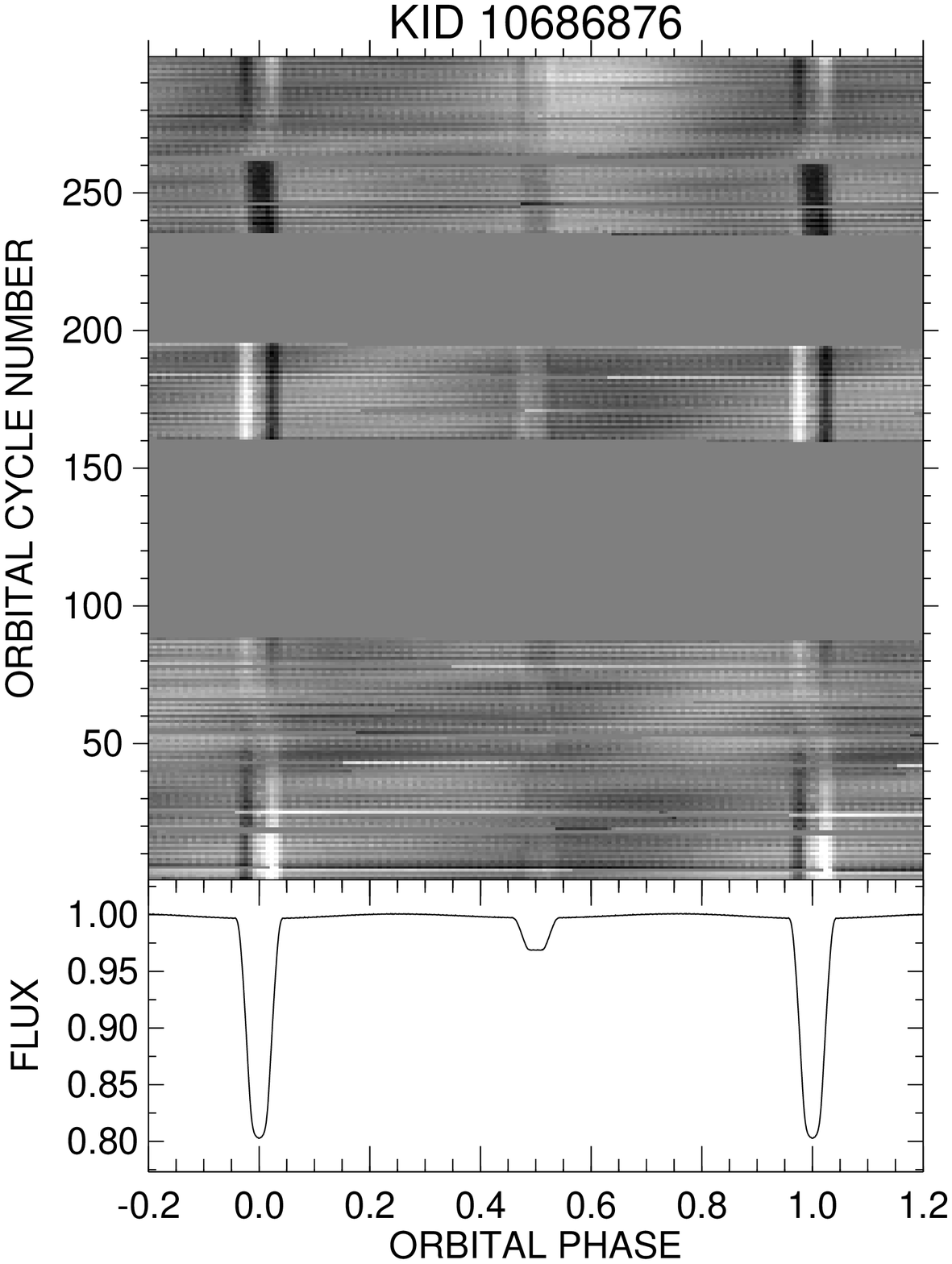}
\figsetgrpnote{The lower panel shows a mean, normalized light curve formed 
by binning in orbital phase.  The top panel shows the 
flux differences as a function of orbital phase and 
cycle number, represented as a gray scale diagram (range $\pm 0.3\%$). 
 }
\figsetgrpend

\figsetgrpstart
\figsetgrpnum{1.39}
\figsetgrptitle{g39}
\figsetplot{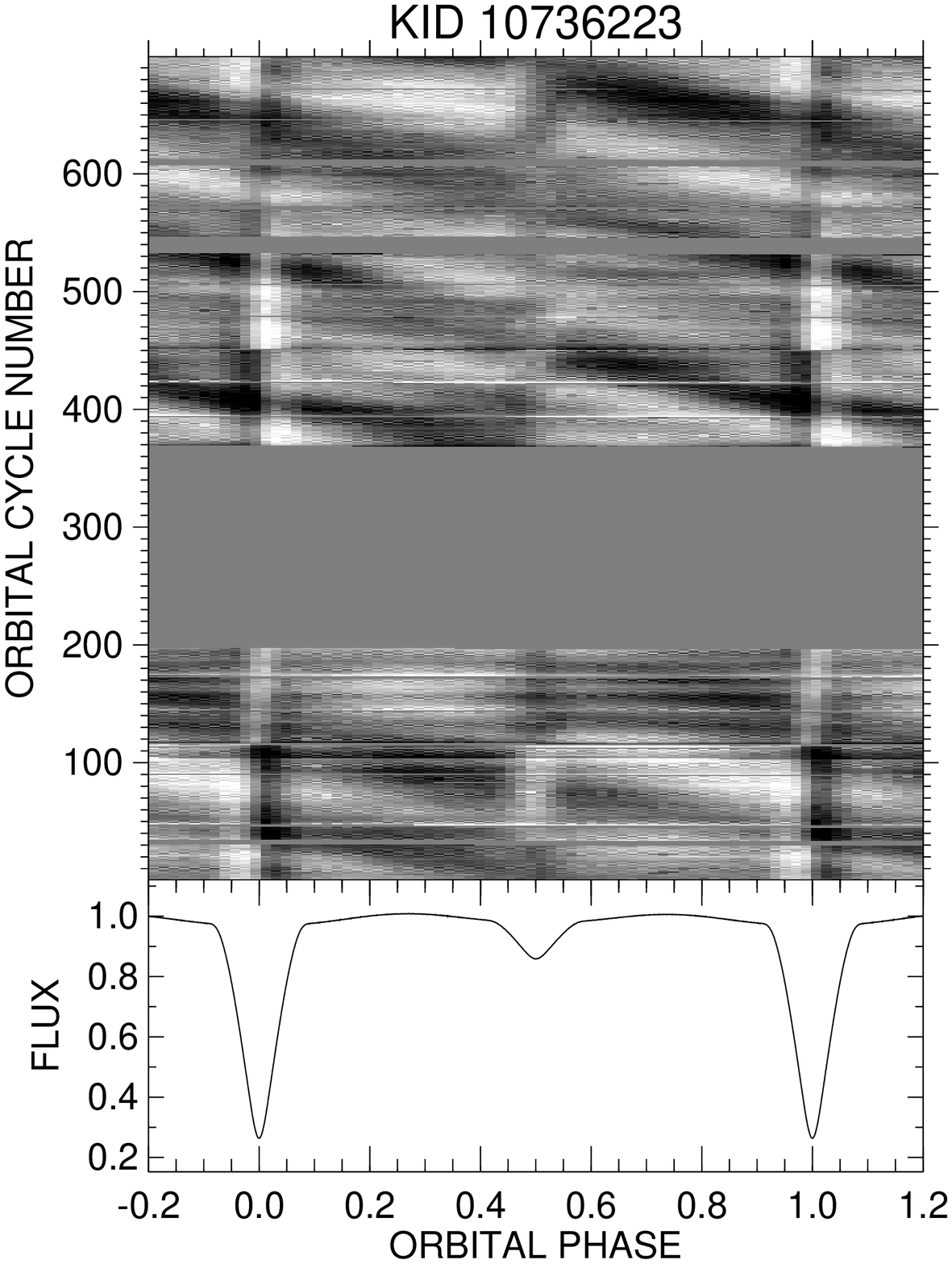}
\figsetgrpnote{The lower panel shows a mean, normalized light curve formed 
by binning in orbital phase.  The top panel shows the 
flux differences as a function of orbital phase and 
cycle number, represented as a gray scale diagram (range $\pm 0.5\%$). 
 }
\figsetgrpend

\figsetgrpstart
\figsetgrpnum{1.40}
\figsetgrptitle{g40}
\figsetplot{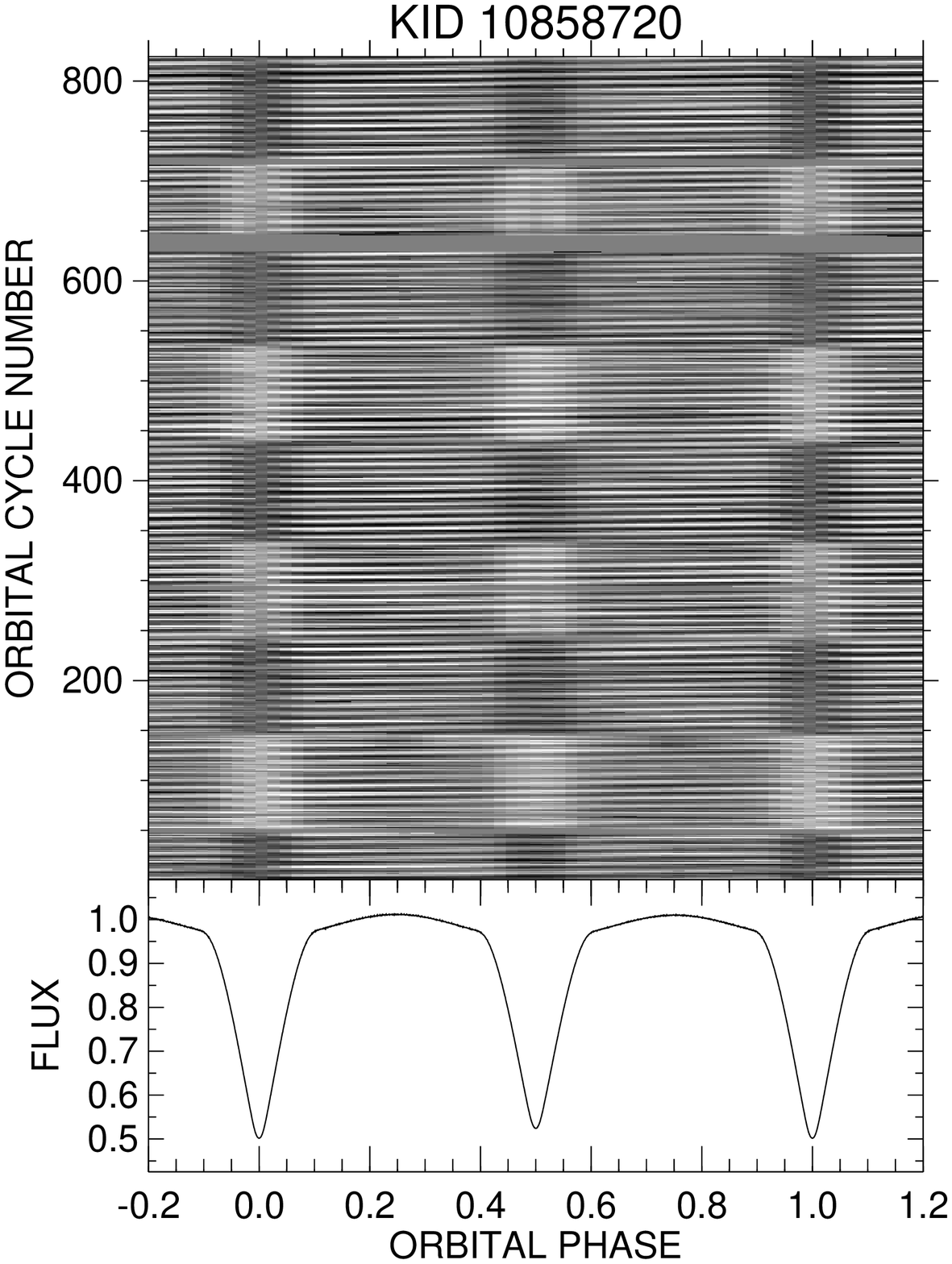}
\figsetgrpnote{The lower panel shows a mean, normalized light curve formed 
by binning in orbital phase.  The top panel shows the 
flux differences as a function of orbital phase and 
cycle number, represented as a gray scale diagram (range $\pm 0.5\%$). 
 }
\figsetgrpend

\figsetgrpstart
\figsetgrpnum{1.41}
\figsetgrptitle{g41}
\figsetplot{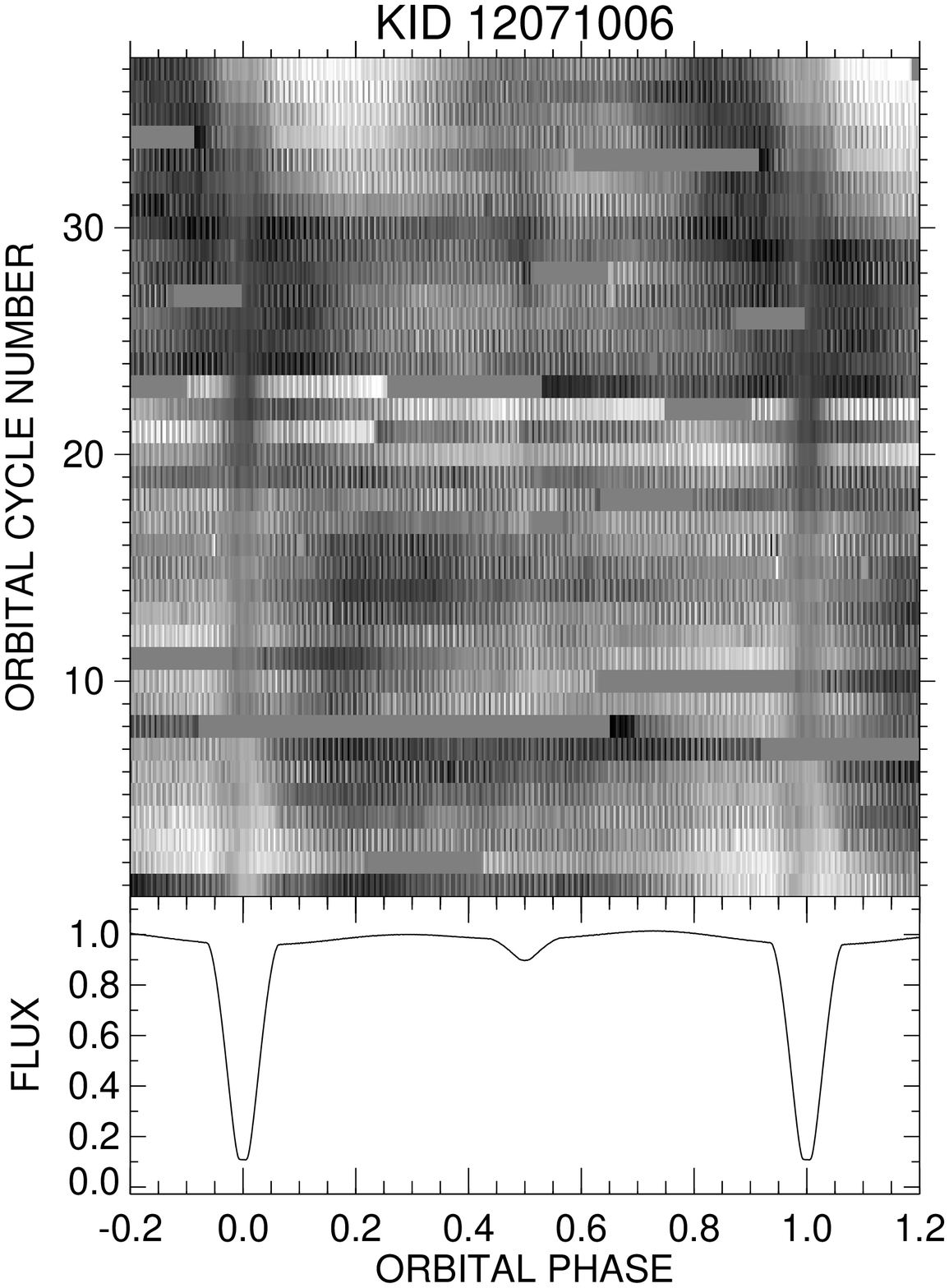}
\figsetgrpnote{The lower panel shows a mean, normalized light curve formed 
by binning in orbital phase.  The top panel shows the 
flux differences as a function of orbital phase and 
cycle number, represented as a gray scale diagram (range $\pm 1\%$). 
 }
\figsetgrpend

\figsetend



\figsetstart
\figsetnum{2}
\figsettitle{Eclipse timing variations}

\figsetgrpstart
\figsetgrpnum{2.1}
\figsetgrptitle{r1}
\figsetplot{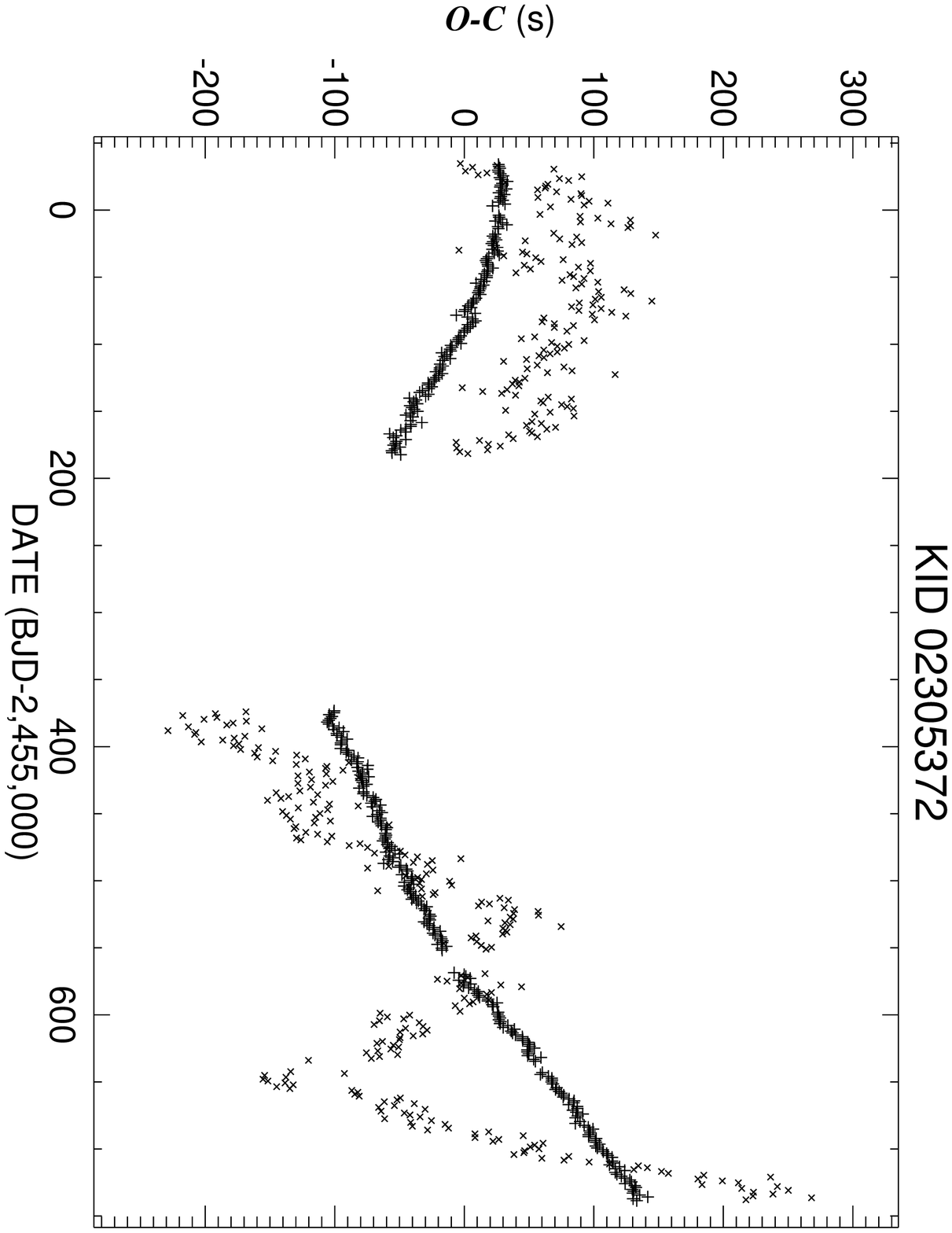}
\figsetgrpnote{The observed minus calculated eclipse times relative to
a linear ephemeris.  The primary and secondary eclipse
times are indicated by $+$ and $\times$ symbols, 
respectively. }
\figsetgrpend

\figsetgrpstart
\figsetgrpnum{2.2}
\figsetgrptitle{r2}
\figsetplot{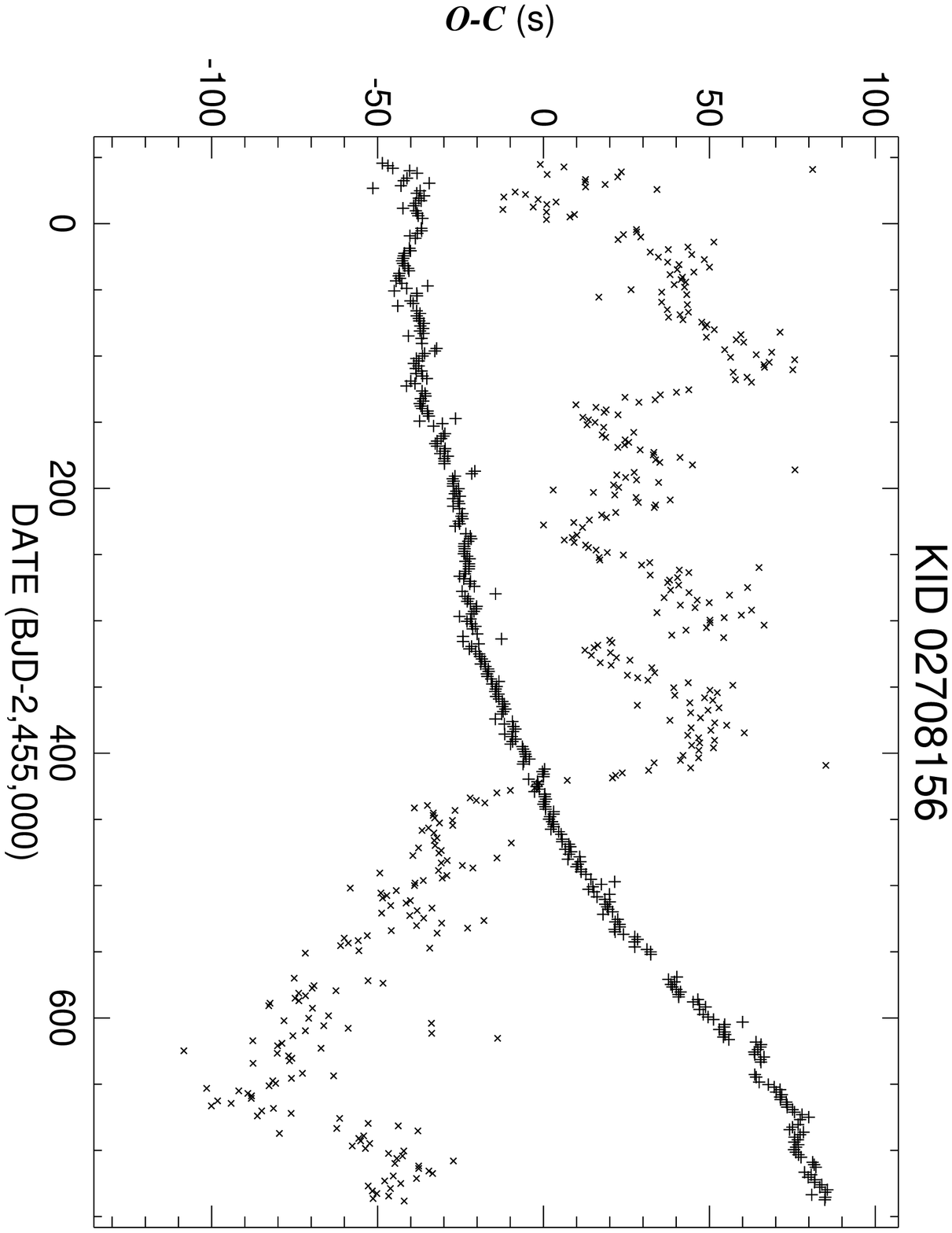}
\figsetgrpnote{The observed minus calculated eclipse times relative to
a linear ephemeris.  The primary and secondary eclipse
times are indicated by $+$ and $\times$ symbols, 
respectively. }
\figsetgrpend

\figsetgrpstart
\figsetgrpnum{2.3}
\figsetgrptitle{r3}
\figsetplot{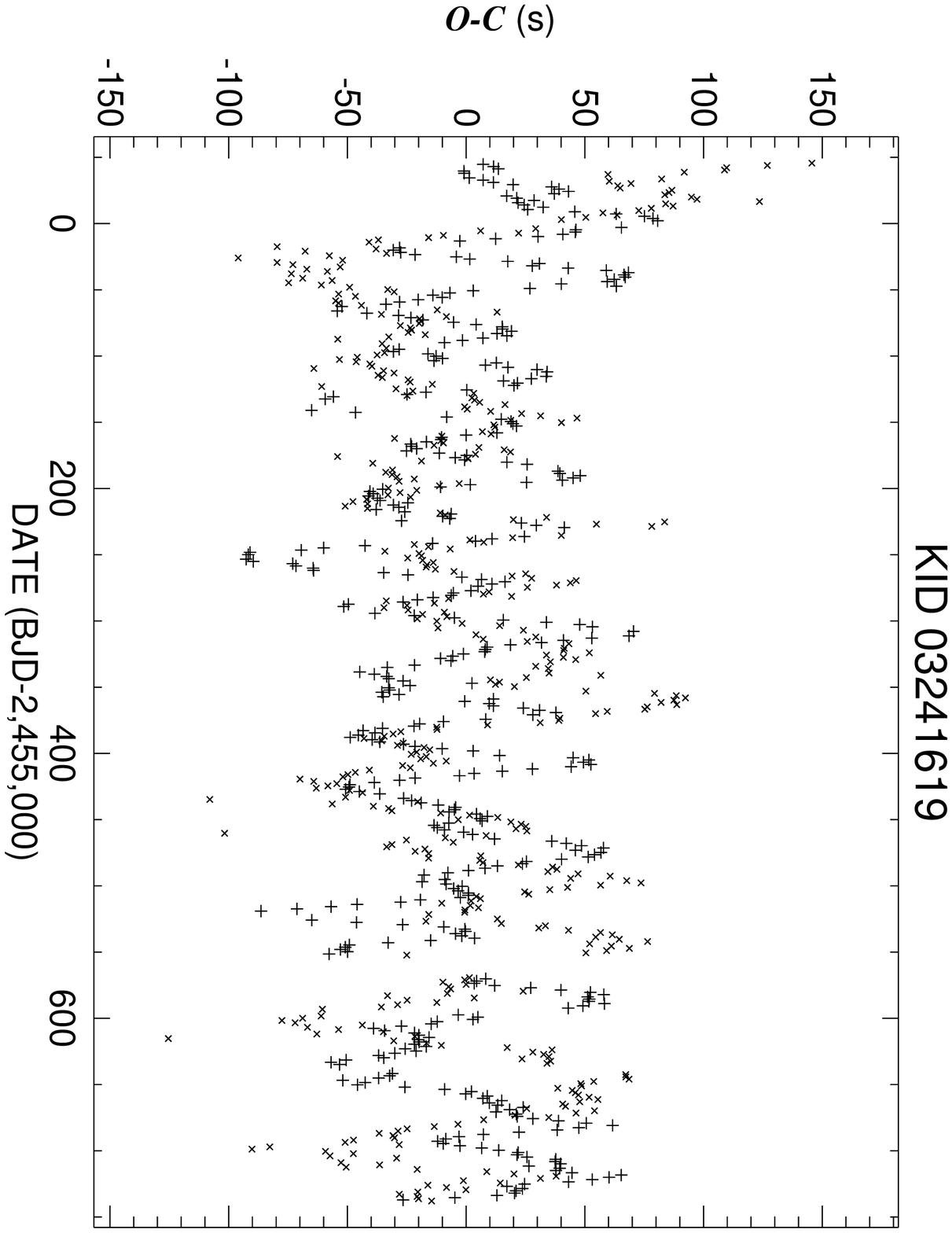}
\figsetgrpnote{The observed minus calculated eclipse times relative to
a linear ephemeris.  The primary and secondary eclipse
times are indicated by $+$ and $\times$ symbols, 
respectively. }
\figsetgrpend

\figsetgrpstart
\figsetgrpnum{2.4}
\figsetgrptitle{r4}
\figsetplot{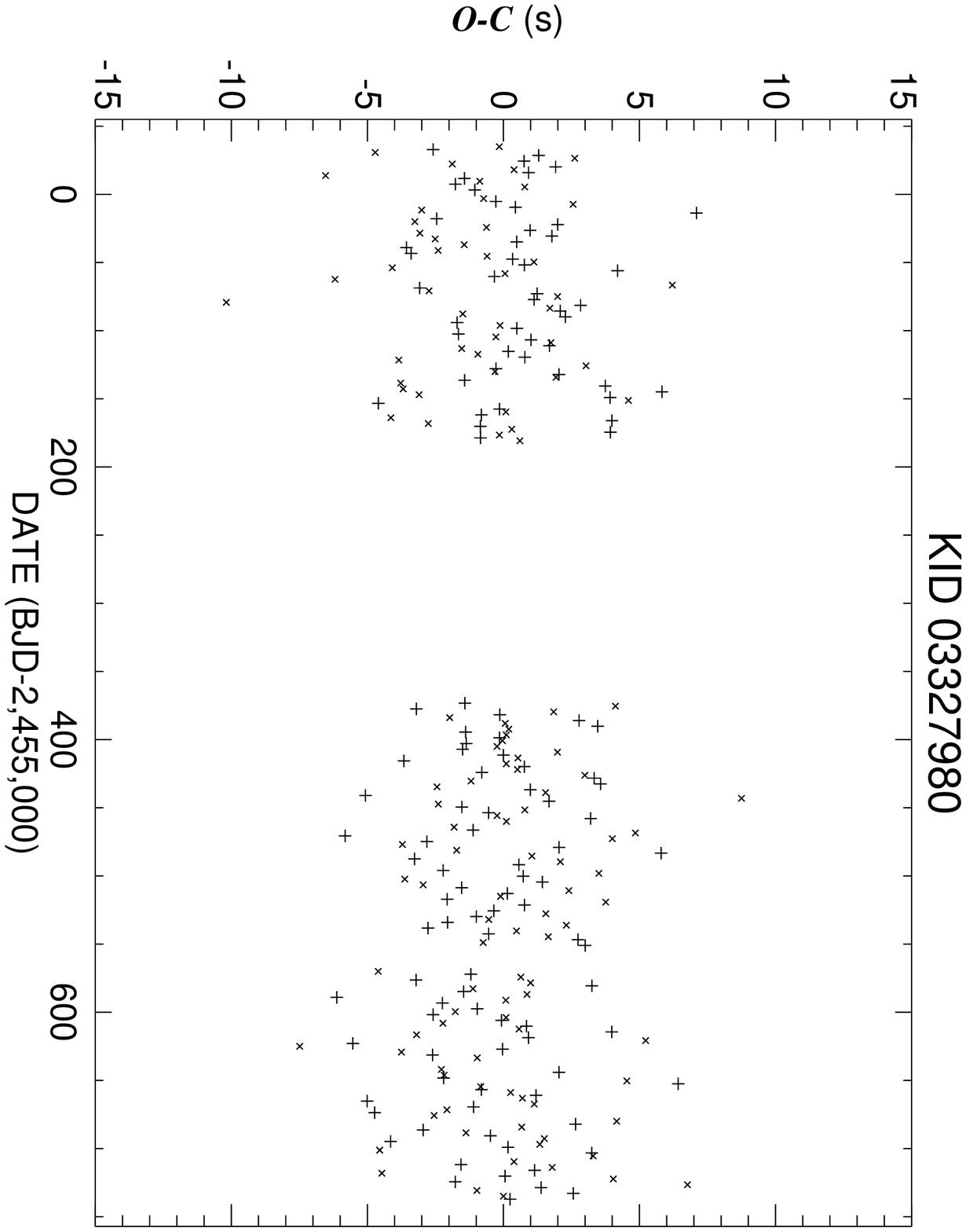}
\figsetgrpnote{The observed minus calculated eclipse times relative to
a linear ephemeris.  The primary and secondary eclipse
times are indicated by $+$ and $\times$ symbols, 
respectively. }
\figsetgrpend

\figsetgrpstart
\figsetgrpnum{2.5}
\figsetgrptitle{r5}
\figsetplot{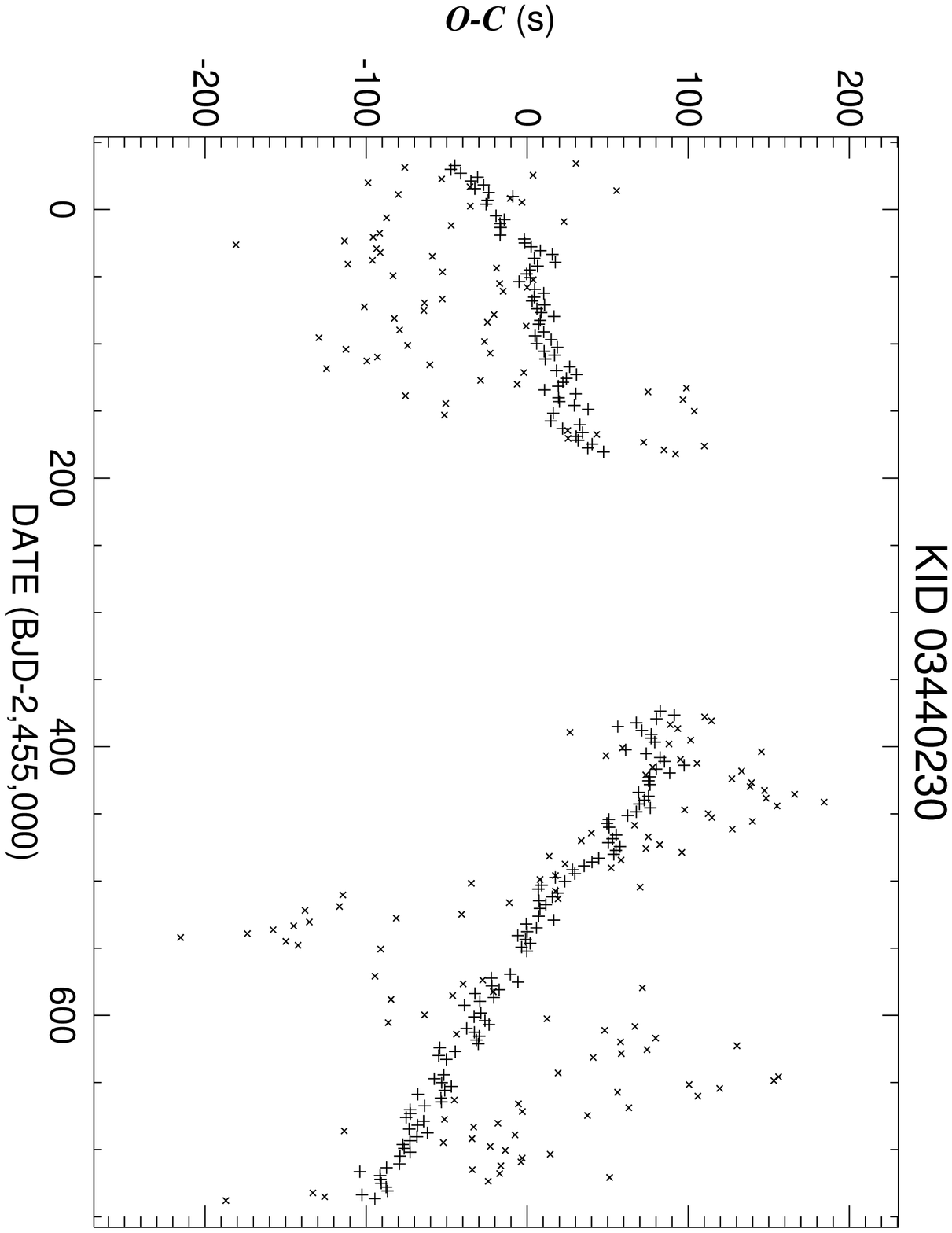}
\figsetgrpnote{The observed minus calculated eclipse times relative to
a linear ephemeris.  The primary and secondary eclipse
times are indicated by $+$ and $\times$ symbols, 
respectively. }
\figsetgrpend

\figsetgrpstart
\figsetgrpnum{2.6}
\figsetgrptitle{r6}
\figsetplot{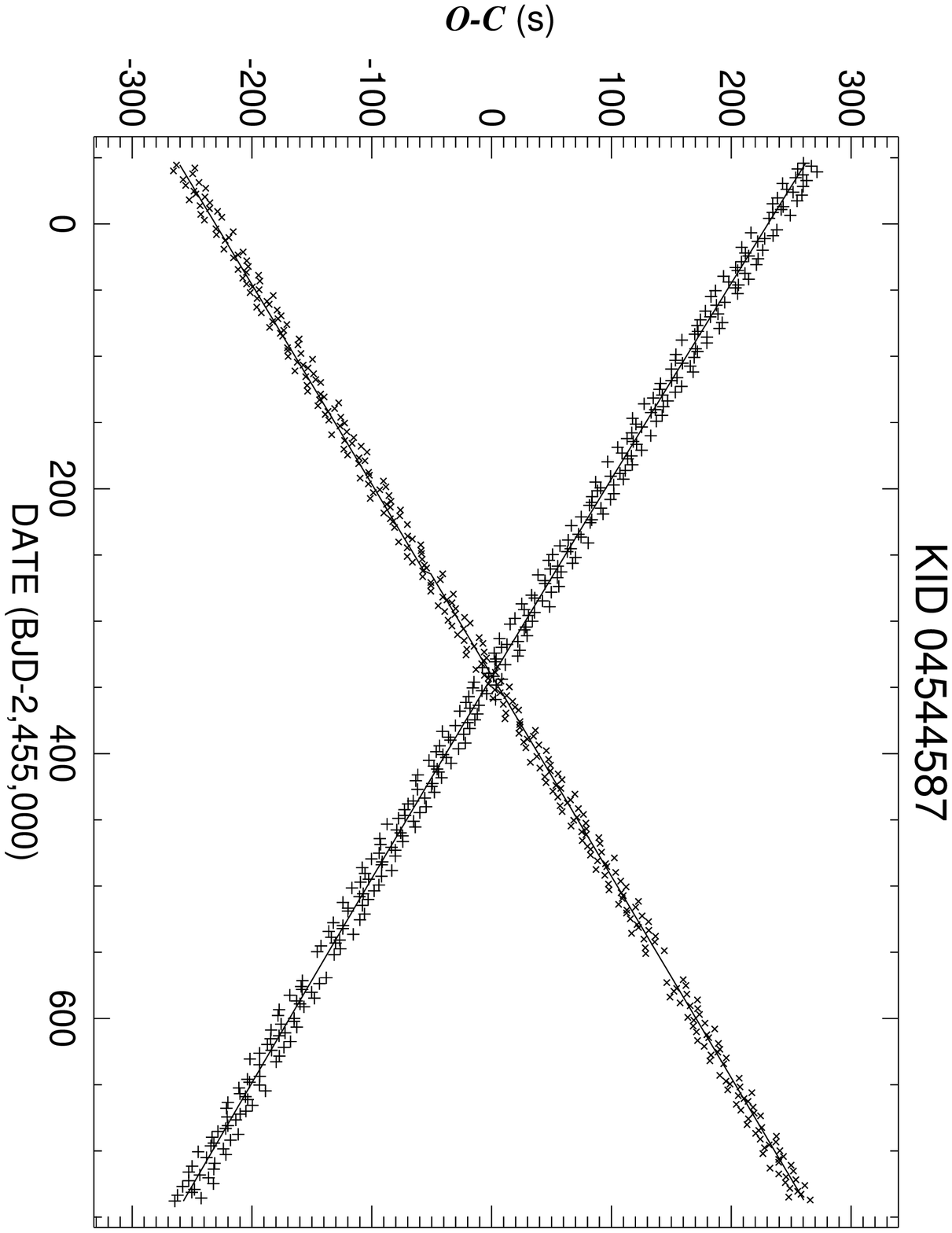}
\figsetgrpnote{The observed minus calculated eclipse times relative to
a linear ephemeris.  The primary and secondary eclipse
times are indicated by $+$ and $\times$ symbols, 
respectively. }
\figsetgrpend

\figsetgrpstart
\figsetgrpnum{2.7}
\figsetgrptitle{r7}
\figsetplot{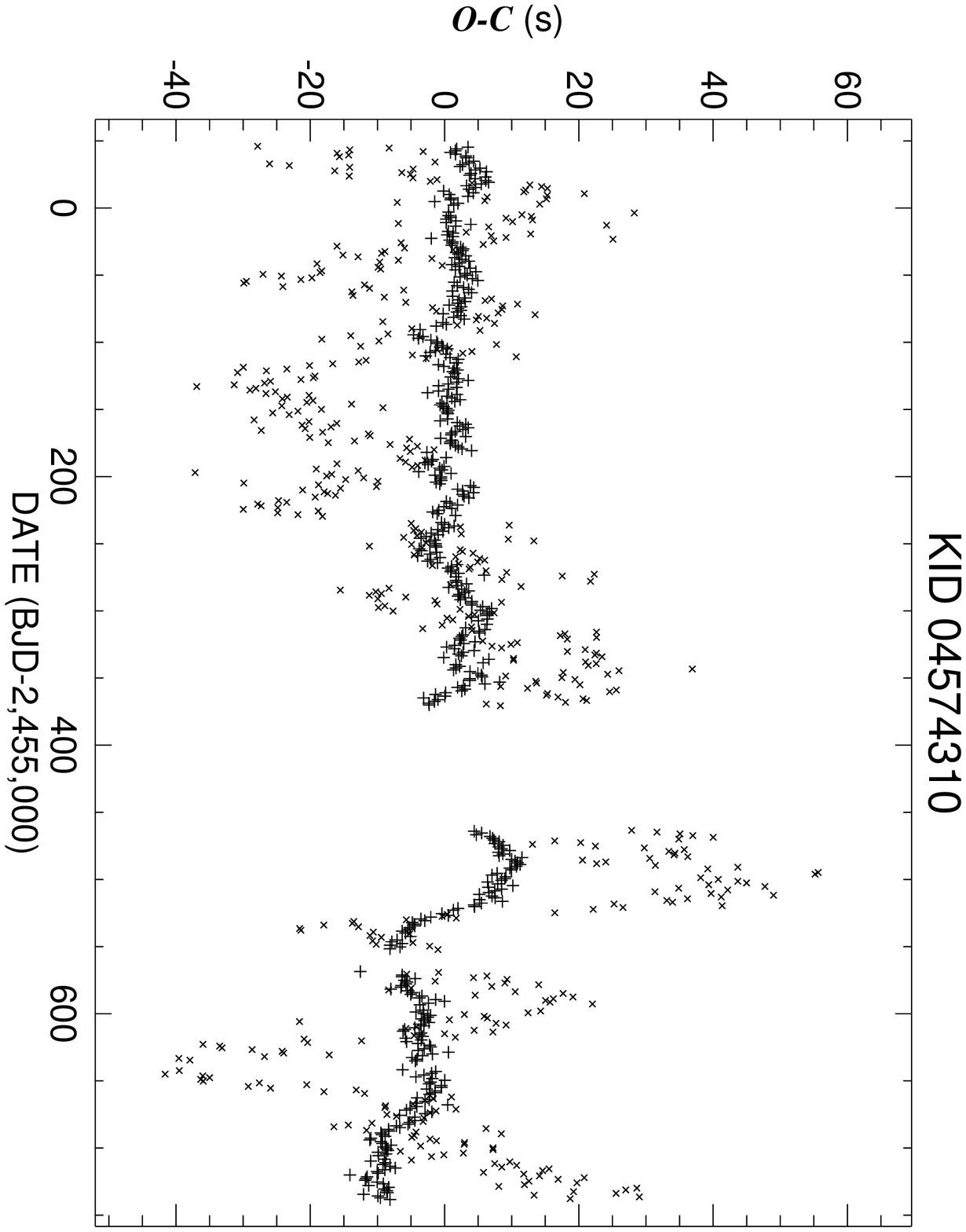}
\figsetgrpnote{The observed minus calculated eclipse times relative to
a linear ephemeris.  The primary and secondary eclipse
times are indicated by $+$ and $\times$ symbols, 
respectively. }
\figsetgrpend

\figsetgrpstart
\figsetgrpnum{2.8}
\figsetgrptitle{r8}
\figsetplot{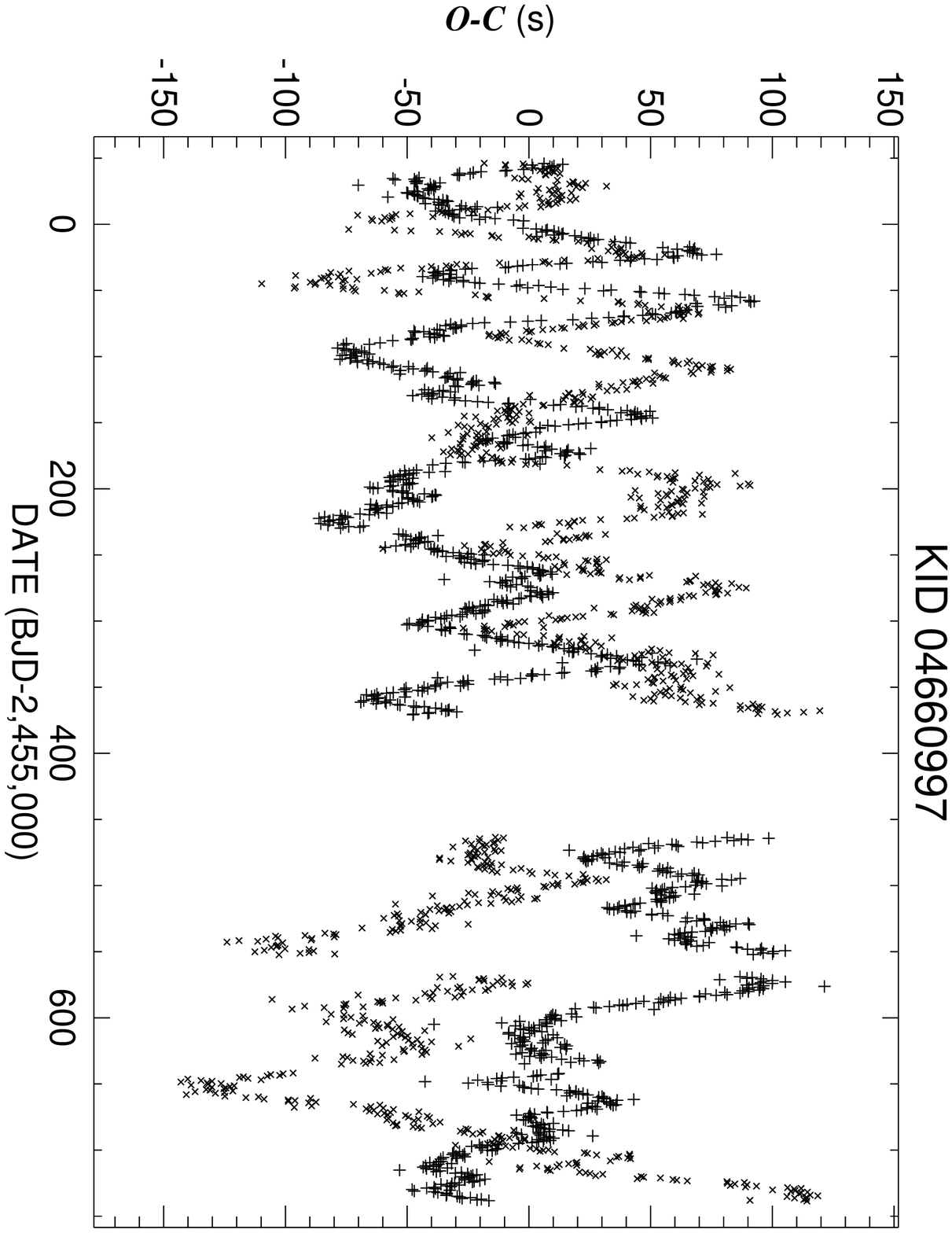}
\figsetgrpnote{The observed minus calculated eclipse times relative to
a linear ephemeris.  The primary and secondary eclipse
times are indicated by $+$ and $\times$ symbols, 
respectively. }
\figsetgrpend

\figsetgrpstart
\figsetgrpnum{2.9}
\figsetgrptitle{r9}
\figsetplot{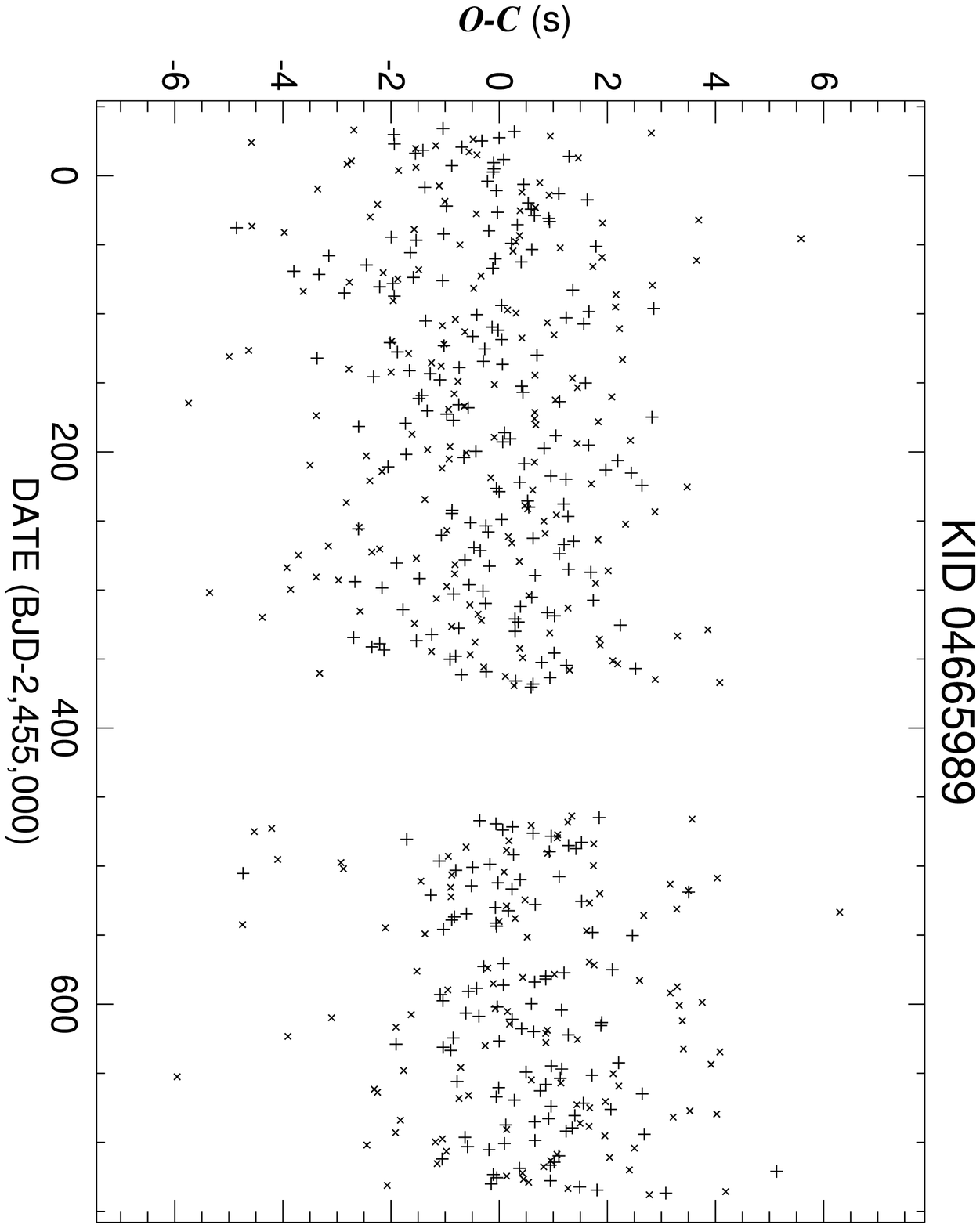}
\figsetgrpnote{The observed minus calculated eclipse times relative to
a linear ephemeris.  The primary and secondary eclipse
times are indicated by $+$ and $\times$ symbols, 
respectively. }
\figsetgrpend

\figsetgrpstart
\figsetgrpnum{2.10}
\figsetgrptitle{r10}
\figsetplot{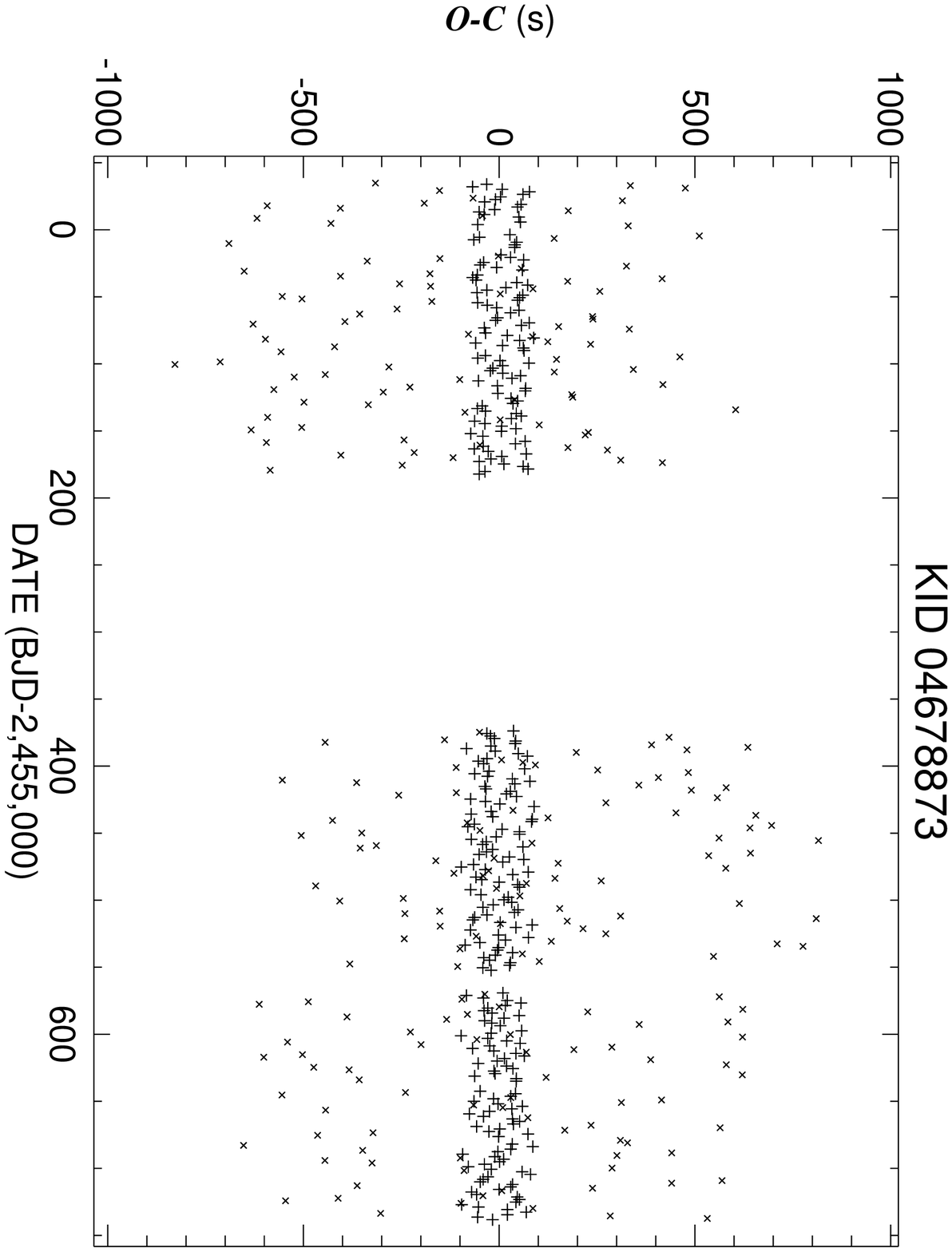}
\figsetgrpnote{The observed minus calculated eclipse times relative to
a linear ephemeris.  The primary and secondary eclipse
times are indicated by $+$ and $\times$ symbols, 
respectively. }
\figsetgrpend

\figsetgrpstart
\figsetgrpnum{2.11}
\figsetgrptitle{r11}
\figsetplot{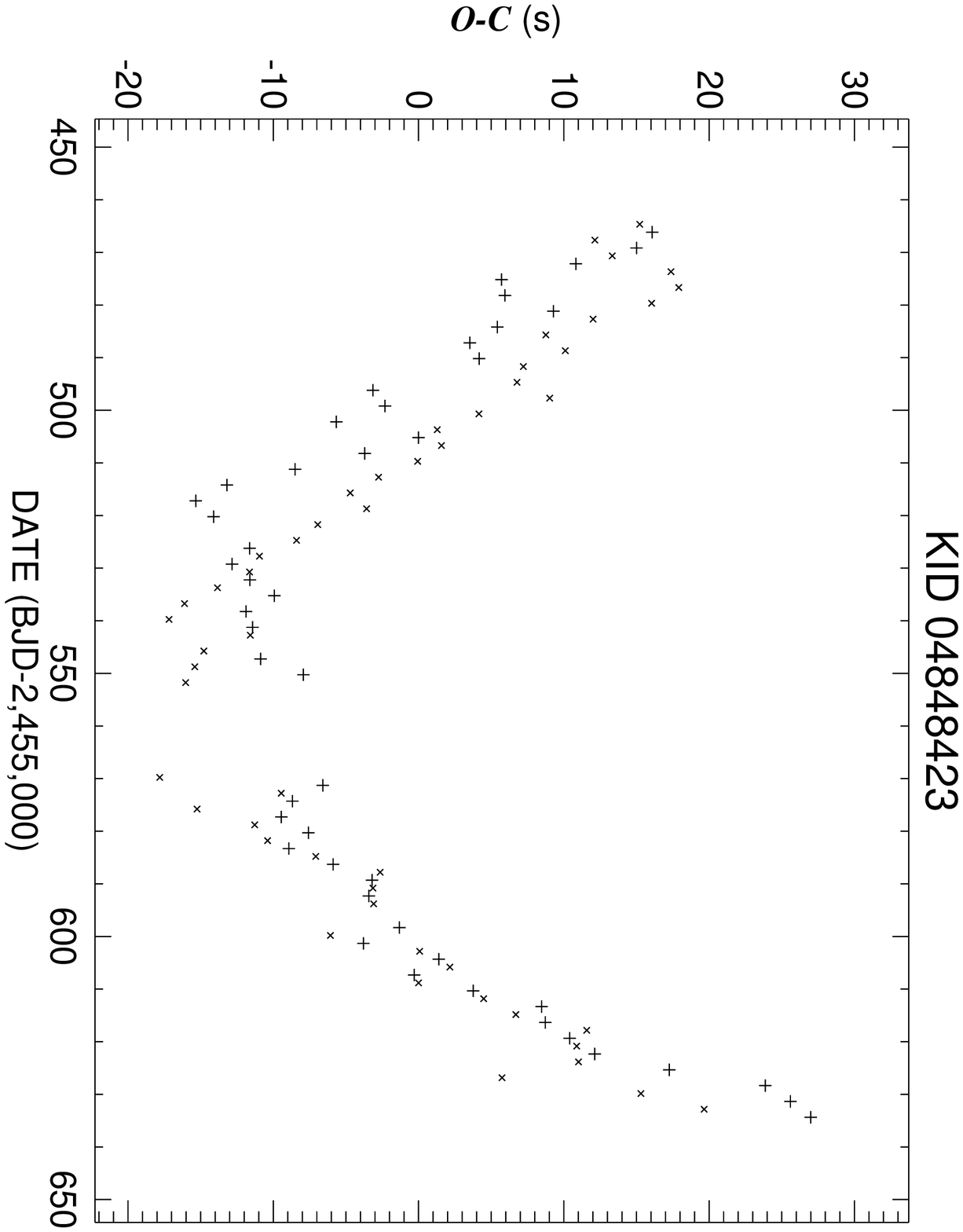}
\figsetgrpnote{The observed minus calculated eclipse times relative to
a linear ephemeris.  The primary and secondary eclipse
times are indicated by $+$ and $\times$ symbols, 
respectively. }
\figsetgrpend

\figsetgrpstart
\figsetgrpnum{2.12}
\figsetgrptitle{r12}
\figsetplot{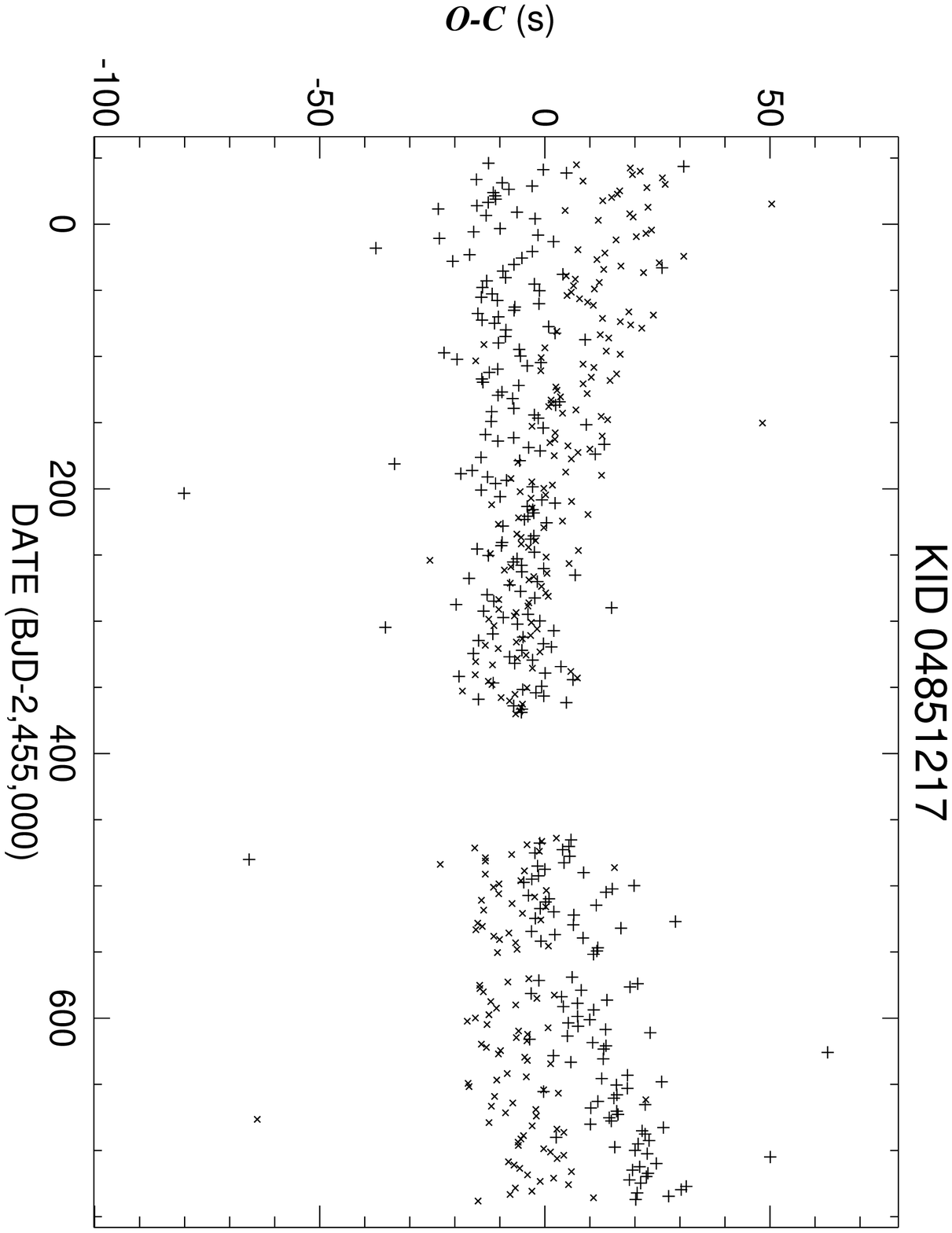}
\figsetgrpnote{The observed minus calculated eclipse times relative to
a linear ephemeris.  The primary and secondary eclipse
times are indicated by $+$ and $\times$ symbols, 
respectively. }
\figsetgrpend

\figsetgrpstart
\figsetgrpnum{2.13}
\figsetgrptitle{r13}
\figsetplot{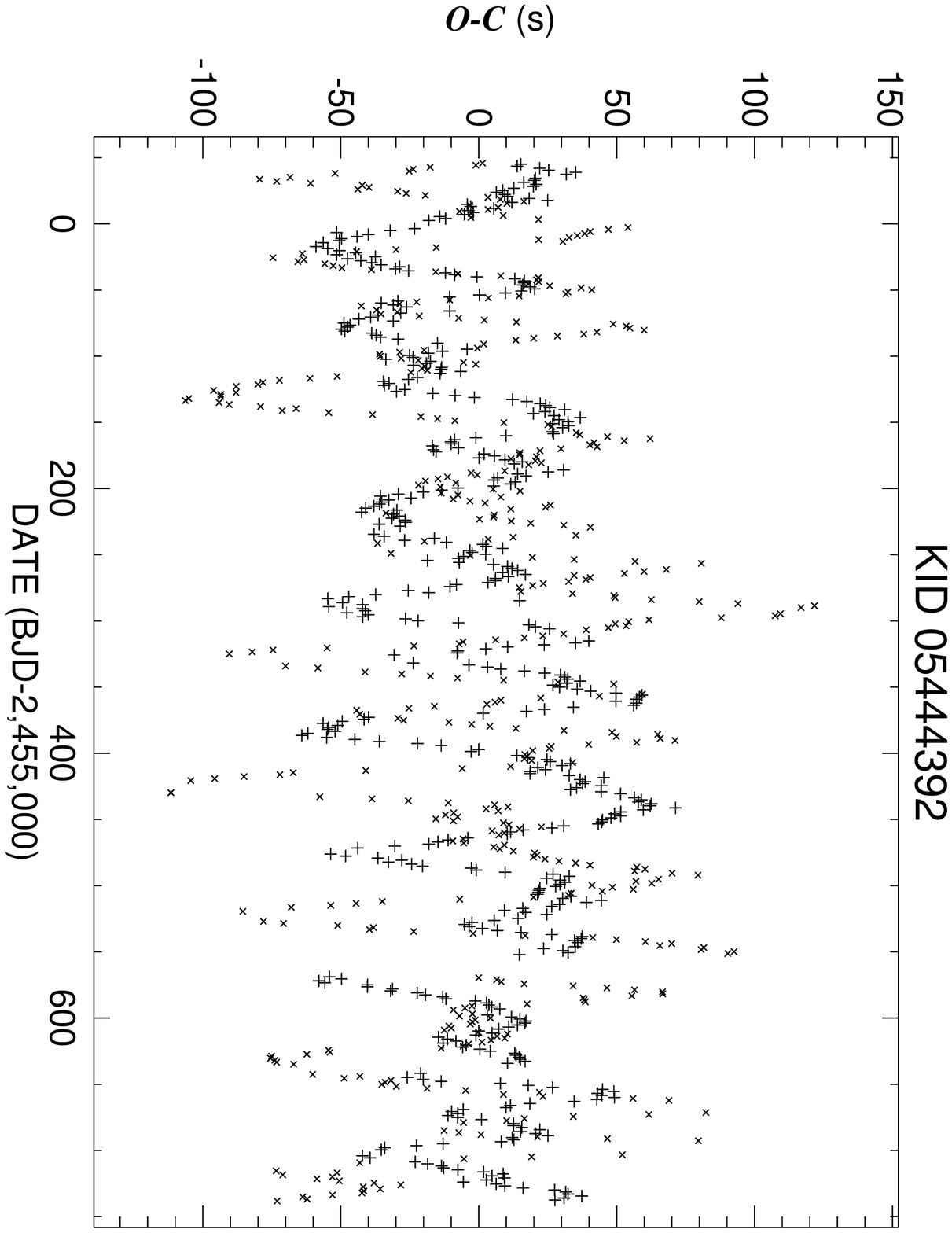}
\figsetgrpnote{The observed minus calculated eclipse times relative to
a linear ephemeris.  The primary and secondary eclipse
times are indicated by $+$ and $\times$ symbols, 
respectively. }
\figsetgrpend

\figsetgrpstart
\figsetgrpnum{2.14}
\figsetgrptitle{r14}
\figsetplot{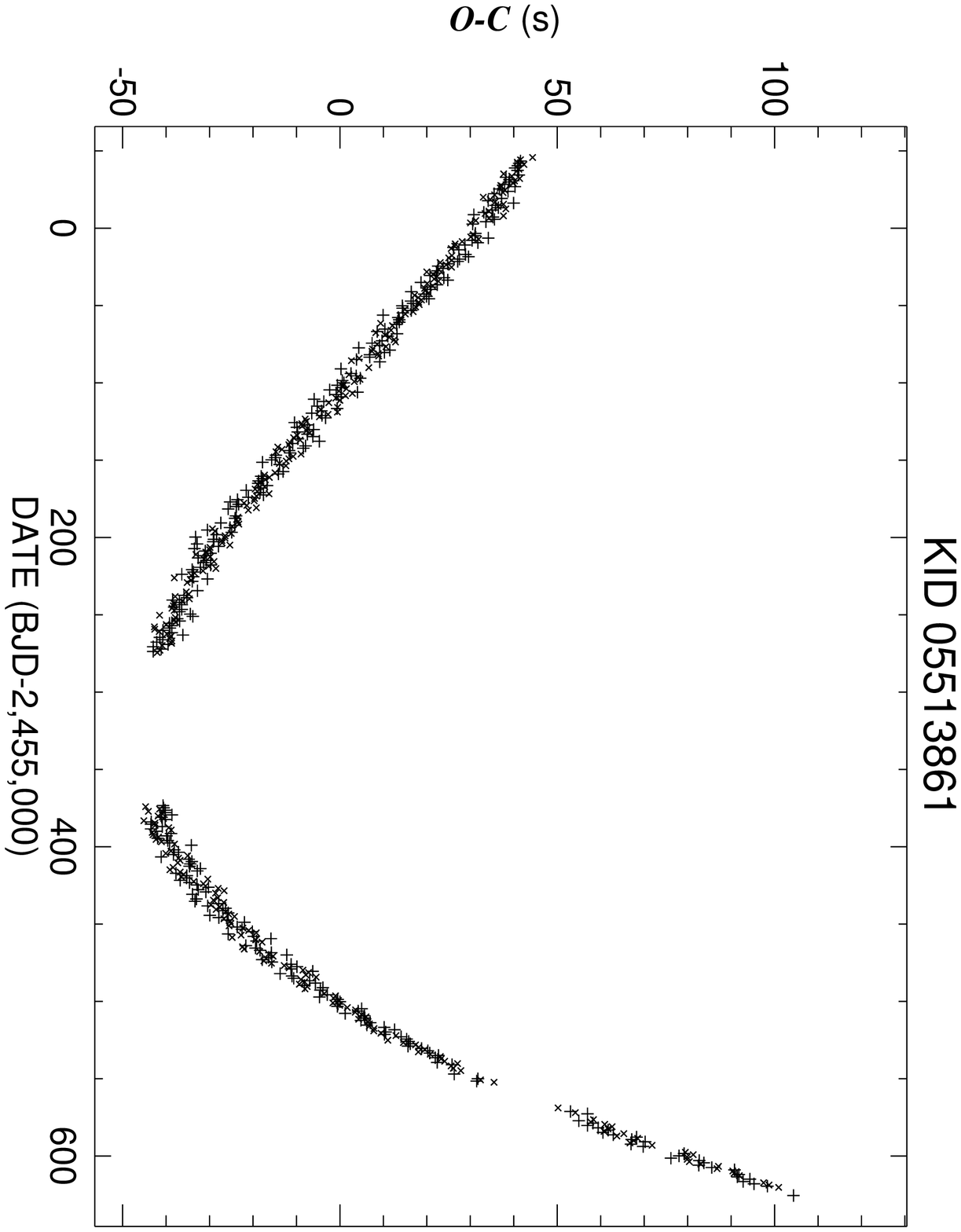}
\figsetgrpnote{The observed minus calculated eclipse times relative to
a linear ephemeris.  The primary and secondary eclipse
times are indicated by $+$ and $\times$ symbols, 
respectively. }
\figsetgrpend

\figsetgrpstart
\figsetgrpnum{2.15}
\figsetgrptitle{r15}
\figsetplot{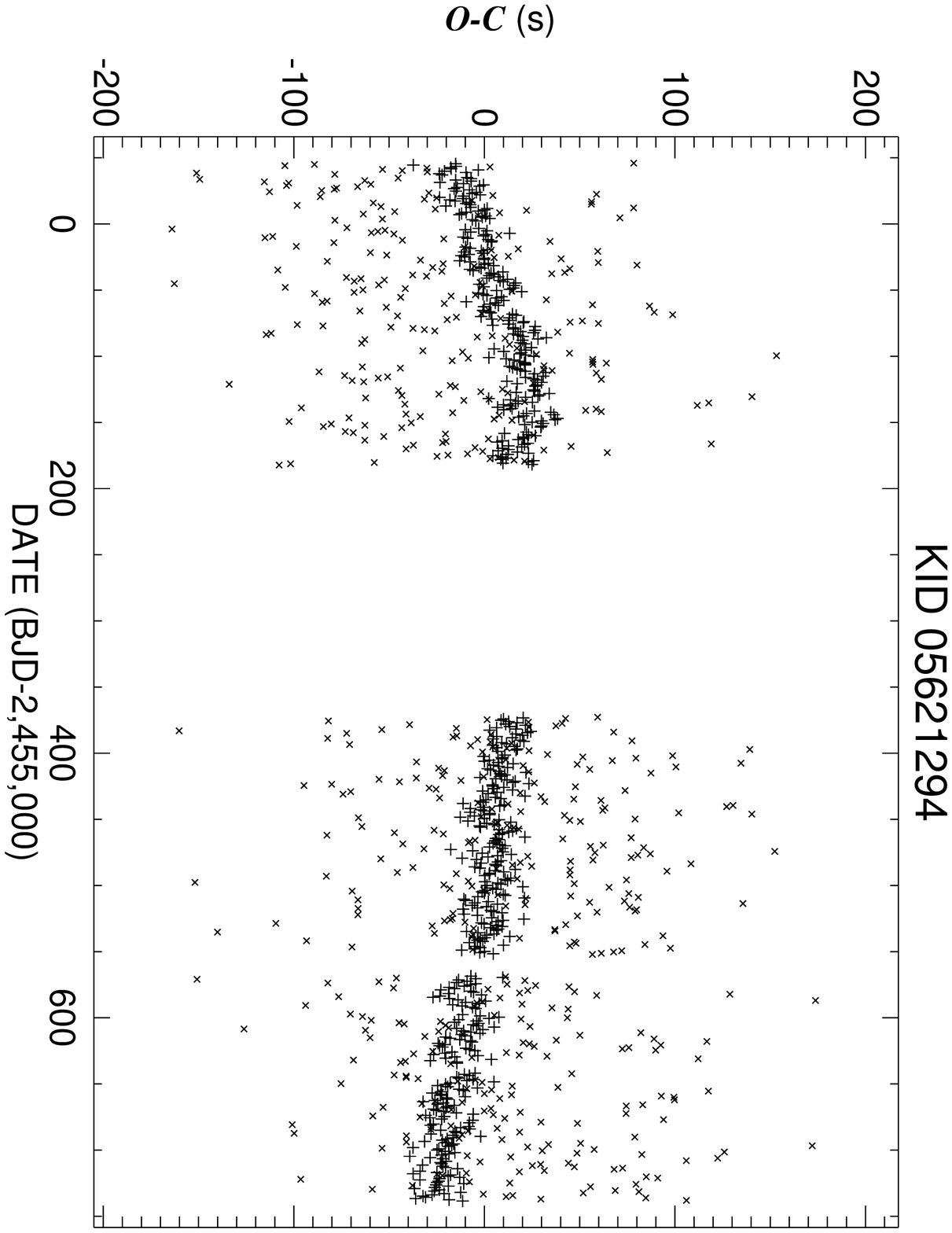}
\figsetgrpnote{The observed minus calculated eclipse times relative to
a linear ephemeris.  The primary and secondary eclipse
times are indicated by $+$ and $\times$ symbols, 
respectively. }
\figsetgrpend

\figsetgrpstart
\figsetgrpnum{2.16}
\figsetgrptitle{r16}
\figsetplot{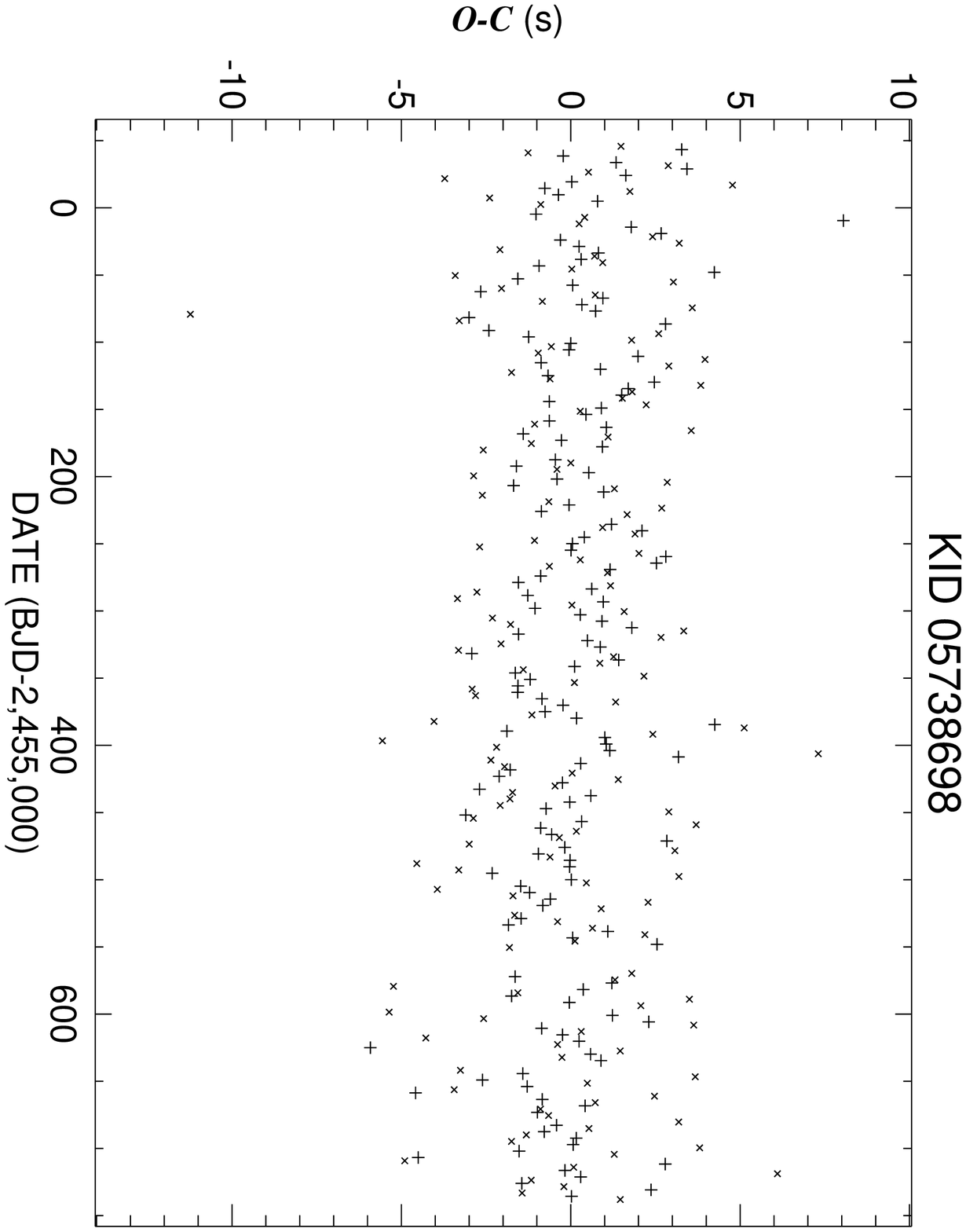}
\figsetgrpnote{The observed minus calculated eclipse times relative to
a linear ephemeris.  The primary and secondary eclipse
times are indicated by $+$ and $\times$ symbols, 
respectively. }
\figsetgrpend

\figsetgrpstart
\figsetgrpnum{2.17}
\figsetgrptitle{r17}
\figsetplot{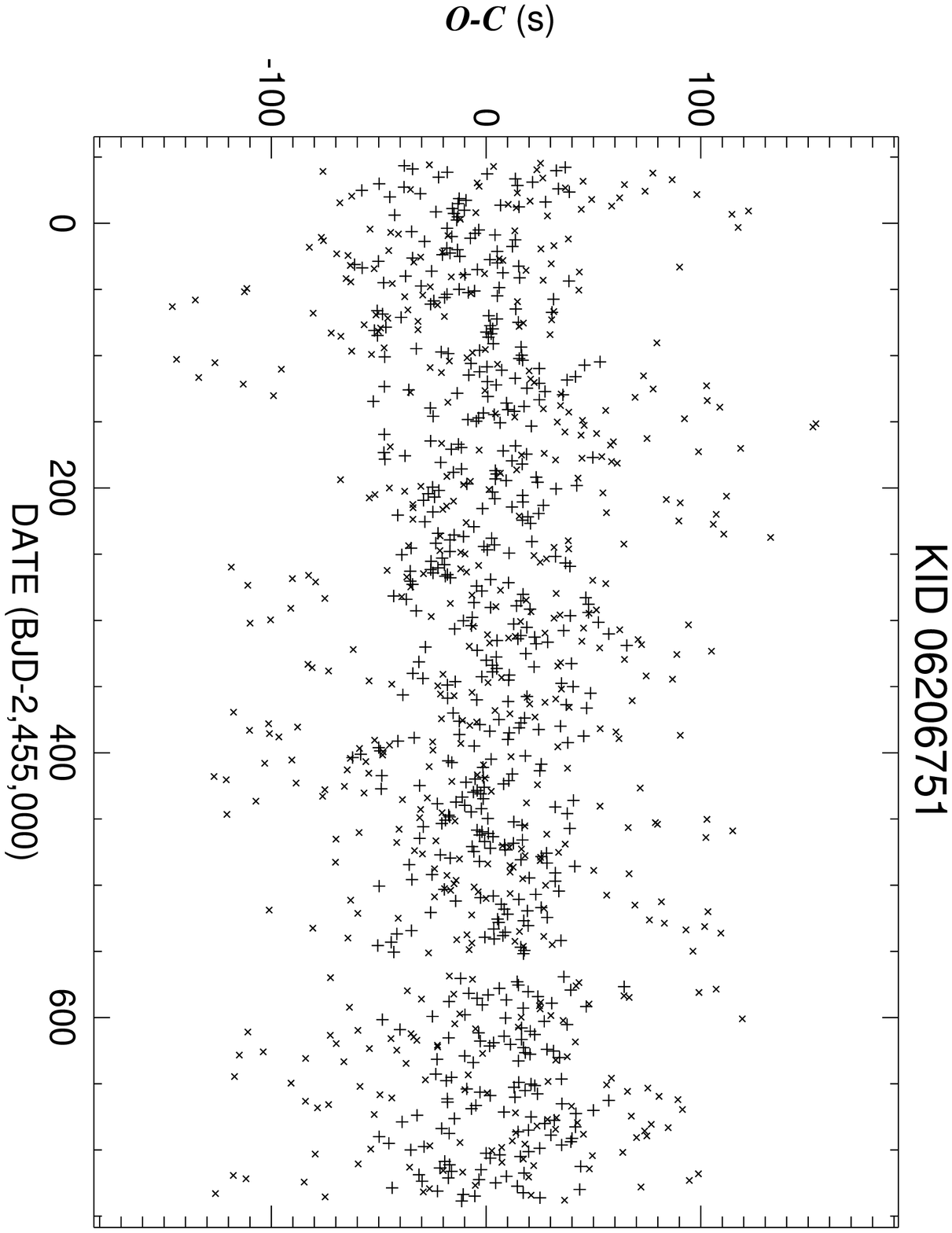}
\figsetgrpnote{The observed minus calculated eclipse times relative to
a linear ephemeris.  The primary and secondary eclipse
times are indicated by $+$ and $\times$ symbols, 
respectively. }
\figsetgrpend

\figsetgrpstart
\figsetgrpnum{2.18}
\figsetgrptitle{r18}
\figsetplot{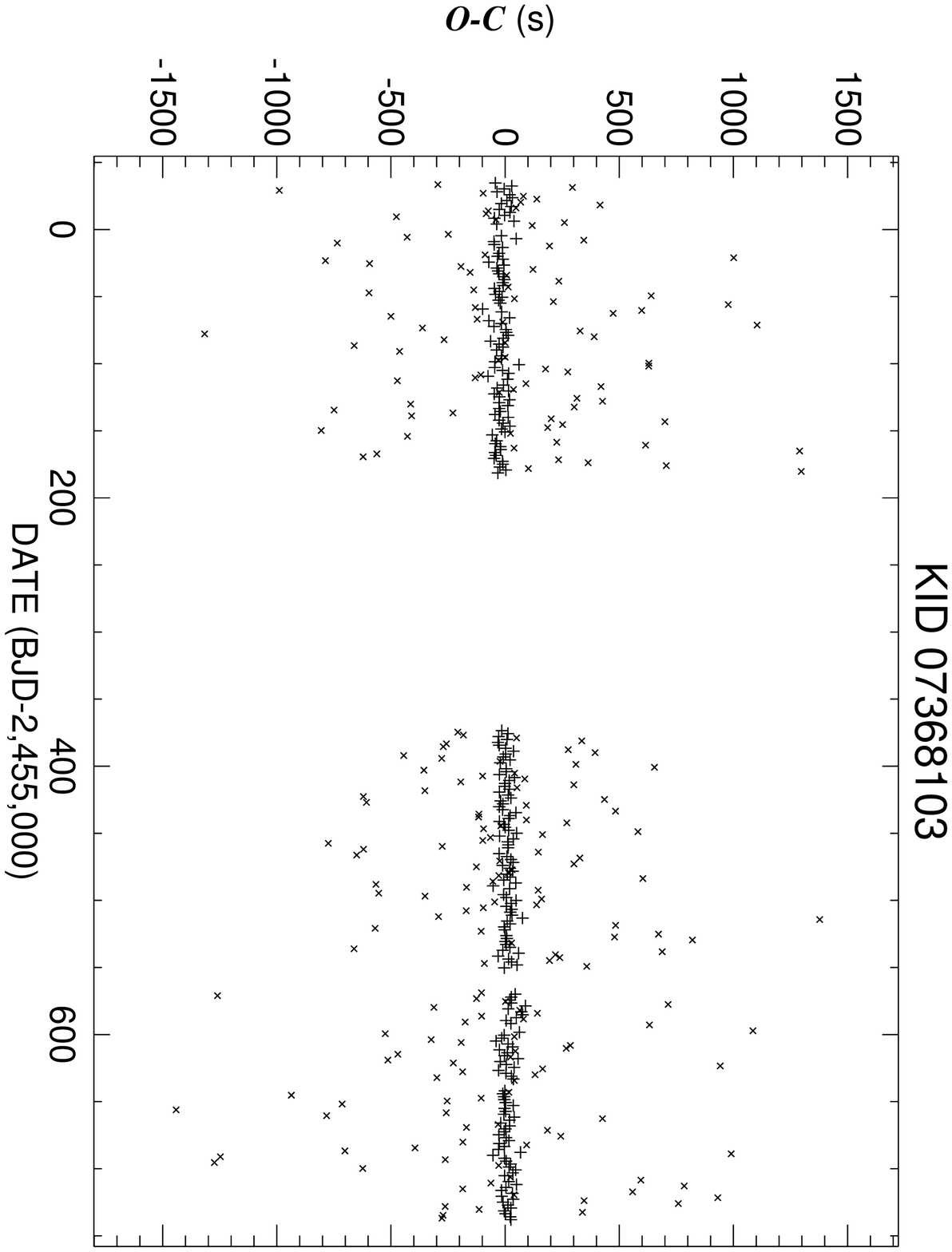}
\figsetgrpnote{The observed minus calculated eclipse times relative to
a linear ephemeris.  The primary and secondary eclipse
times are indicated by $+$ and $\times$ symbols, 
respectively. }
\figsetgrpend

\figsetgrpstart
\figsetgrpnum{2.19}
\figsetgrptitle{r19}
\figsetplot{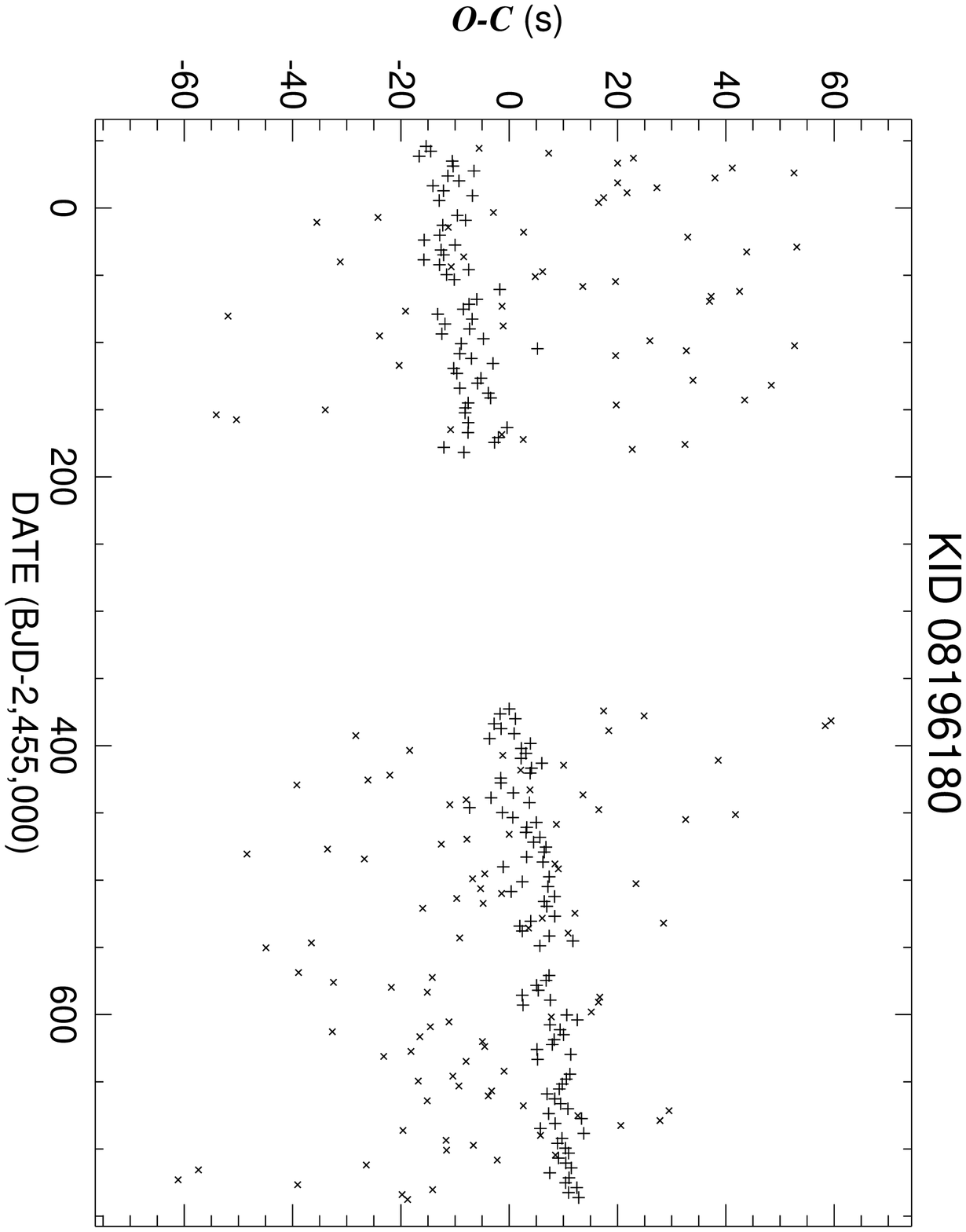}
\figsetgrpnote{The observed minus calculated eclipse times relative to
a linear ephemeris.  The primary and secondary eclipse
times are indicated by $+$ and $\times$ symbols, 
respectively. }
\figsetgrpend

\figsetgrpstart
\figsetgrpnum{2.20}
\figsetgrptitle{r20}
\figsetplot{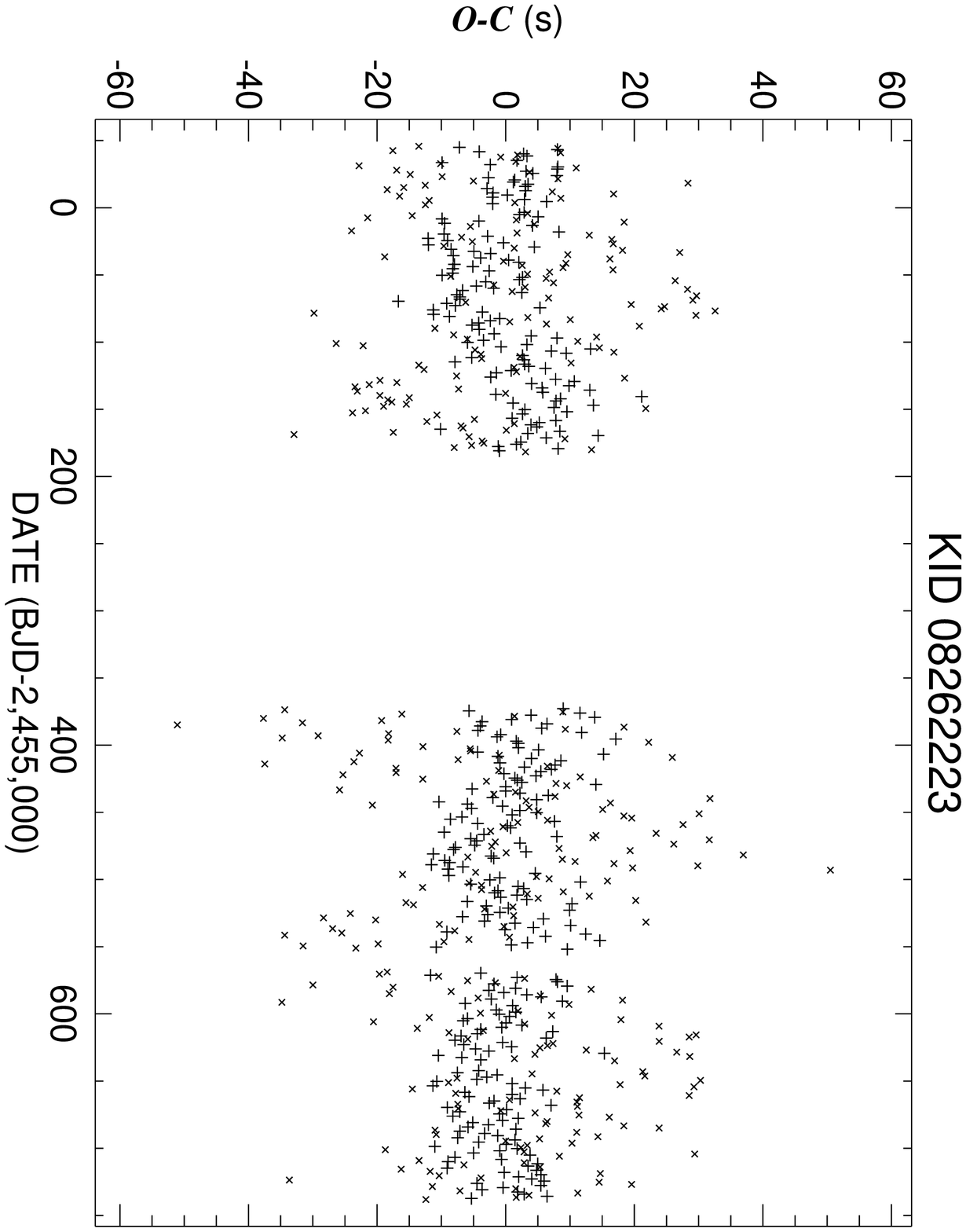}
\figsetgrpnote{The observed minus calculated eclipse times relative to
a linear ephemeris.  The primary and secondary eclipse
times are indicated by $+$ and $\times$ symbols, 
respectively. }
\figsetgrpend

\figsetgrpstart
\figsetgrpnum{2.21}
\figsetgrptitle{r21}
\figsetplot{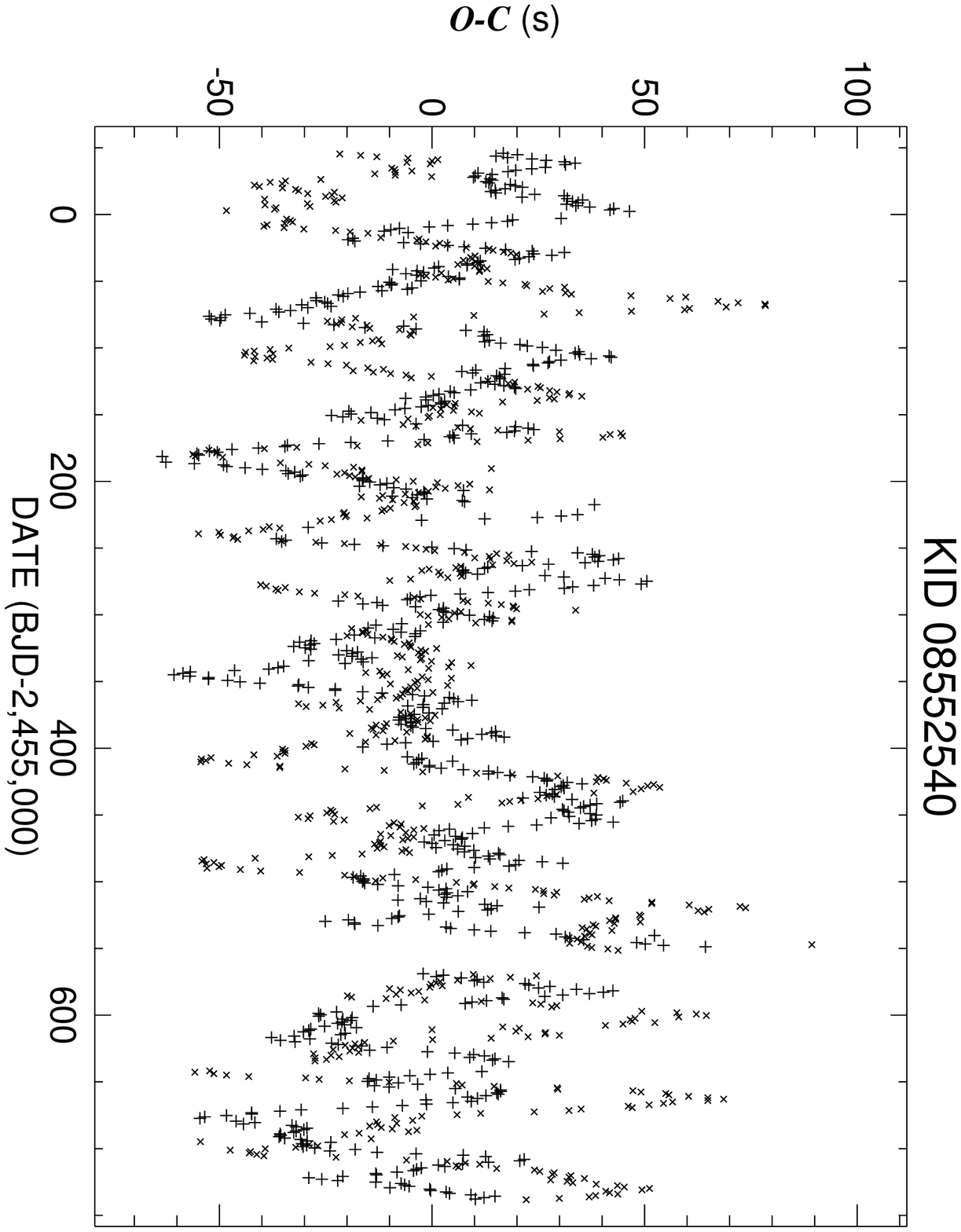}
\figsetgrpnote{The observed minus calculated eclipse times relative to
a linear ephemeris.  The primary and secondary eclipse
times are indicated by $+$ and $\times$ symbols, 
respectively. }
\figsetgrpend

\figsetgrpstart
\figsetgrpnum{2.22}
\figsetgrptitle{r22}
\figsetplot{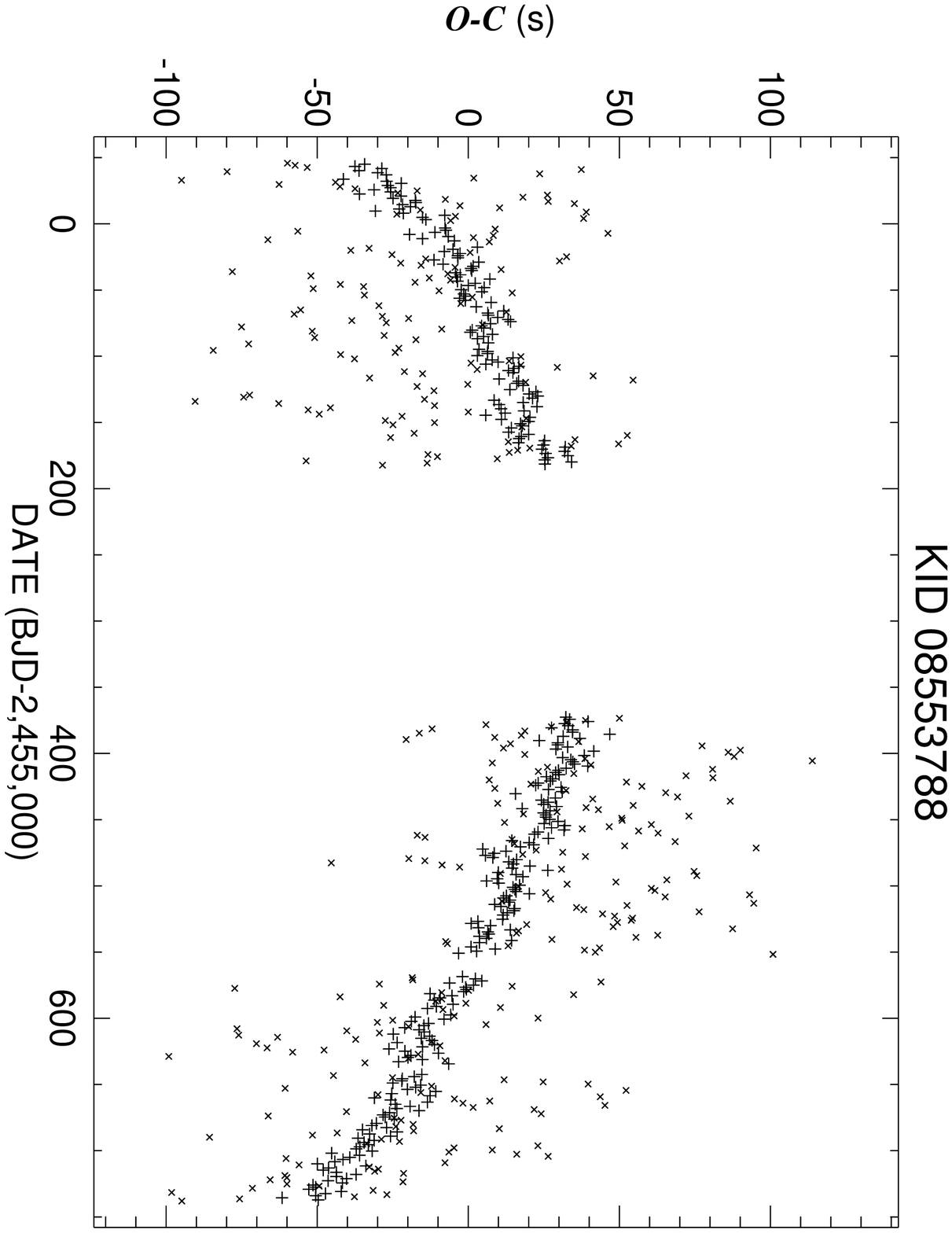}
\figsetgrpnote{The observed minus calculated eclipse times relative to
a linear ephemeris.  The primary and secondary eclipse
times are indicated by $+$ and $\times$ symbols, 
respectively. }
\figsetgrpend

\figsetgrpstart
\figsetgrpnum{2.23}
\figsetgrptitle{r23}
\figsetplot{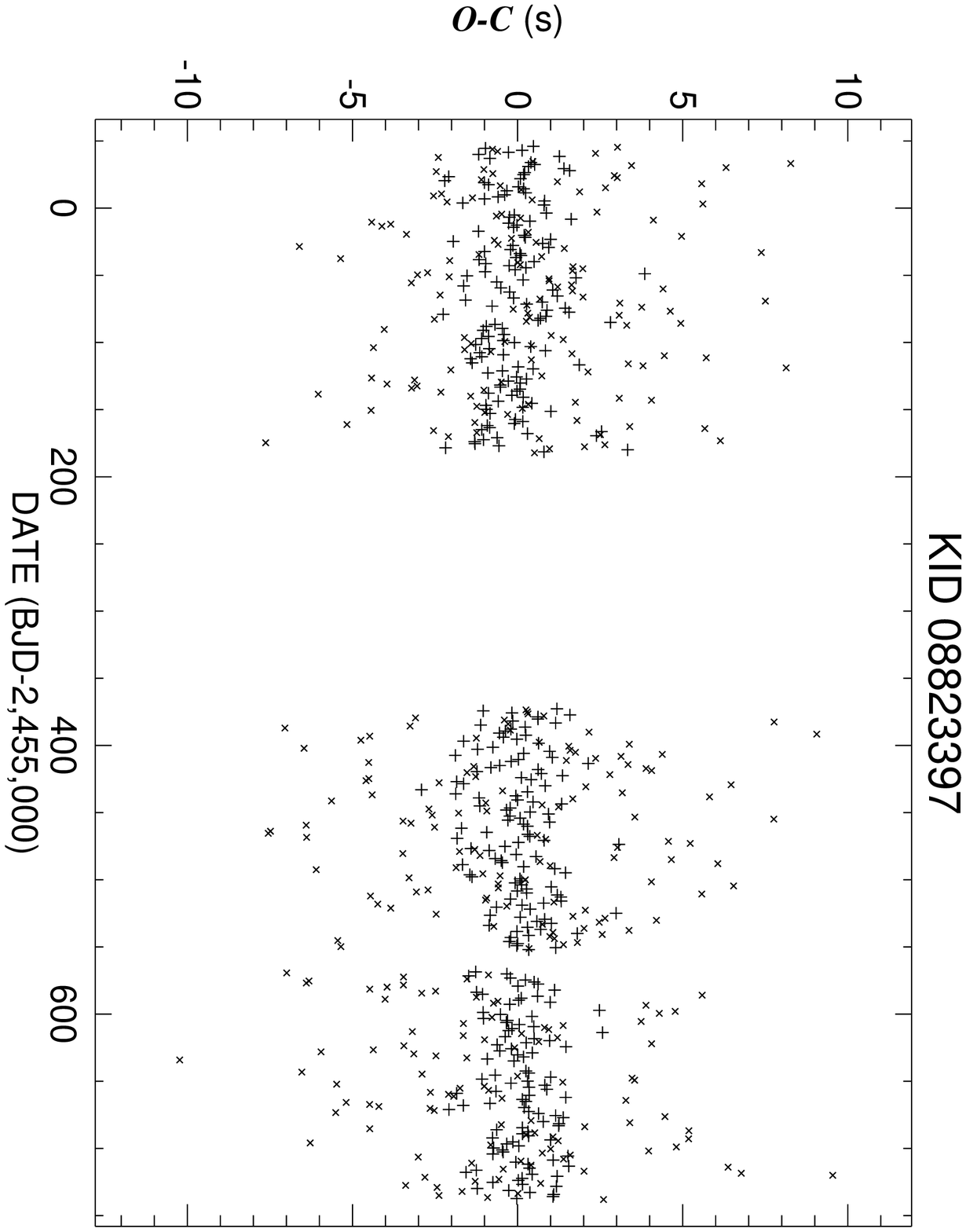}
\figsetgrpnote{The observed minus calculated eclipse times relative to
a linear ephemeris.  The primary and secondary eclipse
times are indicated by $+$ and $\times$ symbols, 
respectively. }
\figsetgrpend

\figsetgrpstart
\figsetgrpnum{2.24}
\figsetgrptitle{r24}
\figsetplot{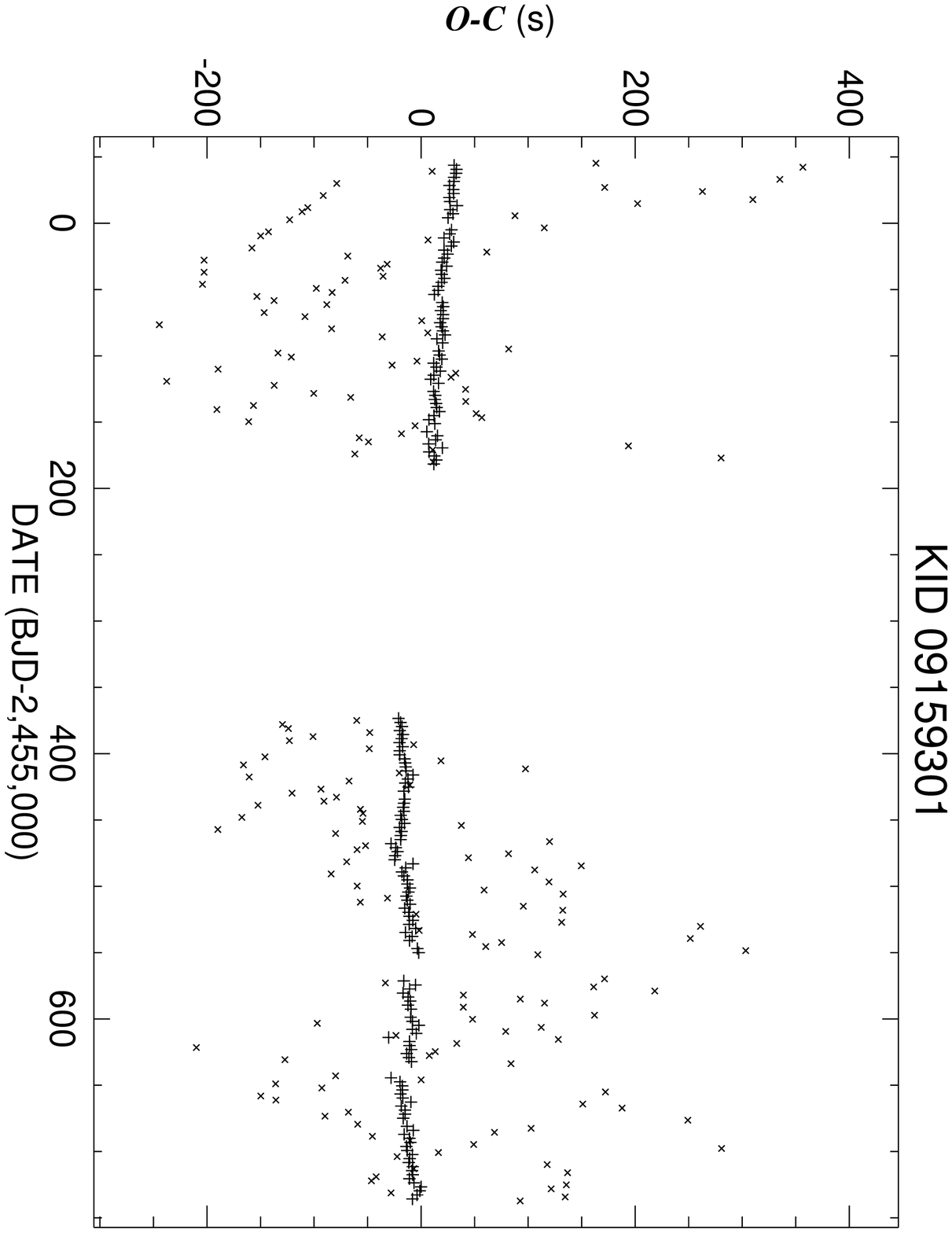}
\figsetgrpnote{The observed minus calculated eclipse times relative to
a linear ephemeris.  The primary and secondary eclipse
times are indicated by $+$ and $\times$ symbols, 
respectively. }
\figsetgrpend

\figsetgrpstart
\figsetgrpnum{2.25}
\figsetgrptitle{r25}
\figsetplot{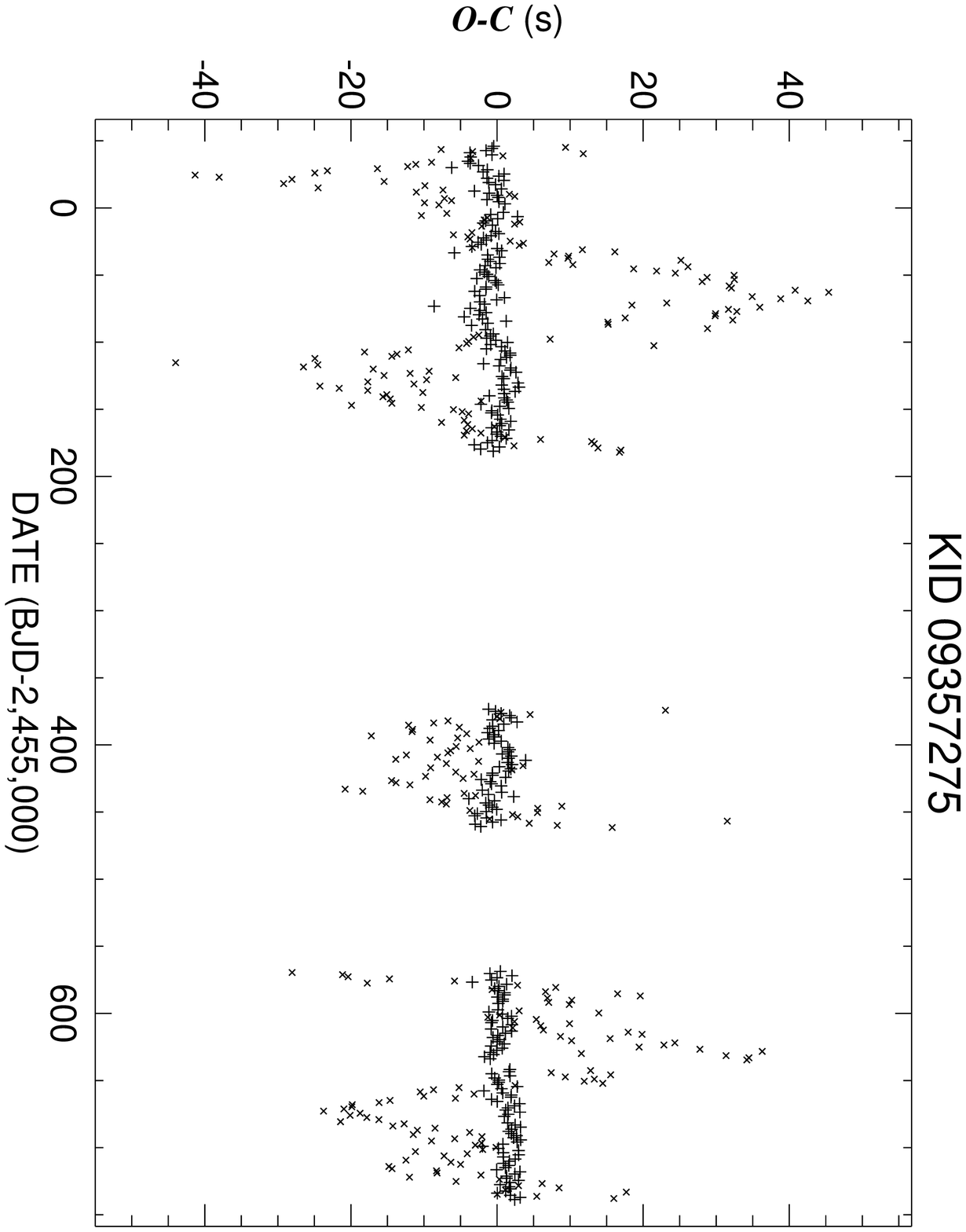}
\figsetgrpnote{The observed minus calculated eclipse times relative to
a linear ephemeris.  The primary and secondary eclipse
times are indicated by $+$ and $\times$ symbols, 
respectively. }
\figsetgrpend

\figsetgrpstart
\figsetgrpnum{2.26}
\figsetgrptitle{r26}
\figsetplot{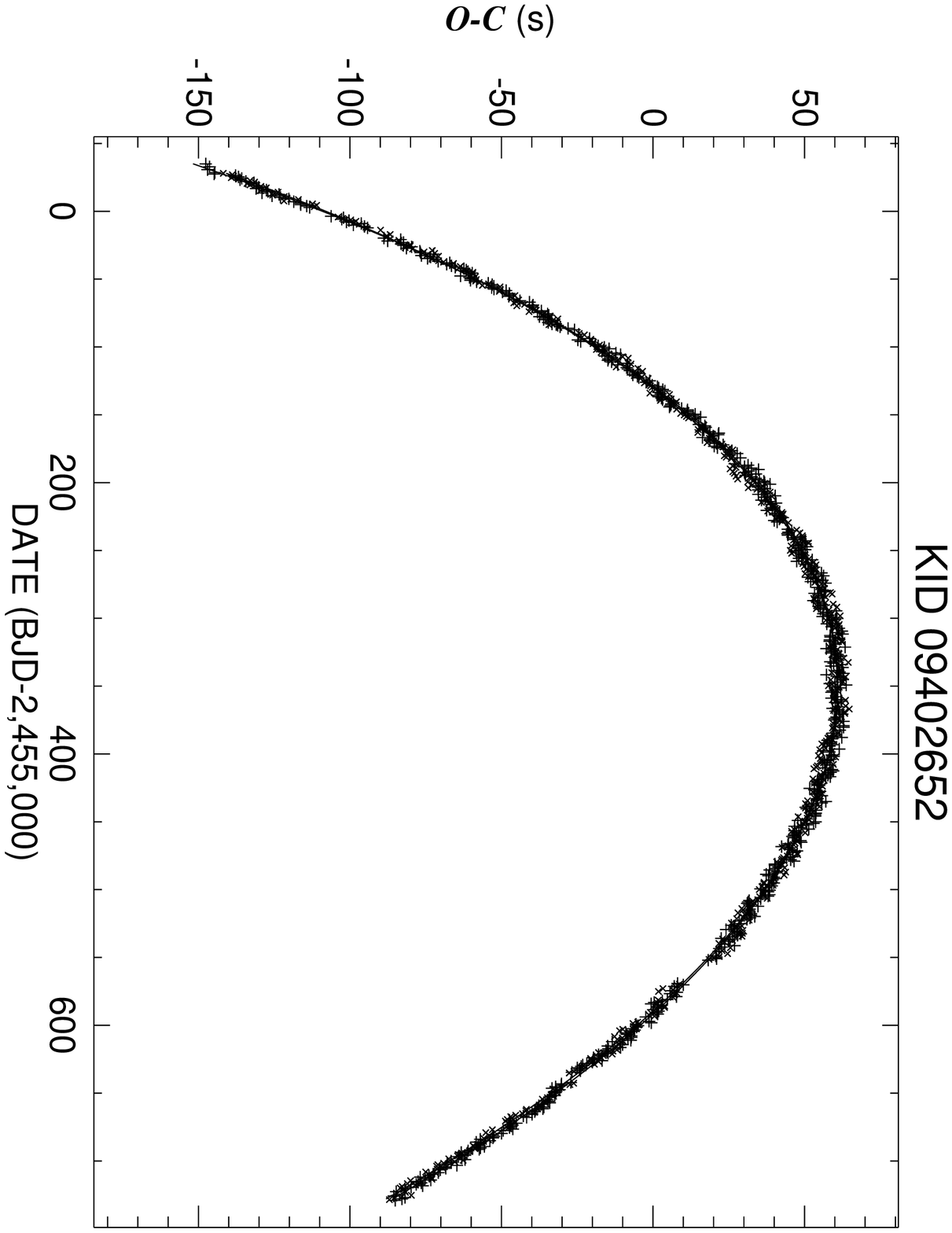}
\figsetgrpnote{The observed minus calculated eclipse times relative to
a linear ephemeris.  The primary and secondary eclipse
times are indicated by $+$ and $\times$ symbols, 
respectively. }
\figsetgrpend

\figsetgrpstart
\figsetgrpnum{2.27}
\figsetgrptitle{r27}
\figsetplot{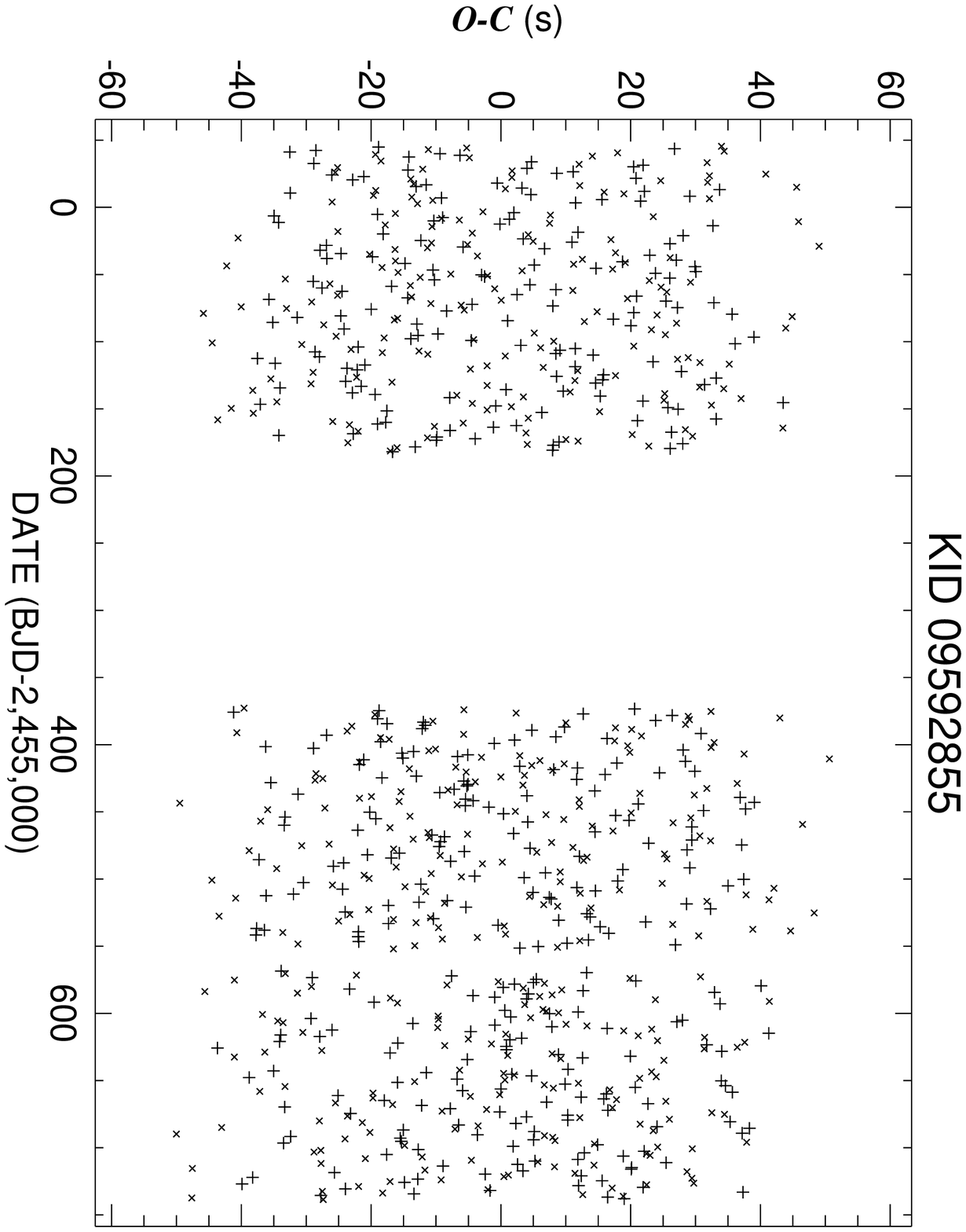}
\figsetgrpnote{The observed minus calculated eclipse times relative to
a linear ephemeris.  The primary and secondary eclipse
times are indicated by $+$ and $\times$ symbols, 
respectively. }
\figsetgrpend

\figsetgrpstart
\figsetgrpnum{2.28}
\figsetgrptitle{r28}
\figsetplot{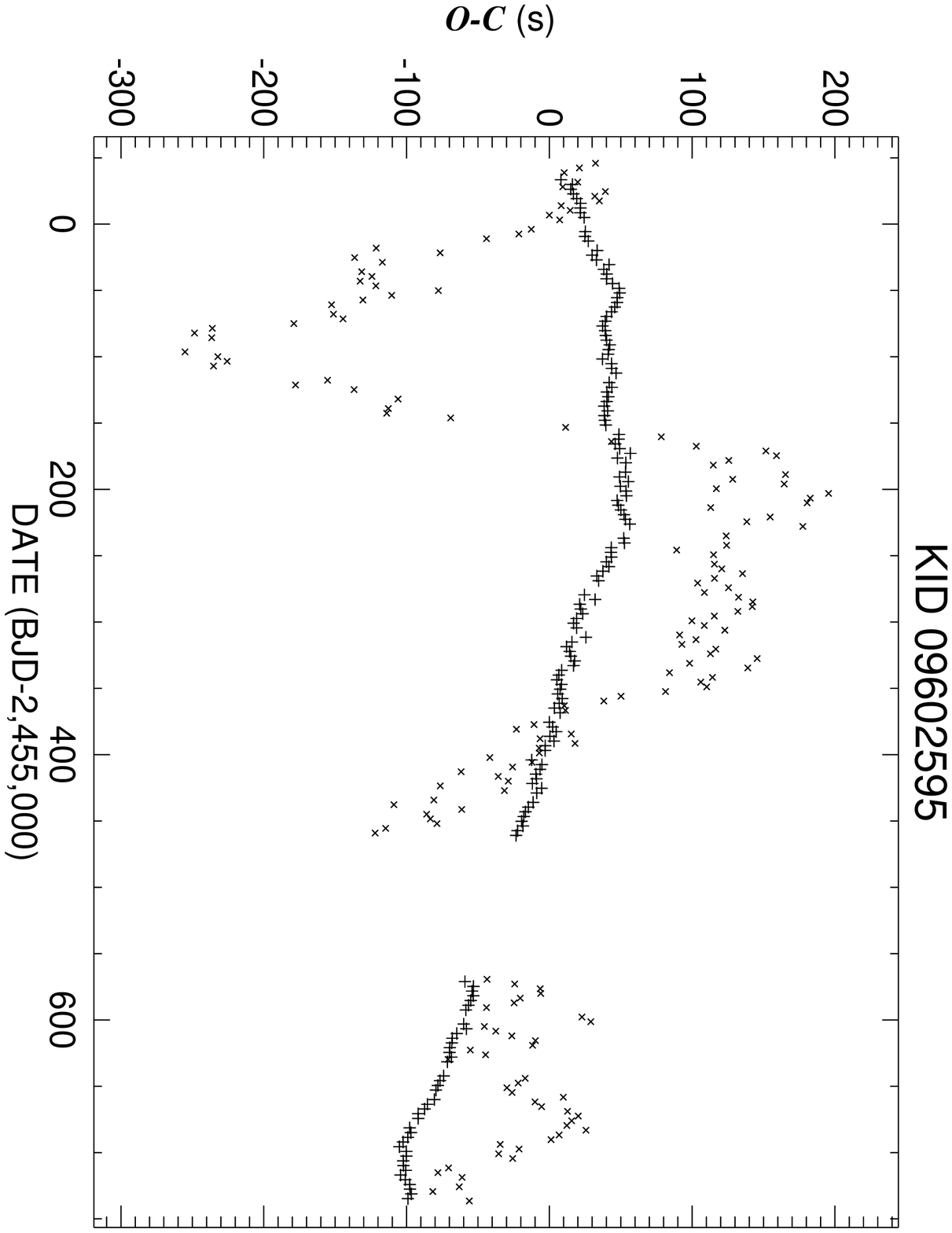}
\figsetgrpnote{The observed minus calculated eclipse times relative to
a linear ephemeris.  The primary and secondary eclipse
times are indicated by $+$ and $\times$ symbols, 
respectively. }
\figsetgrpend

\figsetgrpstart
\figsetgrpnum{2.29}
\figsetgrptitle{r29}
\figsetplot{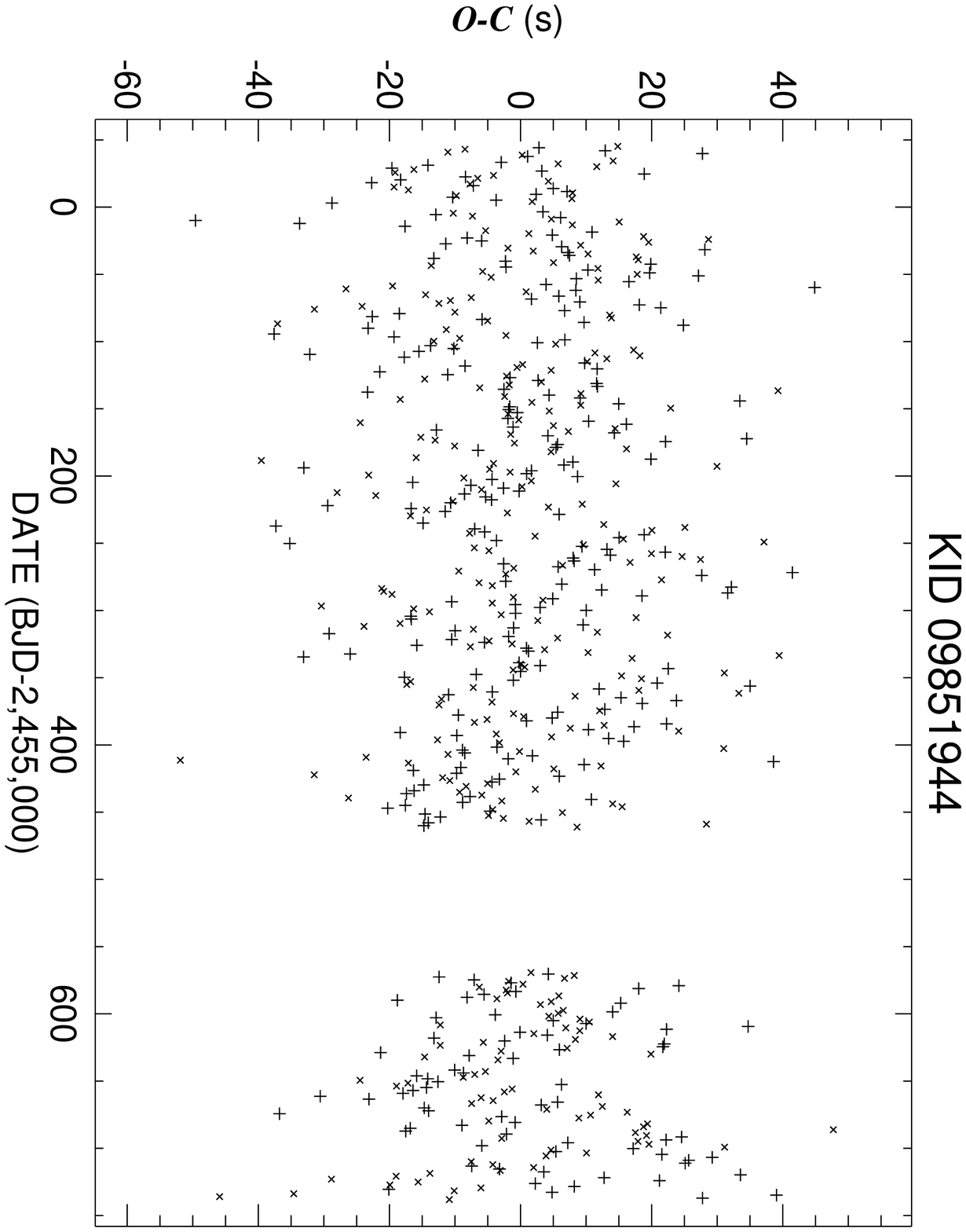}
\figsetgrpnote{The observed minus calculated eclipse times relative to
a linear ephemeris.  The primary and secondary eclipse
times are indicated by $+$ and $\times$ symbols, 
respectively. }
\figsetgrpend

\figsetgrpstart
\figsetgrpnum{2.30}
\figsetgrptitle{r30}
\figsetplot{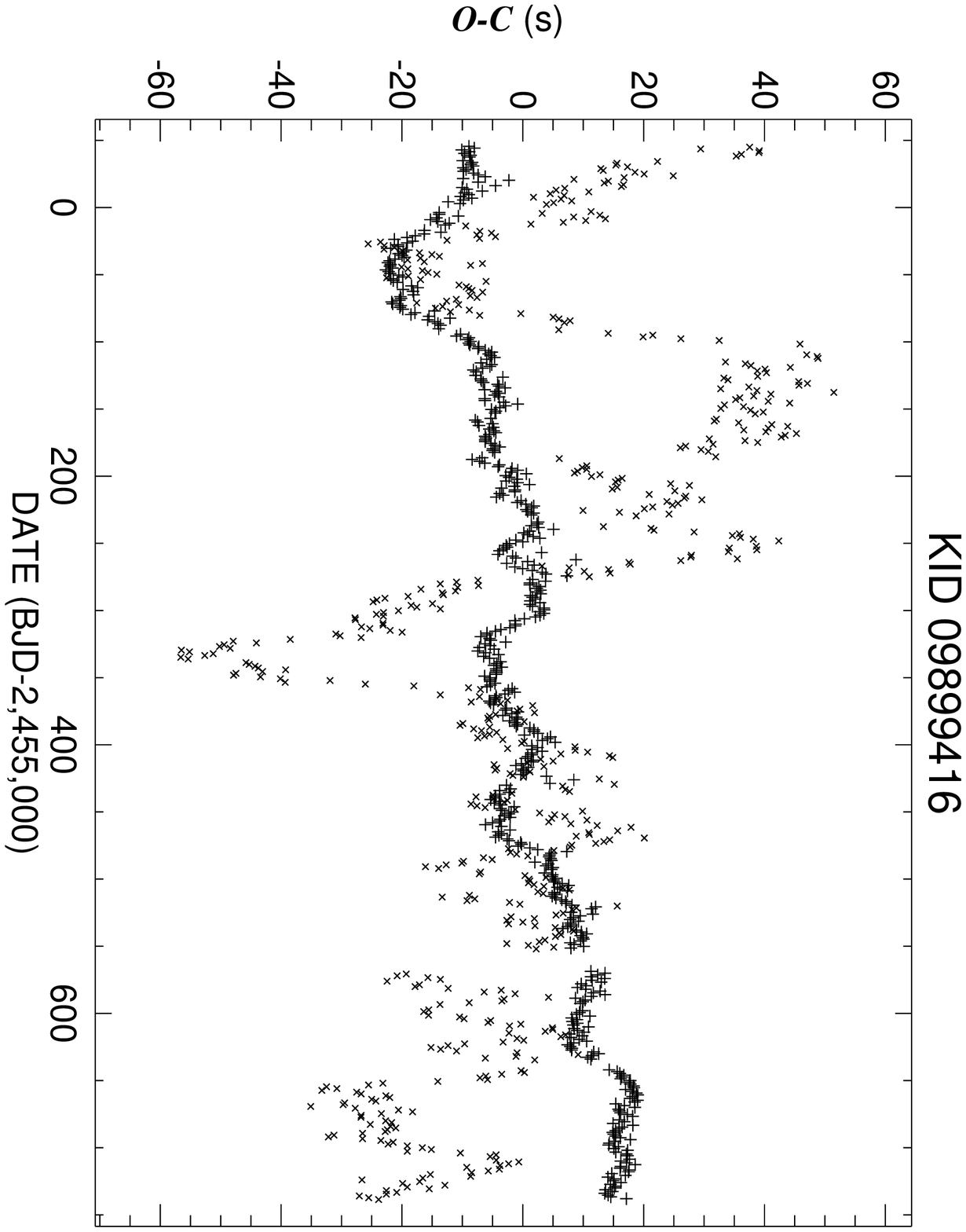}
\figsetgrpnote{The observed minus calculated eclipse times relative to
a linear ephemeris.  The primary and secondary eclipse
times are indicated by $+$ and $\times$ symbols, 
respectively. }
\figsetgrpend

\figsetgrpstart
\figsetgrpnum{2.31}
\figsetgrptitle{r31}
\figsetplot{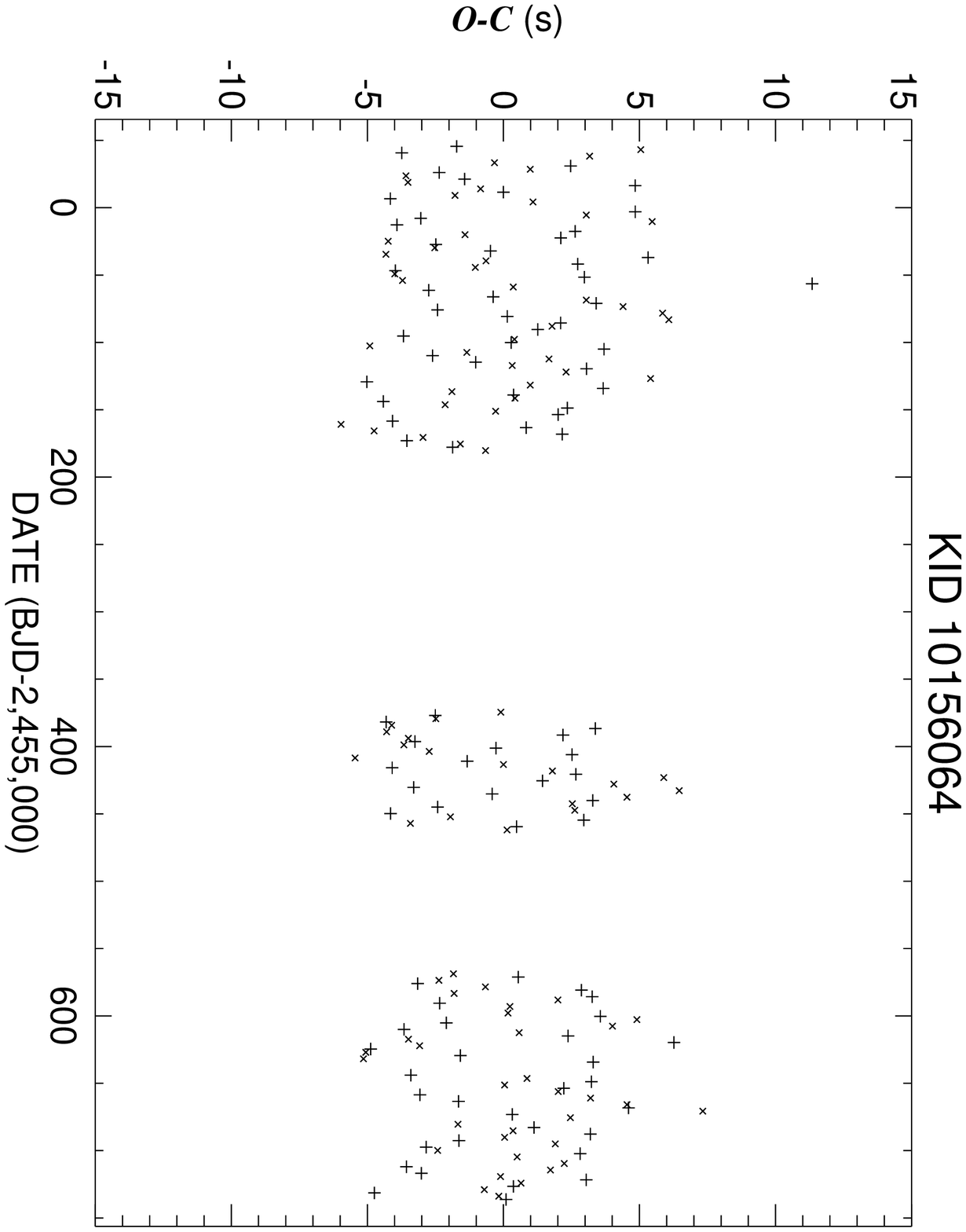}
\figsetgrpnote{The observed minus calculated eclipse times relative to
a linear ephemeris.  The primary and secondary eclipse
times are indicated by $+$ and $\times$ symbols, 
respectively. }
\figsetgrpend

\figsetgrpstart
\figsetgrpnum{2.32}
\figsetgrptitle{r32}
\figsetplot{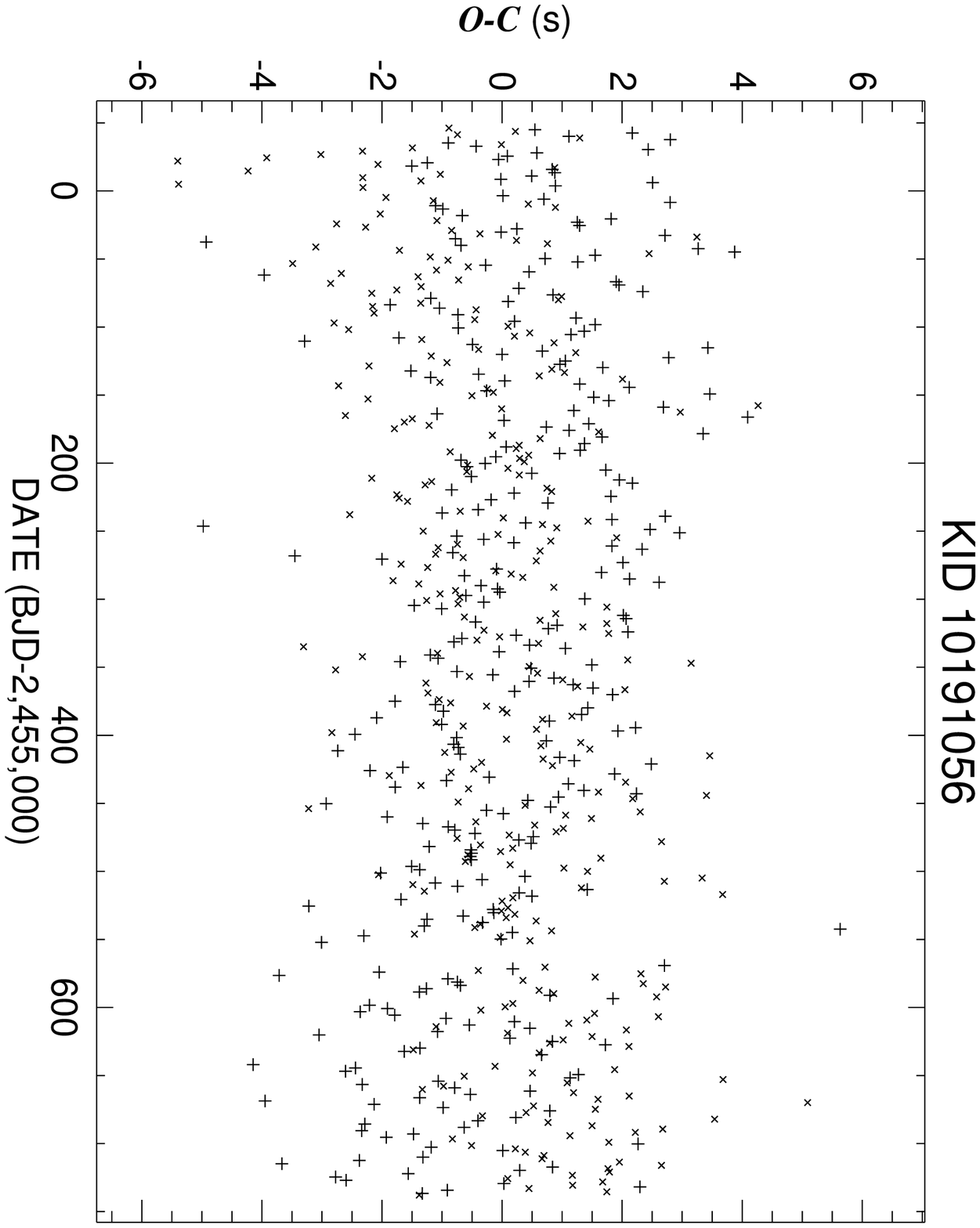}
\figsetgrpnote{The observed minus calculated eclipse times relative to
a linear ephemeris.  The primary and secondary eclipse
times are indicated by $+$ and $\times$ symbols, 
respectively. }
\figsetgrpend

\figsetgrpstart
\figsetgrpnum{2.33}
\figsetgrptitle{r33}
\figsetplot{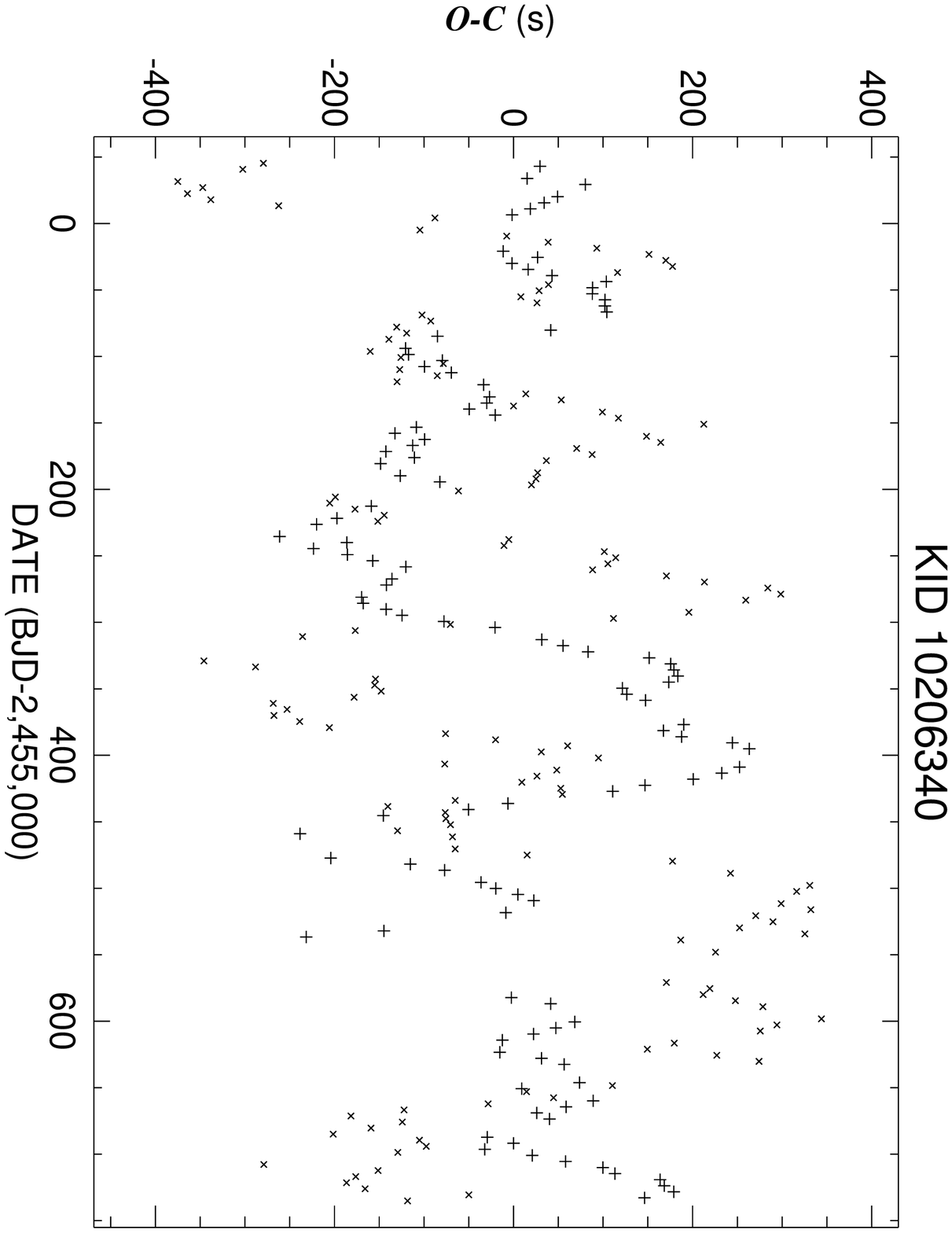}
\figsetgrpnote{The observed minus calculated eclipse times relative to
a linear ephemeris.  The primary and secondary eclipse
times are indicated by $+$ and $\times$ symbols, 
respectively. }
\figsetgrpend

\figsetgrpstart
\figsetgrpnum{2.34}
\figsetgrptitle{r34}
\figsetplot{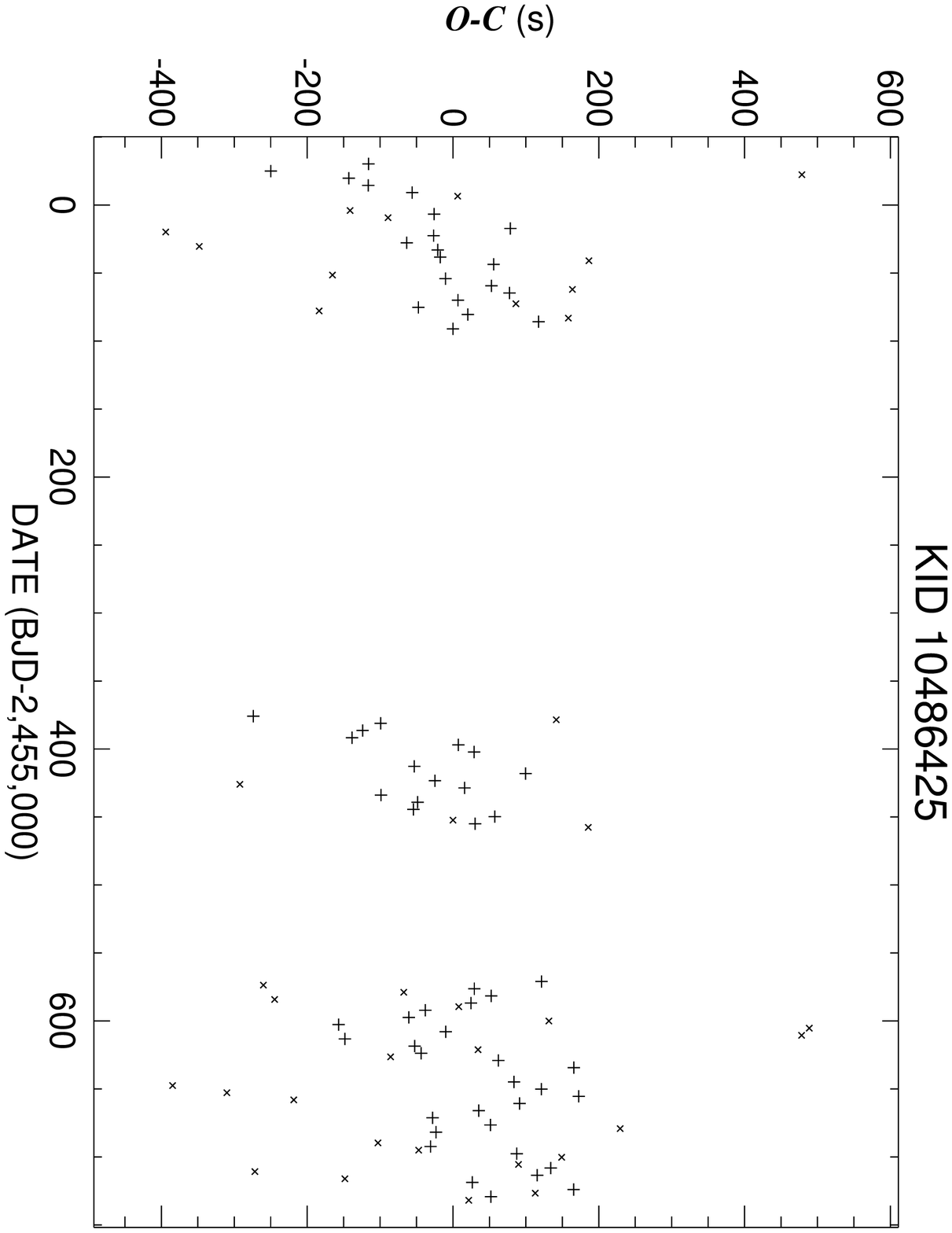}
\figsetgrpnote{The observed minus calculated eclipse times relative to
a linear ephemeris.  The primary and secondary eclipse
times are indicated by $+$ and $\times$ symbols, 
respectively. }
\figsetgrpend

\figsetgrpstart
\figsetgrpnum{2.35}
\figsetgrptitle{r35}
\figsetplot{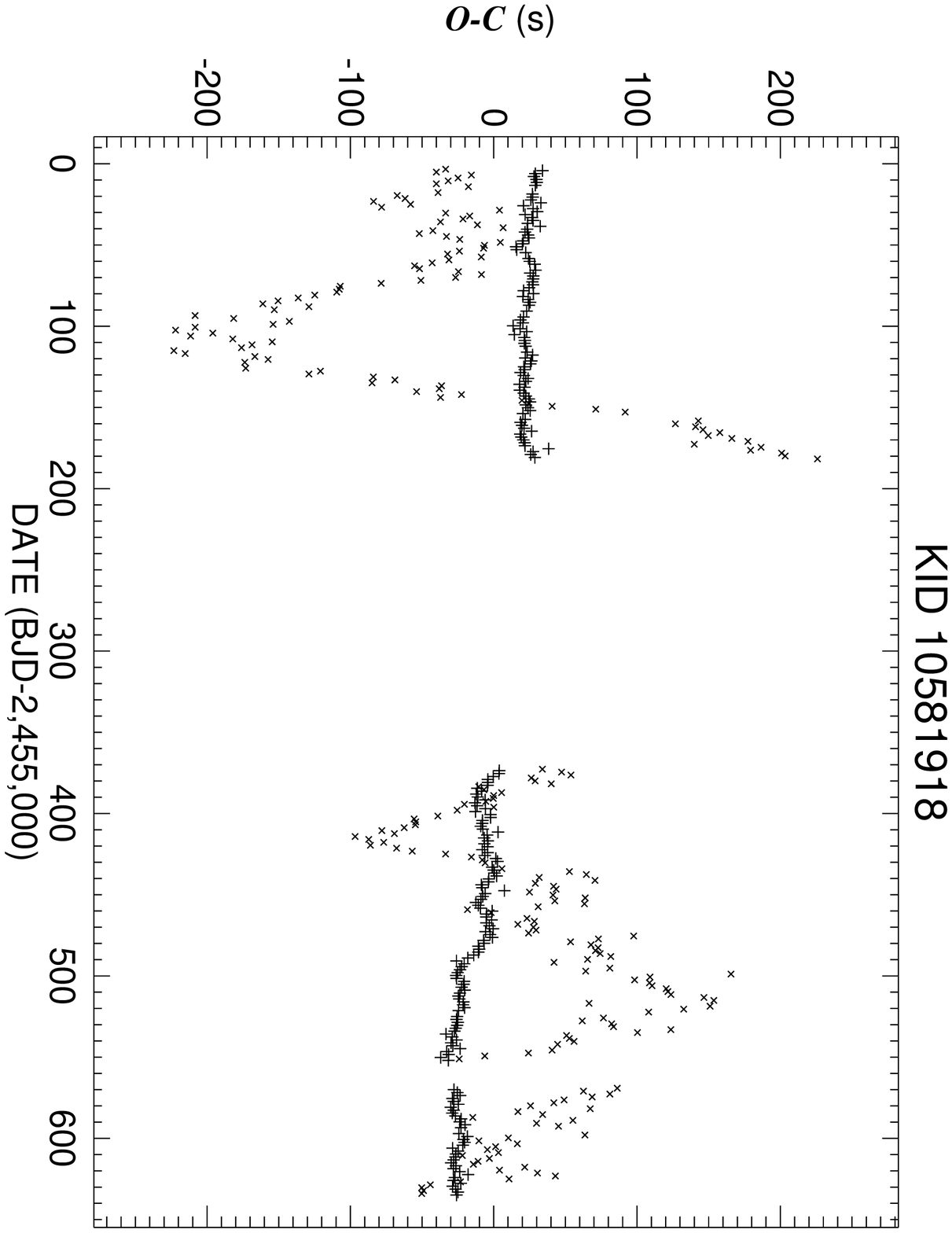}
\figsetgrpnote{The observed minus calculated eclipse times relative to
a linear ephemeris.  The primary and secondary eclipse
times are indicated by $+$ and $\times$ symbols, 
respectively. }
\figsetgrpend

\figsetgrpstart
\figsetgrpnum{2.36}
\figsetgrptitle{r36}
\figsetplot{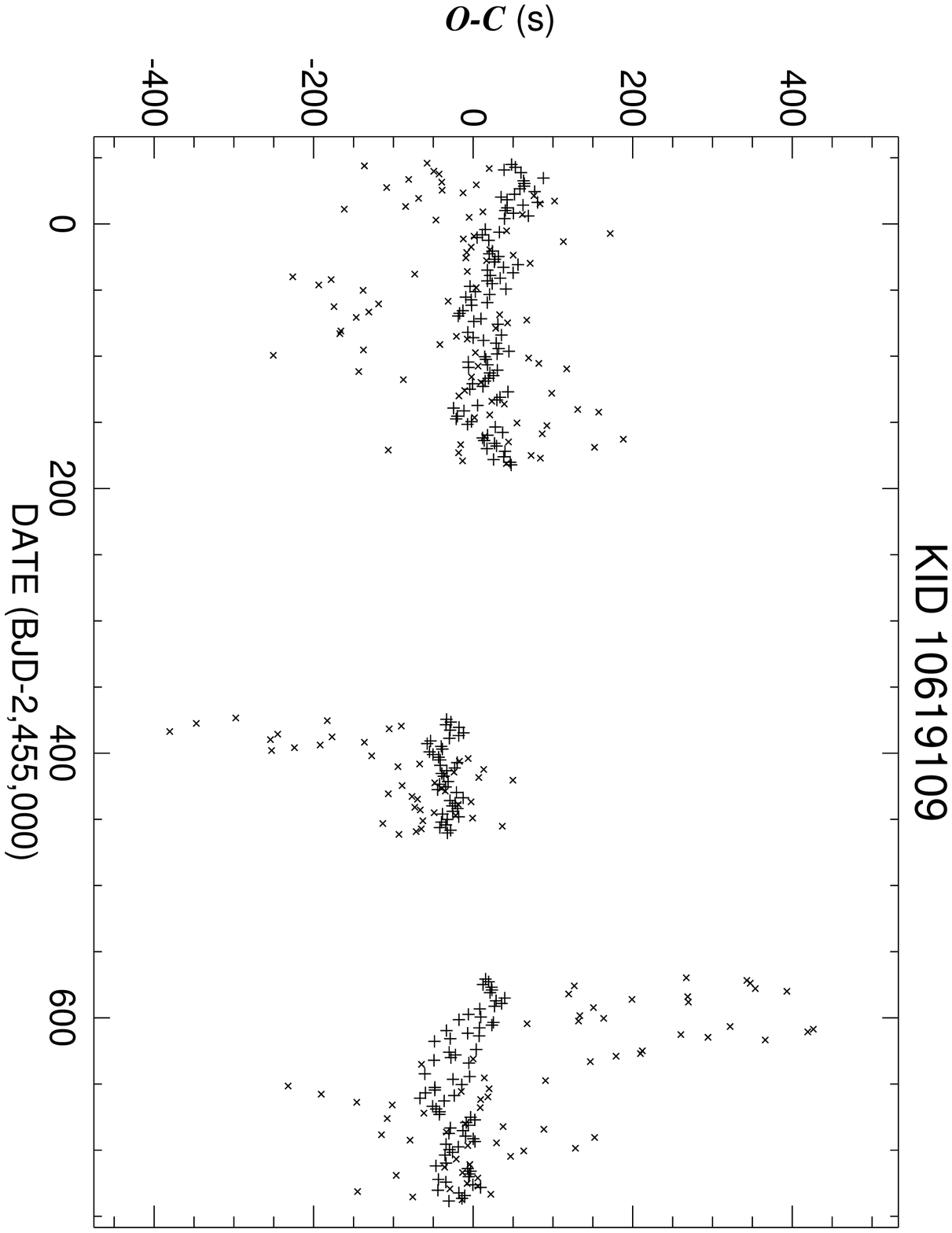}
\figsetgrpnote{The observed minus calculated eclipse times relative to
a linear ephemeris.  The primary and secondary eclipse
times are indicated by $+$ and $\times$ symbols, 
respectively. }
\figsetgrpend

\figsetgrpstart
\figsetgrpnum{2.37}
\figsetgrptitle{r37}
\figsetplot{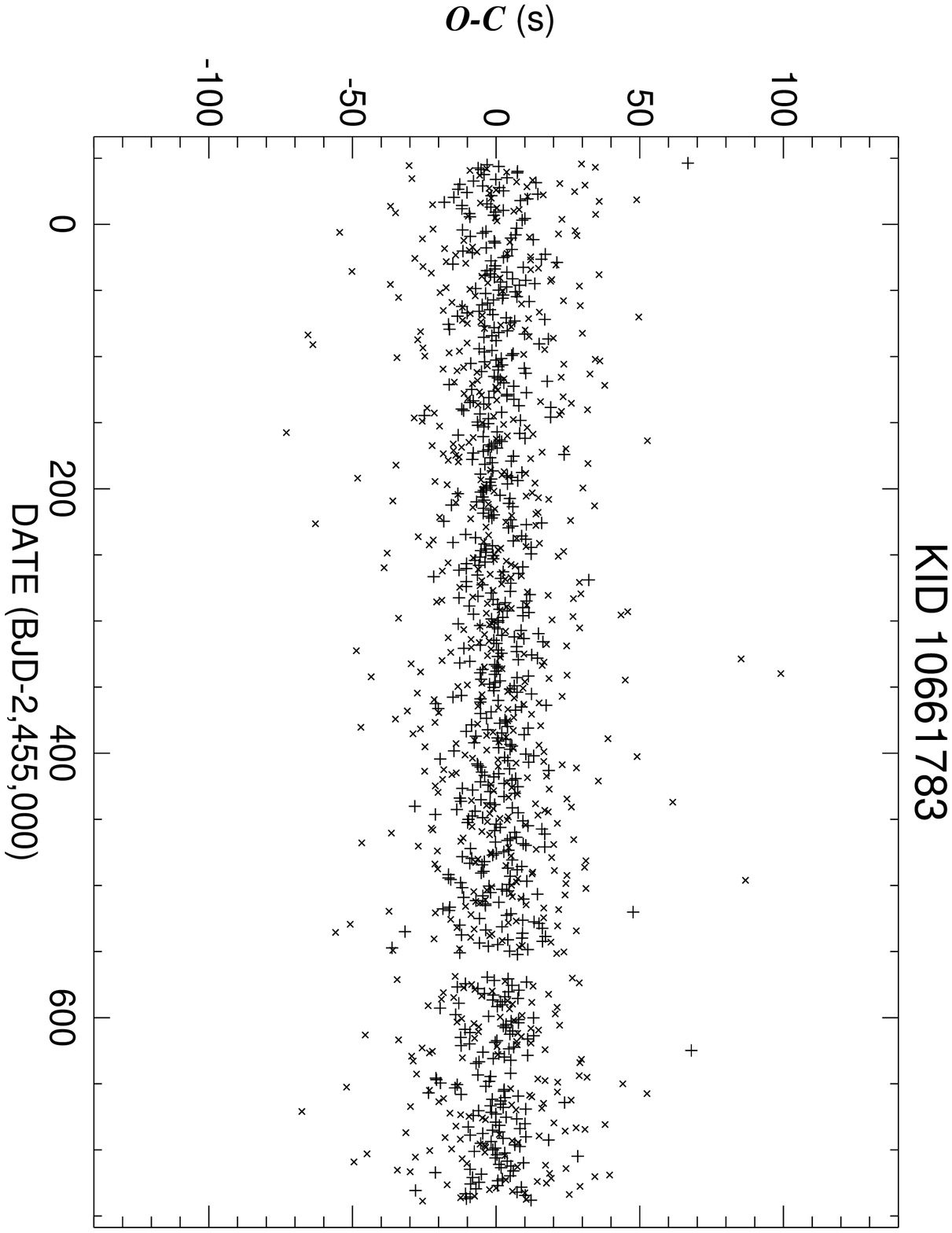}
\figsetgrpnote{The observed minus calculated eclipse times relative to
a linear ephemeris.  The primary and secondary eclipse
times are indicated by $+$ and $\times$ symbols, 
respectively. }
\figsetgrpend

\figsetgrpstart
\figsetgrpnum{2.38}
\figsetgrptitle{r38}
\figsetplot{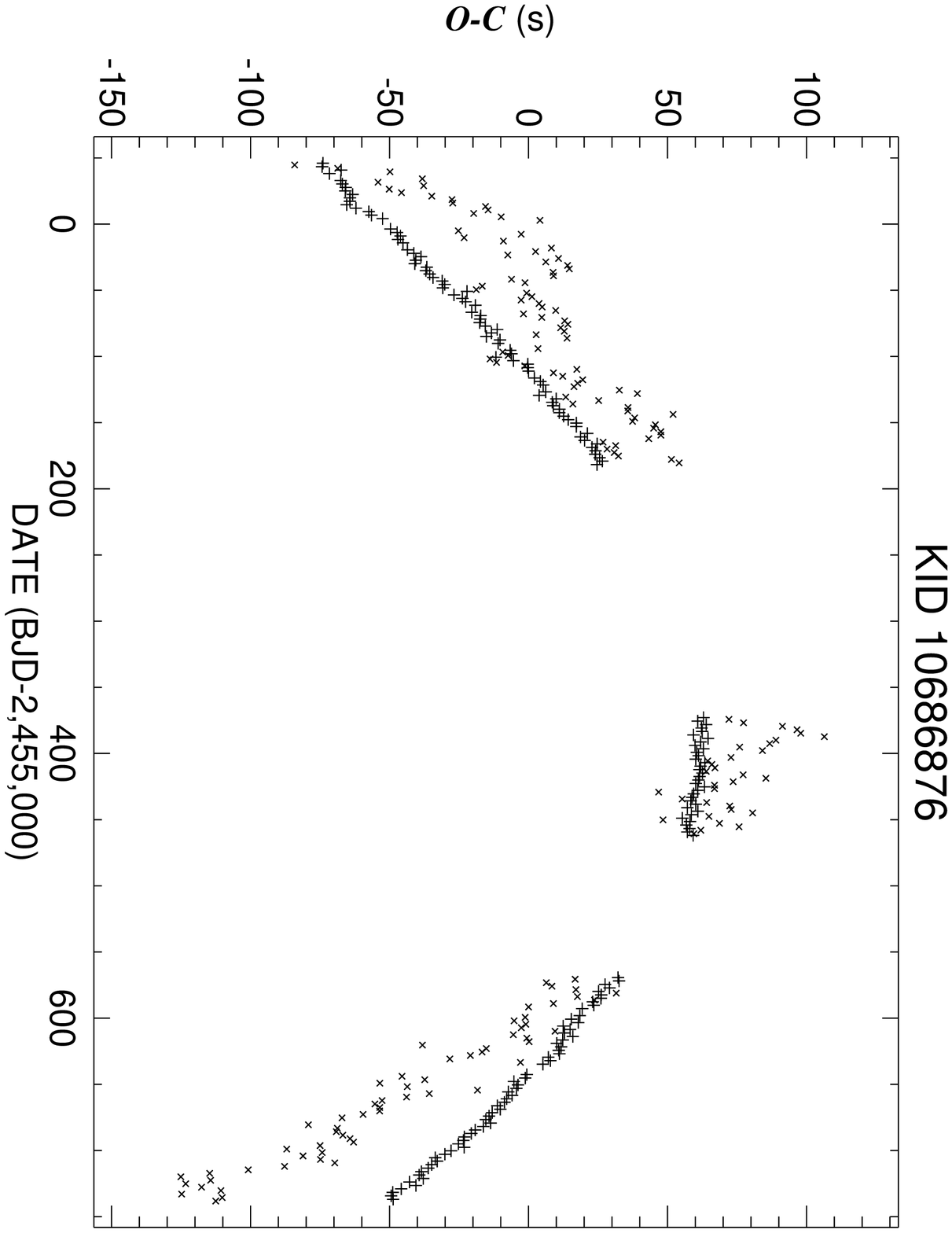}
\figsetgrpnote{The observed minus calculated eclipse times relative to
a linear ephemeris.  The primary and secondary eclipse
times are indicated by $+$ and $\times$ symbols, 
respectively. }
\figsetgrpend

\figsetgrpstart
\figsetgrpnum{2.39}
\figsetgrptitle{r39}
\figsetplot{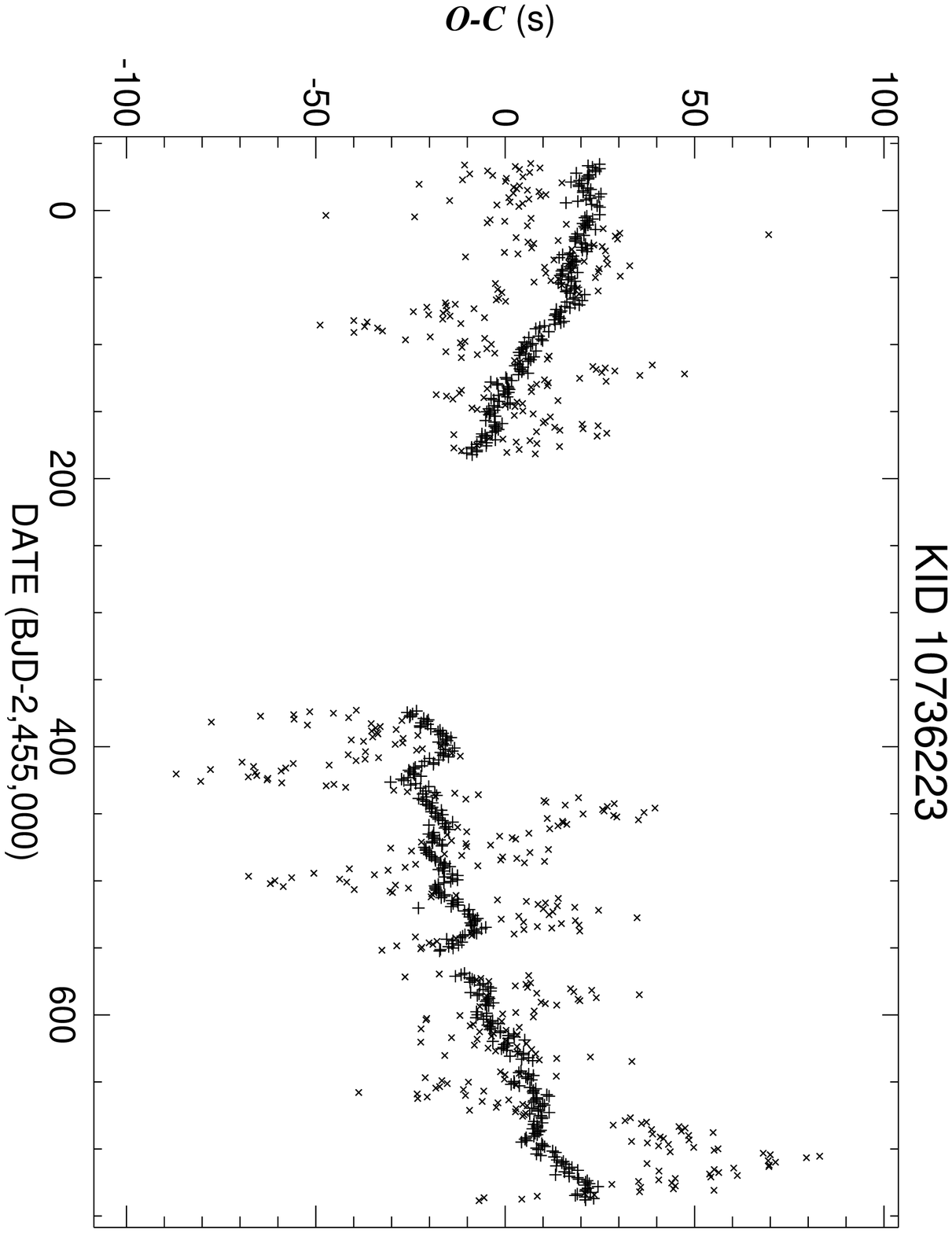}
\figsetgrpnote{The observed minus calculated eclipse times relative to
a linear ephemeris.  The primary and secondary eclipse
times are indicated by $+$ and $\times$ symbols, 
respectively. }
\figsetgrpend

\figsetgrpstart
\figsetgrpnum{2.40}
\figsetgrptitle{r40}
\figsetplot{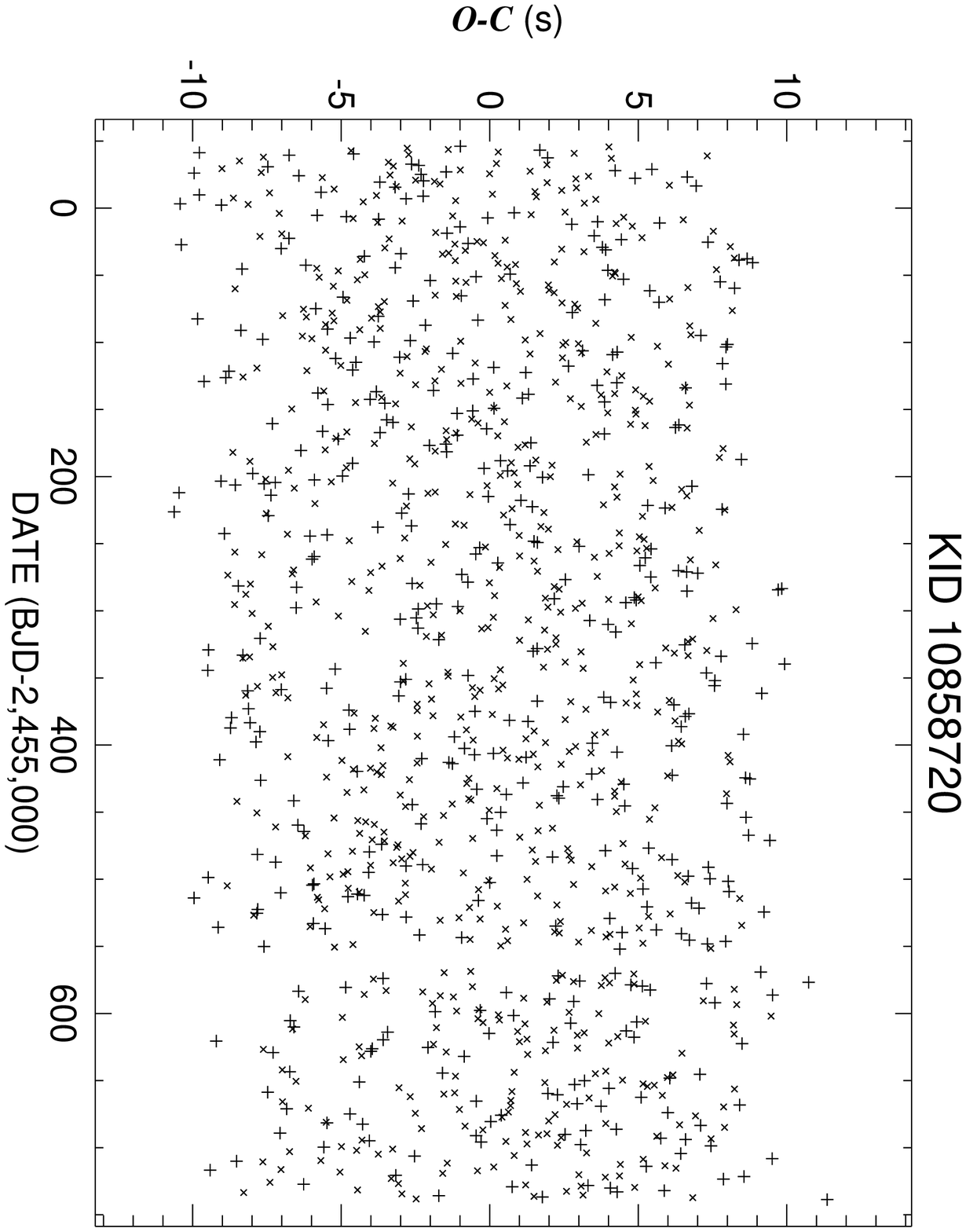}
\figsetgrpnote{The observed minus calculated eclipse times relative to
a linear ephemeris.  The primary and secondary eclipse
times are indicated by $+$ and $\times$ symbols, 
respectively. }
\figsetgrpend

\figsetgrpstart
\figsetgrpnum{2.41}
\figsetgrptitle{r41}
\figsetplot{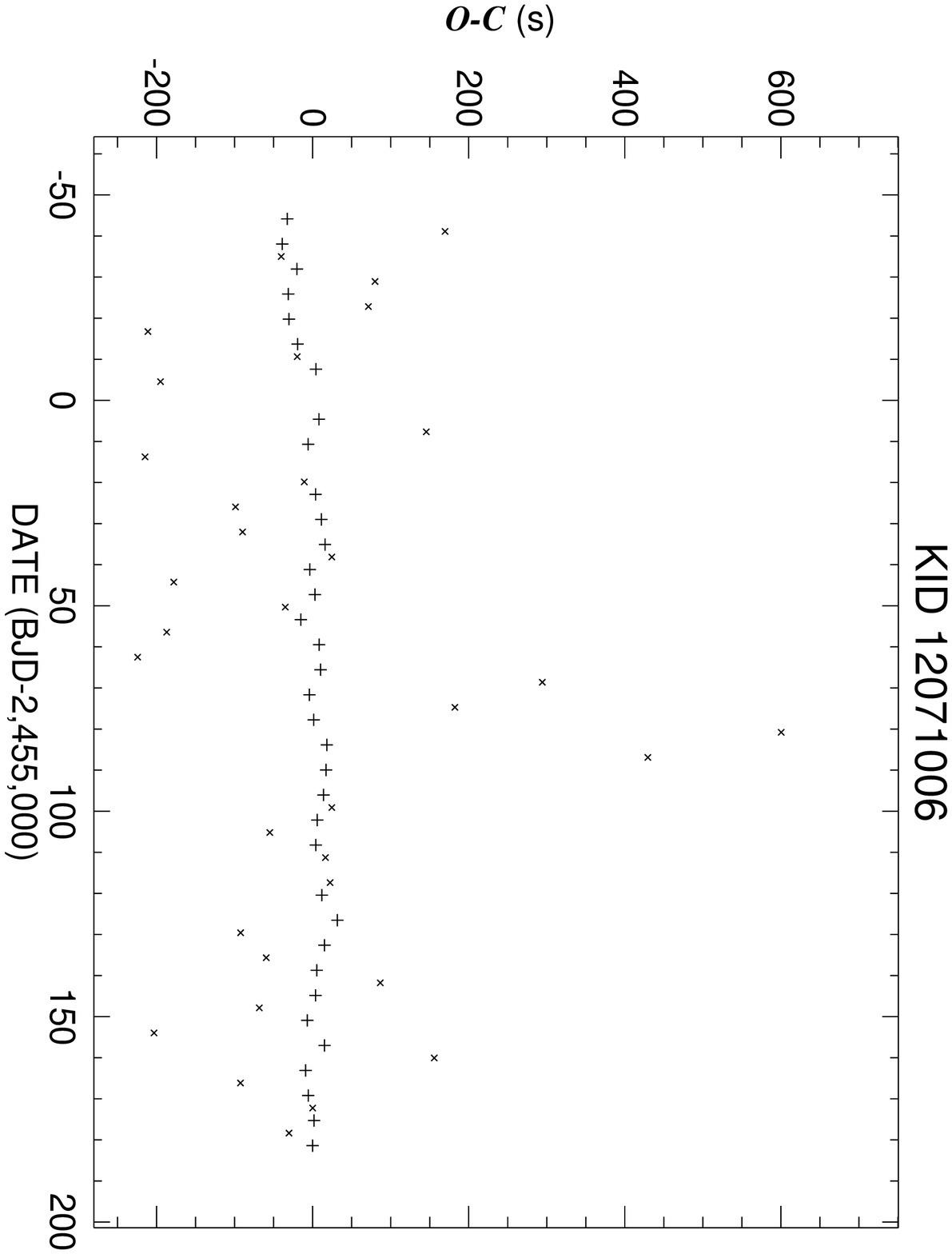}
\figsetgrpnote{The observed minus calculated eclipse times relative to
a linear ephemeris.  The primary and secondary eclipse
times are indicated by $+$ and $\times$ symbols, 
respectively. }
\figsetgrpend

\figsetend



\clearpage
\setcounter{figure}{0}
\renewcommand{\thefigure}{\arabic{figure}.1}
\begin{figure}
\plotone{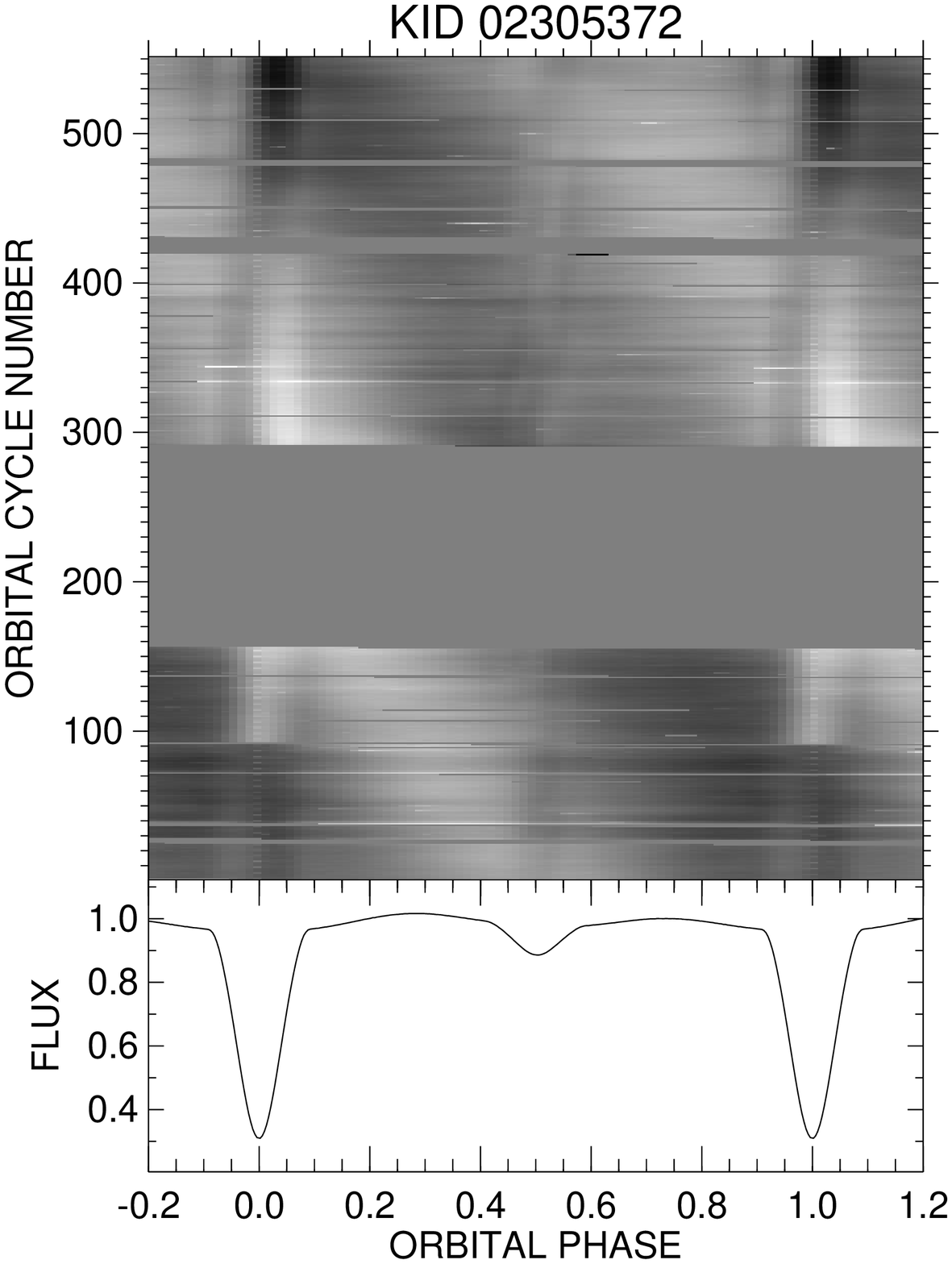}
\caption{The lower panel shows a mean, normalized light curve formed 
by binning in orbital phase.  The top panel shows the 
flux differences as a function of orbital phase and 
cycle number, represented as a gray scale diagram (range $\pm 3\%$). 
 }
\end{figure}

\clearpage
\setcounter{figure}{0}
\renewcommand{\thefigure}{\arabic{figure}.2}
\begin{figure}
\plotone{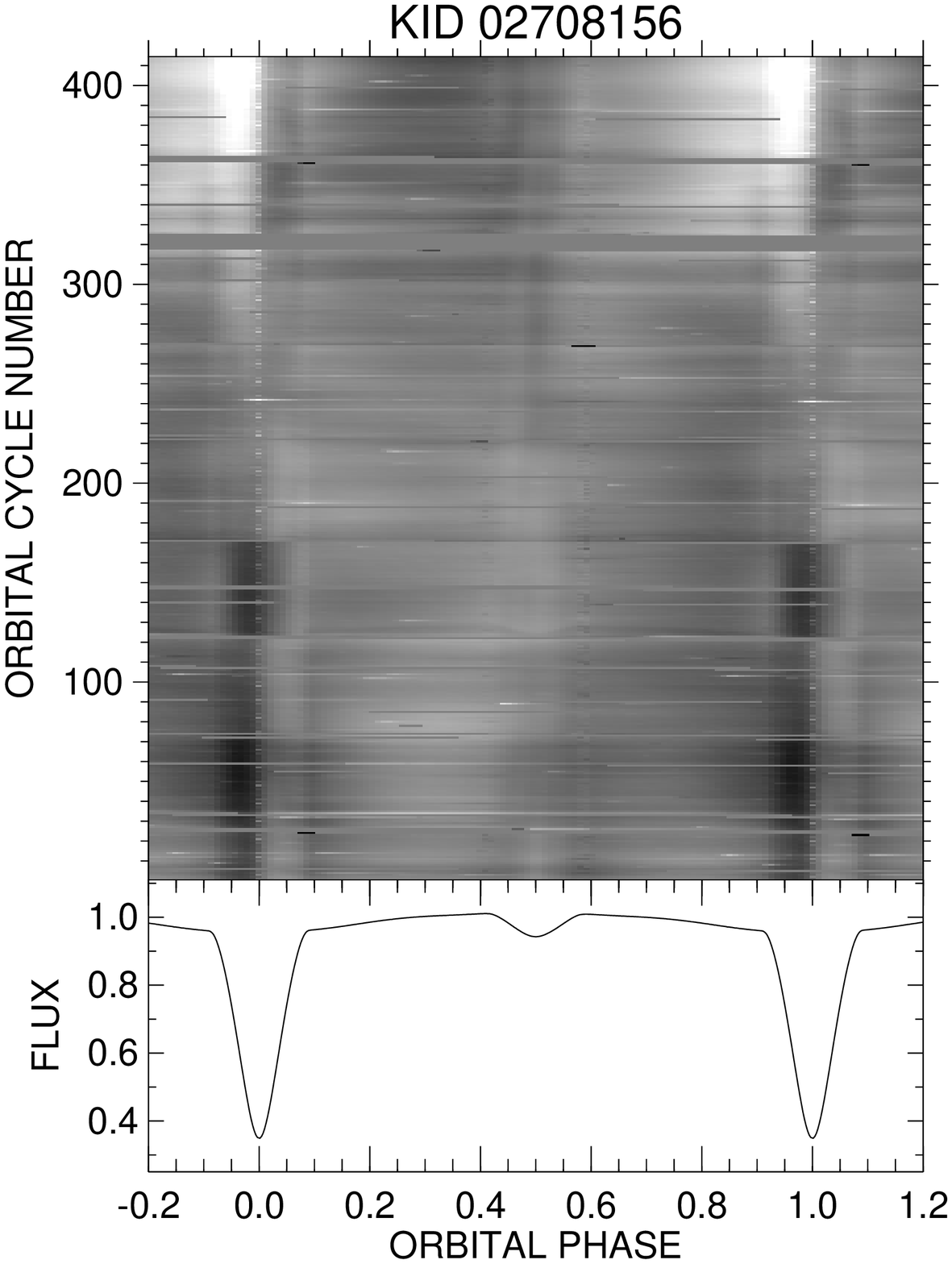}
\caption{The lower panel shows a mean, normalized light curve formed 
by binning in orbital phase.  The top panel shows the 
flux differences as a function of orbital phase and 
cycle number, represented as a gray scale diagram (range $\pm 1\%$). 
 }
\end{figure}

\clearpage
\setcounter{figure}{0}
\renewcommand{\thefigure}{\arabic{figure}.3}
\begin{figure}
\plotone{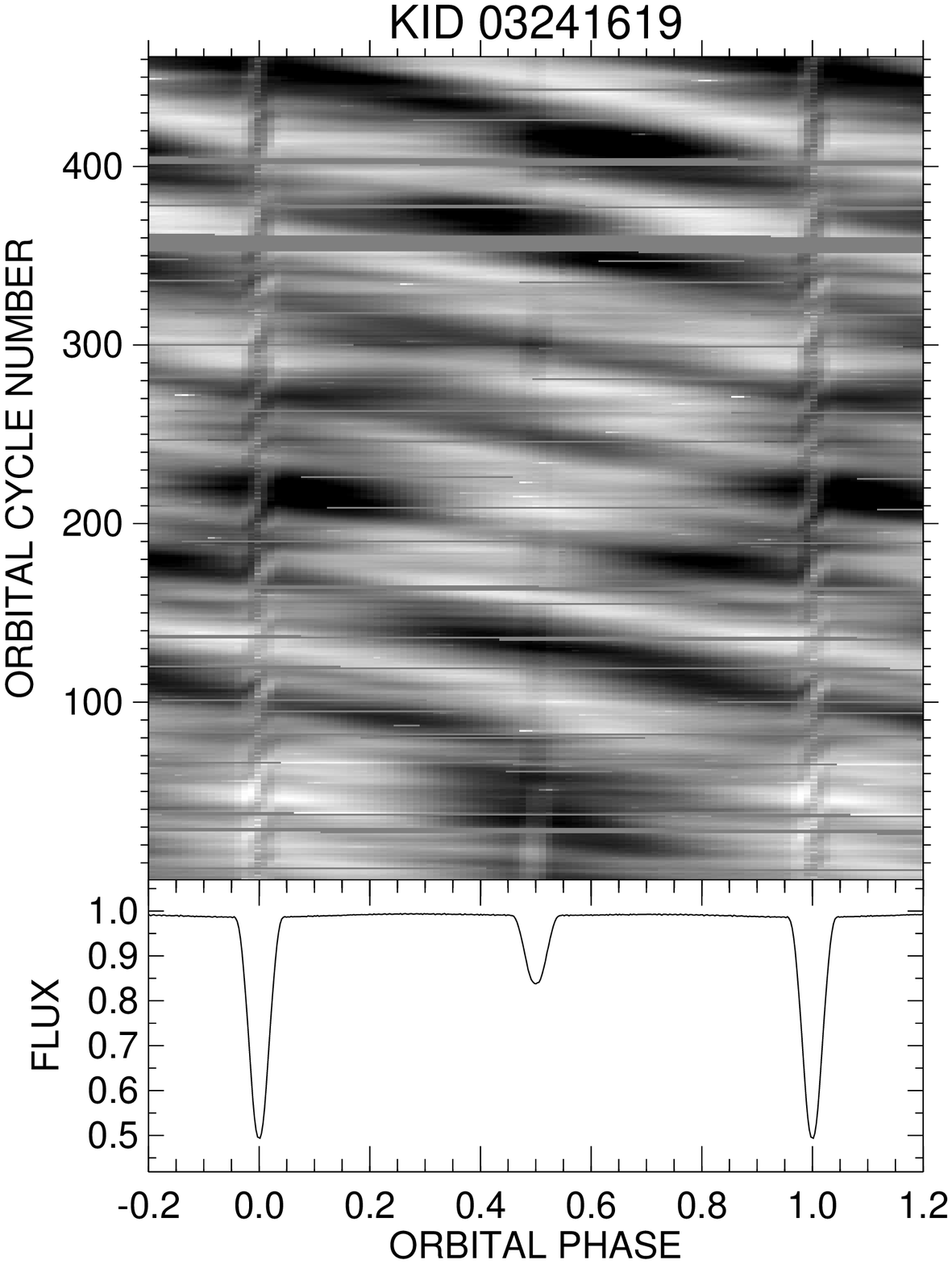}
\caption{The lower panel shows a mean, normalized light curve formed 
by binning in orbital phase.  The top panel shows the 
flux differences as a function of orbital phase and 
cycle number, represented as a gray scale diagram (range $\pm 3\%$). 
 }
\end{figure}

\clearpage
\setcounter{figure}{0}
\renewcommand{\thefigure}{\arabic{figure}.4}
\begin{figure}
\plotone{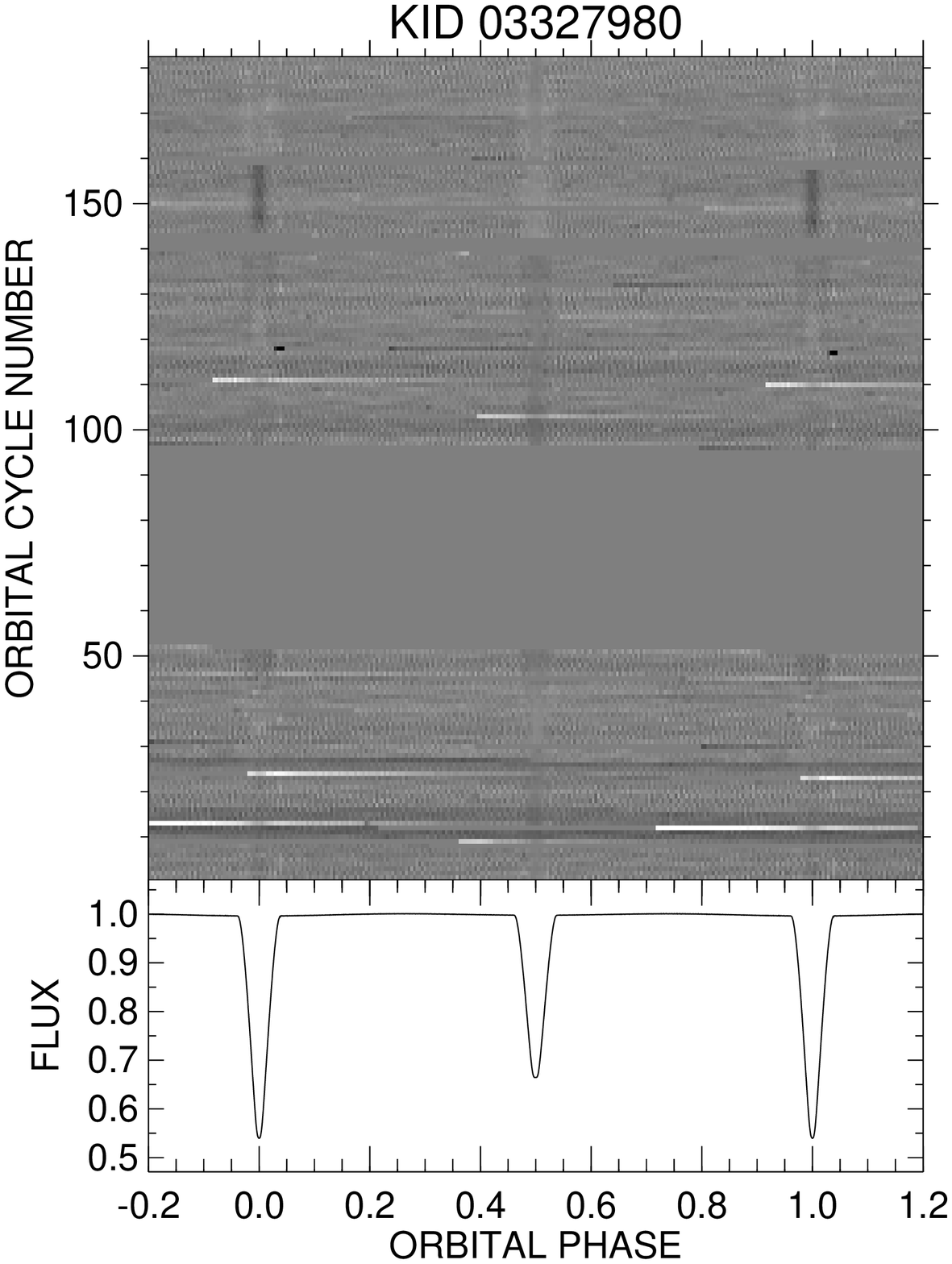}
\caption{The lower panel shows a mean, normalized light curve formed 
by binning in orbital phase.  The top panel shows the 
flux differences as a function of orbital phase and 
cycle number, represented as a gray scale diagram (range $\pm 0.5\%$). 
 }
\end{figure}

\clearpage
\setcounter{figure}{0}
\renewcommand{\thefigure}{\arabic{figure}.5}
\begin{figure}
\plotone{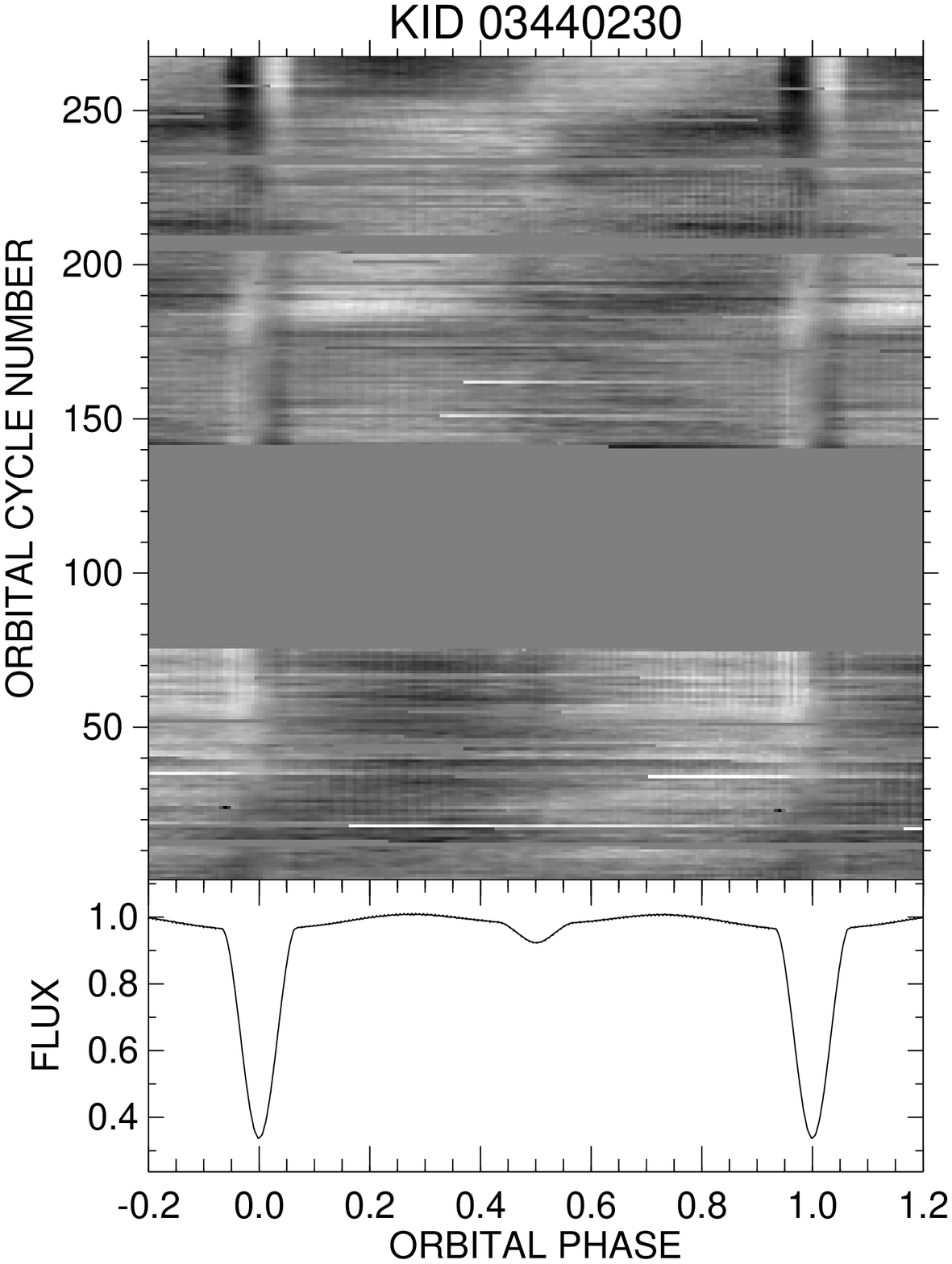}
\caption{The lower panel shows a mean, normalized light curve formed 
by binning in orbital phase.  The top panel shows the 
flux differences as a function of orbital phase and 
cycle number, represented as a gray scale diagram (range $\pm 1\%$). 
 }
\end{figure}

\clearpage
\setcounter{figure}{0}
\renewcommand{\thefigure}{\arabic{figure}.6}
\begin{figure}
\plotone{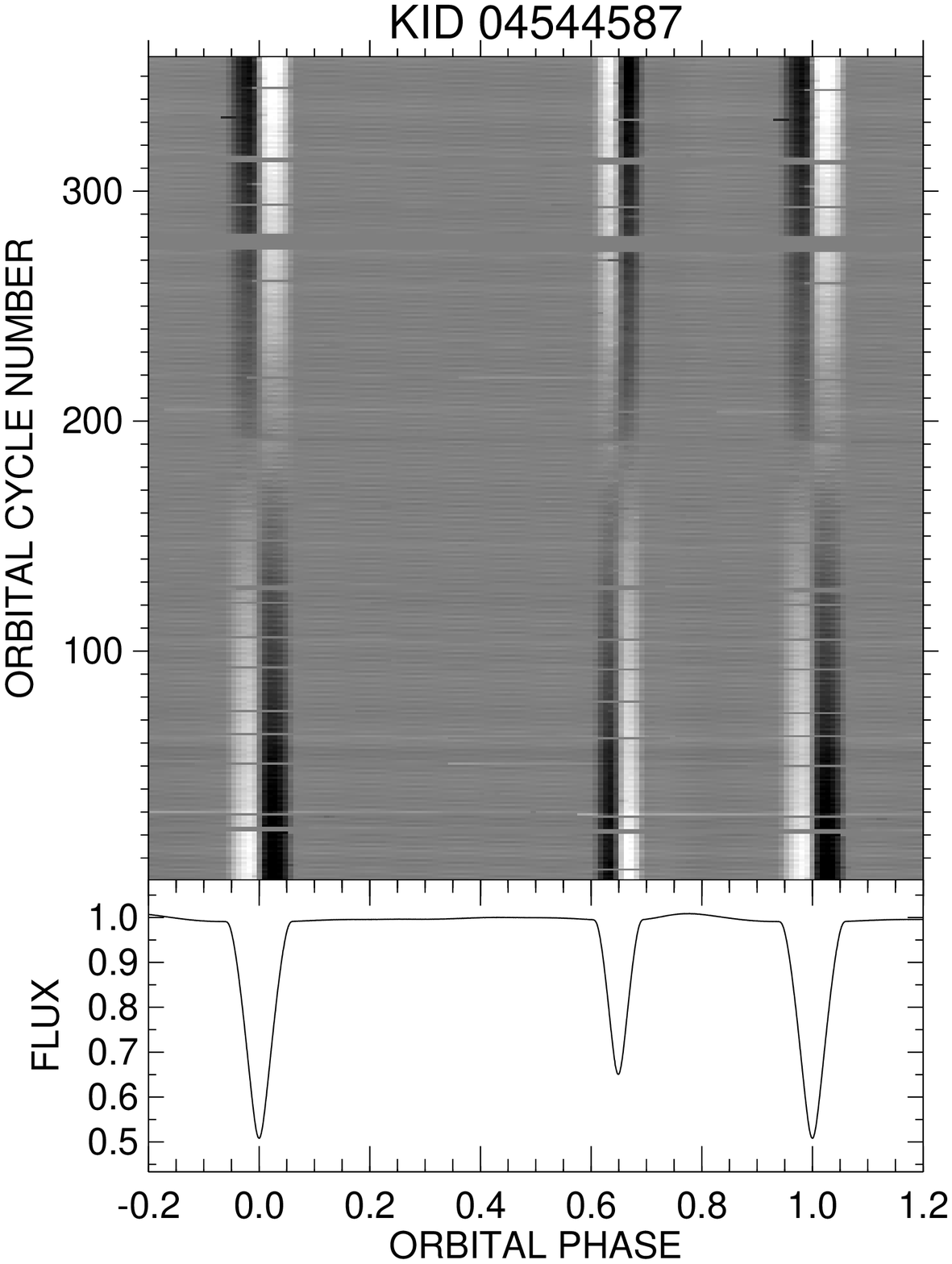}
\caption{The lower panel shows a mean, normalized light curve formed 
by binning in orbital phase.  The top panel shows the 
flux differences as a function of orbital phase and 
cycle number, represented as a gray scale diagram (range $\pm 1.5\%$). 
 }
\end{figure}

\clearpage
\setcounter{figure}{0}
\renewcommand{\thefigure}{\arabic{figure}.7}
\begin{figure}
\plotone{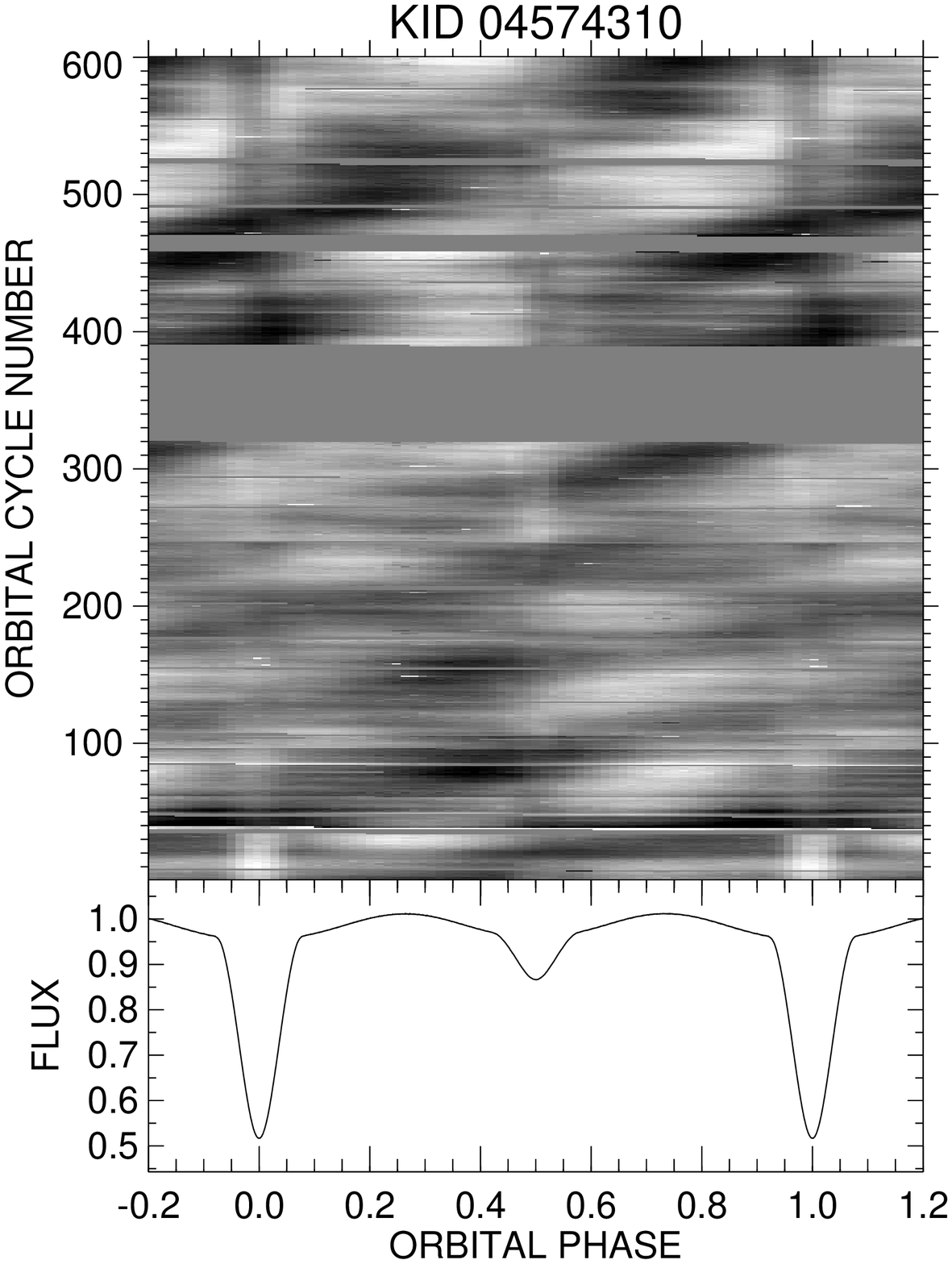}
\caption{The lower panel shows a mean, normalized light curve formed 
by binning in orbital phase.  The top panel shows the 
flux differences as a function of orbital phase and 
cycle number, represented as a gray scale diagram (range $\pm 0.5\%$). 
 }
\end{figure}

\clearpage
\setcounter{figure}{0}
\renewcommand{\thefigure}{\arabic{figure}.8}
\begin{figure}
\plotone{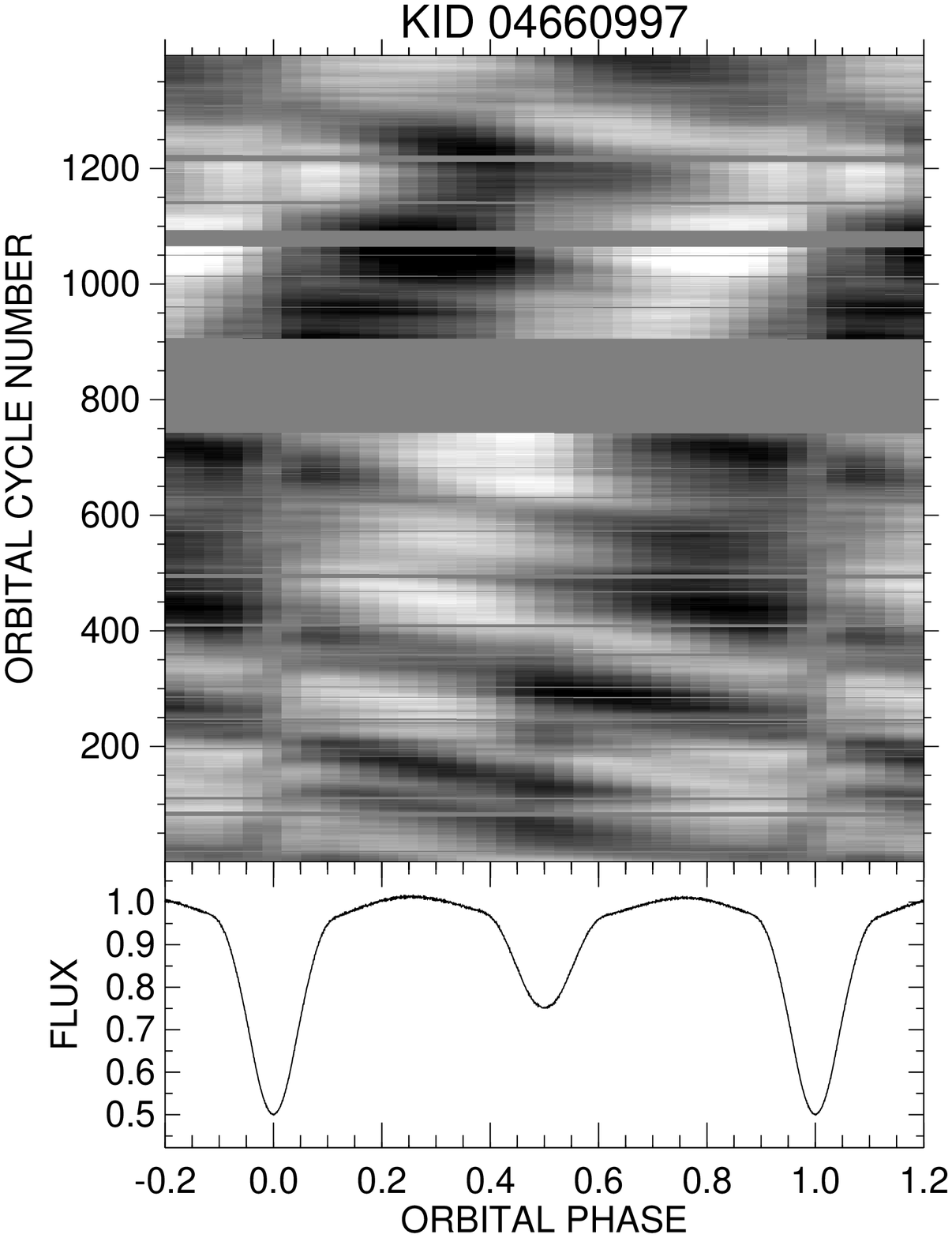}
\caption{The lower panel shows a mean, normalized light curve formed 
by binning in orbital phase.  The top panel shows the 
flux differences as a function of orbital phase and 
cycle number, represented as a gray scale diagram (range $\pm 4\%$). 
 }
\end{figure}

\clearpage
\setcounter{figure}{0}
\renewcommand{\thefigure}{\arabic{figure}.9}
\begin{figure}
\plotone{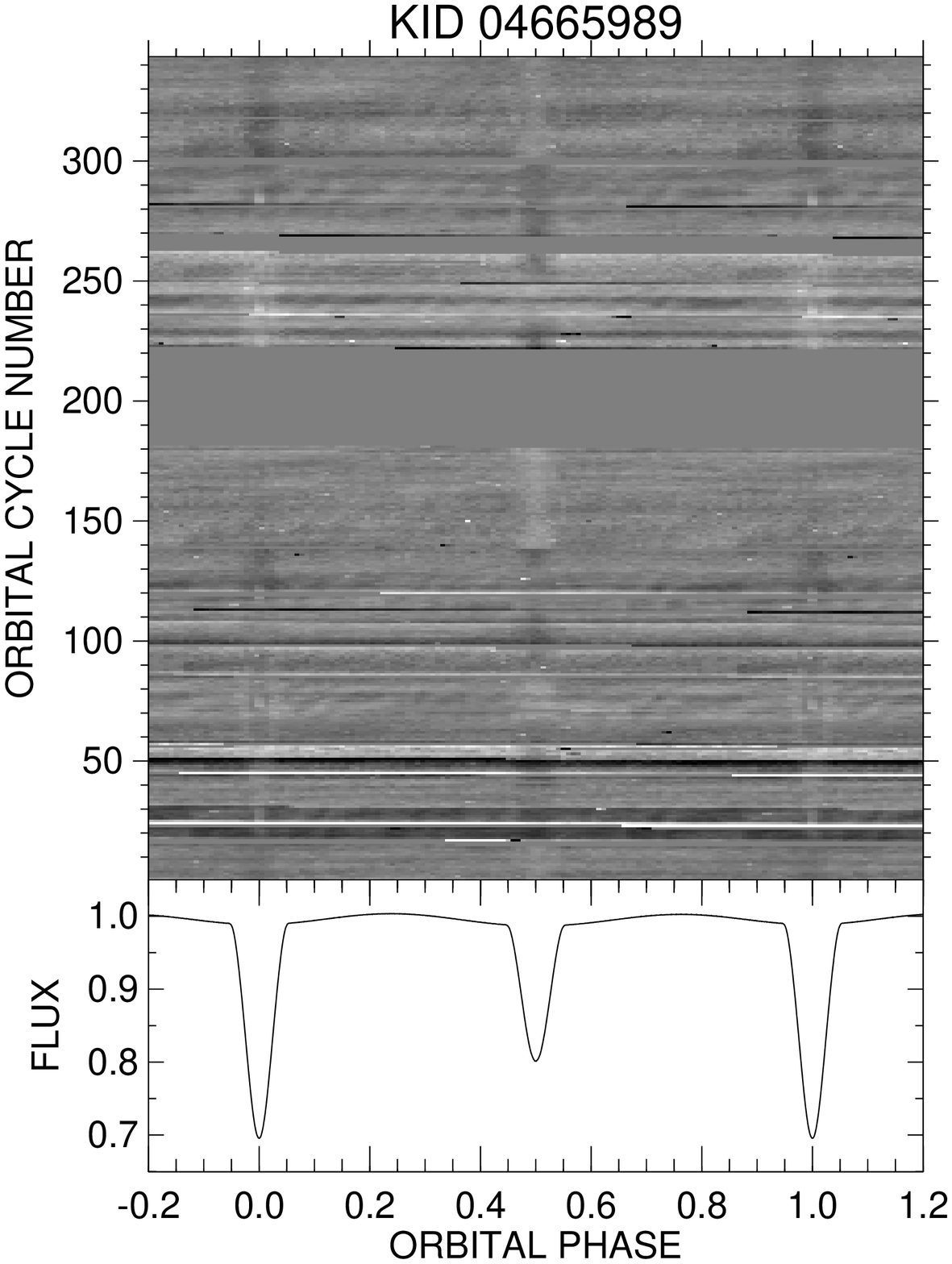}
\caption{The lower panel shows a mean, normalized light curve formed 
by binning in orbital phase.  The top panel shows the 
flux differences as a function of orbital phase and 
cycle number, represented as a gray scale diagram (range $\pm 0.2\%$). 
 }
\end{figure}

\clearpage
\setcounter{figure}{0}
\renewcommand{\thefigure}{\arabic{figure}.10}
\begin{figure}
\plotone{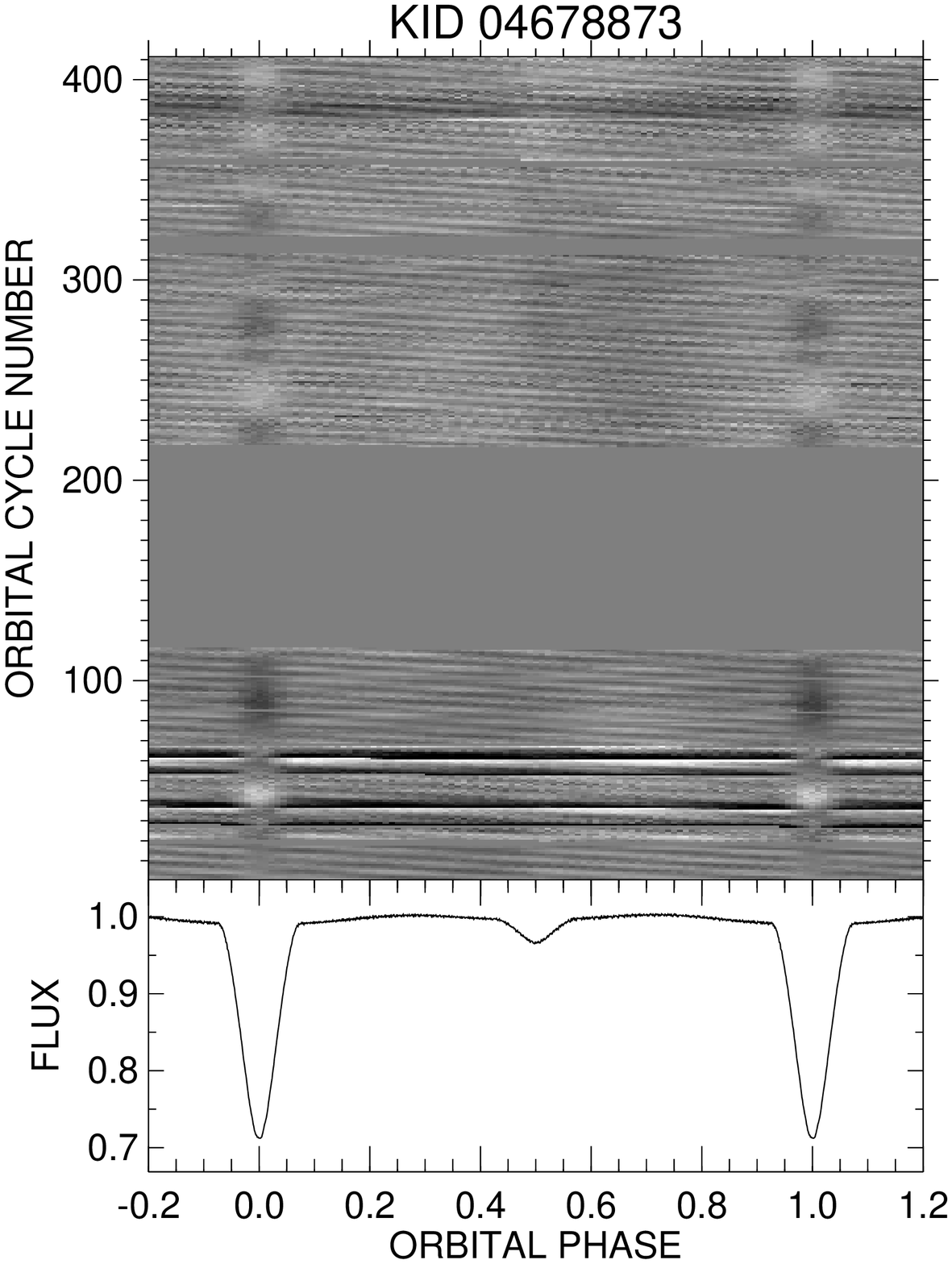}
\caption{The lower panel shows a mean, normalized light curve formed 
by binning in orbital phase.  The top panel shows the 
flux differences as a function of orbital phase and 
cycle number, represented as a gray scale diagram (range $\pm 2\%$). 
 }
\end{figure}

\clearpage
\setcounter{figure}{0}
\renewcommand{\thefigure}{\arabic{figure}.11}
\begin{figure}
\plotone{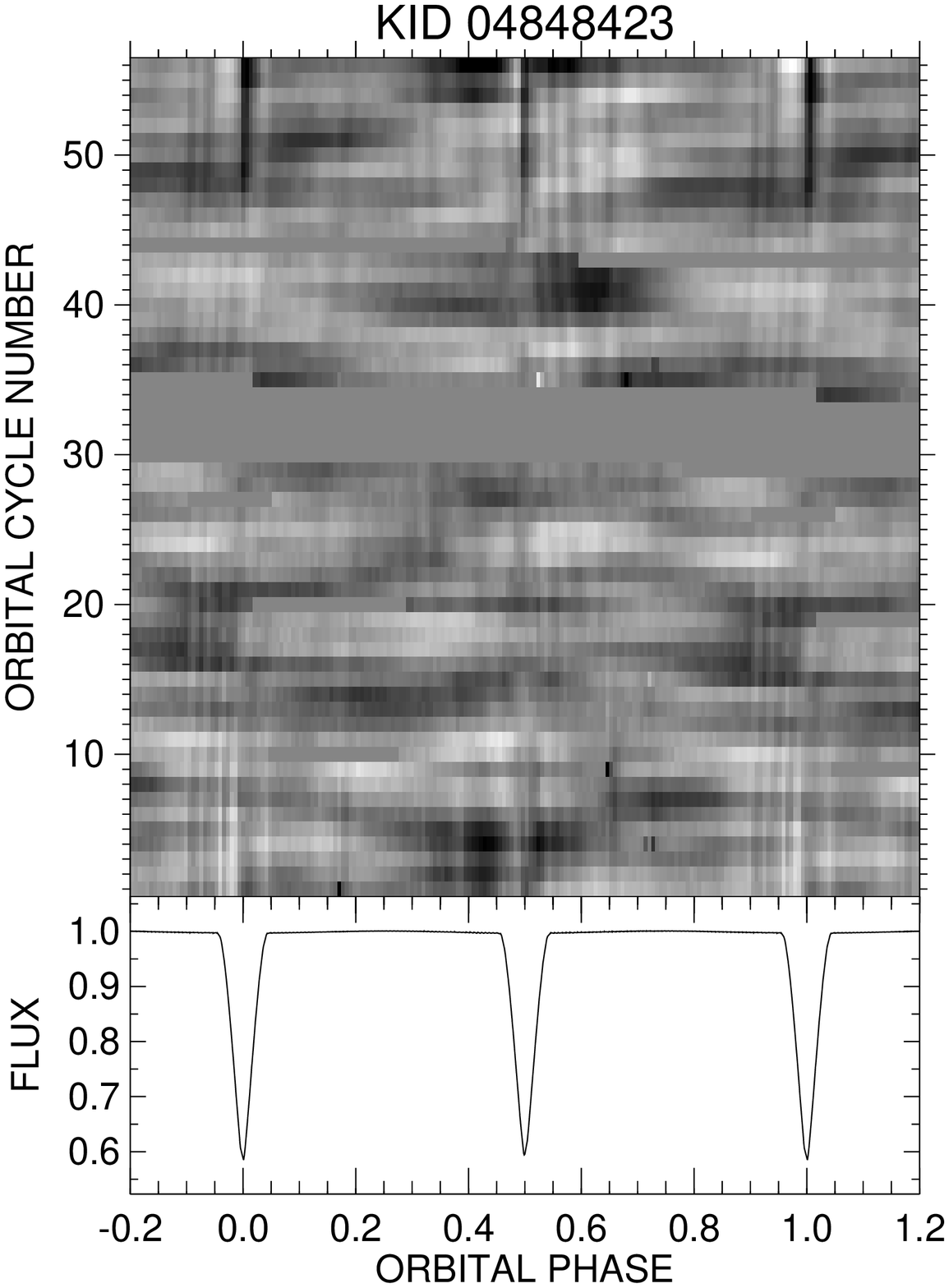}
\caption{The lower panel shows a mean, normalized light curve formed 
by binning in orbital phase.  The top panel shows the 
flux differences as a function of orbital phase and 
cycle number, represented as a gray scale diagram (range $\pm 0.2\%$). 
 }
\end{figure}

\clearpage
\setcounter{figure}{0}
\renewcommand{\thefigure}{\arabic{figure}.12}
\begin{figure}
\plotone{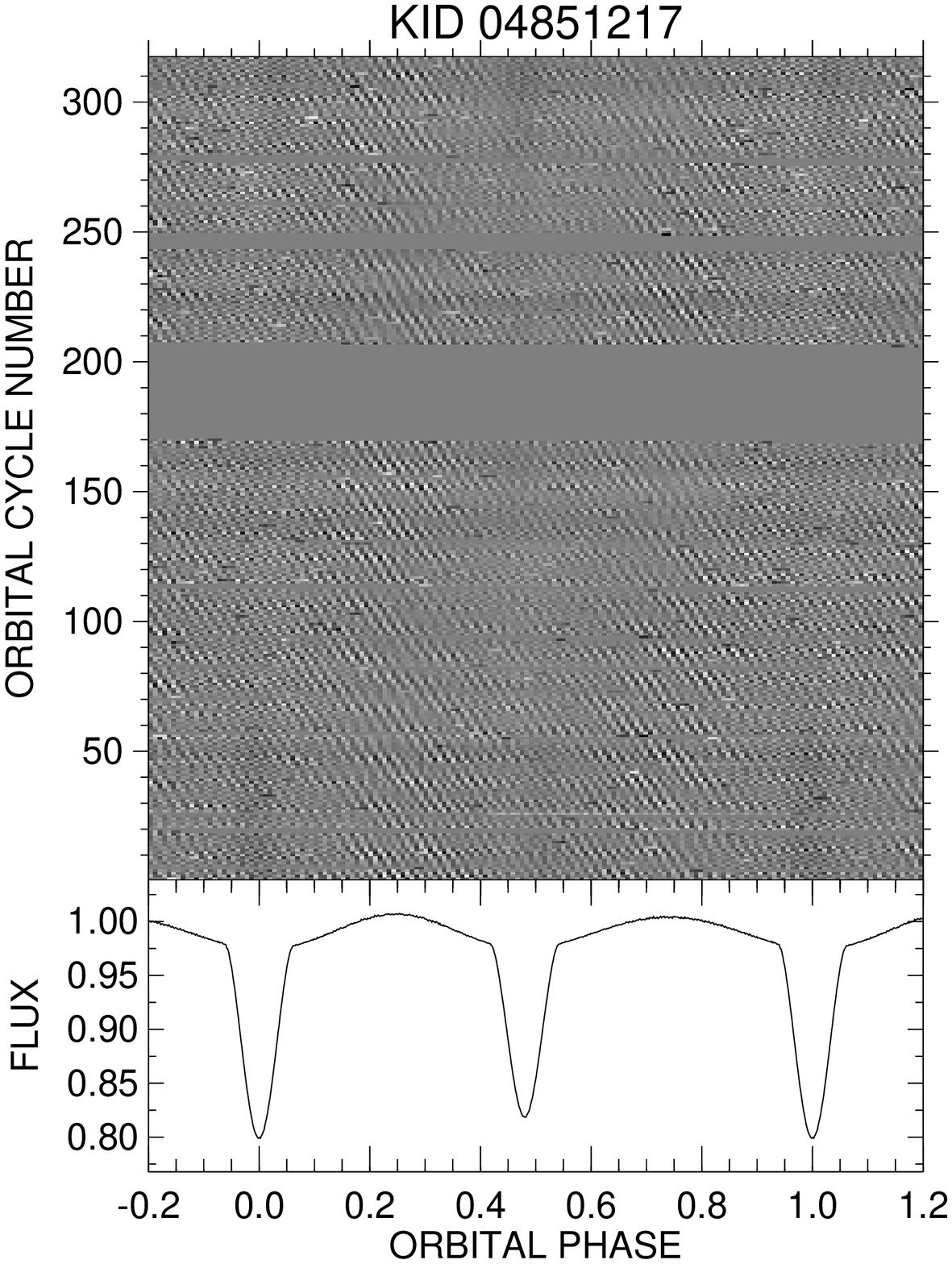}
\caption{The lower panel shows a mean, normalized light curve formed 
by binning in orbital phase.  The top panel shows the 
flux differences as a function of orbital phase and 
cycle number, represented as a gray scale diagram (range $\pm 1\%$). 
 }
\end{figure}

\clearpage
\setcounter{figure}{0}
\renewcommand{\thefigure}{\arabic{figure}.13}
\begin{figure}
\plotone{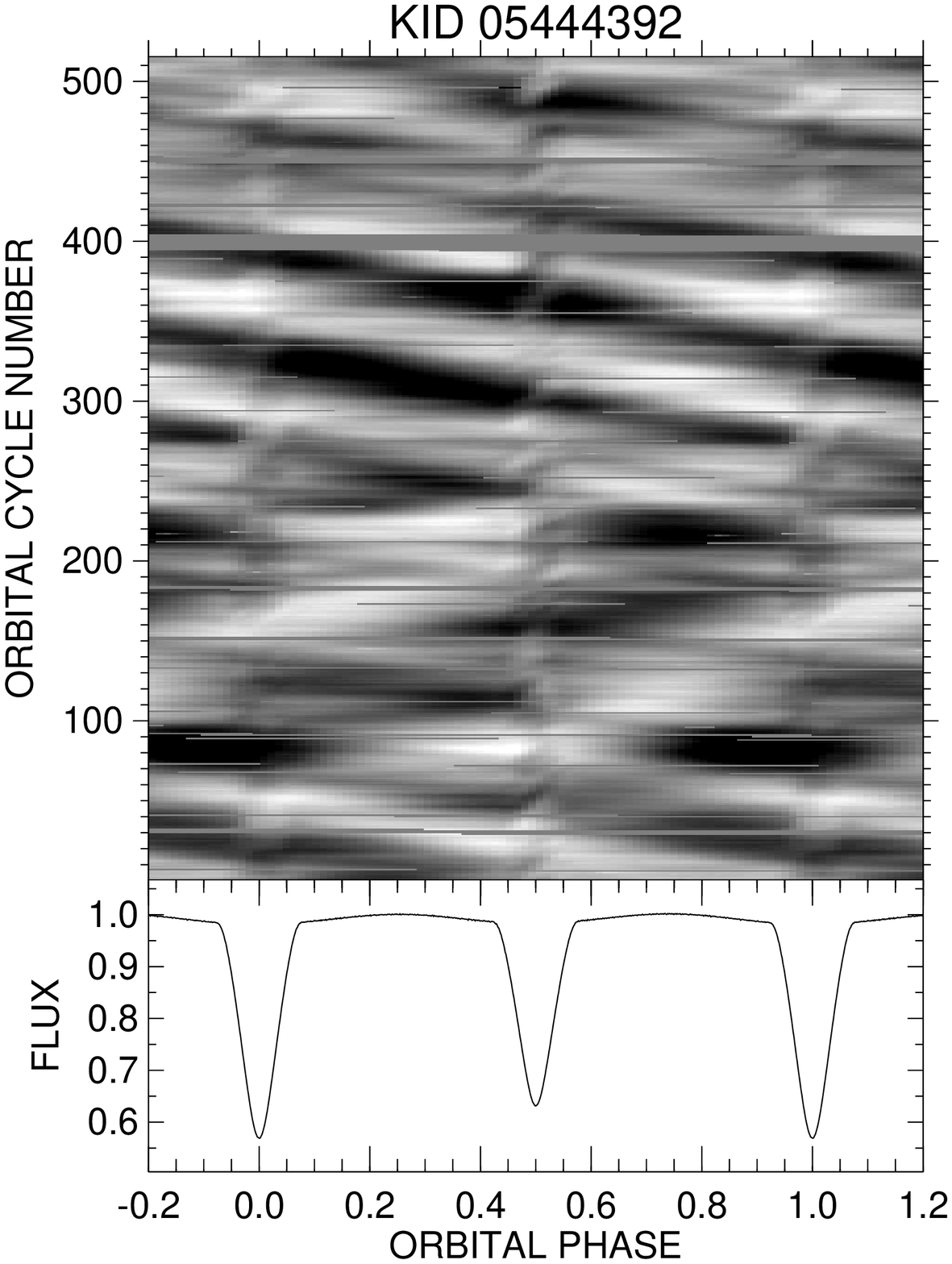}
\caption{The lower panel shows a mean, normalized light curve formed 
by binning in orbital phase.  The top panel shows the 
flux differences as a function of orbital phase and 
cycle number, represented as a gray scale diagram (range $\pm 2\%$). 
 }
\end{figure}

\clearpage
\setcounter{figure}{0}
\renewcommand{\thefigure}{\arabic{figure}.14}
\begin{figure}
\plotone{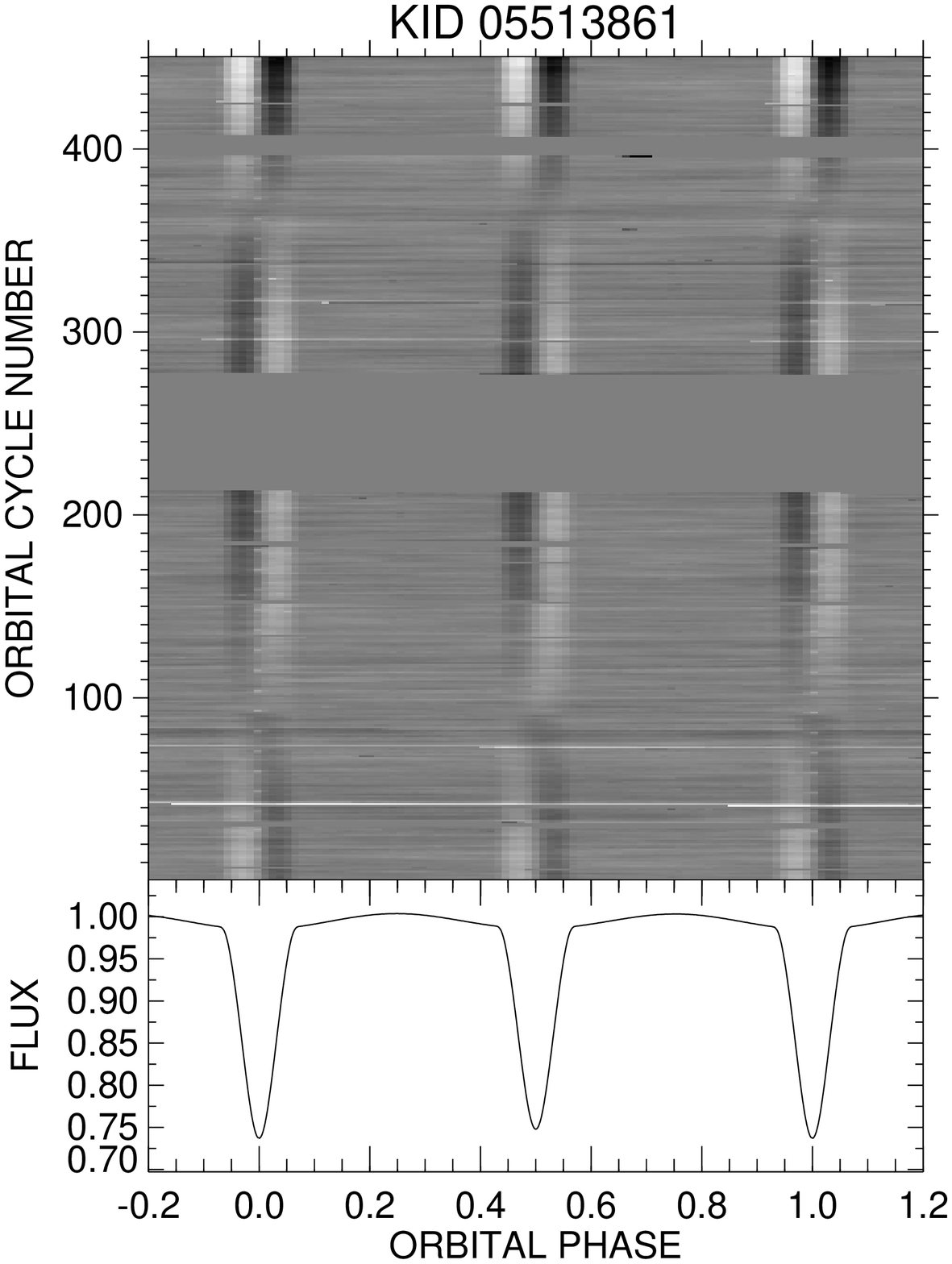}
\caption{The lower panel shows a mean, normalized light curve formed 
by binning in orbital phase.  The top panel shows the 
flux differences as a function of orbital phase and 
cycle number, represented as a gray scale diagram (range $\pm 0.5\%$). 
 }
\end{figure}

\clearpage
\setcounter{figure}{0}
\renewcommand{\thefigure}{\arabic{figure}.15}
\begin{figure}
\plotone{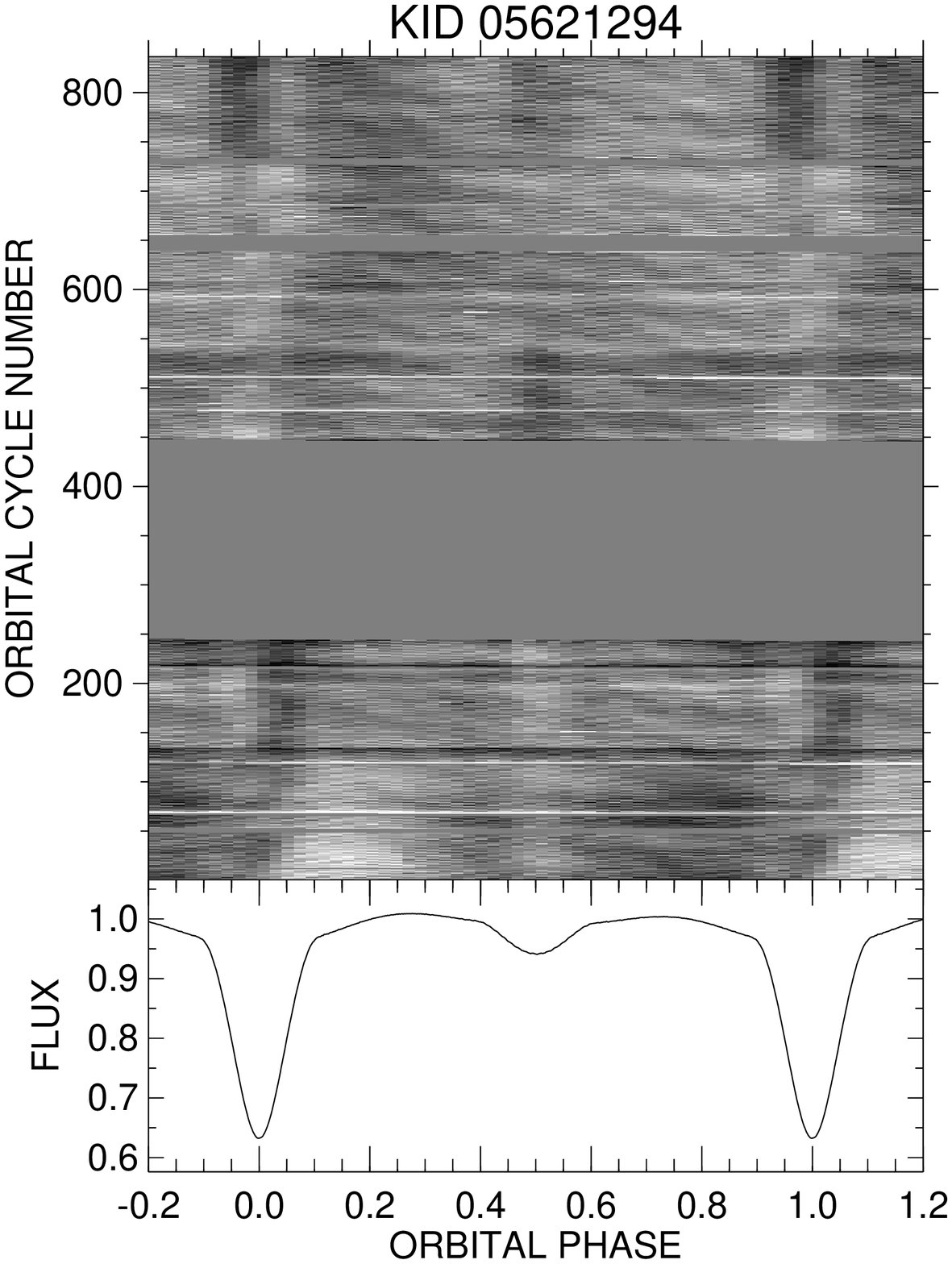}
\caption{The lower panel shows a mean, normalized light curve formed 
by binning in orbital phase.  The top panel shows the 
flux differences as a function of orbital phase and 
cycle number, represented as a gray scale diagram (range $\pm 0.5\%$). 
 }
\end{figure}

\clearpage
\setcounter{figure}{0}
\renewcommand{\thefigure}{\arabic{figure}.16}
\begin{figure}
\plotone{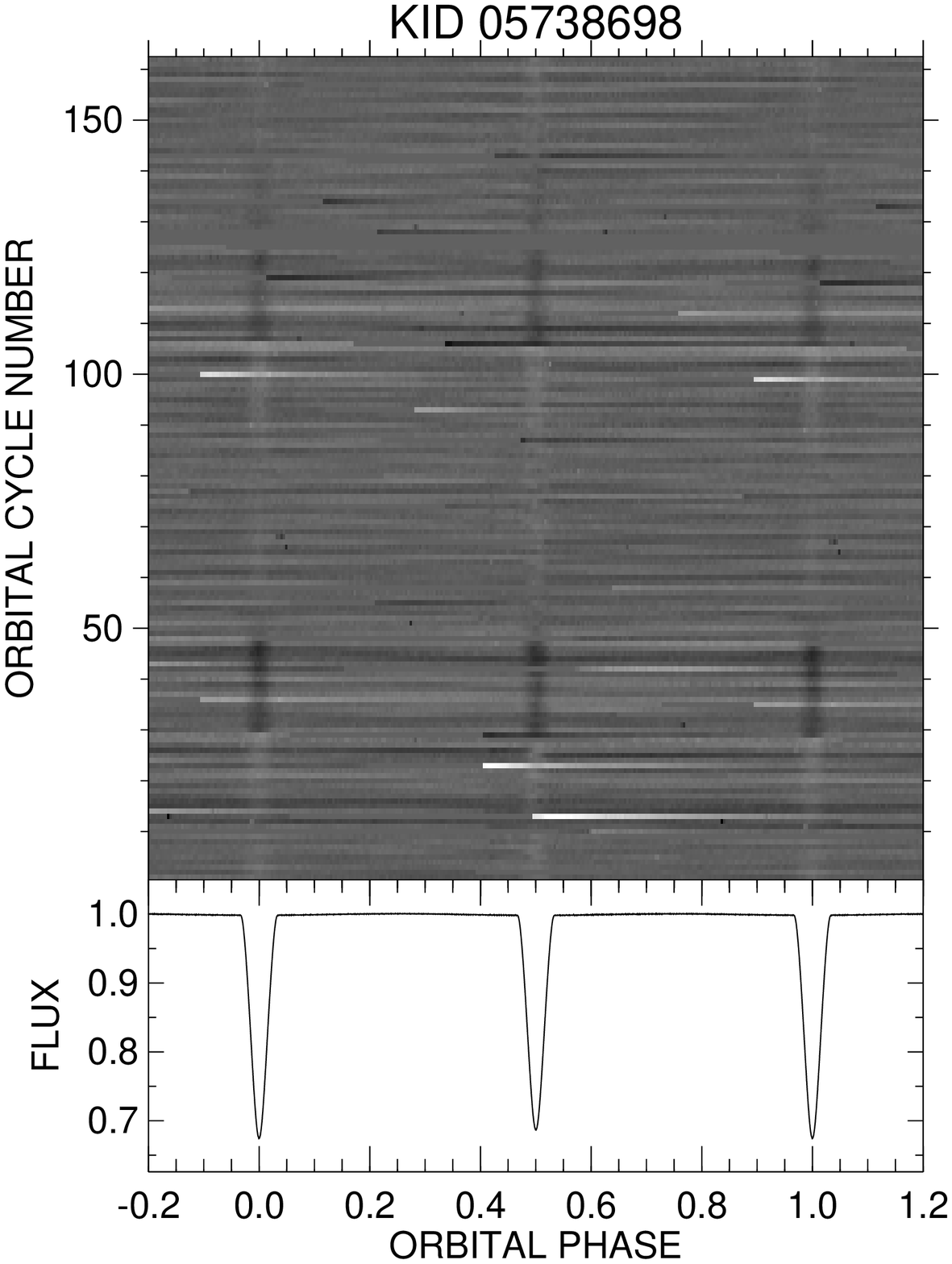}
\caption{The lower panel shows a mean, normalized light curve formed 
by binning in orbital phase.  The top panel shows the 
flux differences as a function of orbital phase and 
cycle number, represented as a gray scale diagram (range $\pm 0.5\%$). 
 }
\end{figure}

\clearpage
\setcounter{figure}{0}
\renewcommand{\thefigure}{\arabic{figure}.17}
\begin{figure}
\plotone{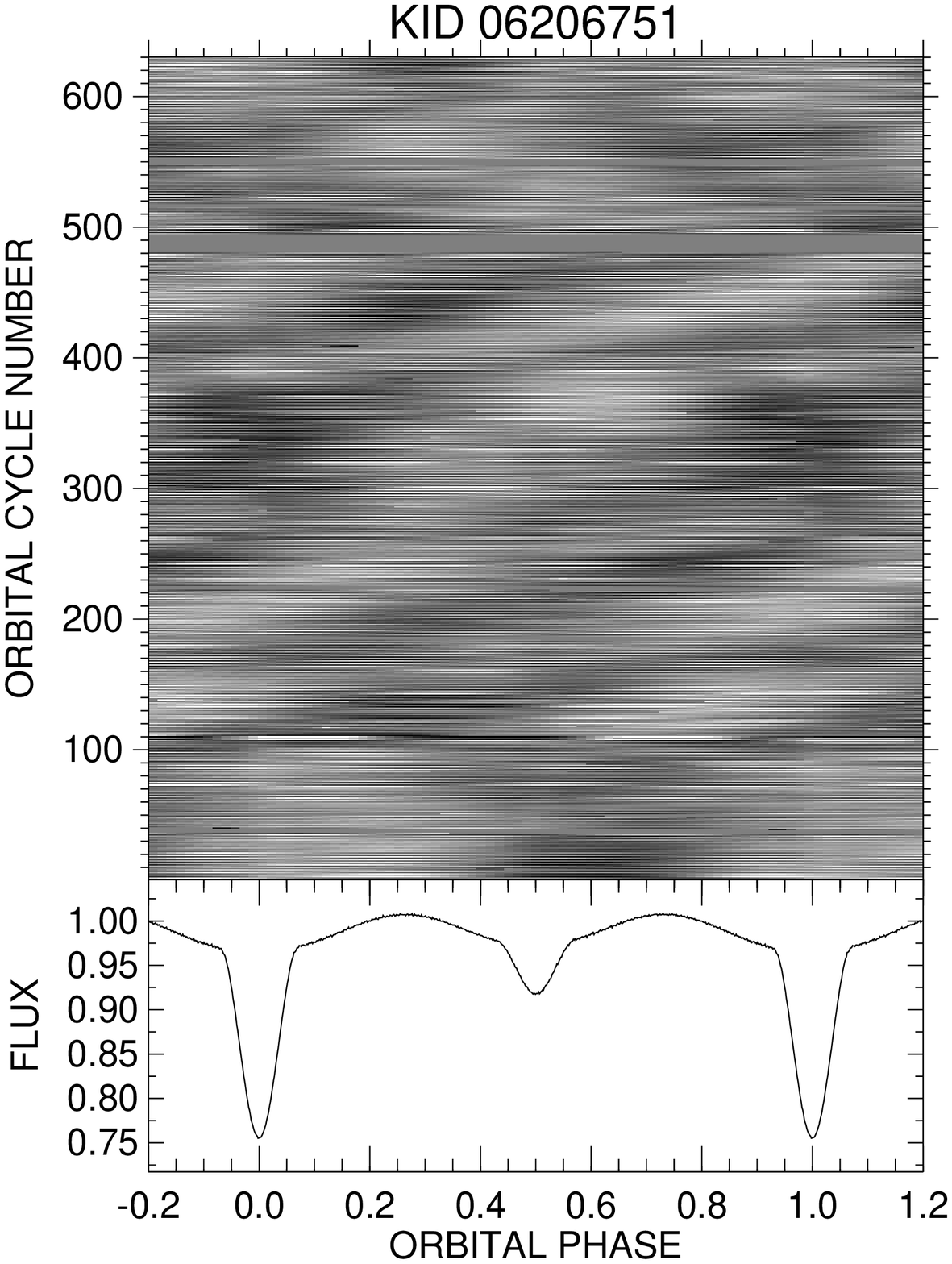}
\caption{The lower panel shows a mean, normalized light curve formed 
by binning in orbital phase.  The top panel shows the 
flux differences as a function of orbital phase and 
cycle number, represented as a gray scale diagram (range $\pm 1.5\%$). 
 }
\end{figure}

\clearpage
\setcounter{figure}{0}
\renewcommand{\thefigure}{\arabic{figure}.18}
\begin{figure}
\plotone{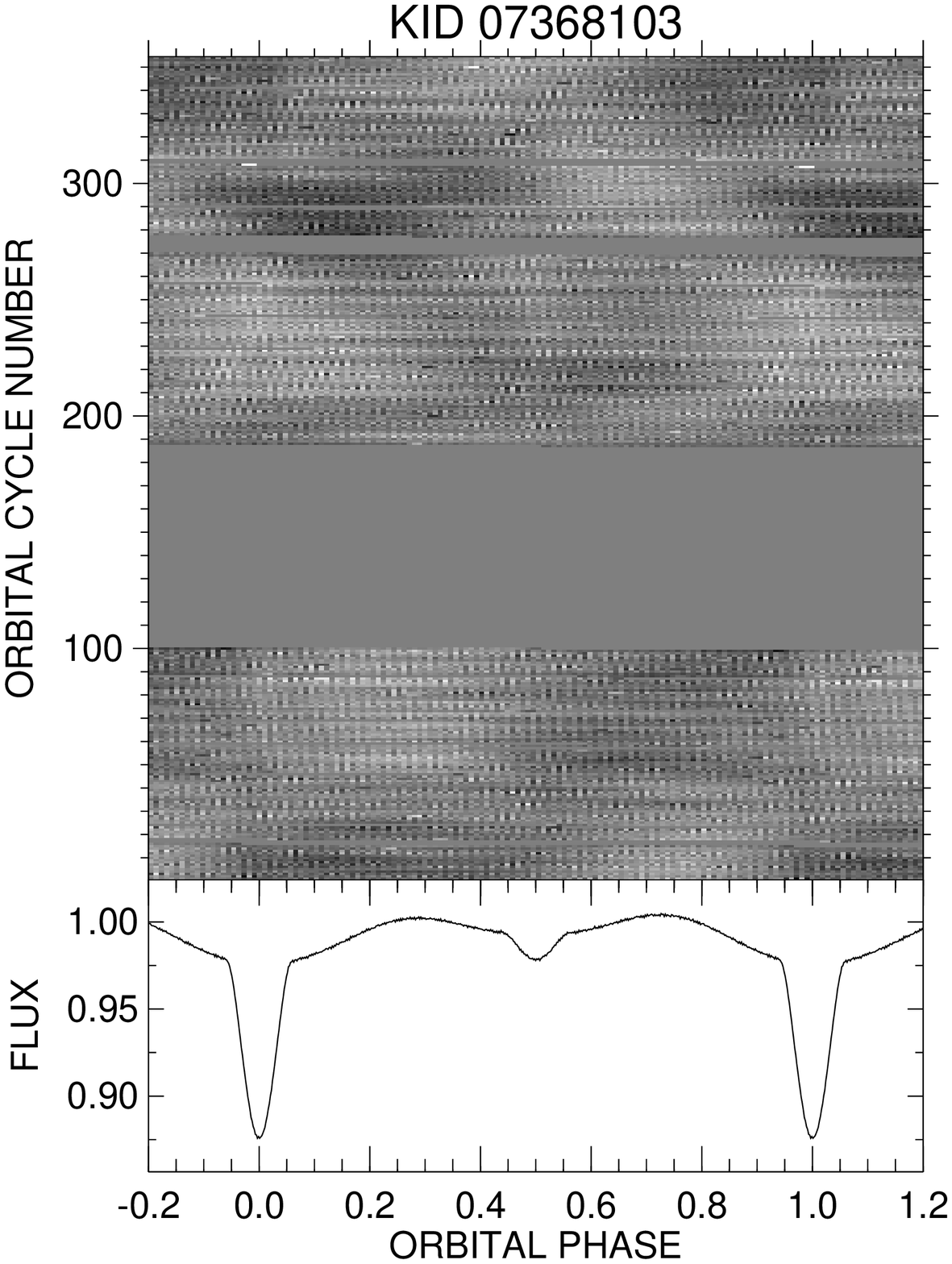}
\caption{The lower panel shows a mean, normalized light curve formed 
by binning in orbital phase.  The top panel shows the 
flux differences as a function of orbital phase and 
cycle number, represented as a gray scale diagram (range $\pm 1\%$). 
 }
\end{figure}

\clearpage
\setcounter{figure}{0}
\renewcommand{\thefigure}{\arabic{figure}.19}
\begin{figure}
\plotone{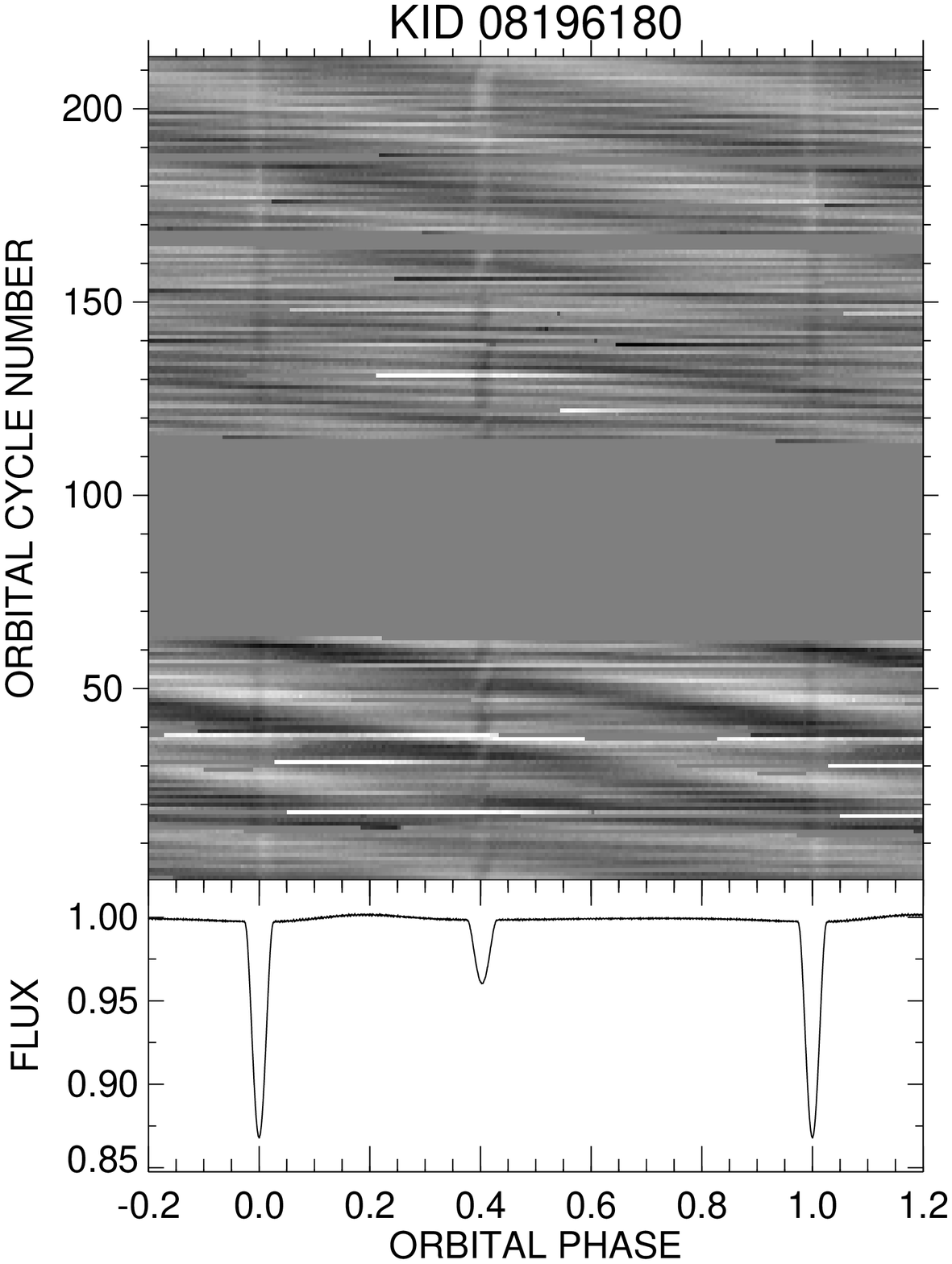}
\caption{The lower panel shows a mean, normalized light curve formed 
by binning in orbital phase.  The top panel shows the 
flux differences as a function of orbital phase and 
cycle number, represented as a gray scale diagram (range $\pm 0.5\%$). 
 }
\end{figure}

\clearpage
\setcounter{figure}{0}
\renewcommand{\thefigure}{\arabic{figure}.20}
\begin{figure}
\plotone{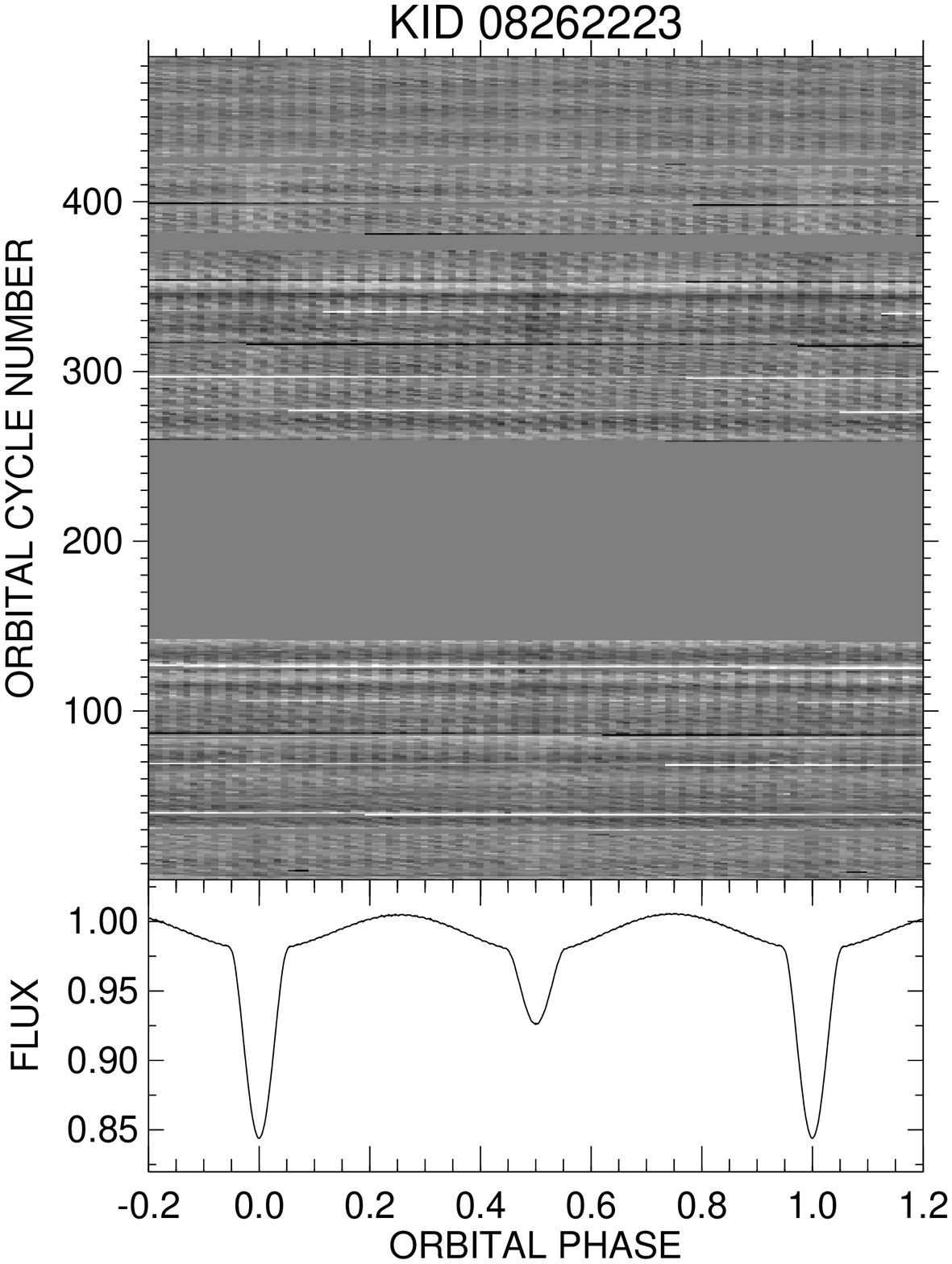}
\caption{The lower panel shows a mean, normalized light curve formed 
by binning in orbital phase.  The top panel shows the 
flux differences as a function of orbital phase and 
cycle number, represented as a gray scale diagram (range $\pm 0.3\%$). 
 }
\end{figure}

\clearpage
\setcounter{figure}{0}
\renewcommand{\thefigure}{\arabic{figure}.21}
\begin{figure}
\plotone{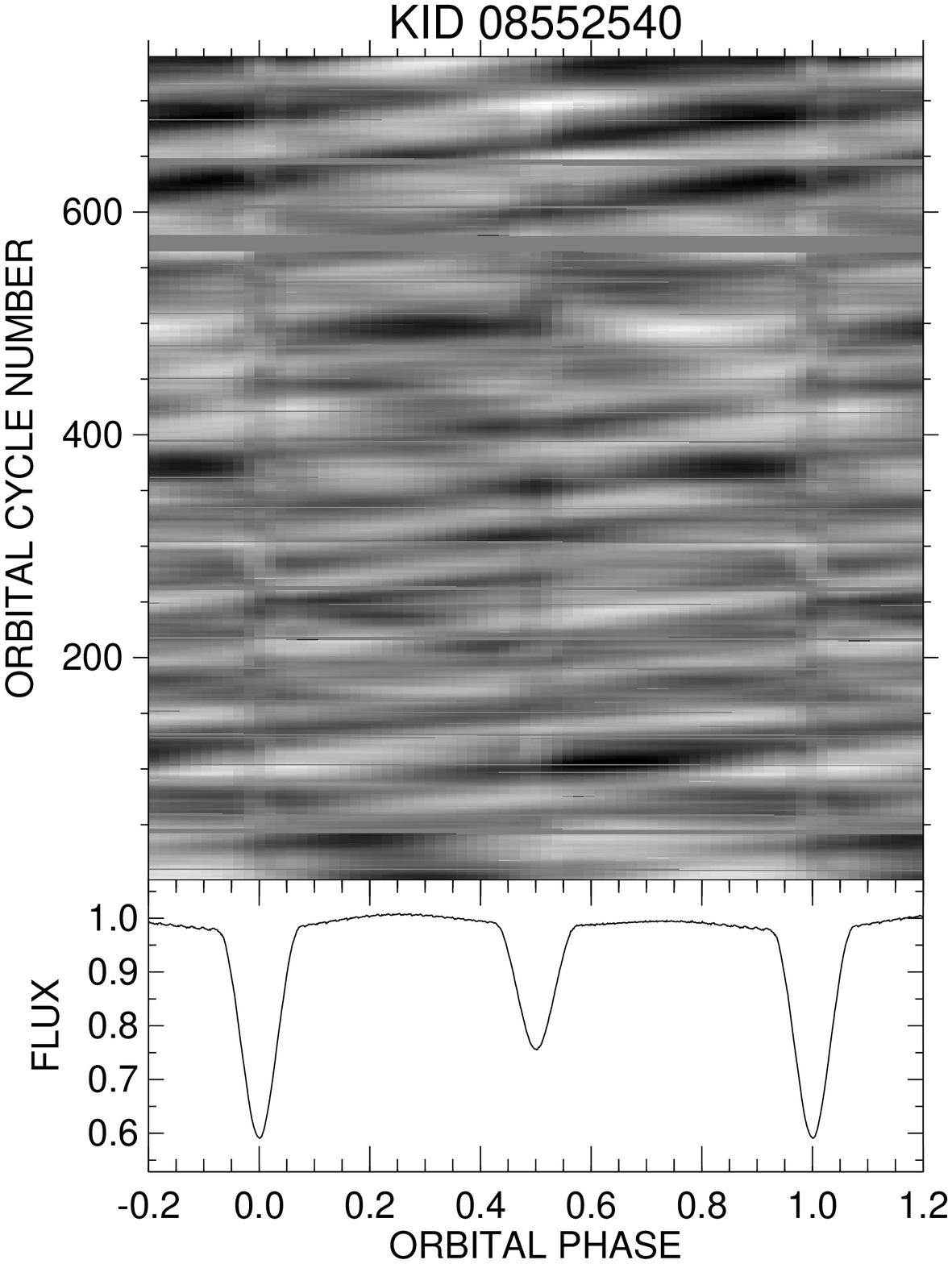}
\caption{The lower panel shows a mean, normalized light curve formed 
by binning in orbital phase.  The top panel shows the 
flux differences as a function of orbital phase and 
cycle number, represented as a gray scale diagram (range $\pm 3\%$). 
 }
\end{figure}

\clearpage
\setcounter{figure}{0}
\renewcommand{\thefigure}{\arabic{figure}.22}
\begin{figure}
\plotone{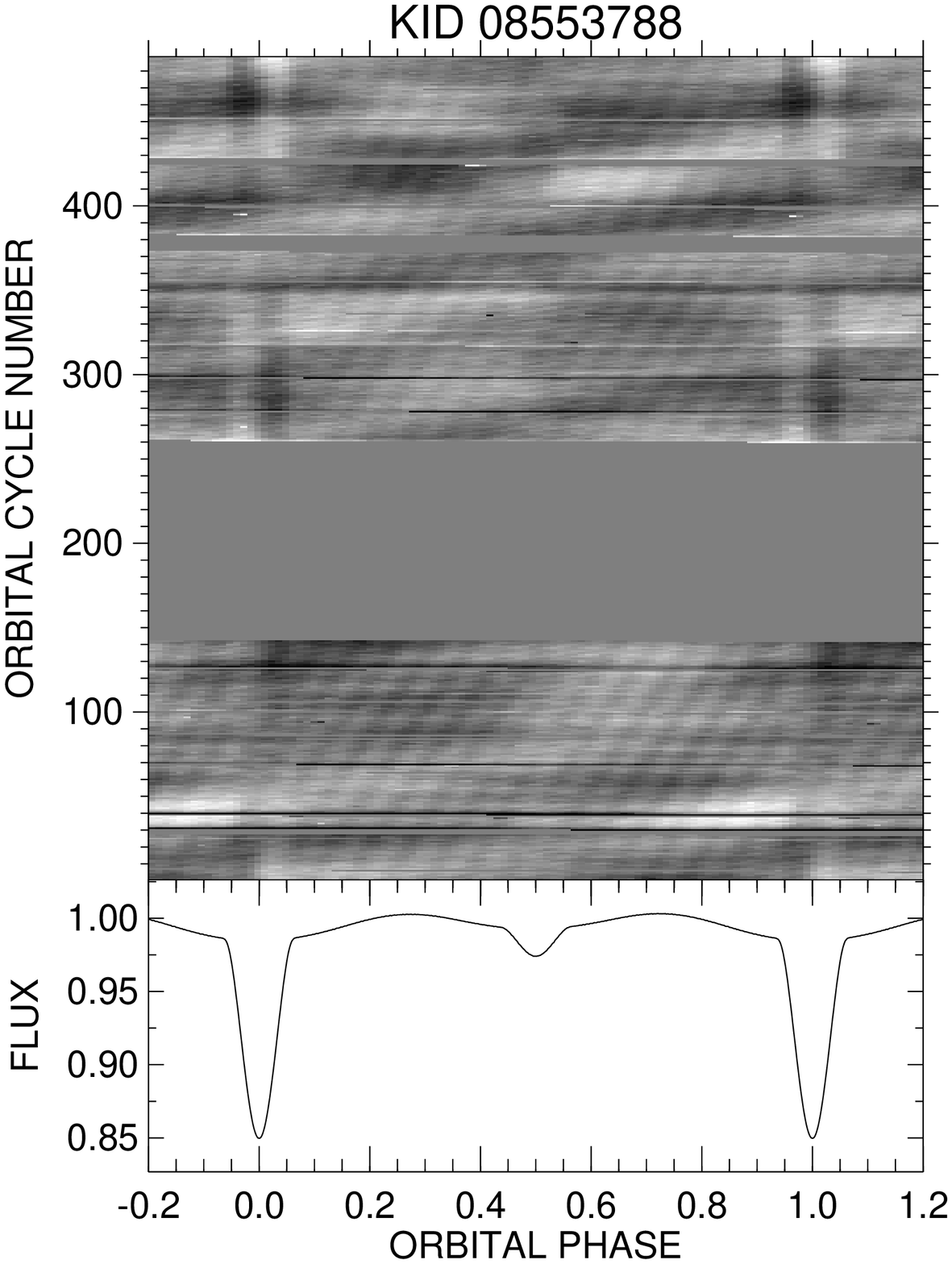}
\caption{The lower panel shows a mean, normalized light curve formed 
by binning in orbital phase.  The top panel shows the 
flux differences as a function of orbital phase and 
cycle number, represented as a gray scale diagram (range $\pm 0.3\%$). 
 }
\end{figure}

\clearpage
\setcounter{figure}{0}
\renewcommand{\thefigure}{\arabic{figure}.23}
\begin{figure}
\plotone{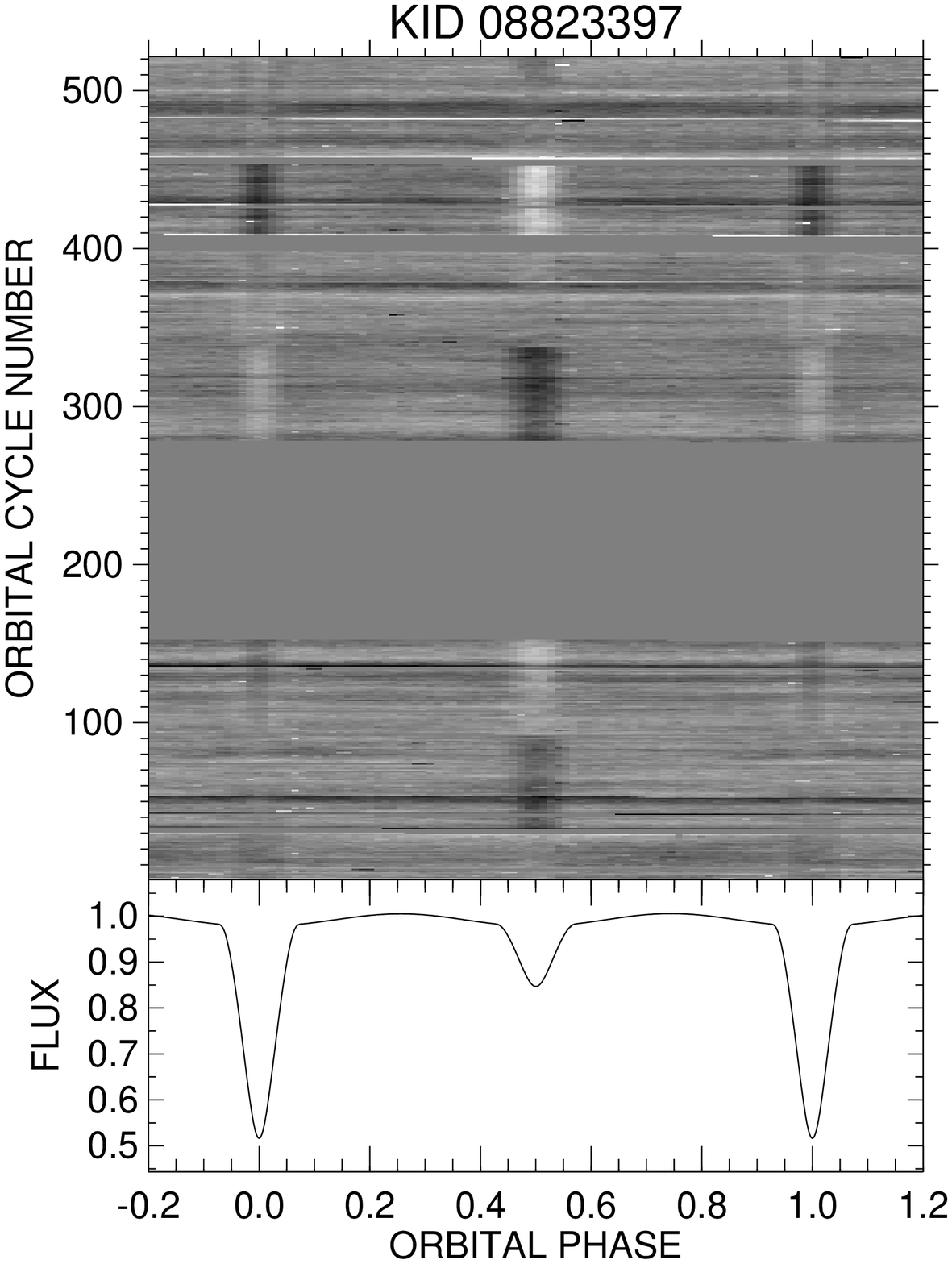}
\caption{The lower panel shows a mean, normalized light curve formed 
by binning in orbital phase.  The top panel shows the 
flux differences as a function of orbital phase and 
cycle number, represented as a gray scale diagram (range $\pm 0.2\%$). 
 }
\end{figure}

\clearpage
\setcounter{figure}{0}
\renewcommand{\thefigure}{\arabic{figure}.24}
\begin{figure}
\plotone{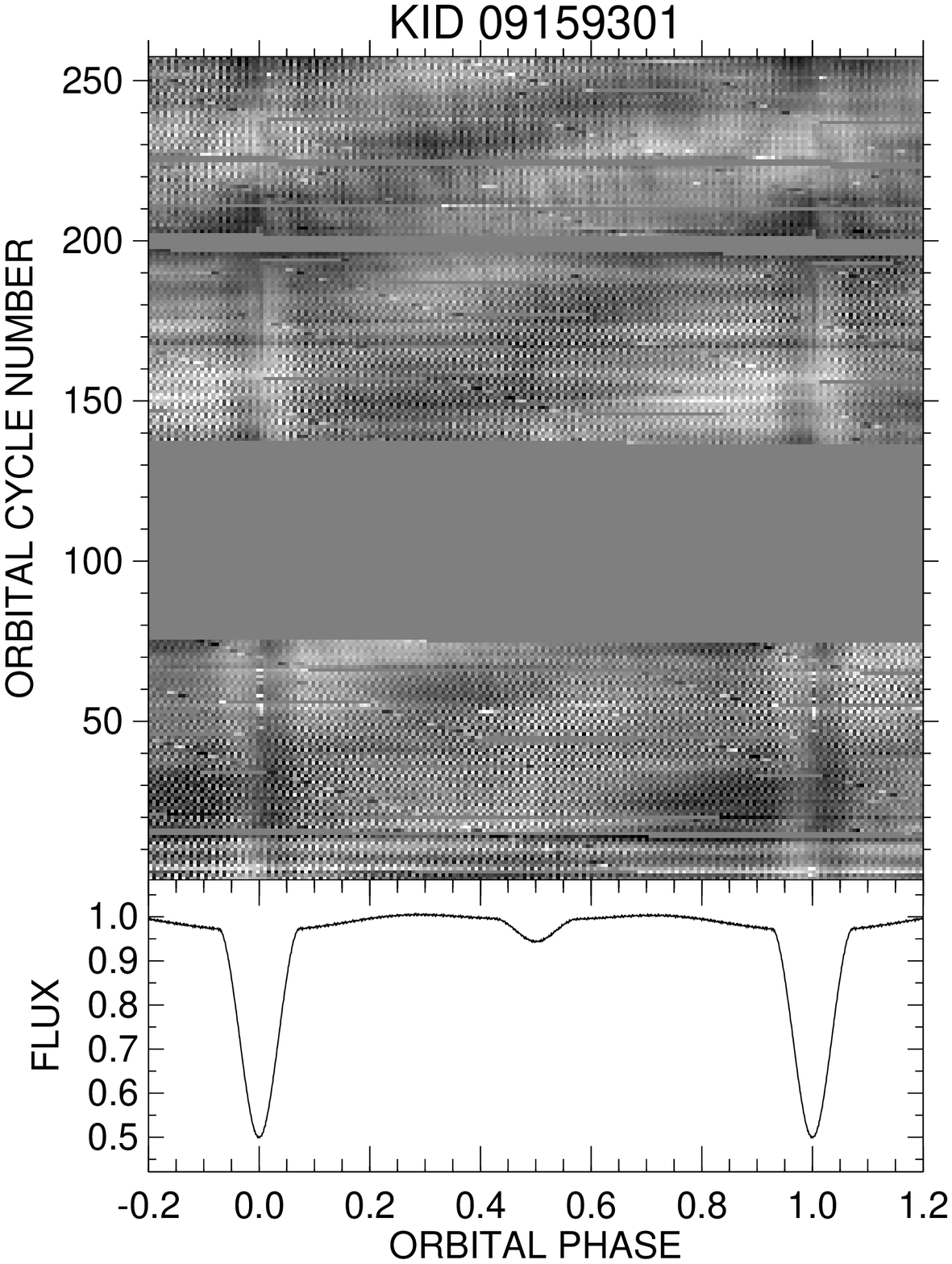}
\caption{The lower panel shows a mean, normalized light curve formed 
by binning in orbital phase.  The top panel shows the 
flux differences as a function of orbital phase and 
cycle number, represented as a gray scale diagram (range $\pm 0.4\%$). 
 }
\end{figure}

\clearpage
\setcounter{figure}{0}
\renewcommand{\thefigure}{\arabic{figure}.25}
\begin{figure}
\plotone{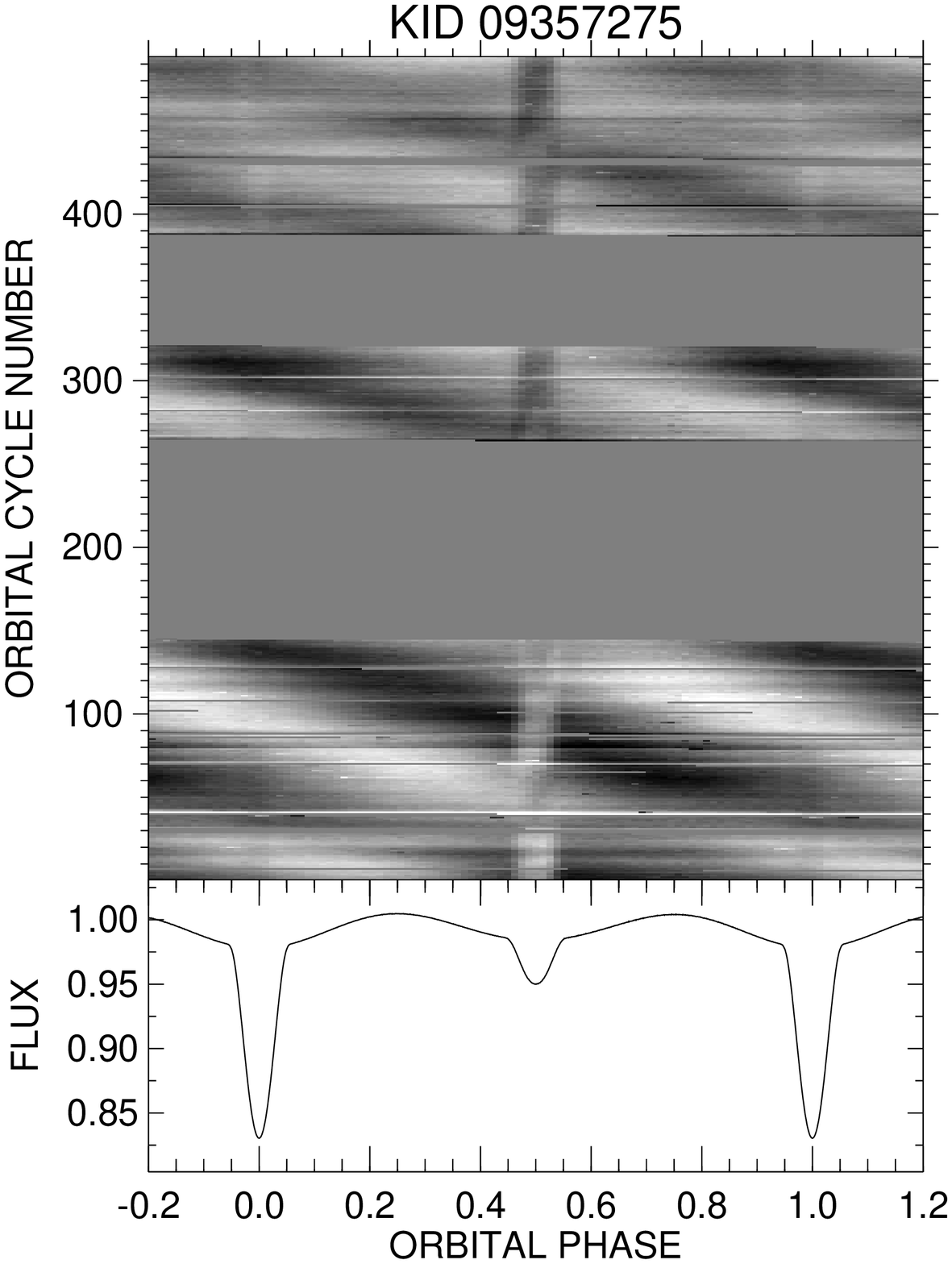}
\caption{The lower panel shows a mean, normalized light curve formed 
by binning in orbital phase.  The top panel shows the 
flux differences as a function of orbital phase and 
cycle number, represented as a gray scale diagram (range $\pm 0.2\%$). 
 }
\end{figure}

\clearpage
\setcounter{figure}{0}
\renewcommand{\thefigure}{\arabic{figure}.26}
\begin{figure}
\plotone{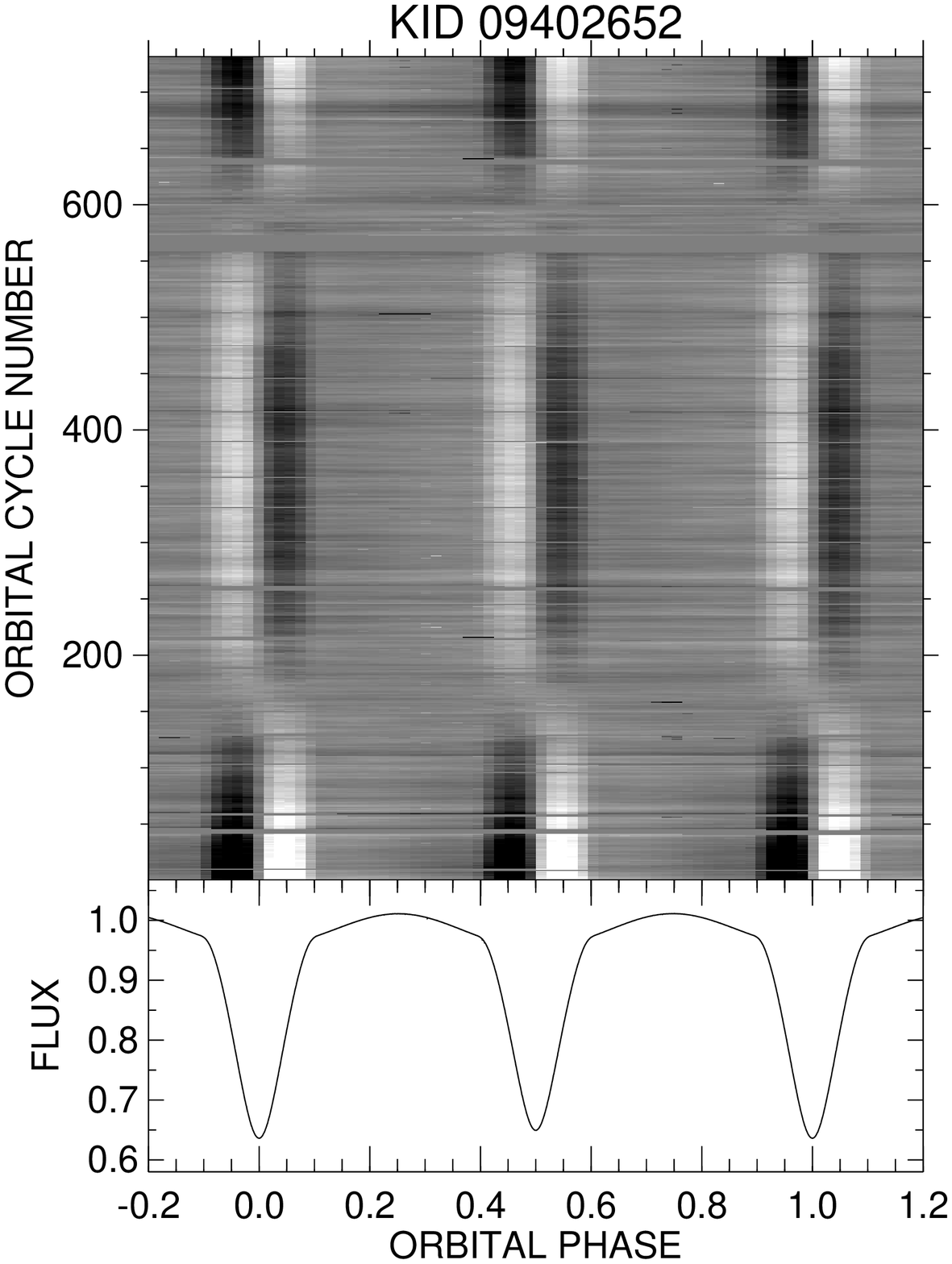}
\caption{The lower panel shows a mean, normalized light curve formed 
by binning in orbital phase.  The top panel shows the 
flux differences as a function of orbital phase and 
cycle number, represented as a gray scale diagram (range $\pm 0.5\%$). 
 }
\end{figure}

\clearpage
\setcounter{figure}{0}
\renewcommand{\thefigure}{\arabic{figure}.27}
\begin{figure}
\plotone{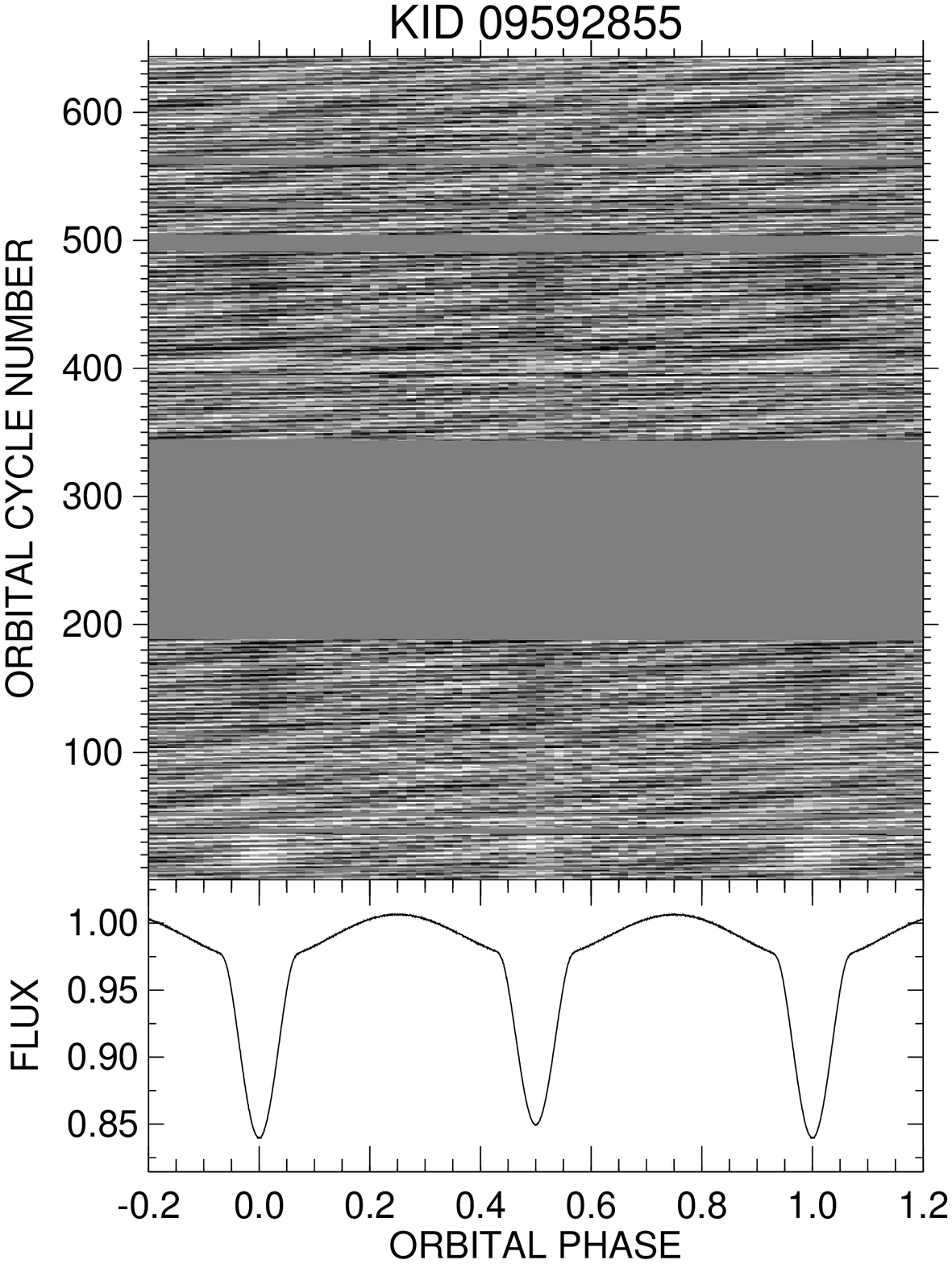}
\caption{The lower panel shows a mean, normalized light curve formed 
by binning in orbital phase.  The top panel shows the 
flux differences as a function of orbital phase and 
cycle number, represented as a gray scale diagram (range $\pm 0.3\%$). 
 }
\end{figure}

\clearpage
\setcounter{figure}{0}
\renewcommand{\thefigure}{\arabic{figure}.28}
\begin{figure}
\plotone{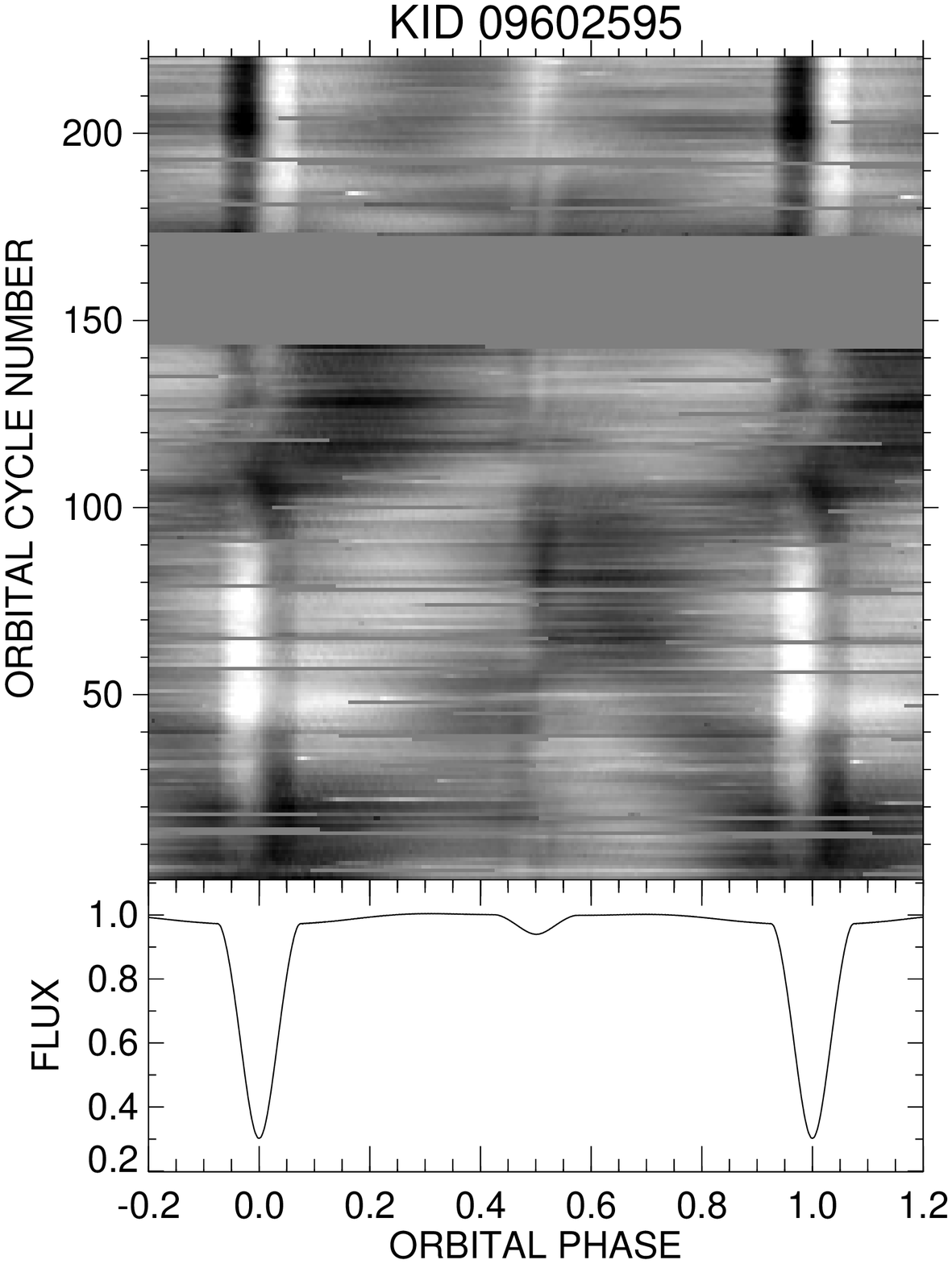}
\caption{The lower panel shows a mean, normalized light curve formed 
by binning in orbital phase.  The top panel shows the 
flux differences as a function of orbital phase and 
cycle number, represented as a gray scale diagram (range $\pm 0.5\%$). 
 }
\end{figure}

\clearpage
\setcounter{figure}{0}
\renewcommand{\thefigure}{\arabic{figure}.29}
\begin{figure}
\plotone{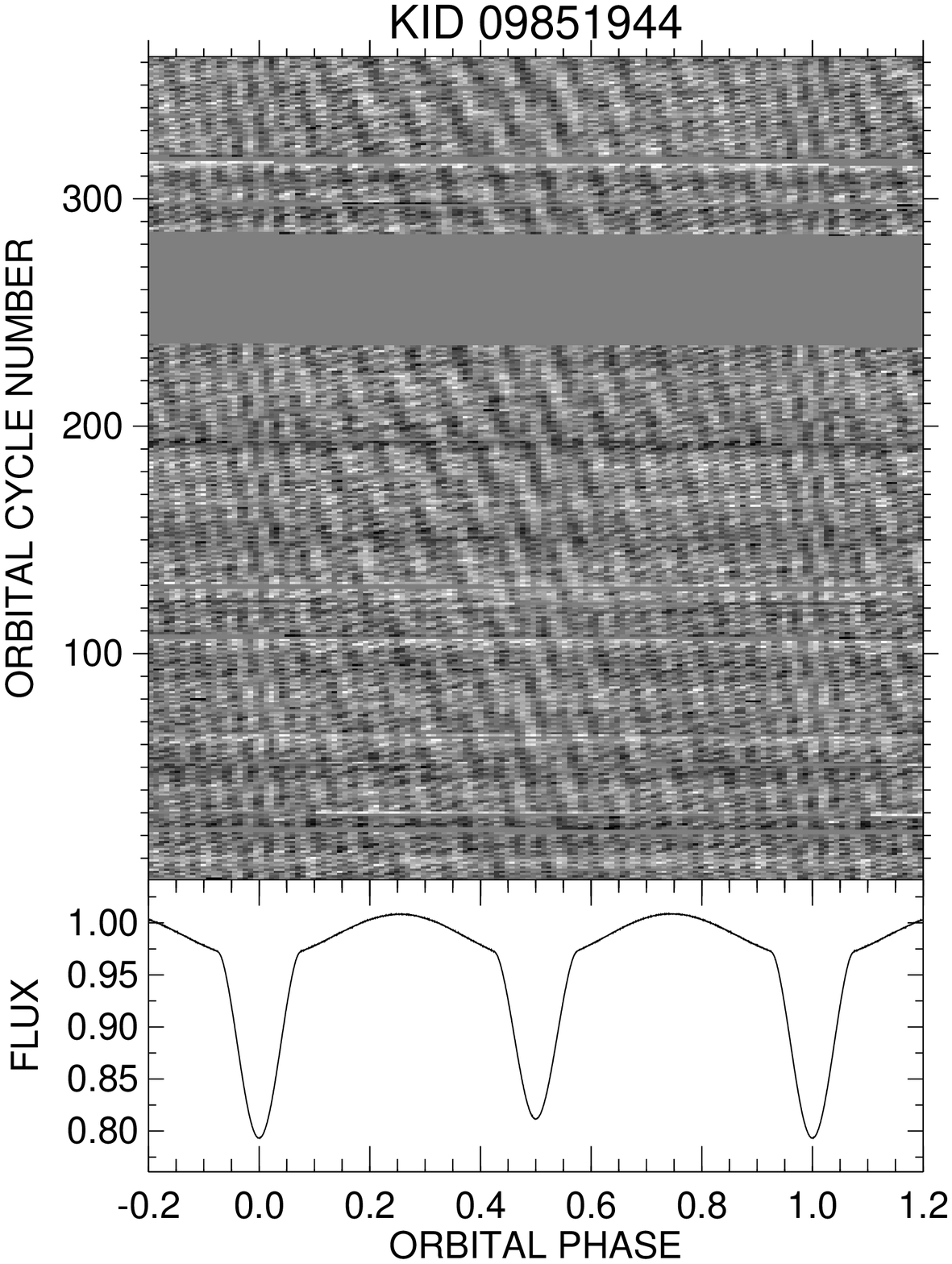}
\caption{The lower panel shows a mean, normalized light curve formed 
by binning in orbital phase.  The top panel shows the 
flux differences as a function of orbital phase and 
cycle number, represented as a gray scale diagram (range $\pm 0.3\%$). 
 }
\end{figure}

\clearpage
\setcounter{figure}{0}
\renewcommand{\thefigure}{\arabic{figure}.30}
\begin{figure}
\plotone{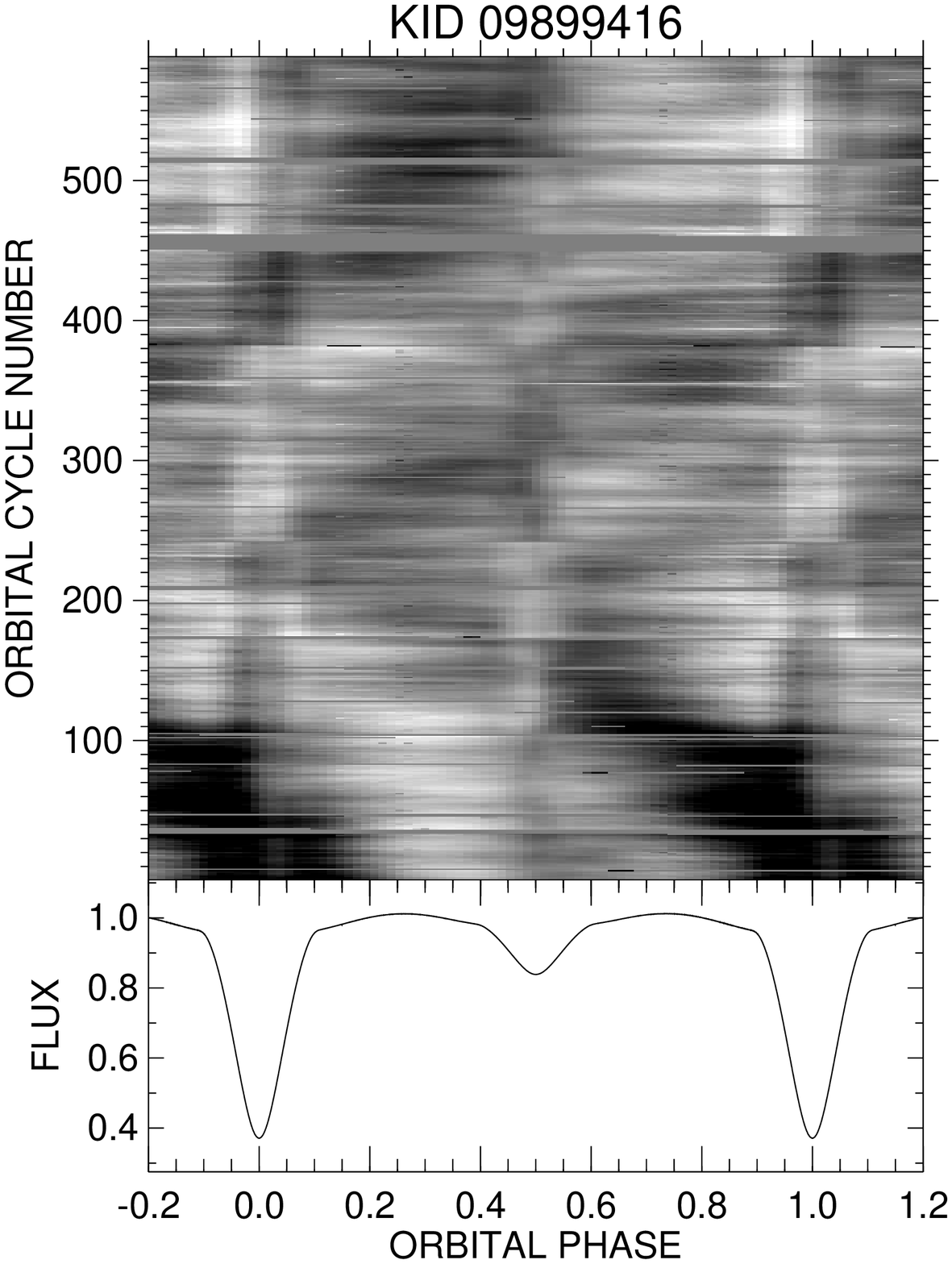}
\caption{The lower panel shows a mean, normalized light curve formed 
by binning in orbital phase.  The top panel shows the 
flux differences as a function of orbital phase and 
cycle number, represented as a gray scale diagram (range $\pm 0.4\%$). 
 }
\end{figure}

\clearpage
\setcounter{figure}{0}
\renewcommand{\thefigure}{\arabic{figure}.31}
\begin{figure}
\plotone{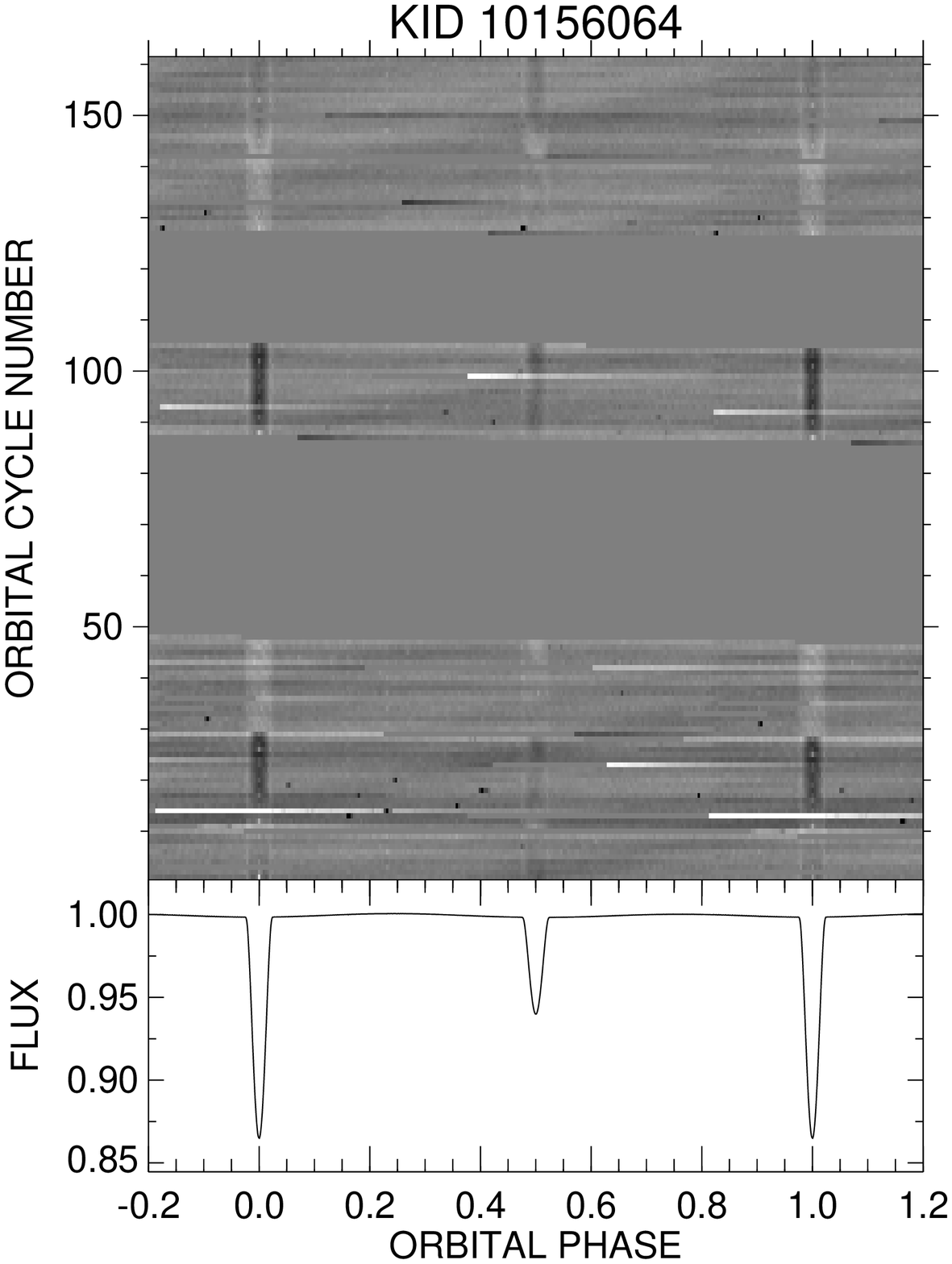}
\caption{The lower panel shows a mean, normalized light curve formed 
by binning in orbital phase.  The top panel shows the 
flux differences as a function of orbital phase and 
cycle number, represented as a gray scale diagram (range $\pm 0.1\%$). 
 }
\end{figure}

\clearpage
\setcounter{figure}{0}
\renewcommand{\thefigure}{\arabic{figure}.32}
\begin{figure}
\plotone{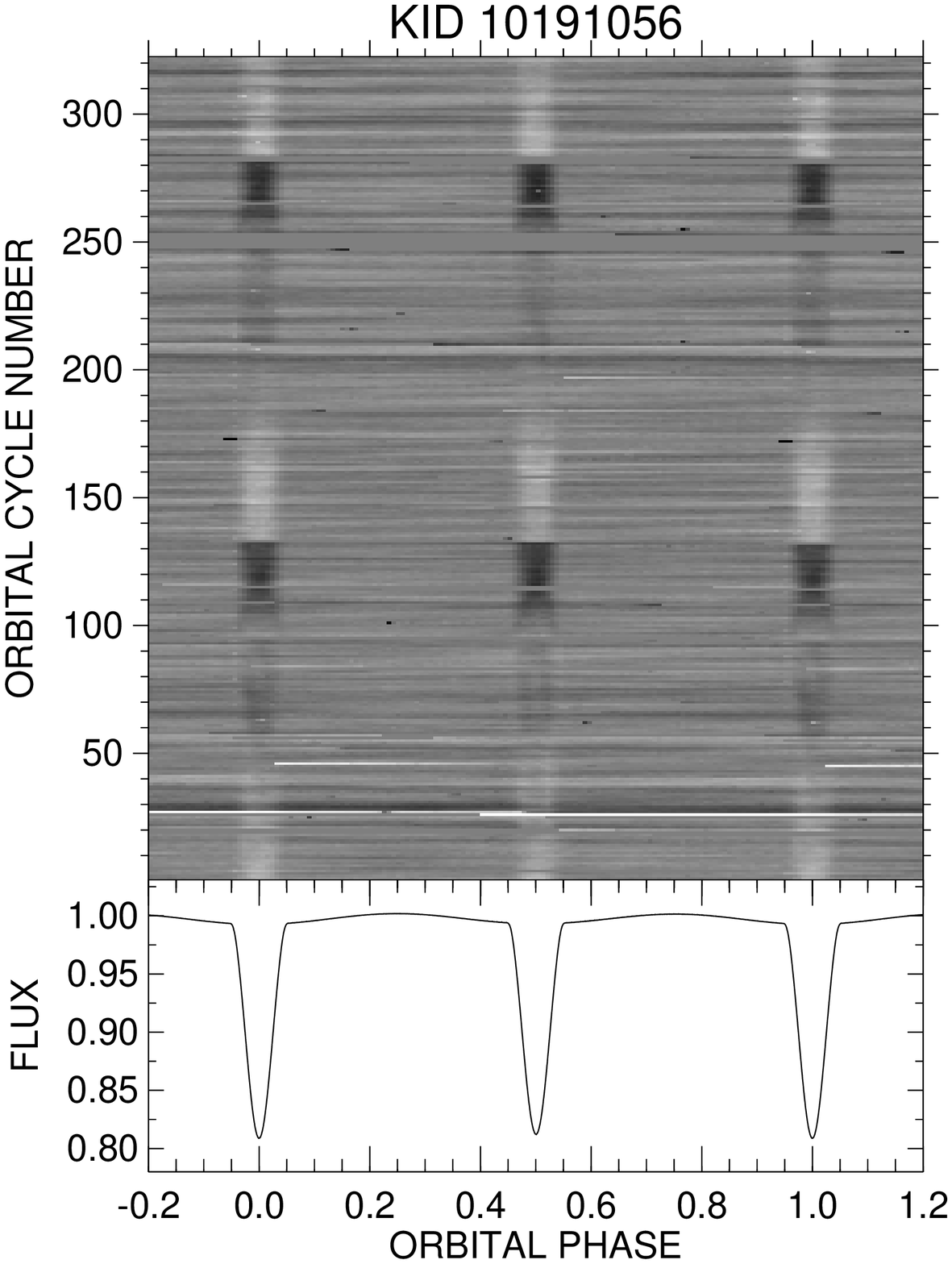}
\caption{The lower panel shows a mean, normalized light curve formed 
by binning in orbital phase.  The top panel shows the 
flux differences as a function of orbital phase and 
cycle number, represented as a gray scale diagram (range $\pm 0.2\%$). 
 }
\end{figure}

\clearpage
\setcounter{figure}{0}
\renewcommand{\thefigure}{\arabic{figure}.33}
\begin{figure}
\plotone{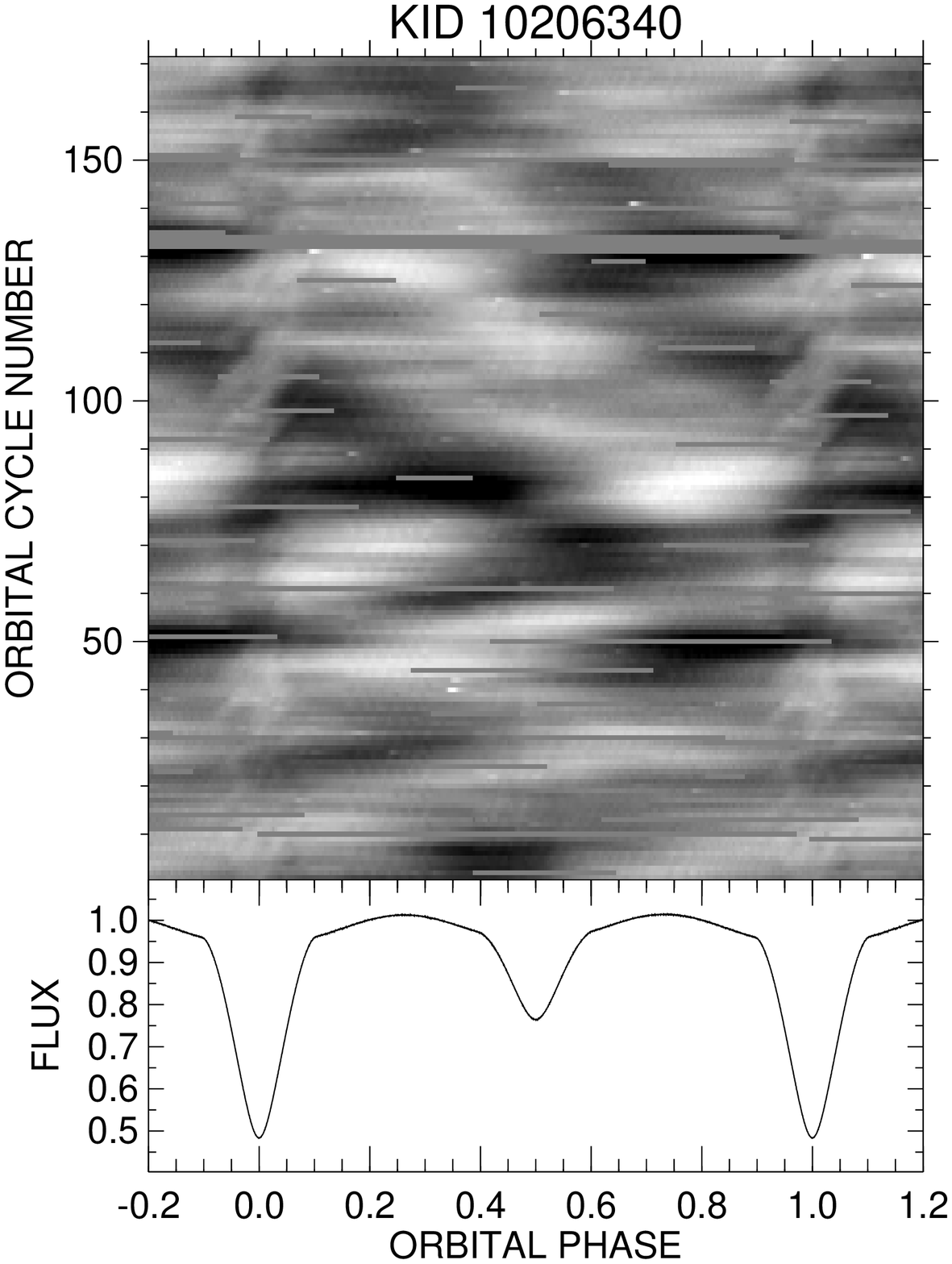}
\caption{The lower panel shows a mean, normalized light curve formed 
by binning in orbital phase.  The top panel shows the 
flux differences as a function of orbital phase and 
cycle number, represented as a gray scale diagram (range $\pm 2\%$). 
 }
\end{figure}

\clearpage
\setcounter{figure}{0}
\renewcommand{\thefigure}{\arabic{figure}.34}
\begin{figure}
\plotone{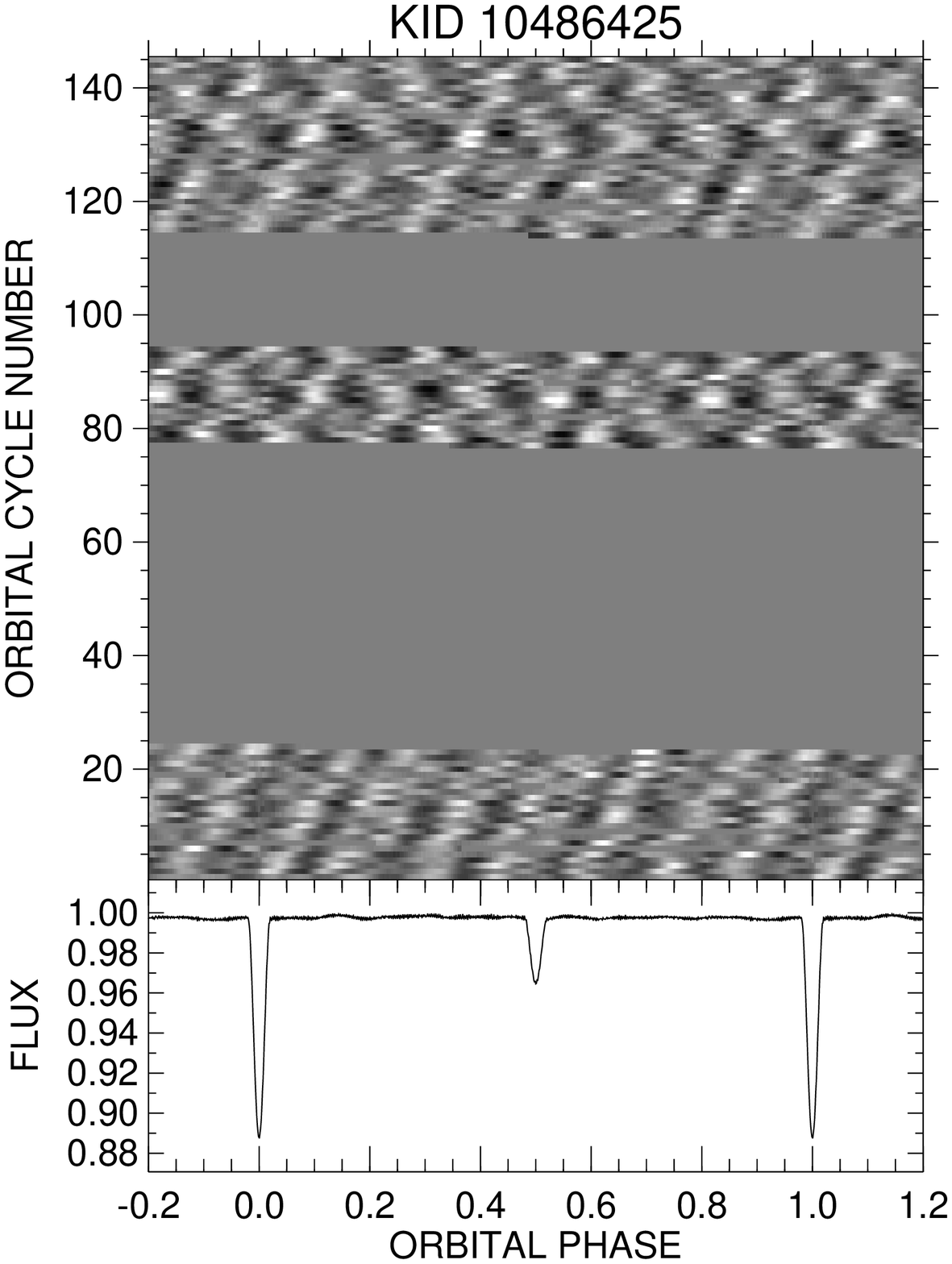}
\caption{The lower panel shows a mean, normalized light curve formed 
by binning in orbital phase.  The top panel shows the 
flux differences as a function of orbital phase and 
cycle number, represented as a gray scale diagram (range $\pm 1\%$). 
 }
\end{figure}

\clearpage
\setcounter{figure}{0}
\renewcommand{\thefigure}{\arabic{figure}.35}
\begin{figure}
\plotone{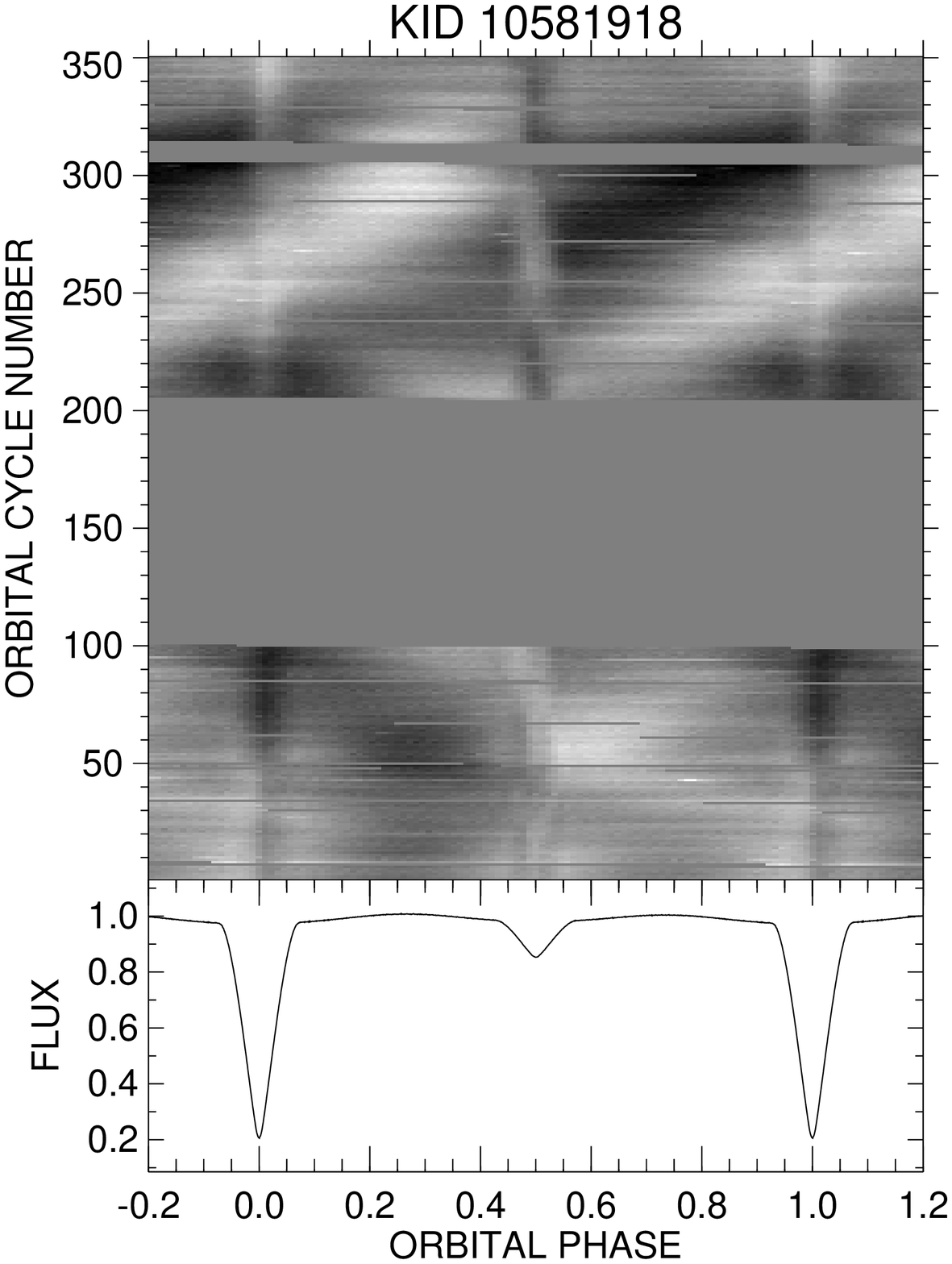}
\caption{The lower panel shows a mean, normalized light curve formed 
by binning in orbital phase.  The top panel shows the 
flux differences as a function of orbital phase and 
cycle number, represented as a gray scale diagram (range $\pm 2\%$). 
 }
\end{figure}

\clearpage
\setcounter{figure}{0}
\renewcommand{\thefigure}{\arabic{figure}.36}
\begin{figure}
\plotone{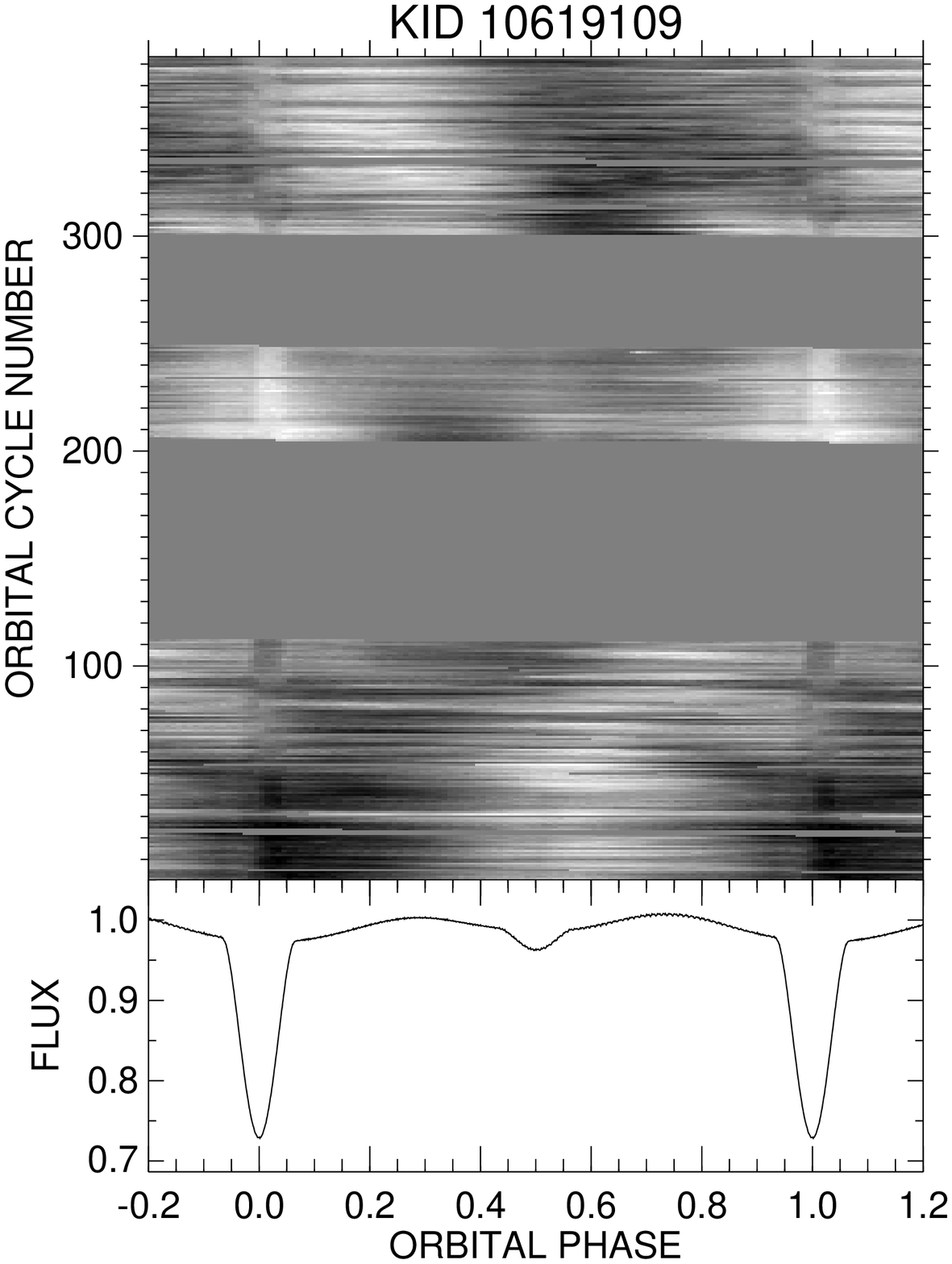}
\caption{The lower panel shows a mean, normalized light curve formed 
by binning in orbital phase.  The top panel shows the 
flux differences as a function of orbital phase and 
cycle number, represented as a gray scale diagram (range $\pm 1\%$). 
 }
\end{figure}

\clearpage
\setcounter{figure}{0}
\renewcommand{\thefigure}{\arabic{figure}.37}
\begin{figure}
\plotone{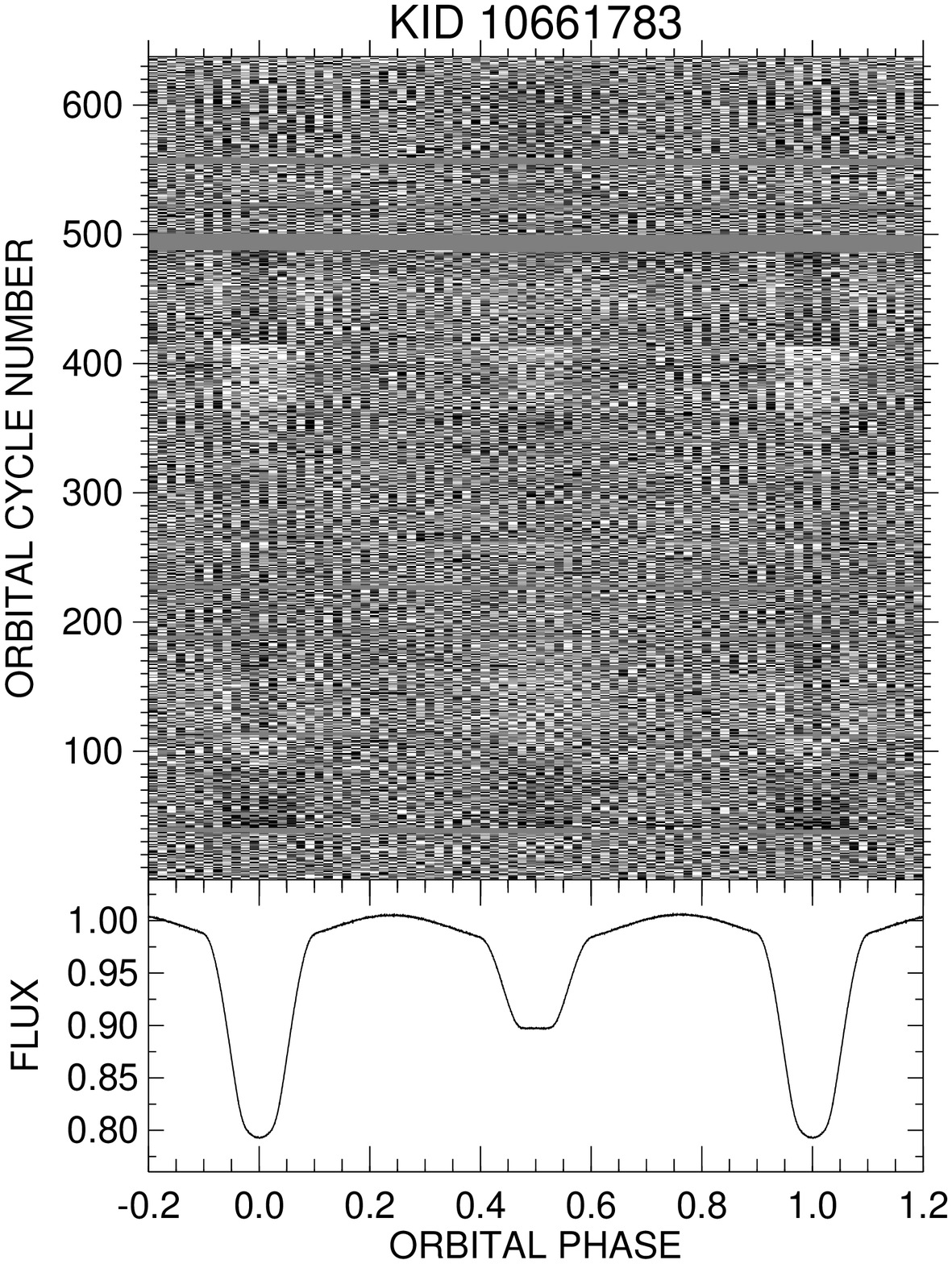}
\caption{The lower panel shows a mean, normalized light curve formed 
by binning in orbital phase.  The top panel shows the 
flux differences as a function of orbital phase and 
cycle number, represented as a gray scale diagram (range $\pm 0.2\%$). 
 }
\end{figure}

\clearpage
\setcounter{figure}{0}
\renewcommand{\thefigure}{\arabic{figure}.38}
\begin{figure}
\plotone{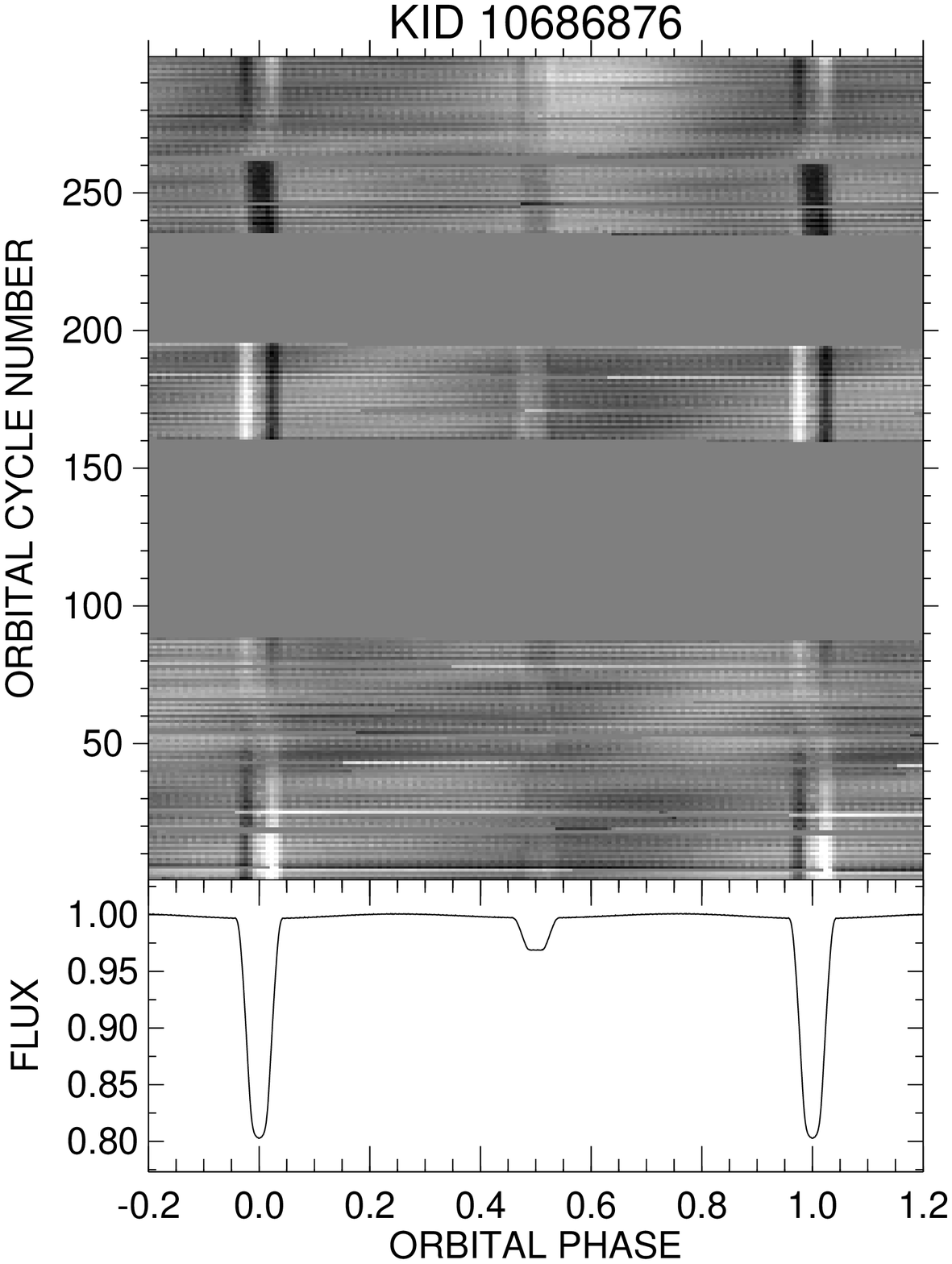}
\caption{The lower panel shows a mean, normalized light curve formed 
by binning in orbital phase.  The top panel shows the 
flux differences as a function of orbital phase and 
cycle number, represented as a gray scale diagram (range $\pm 0.3\%$). 
 }
\end{figure}

\clearpage
\setcounter{figure}{0}
\renewcommand{\thefigure}{\arabic{figure}.39}
\begin{figure}
\plotone{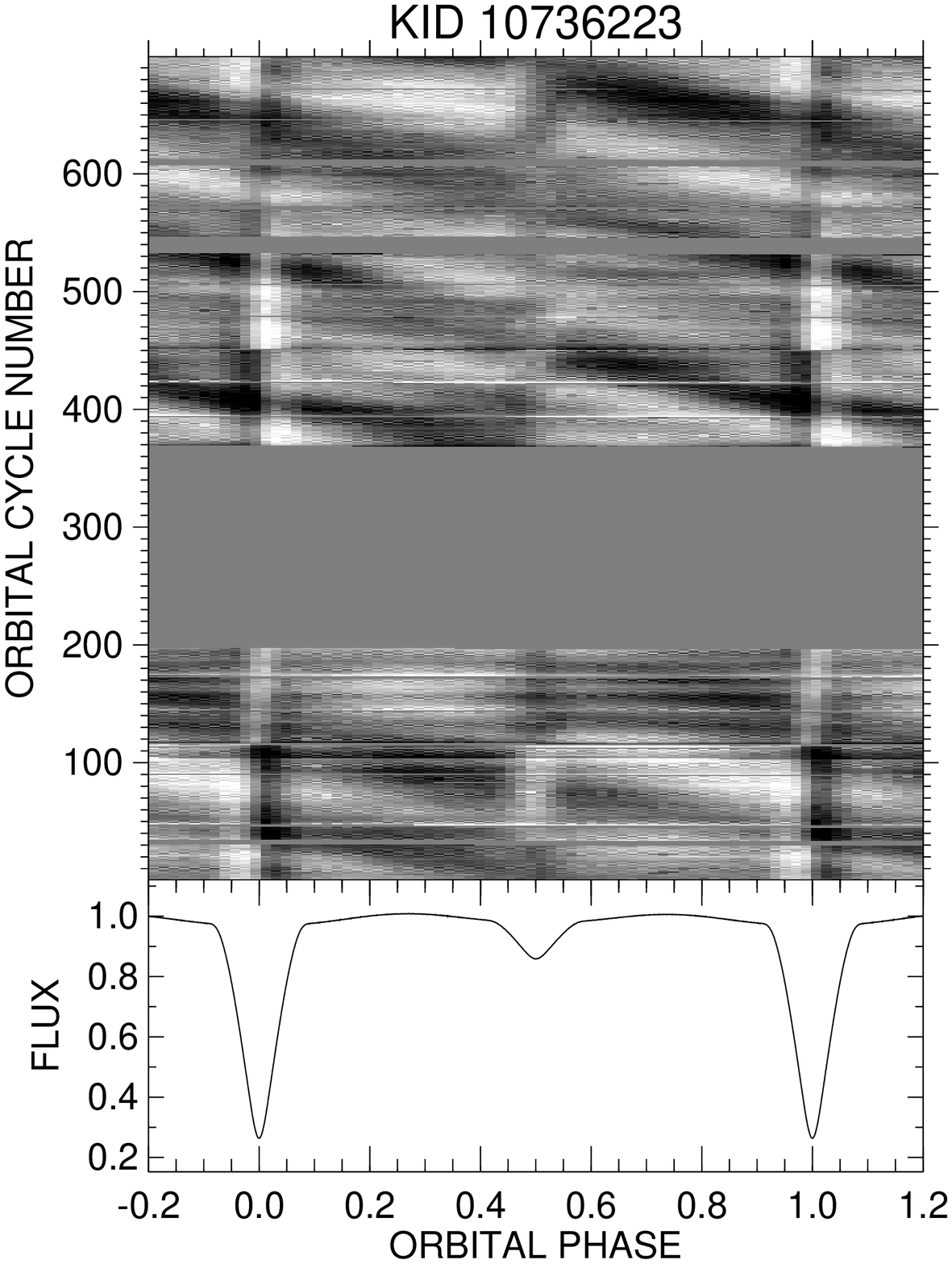}
\caption{The lower panel shows a mean, normalized light curve formed 
by binning in orbital phase.  The top panel shows the 
flux differences as a function of orbital phase and 
cycle number, represented as a gray scale diagram (range $\pm 0.5\%$). 
 }
\end{figure}

\clearpage
\setcounter{figure}{0}
\renewcommand{\thefigure}{\arabic{figure}.40}
\begin{figure}
\plotone{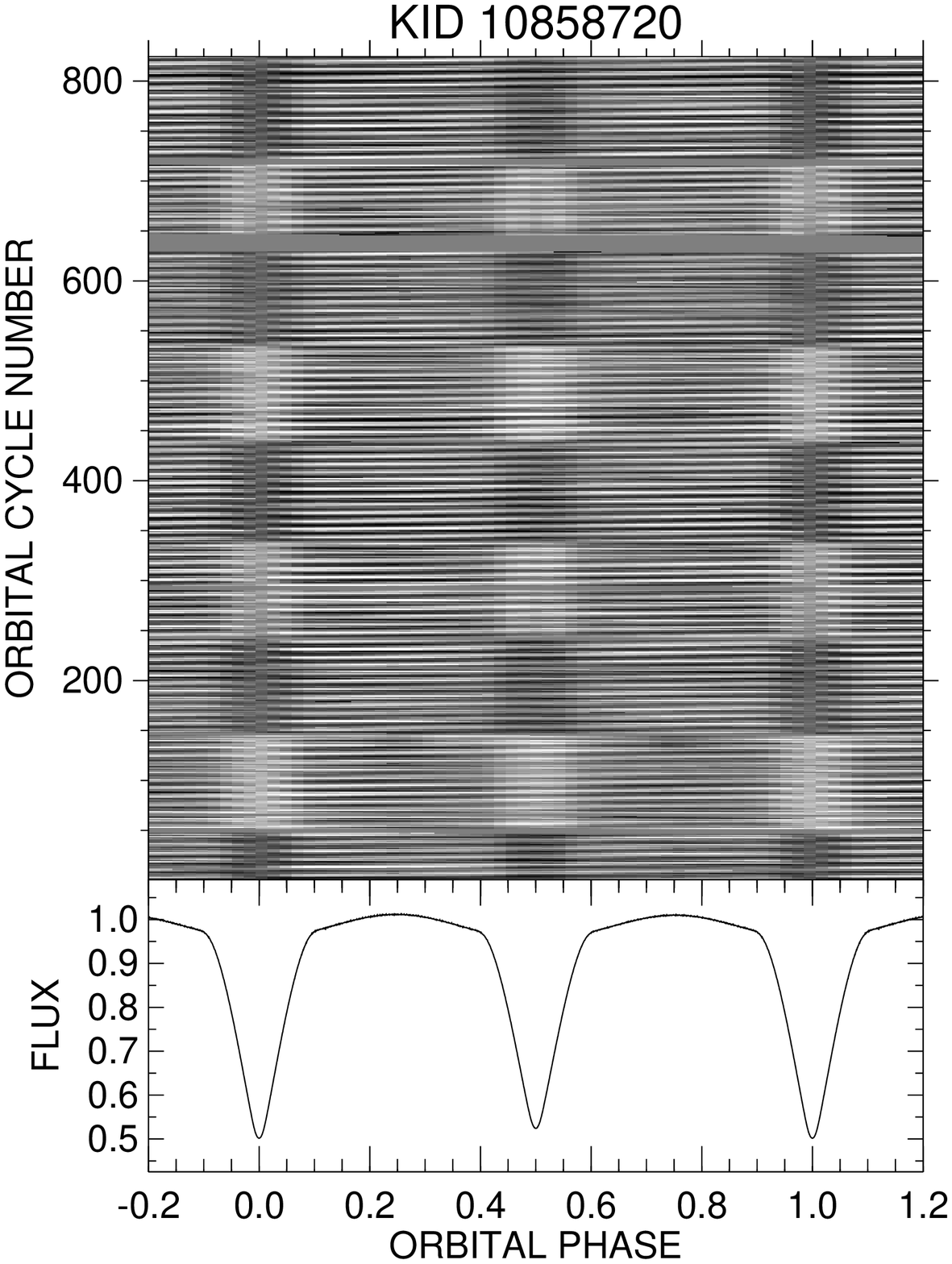}
\caption{The lower panel shows a mean, normalized light curve formed 
by binning in orbital phase.  The top panel shows the 
flux differences as a function of orbital phase and 
cycle number, represented as a gray scale diagram (range $\pm 0.5\%$). 
 }
\end{figure}

\clearpage
\setcounter{figure}{0}
\renewcommand{\thefigure}{\arabic{figure}.41}
\begin{figure}
\plotone{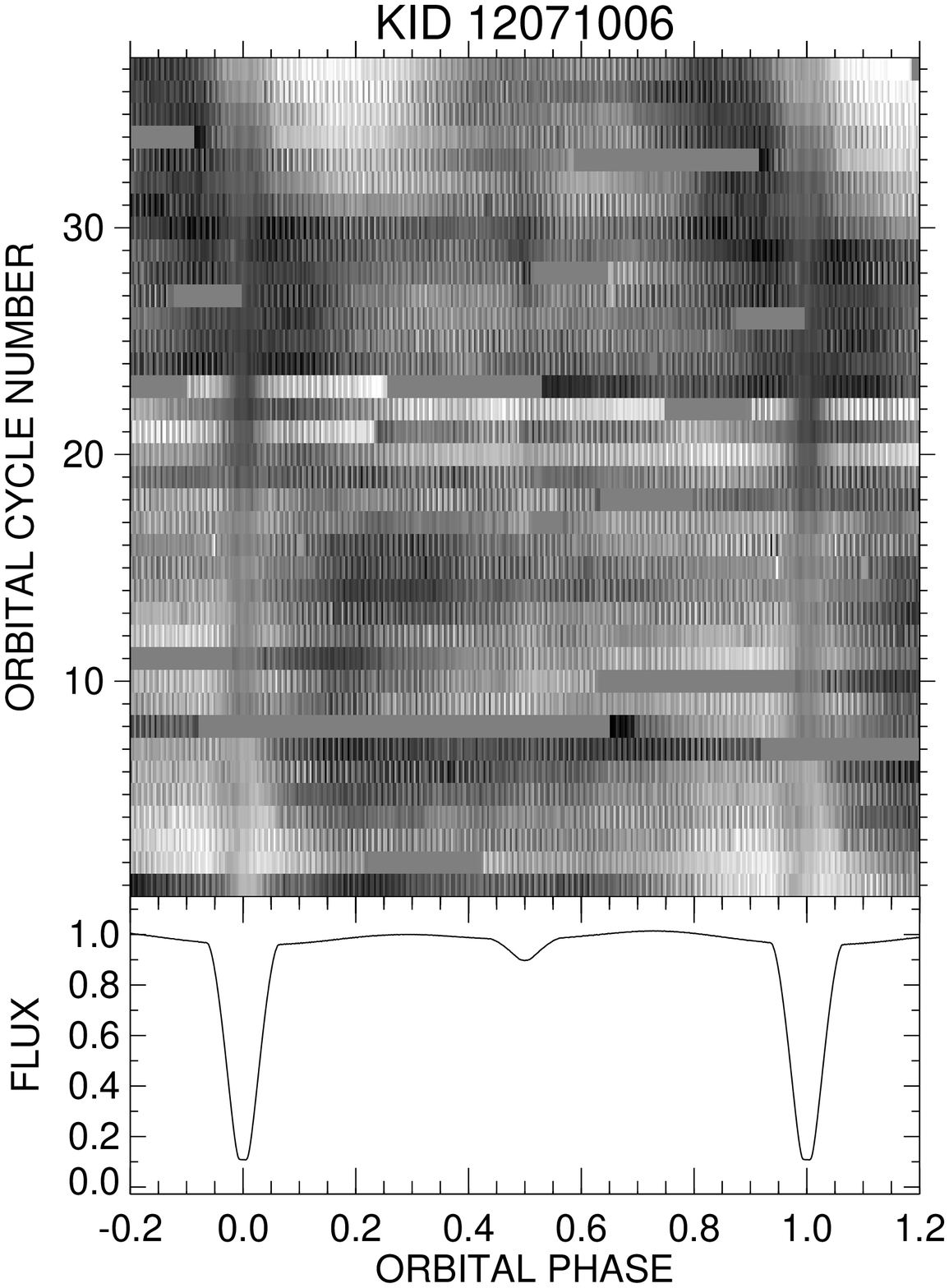}
\caption{The lower panel shows a mean, normalized light curve formed 
by binning in orbital phase.  The top panel shows the 
flux differences as a function of orbital phase and 
cycle number, represented as a gray scale diagram (range $\pm 1\%$). 
 }
\end{figure}

\clearpage
\setcounter{figure}{1}
\renewcommand{\thefigure}{\arabic{figure}.1}
\begin{figure}
\begin{center}
{\includegraphics[angle=90,height=12cm]{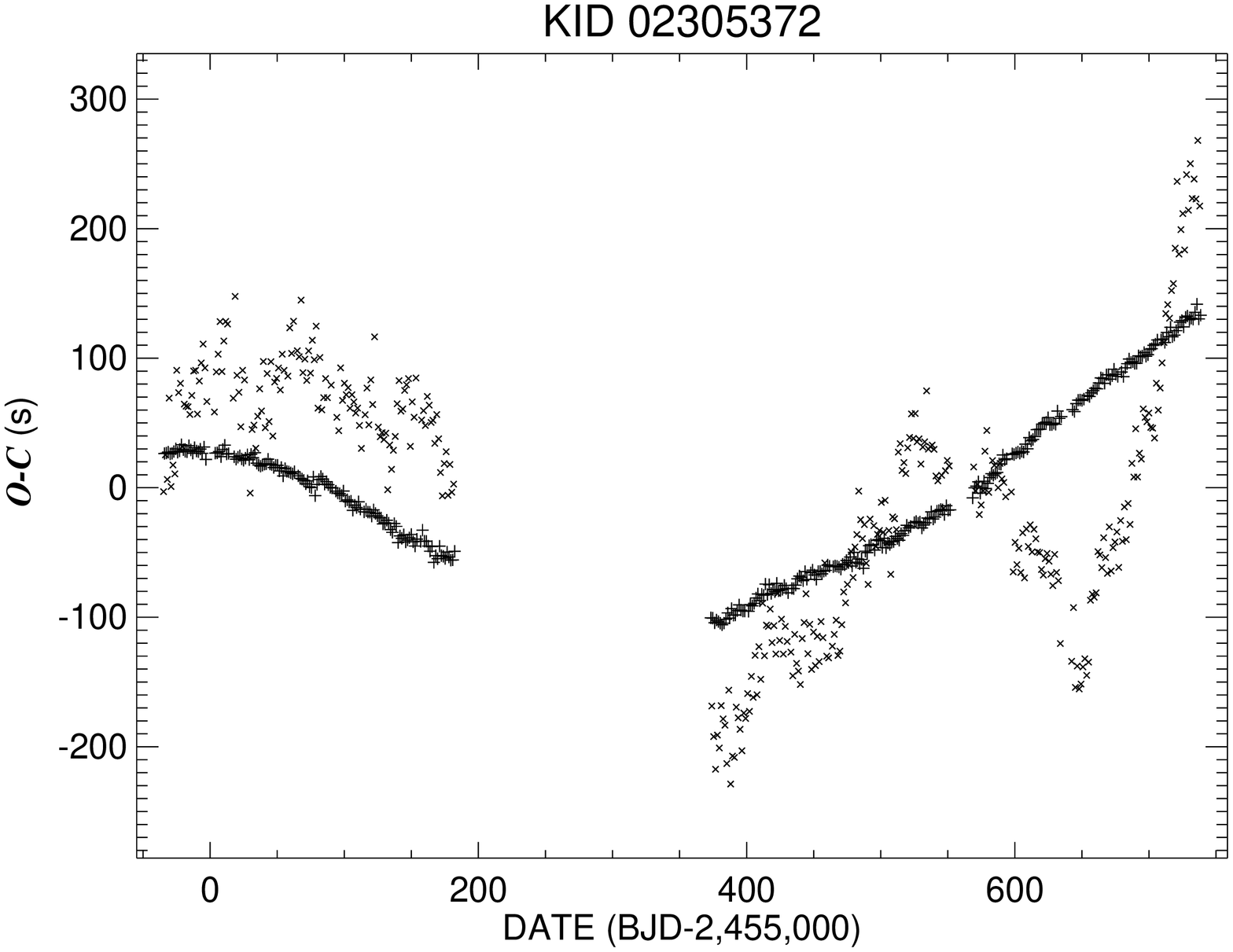}}
\end{center}
\caption{The observed minus calculated eclipse times relative to
a linear ephemeris.  The primary and secondary eclipse
times are indicated by $+$ and $\times$ symbols, 
respectively.}
\end{figure}

\clearpage
\setcounter{figure}{1}
\renewcommand{\thefigure}{\arabic{figure}.2}
\begin{figure}
\begin{center}
{\includegraphics[angle=90,height=12cm]{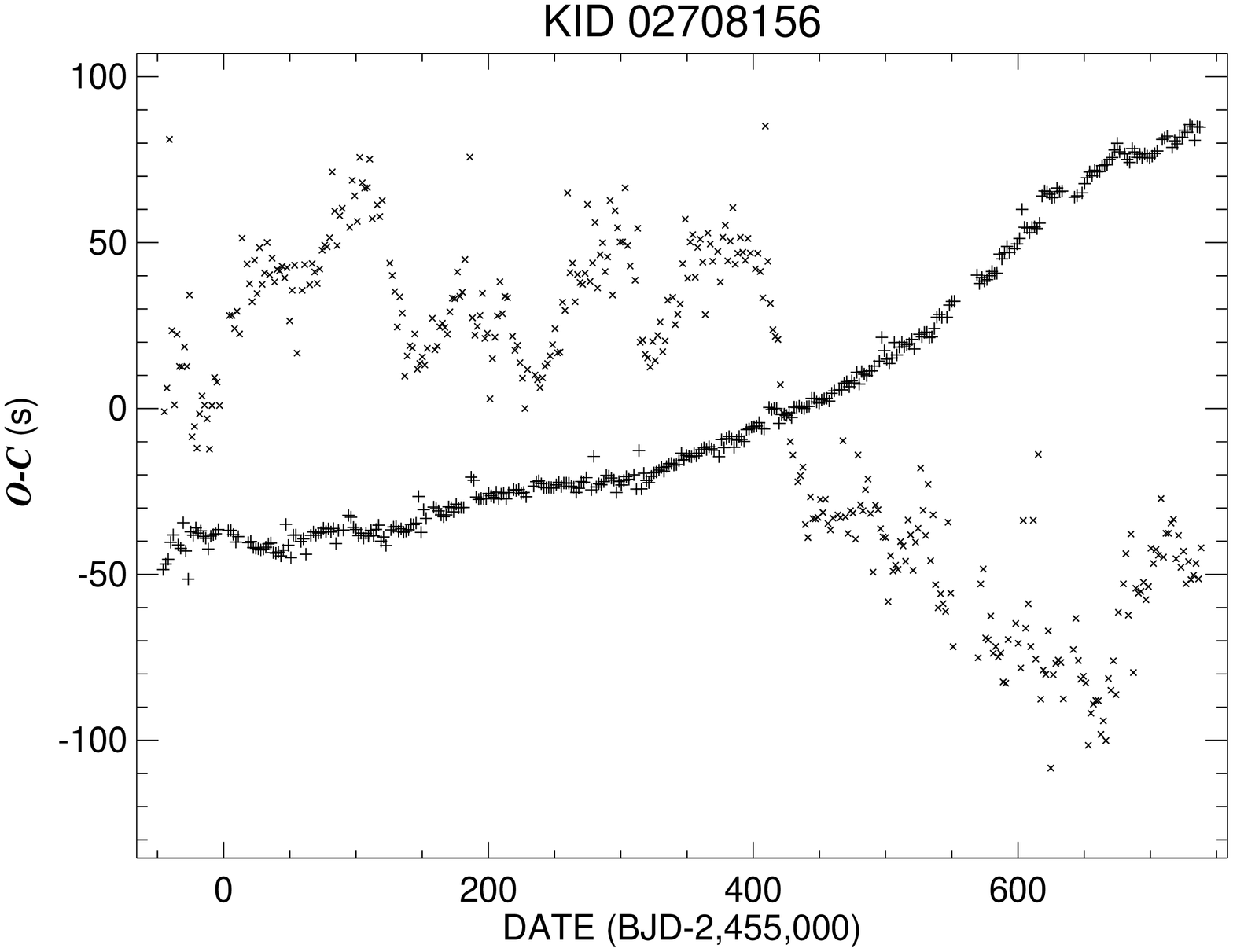}}
\end{center}
\caption{The observed minus calculated eclipse times relative to
a linear ephemeris.  The primary and secondary eclipse
times are indicated by $+$ and $\times$ symbols, 
respectively.}
\end{figure}

\clearpage
\setcounter{figure}{1}
\renewcommand{\thefigure}{\arabic{figure}.3}
\begin{figure}
\begin{center}
{\includegraphics[angle=90,height=12cm]{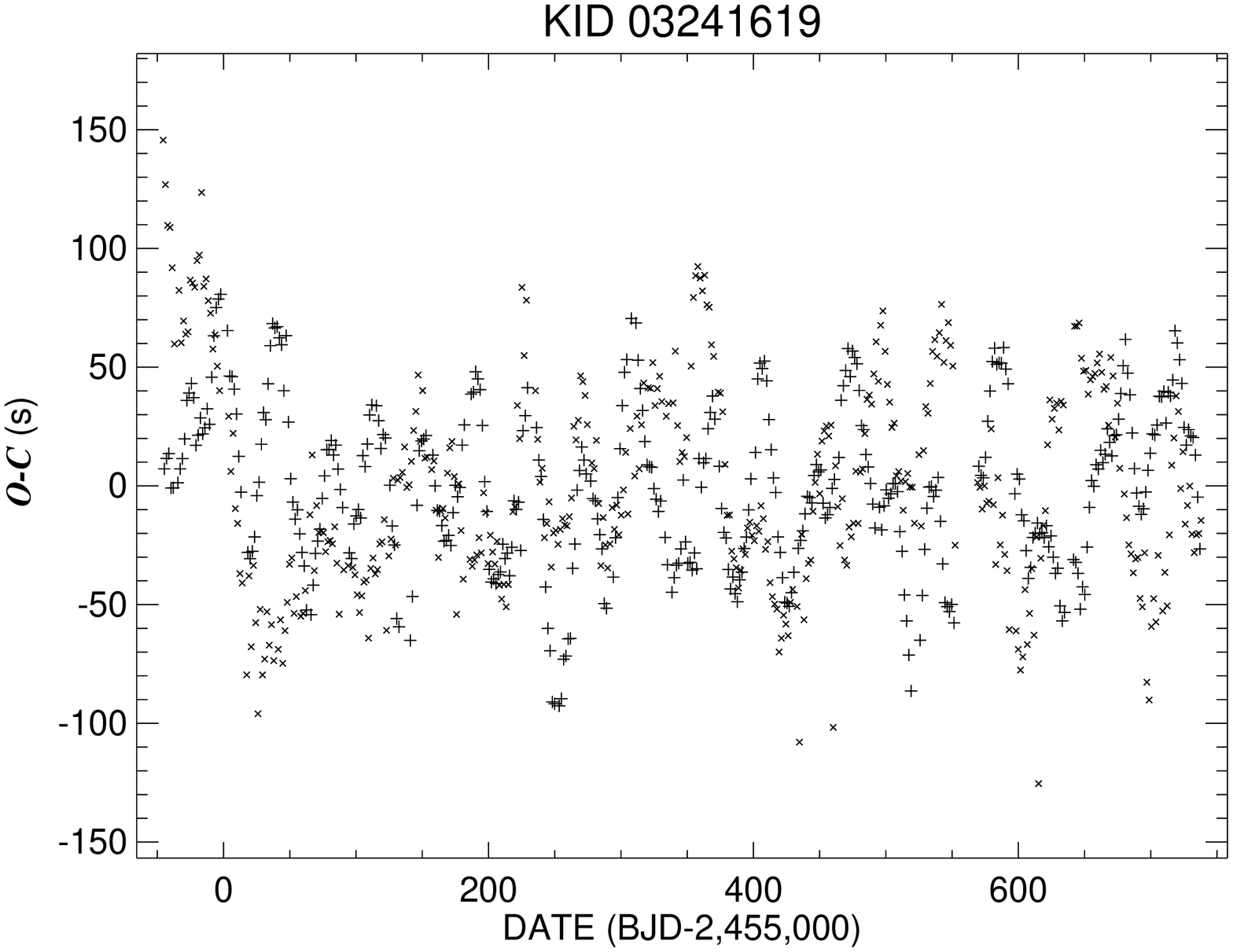}}
\end{center}
\caption{The observed minus calculated eclipse times relative to
a linear ephemeris.  The primary and secondary eclipse
times are indicated by $+$ and $\times$ symbols, 
respectively.}
\end{figure}

\clearpage
\setcounter{figure}{1}
\renewcommand{\thefigure}{\arabic{figure}.4}
\begin{figure}
\begin{center}
{\includegraphics[angle=90,height=12cm]{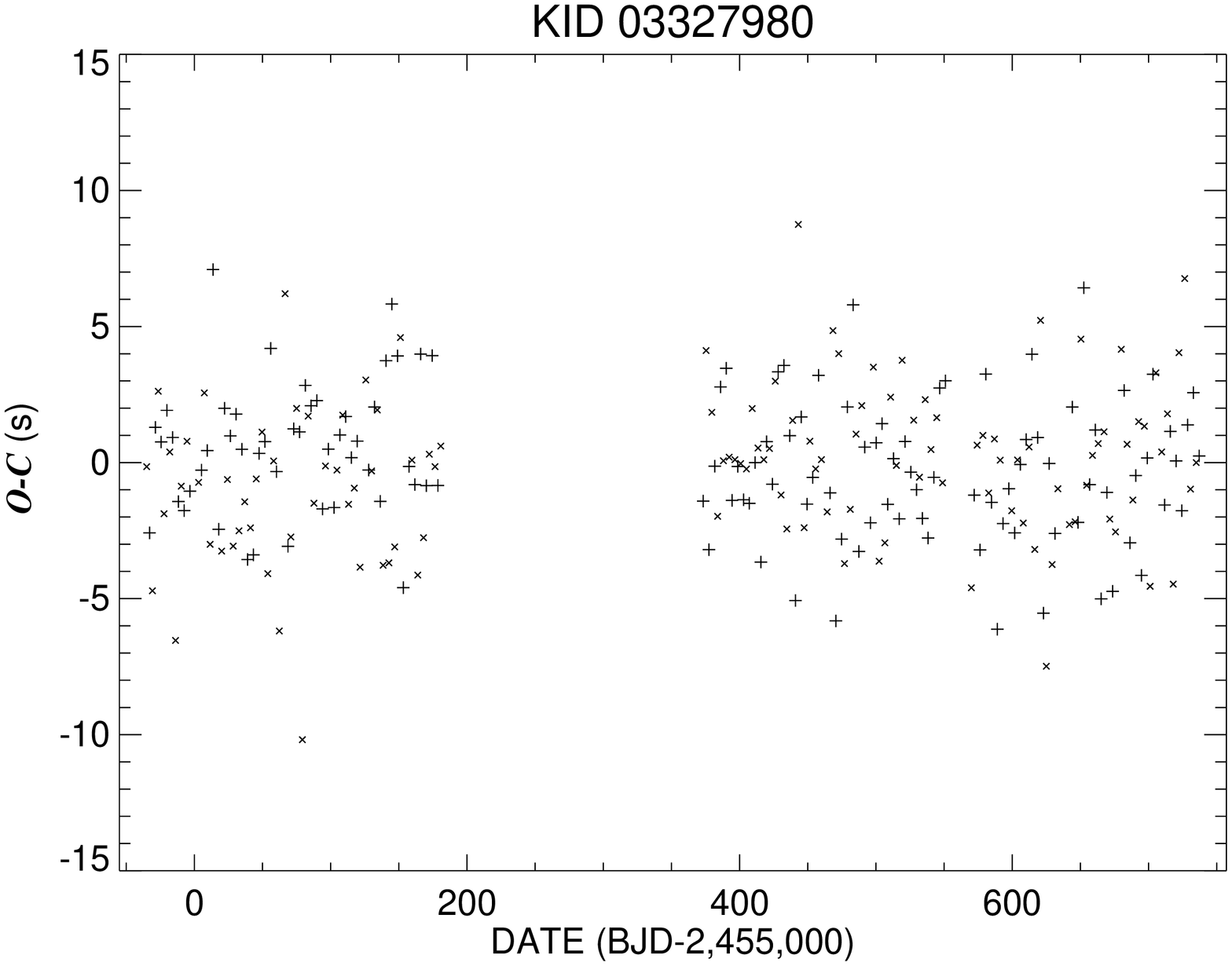}}
\end{center}
\caption{The observed minus calculated eclipse times relative to
a linear ephemeris.  The primary and secondary eclipse
times are indicated by $+$ and $\times$ symbols, 
respectively.}
\end{figure}

\clearpage
\setcounter{figure}{1}
\renewcommand{\thefigure}{\arabic{figure}.5}
\begin{figure}
\begin{center}
{\includegraphics[angle=90,height=12cm]{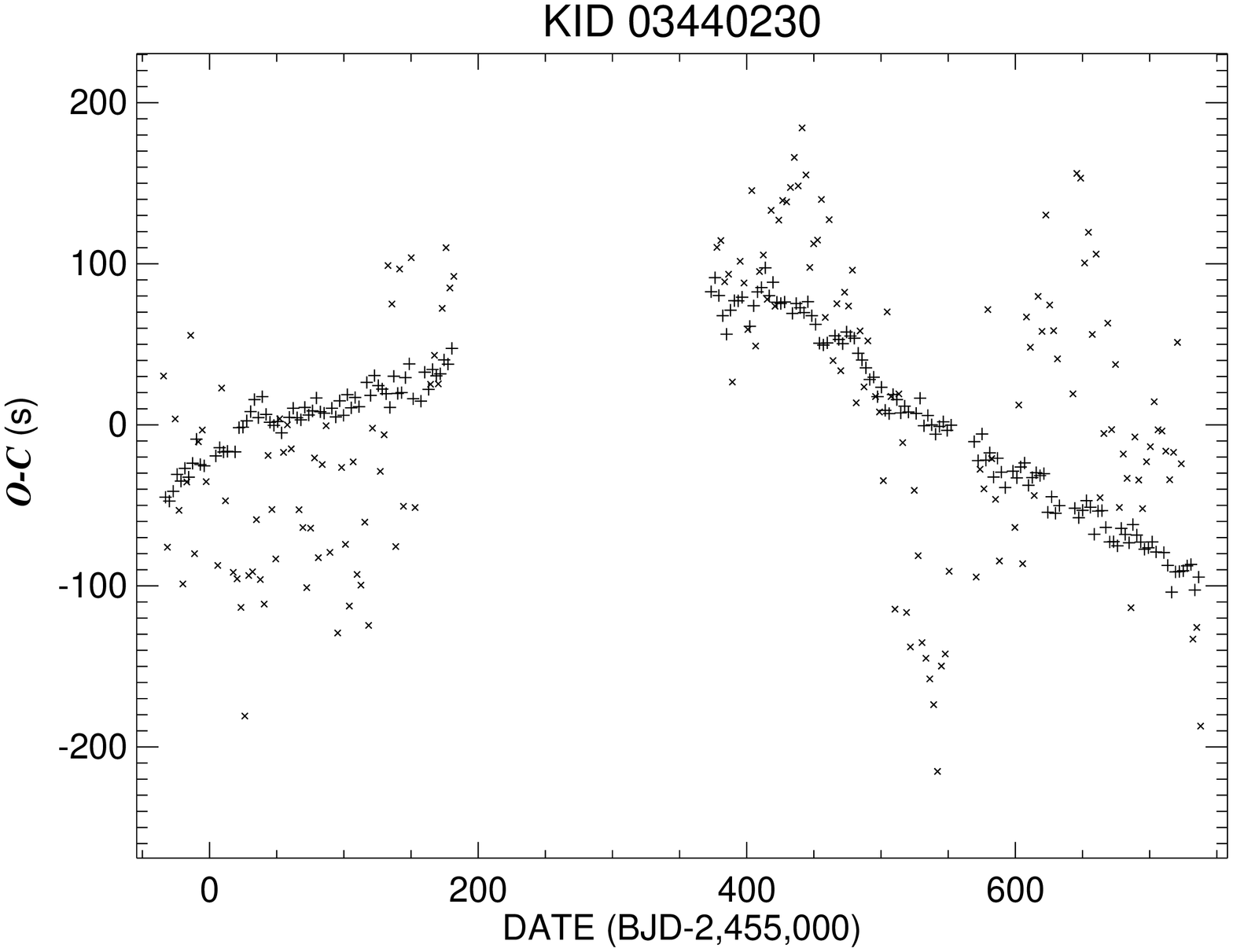}}
\end{center}
\caption{The observed minus calculated eclipse times relative to
a linear ephemeris.  The primary and secondary eclipse
times are indicated by $+$ and $\times$ symbols, 
respectively.}
\end{figure}

\clearpage
\setcounter{figure}{1}
\renewcommand{\thefigure}{\arabic{figure}.6}
\begin{figure}
\begin{center}
{\includegraphics[angle=90,height=12cm]{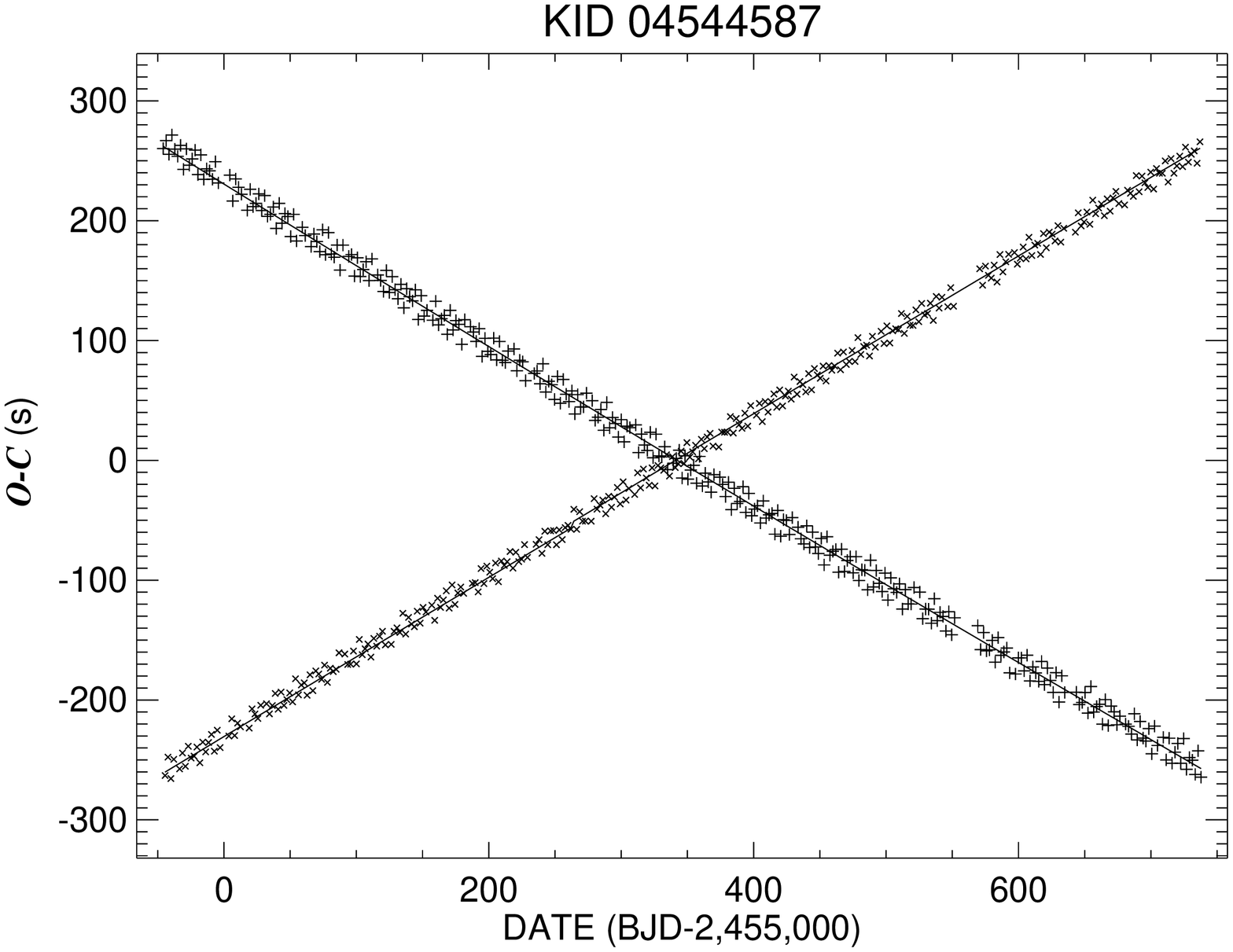}}
\end{center}
\caption{The observed minus calculated eclipse times relative to
a linear ephemeris.  The primary and secondary eclipse
times are indicated by $+$ and $\times$ symbols, 
respectively.}
\end{figure}

\clearpage
\setcounter{figure}{1}
\renewcommand{\thefigure}{\arabic{figure}.7}
\begin{figure}
\begin{center}
{\includegraphics[angle=90,height=12cm]{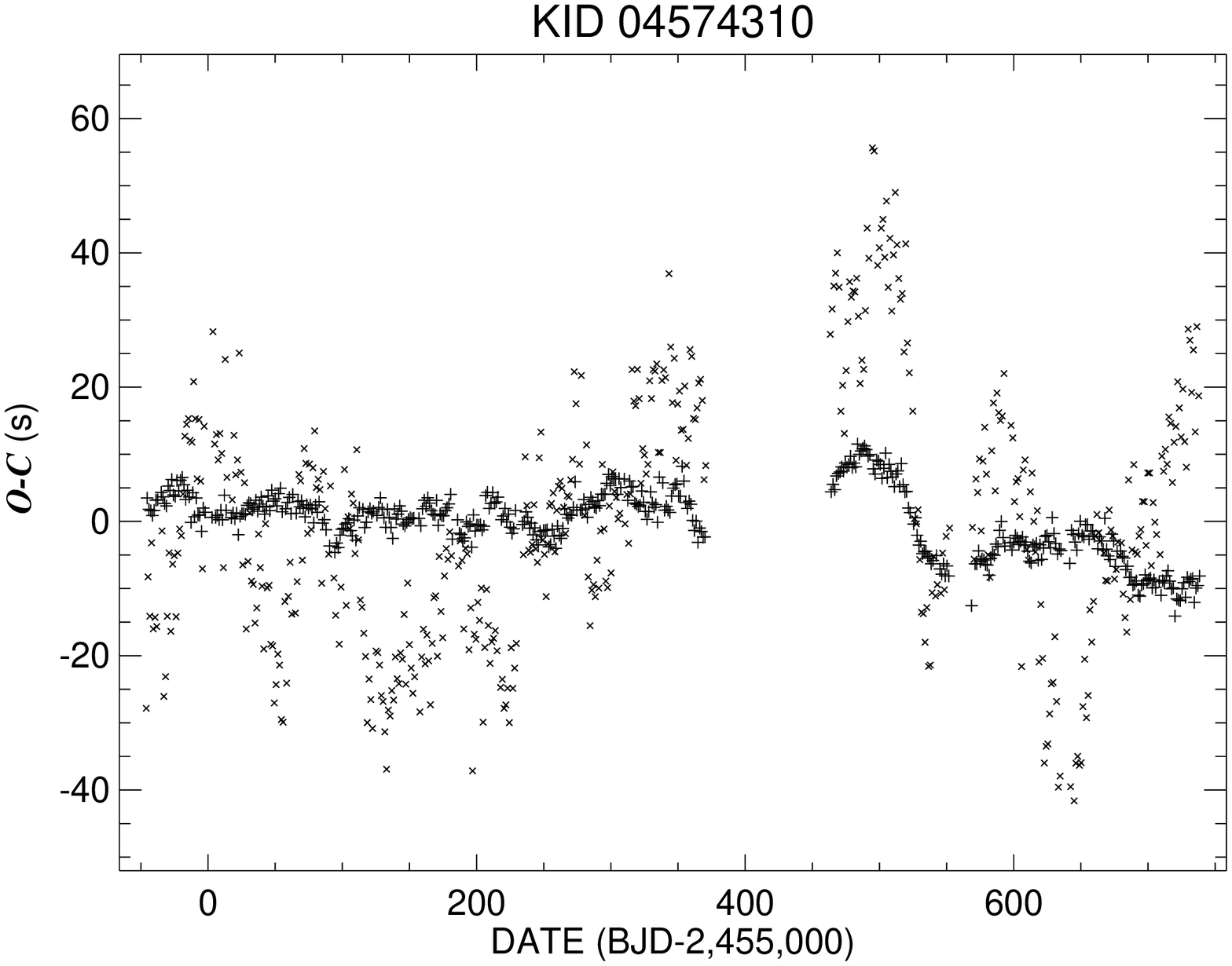}}
\end{center}
\caption{The observed minus calculated eclipse times relative to
a linear ephemeris.  The primary and secondary eclipse
times are indicated by $+$ and $\times$ symbols, 
respectively.}
\end{figure}

\clearpage
\setcounter{figure}{1}
\renewcommand{\thefigure}{\arabic{figure}.8}
\begin{figure}
\begin{center}
{\includegraphics[angle=90,height=12cm]{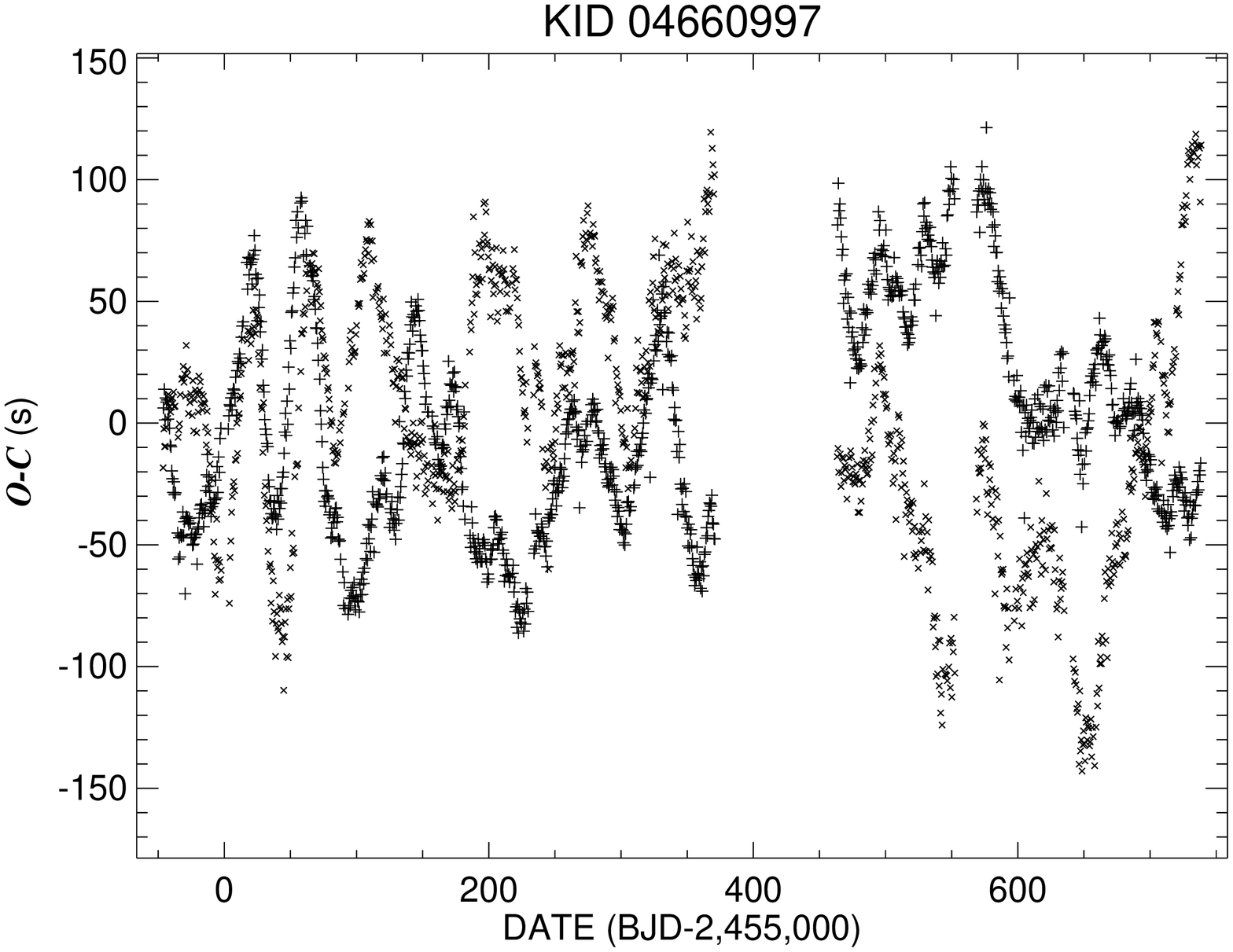}}
\end{center}
\caption{The observed minus calculated eclipse times relative to
a linear ephemeris.  The primary and secondary eclipse
times are indicated by $+$ and $\times$ symbols, 
respectively.}
\end{figure}

\clearpage
\setcounter{figure}{1}
\renewcommand{\thefigure}{\arabic{figure}.9}
\begin{figure}
\begin{center}
{\includegraphics[angle=90,height=12cm]{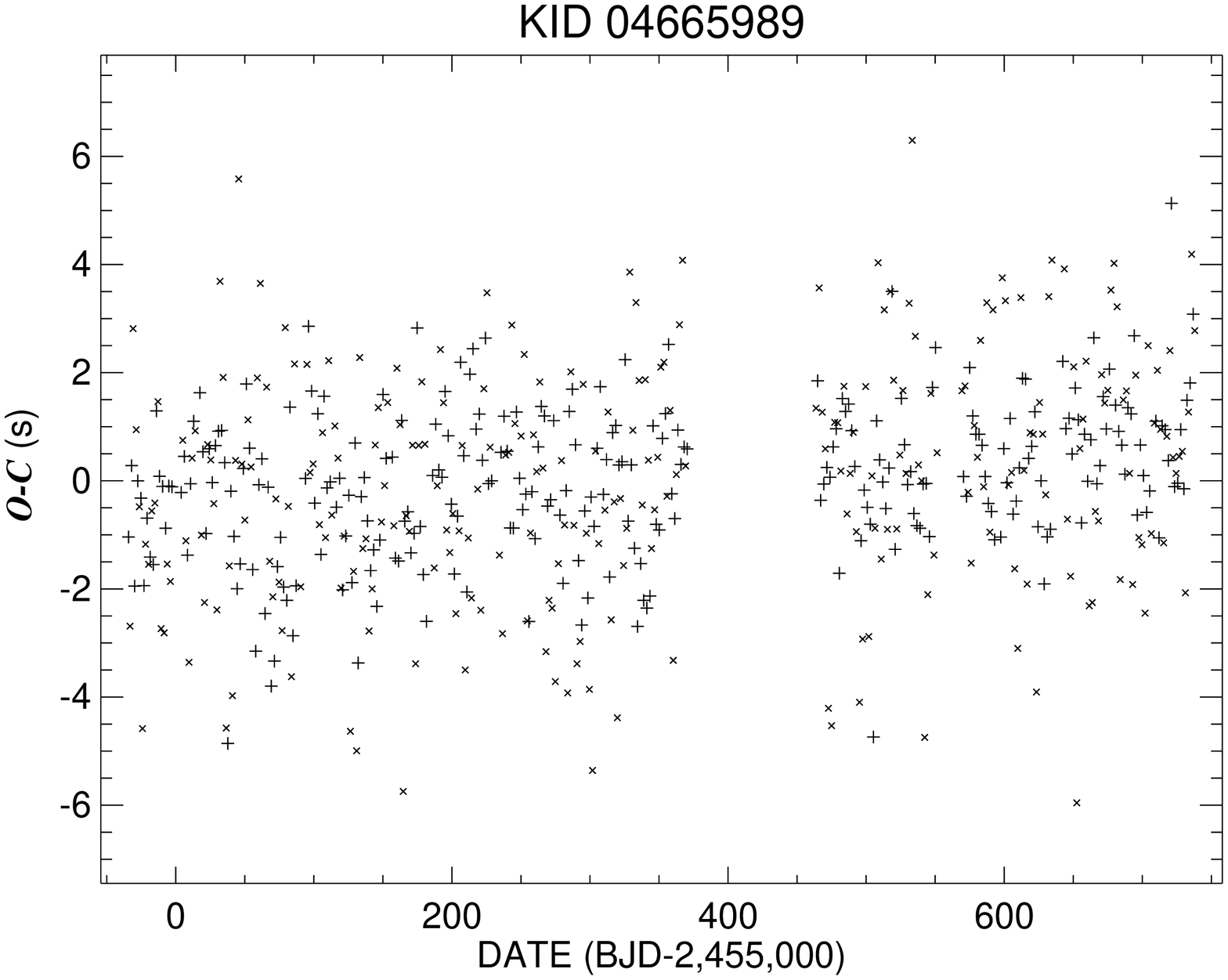}}
\end{center}
\caption{The observed minus calculated eclipse times relative to
a linear ephemeris.  The primary and secondary eclipse
times are indicated by $+$ and $\times$ symbols, 
respectively.}
\end{figure}

\clearpage
\setcounter{figure}{1}
\renewcommand{\thefigure}{\arabic{figure}.10}
\begin{figure}
\begin{center}
{\includegraphics[angle=90,height=12cm]{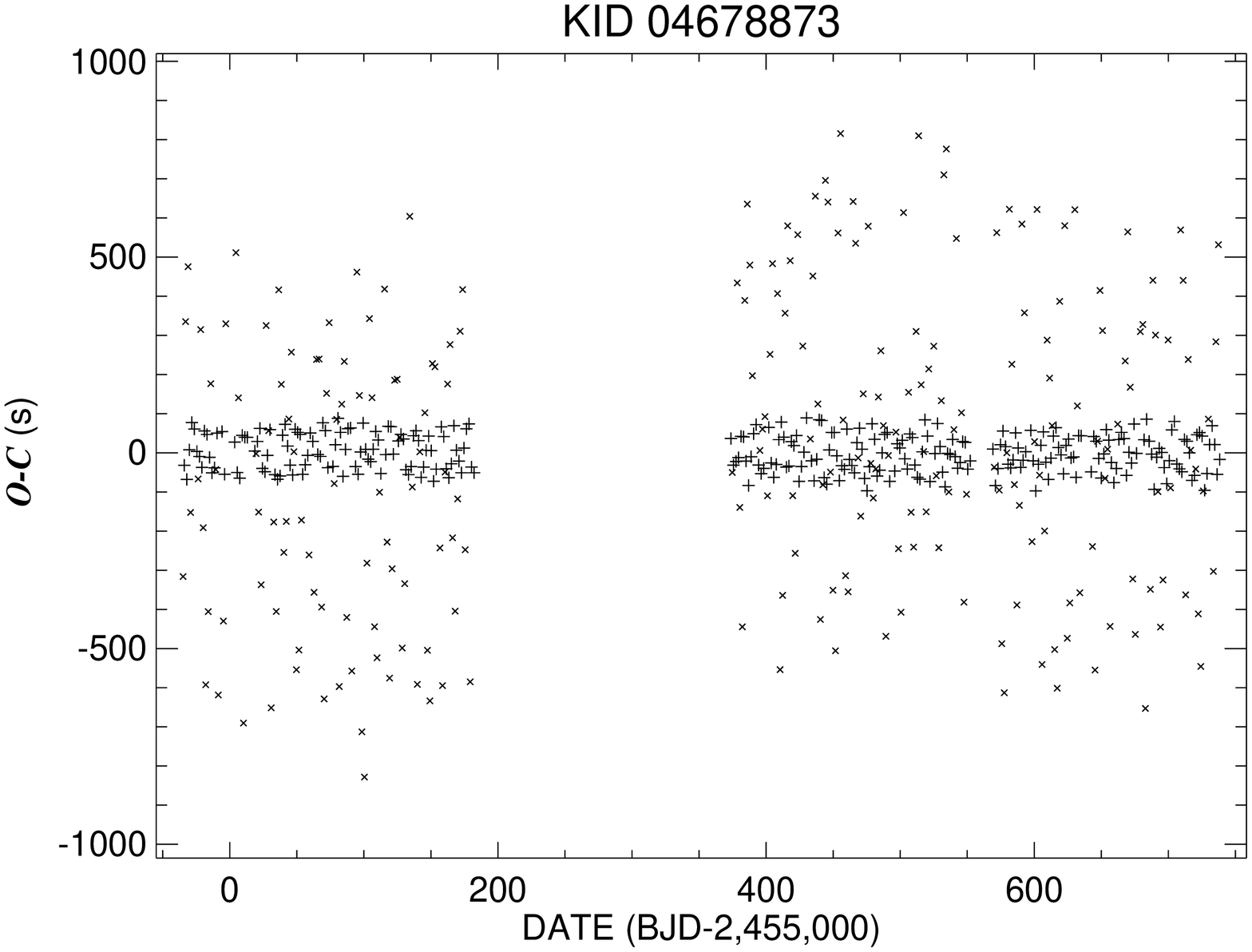}}
\end{center}
\caption{The observed minus calculated eclipse times relative to
a linear ephemeris.  The primary and secondary eclipse
times are indicated by $+$ and $\times$ symbols, 
respectively.}
\end{figure}

\clearpage
\setcounter{figure}{1}
\renewcommand{\thefigure}{\arabic{figure}.11}
\begin{figure}
\begin{center}
{\includegraphics[angle=90,height=12cm]{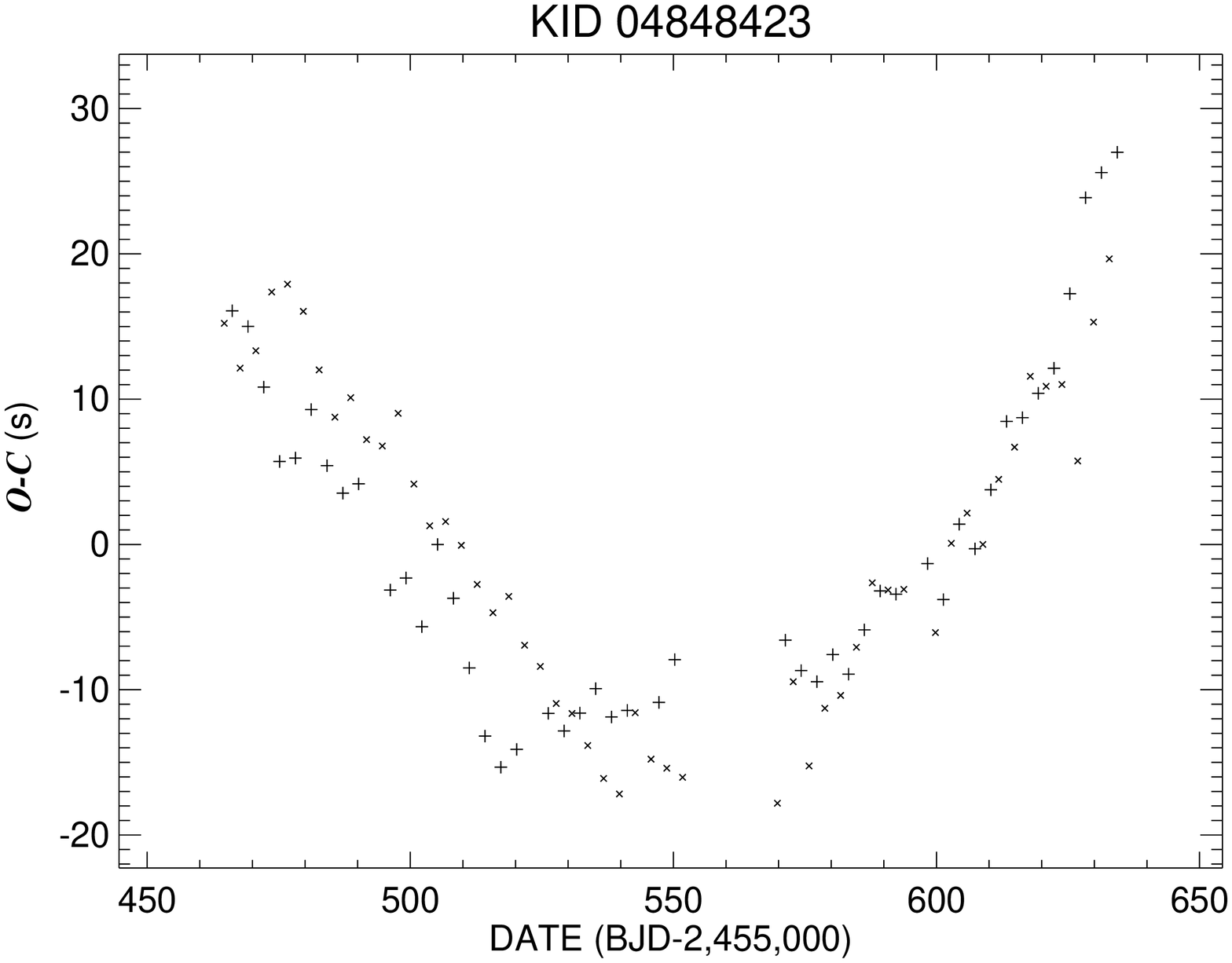}}
\end{center}
\caption{The observed minus calculated eclipse times relative to
a linear ephemeris.  The primary and secondary eclipse
times are indicated by $+$ and $\times$ symbols, 
respectively.}
\end{figure}

\clearpage
\setcounter{figure}{1}
\renewcommand{\thefigure}{\arabic{figure}.12}
\begin{figure}
\begin{center}
{\includegraphics[angle=90,height=12cm]{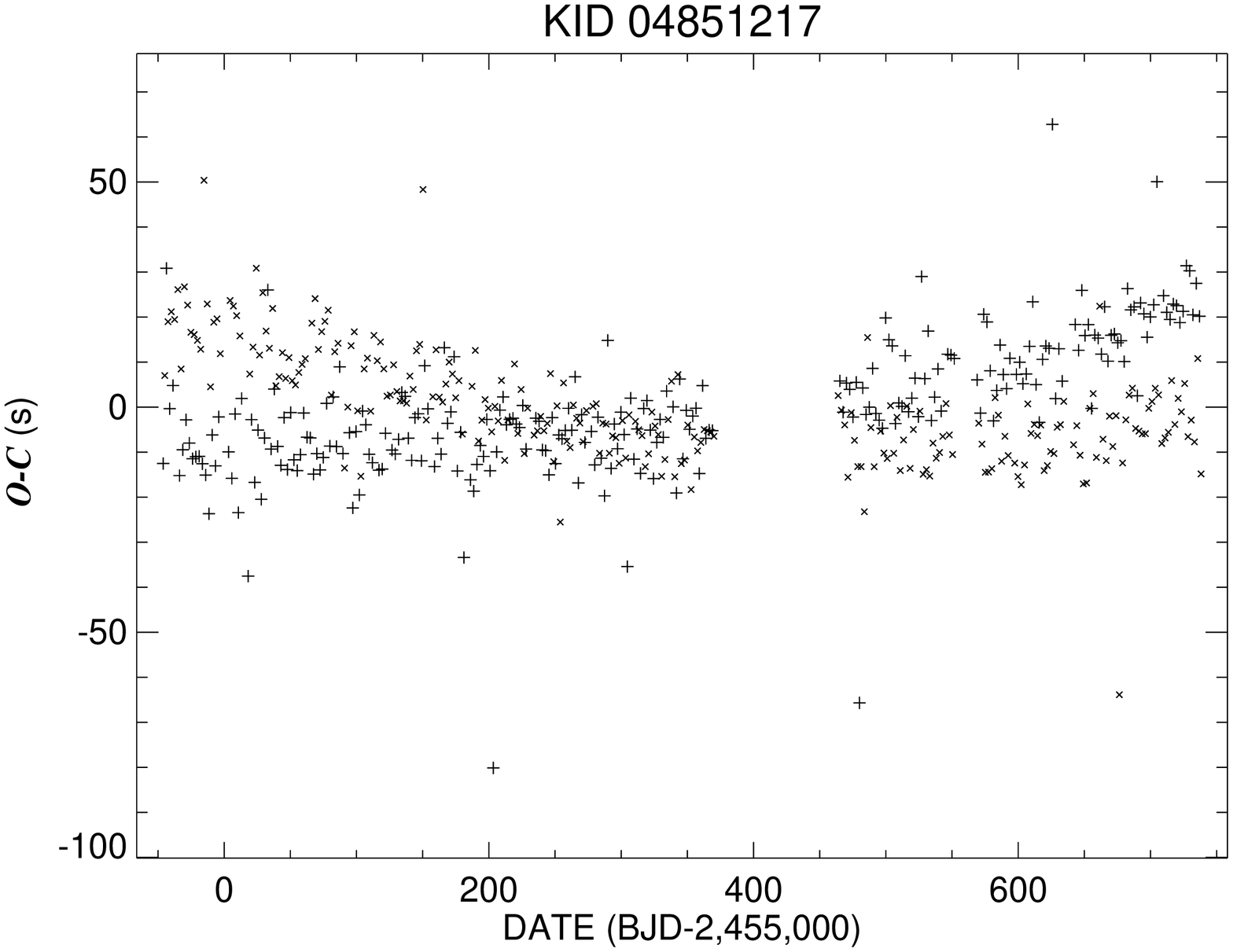}}
\end{center}
\caption{The observed minus calculated eclipse times relative to
a linear ephemeris.  The primary and secondary eclipse
times are indicated by $+$ and $\times$ symbols, 
respectively.}
\end{figure}

\clearpage
\setcounter{figure}{1}
\renewcommand{\thefigure}{\arabic{figure}.13}
\begin{figure}
\begin{center}
{\includegraphics[angle=90,height=12cm]{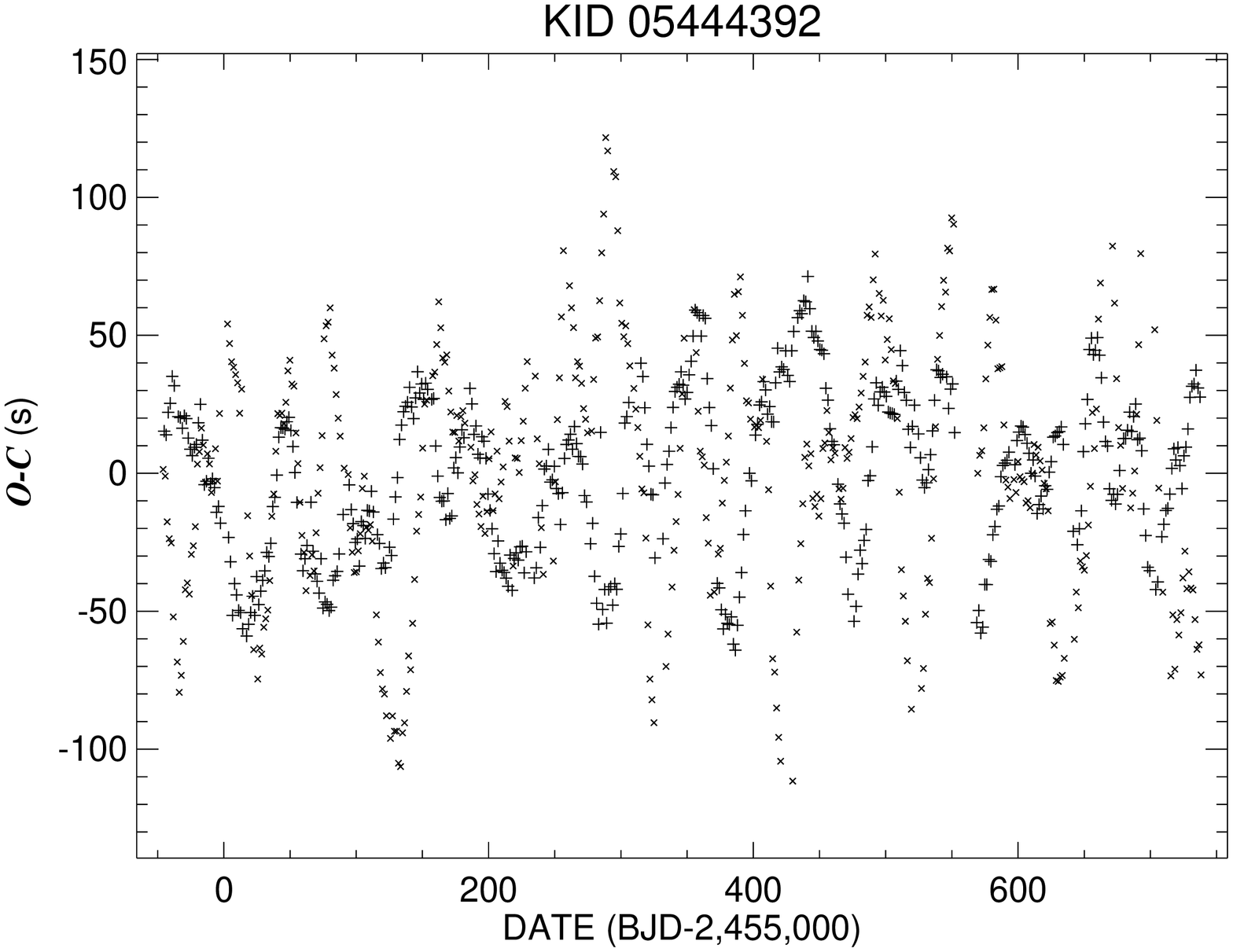}}
\end{center}
\caption{The observed minus calculated eclipse times relative to
a linear ephemeris.  The primary and secondary eclipse
times are indicated by $+$ and $\times$ symbols, 
respectively.}
\end{figure}

\clearpage
\setcounter{figure}{1}
\renewcommand{\thefigure}{\arabic{figure}.14}
\begin{figure}
\begin{center}
{\includegraphics[angle=90,height=12cm]{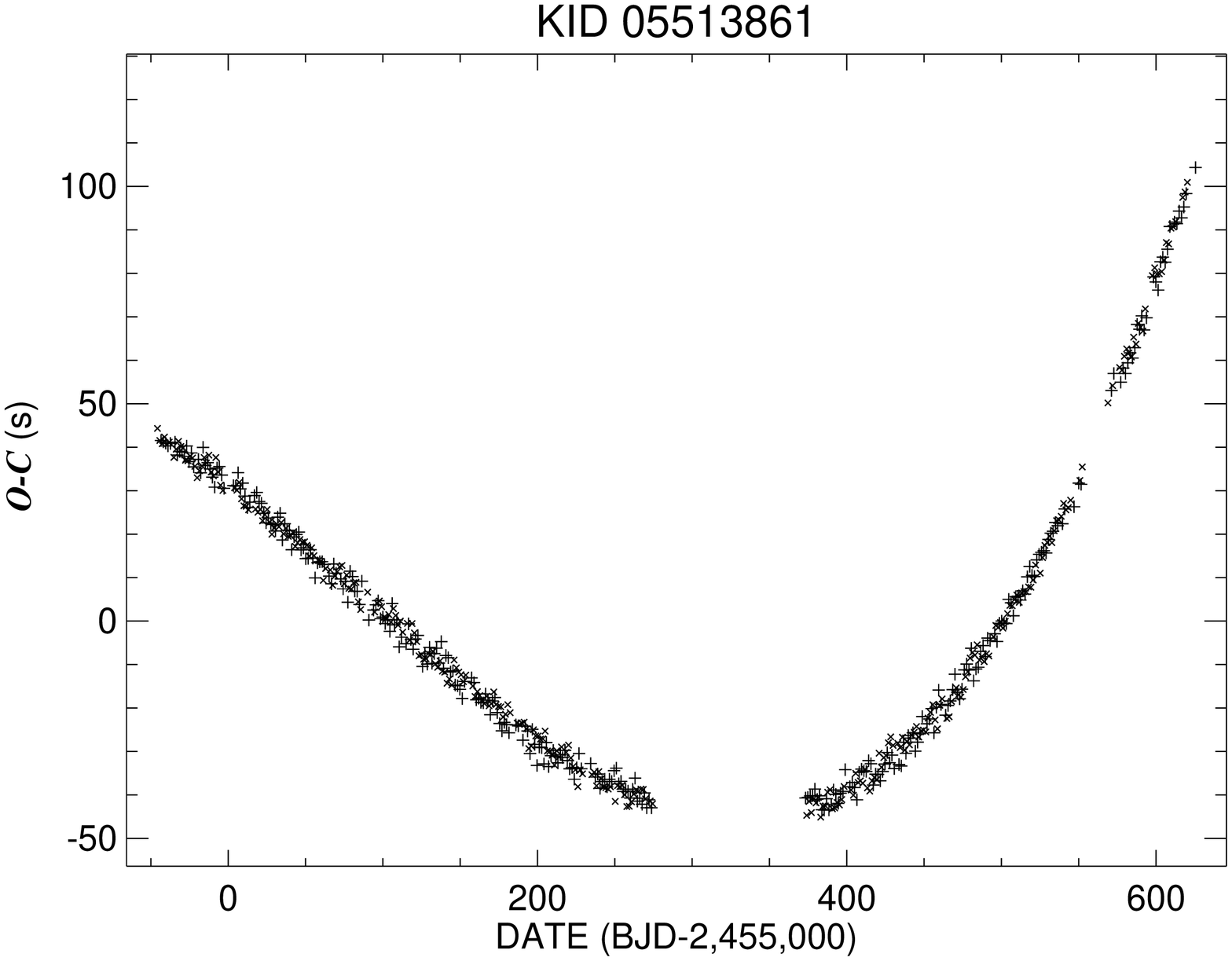}}
\end{center}
\caption{The observed minus calculated eclipse times relative to
a linear ephemeris.  The primary and secondary eclipse
times are indicated by $+$ and $\times$ symbols, 
respectively.}
\end{figure}

\clearpage
\setcounter{figure}{1}
\renewcommand{\thefigure}{\arabic{figure}.15}
\begin{figure}
\begin{center}
{\includegraphics[angle=90,height=12cm]{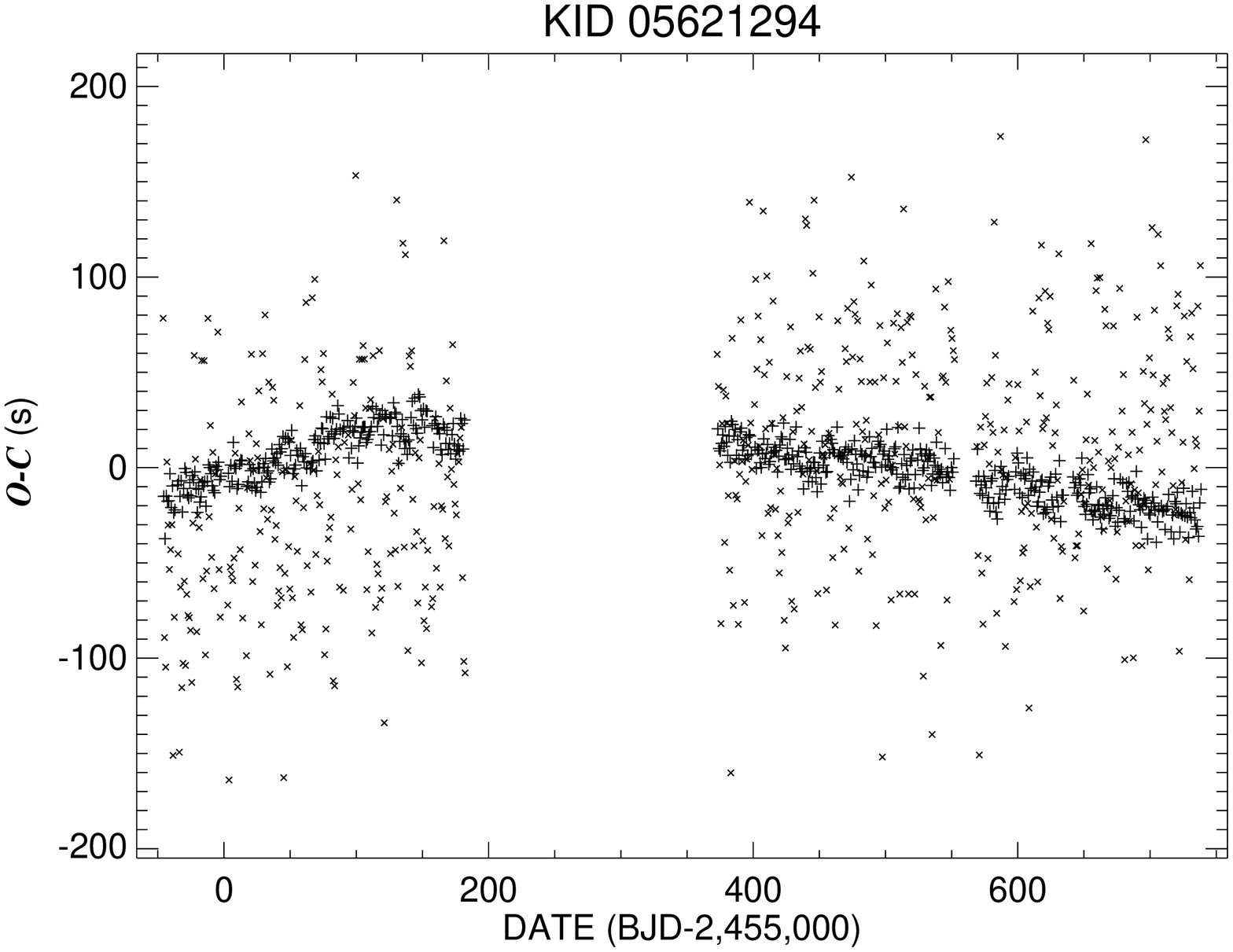}}
\end{center}
\caption{The observed minus calculated eclipse times relative to
a linear ephemeris.  The primary and secondary eclipse
times are indicated by $+$ and $\times$ symbols, 
respectively.}
\end{figure}

\clearpage
\setcounter{figure}{1}
\renewcommand{\thefigure}{\arabic{figure}.16}
\begin{figure}
\begin{center}
{\includegraphics[angle=90,height=12cm]{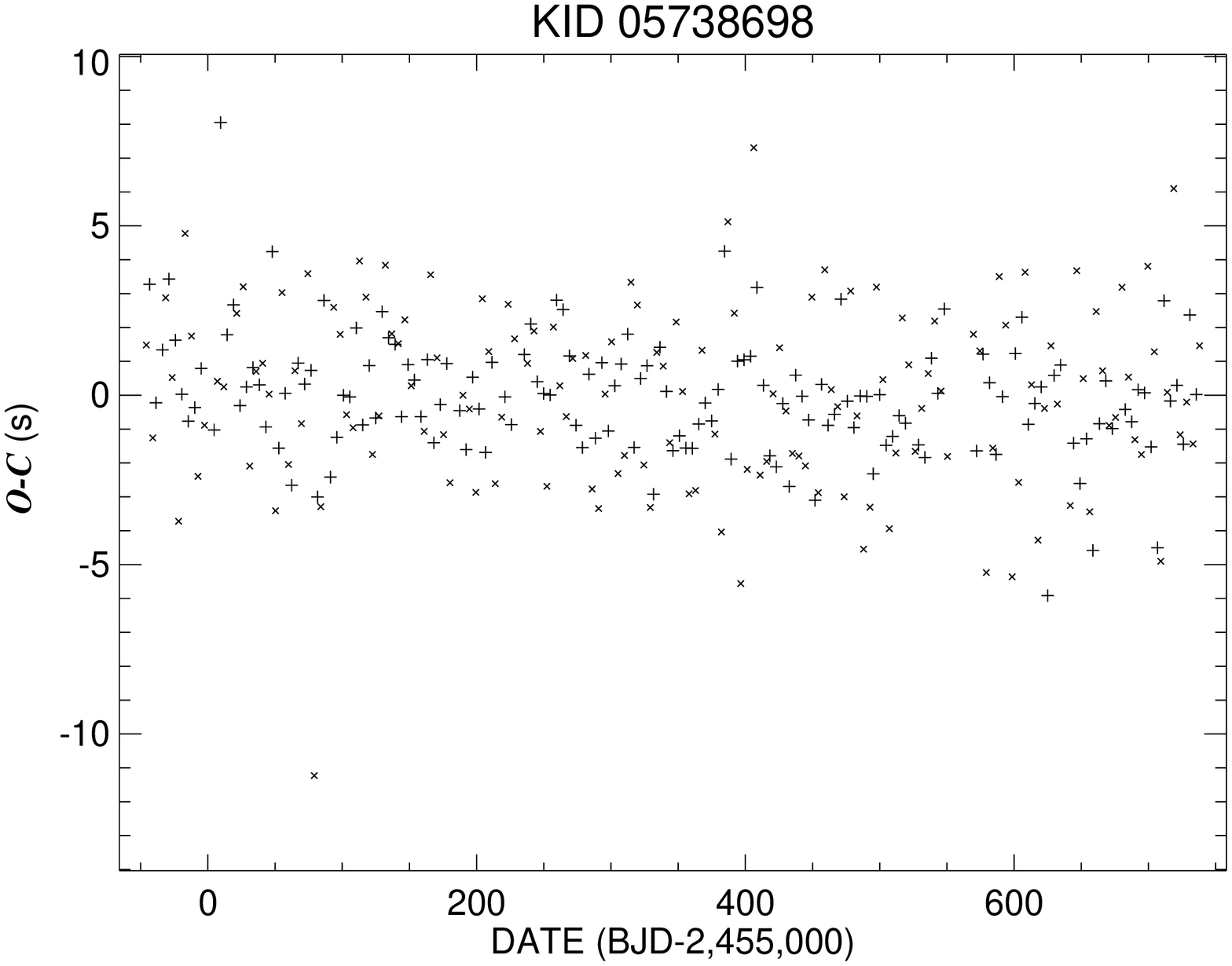}}
\end{center}
\caption{The observed minus calculated eclipse times relative to
a linear ephemeris.  The primary and secondary eclipse
times are indicated by $+$ and $\times$ symbols, 
respectively.}
\end{figure}

\clearpage
\setcounter{figure}{1}
\renewcommand{\thefigure}{\arabic{figure}.17}
\begin{figure}
\begin{center}
{\includegraphics[angle=90,height=12cm]{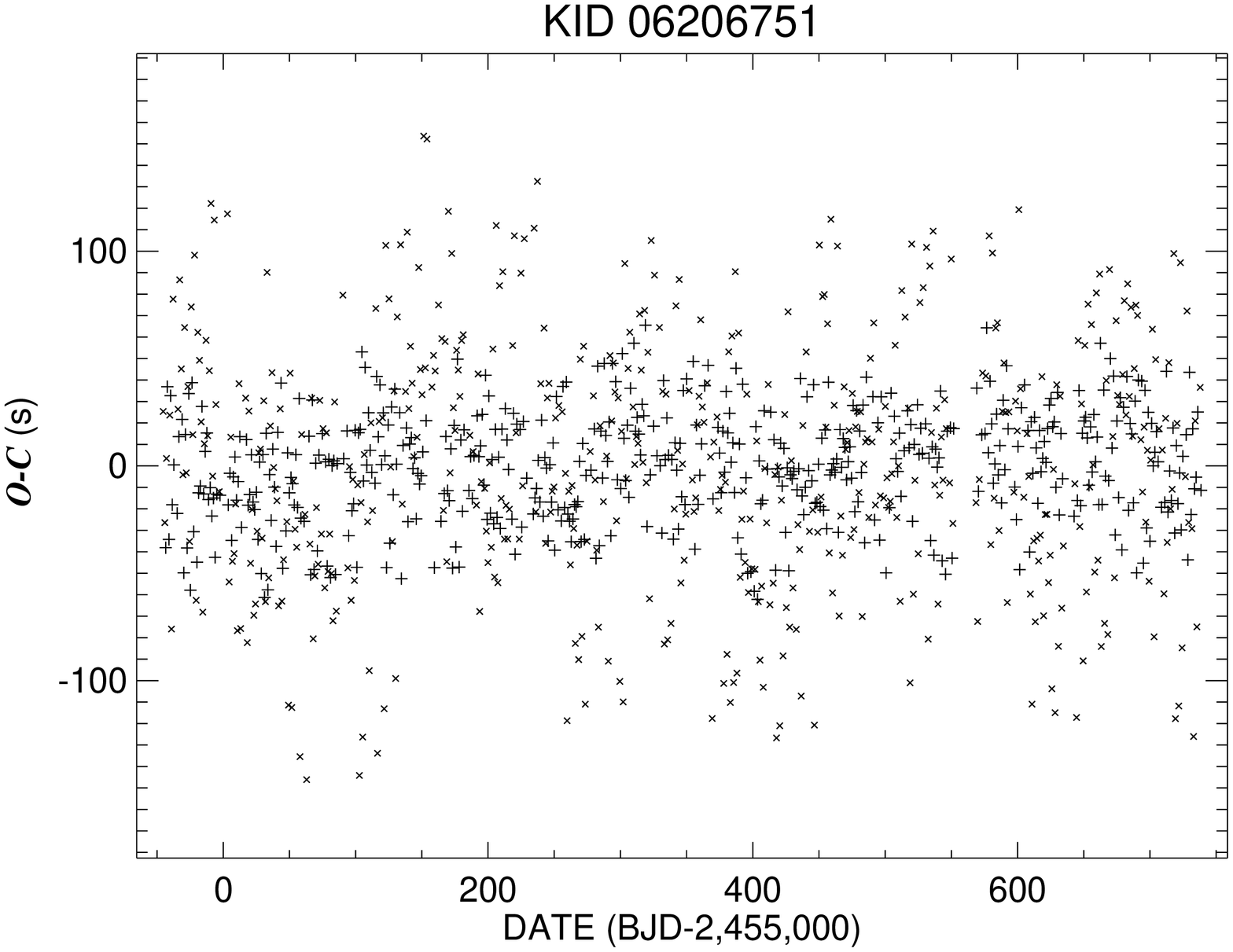}}
\end{center}
\caption{The observed minus calculated eclipse times relative to
a linear ephemeris.  The primary and secondary eclipse
times are indicated by $+$ and $\times$ symbols, 
respectively.}
\end{figure}

\clearpage
\setcounter{figure}{1}
\renewcommand{\thefigure}{\arabic{figure}.18}
\begin{figure}
\begin{center}
{\includegraphics[angle=90,height=12cm]{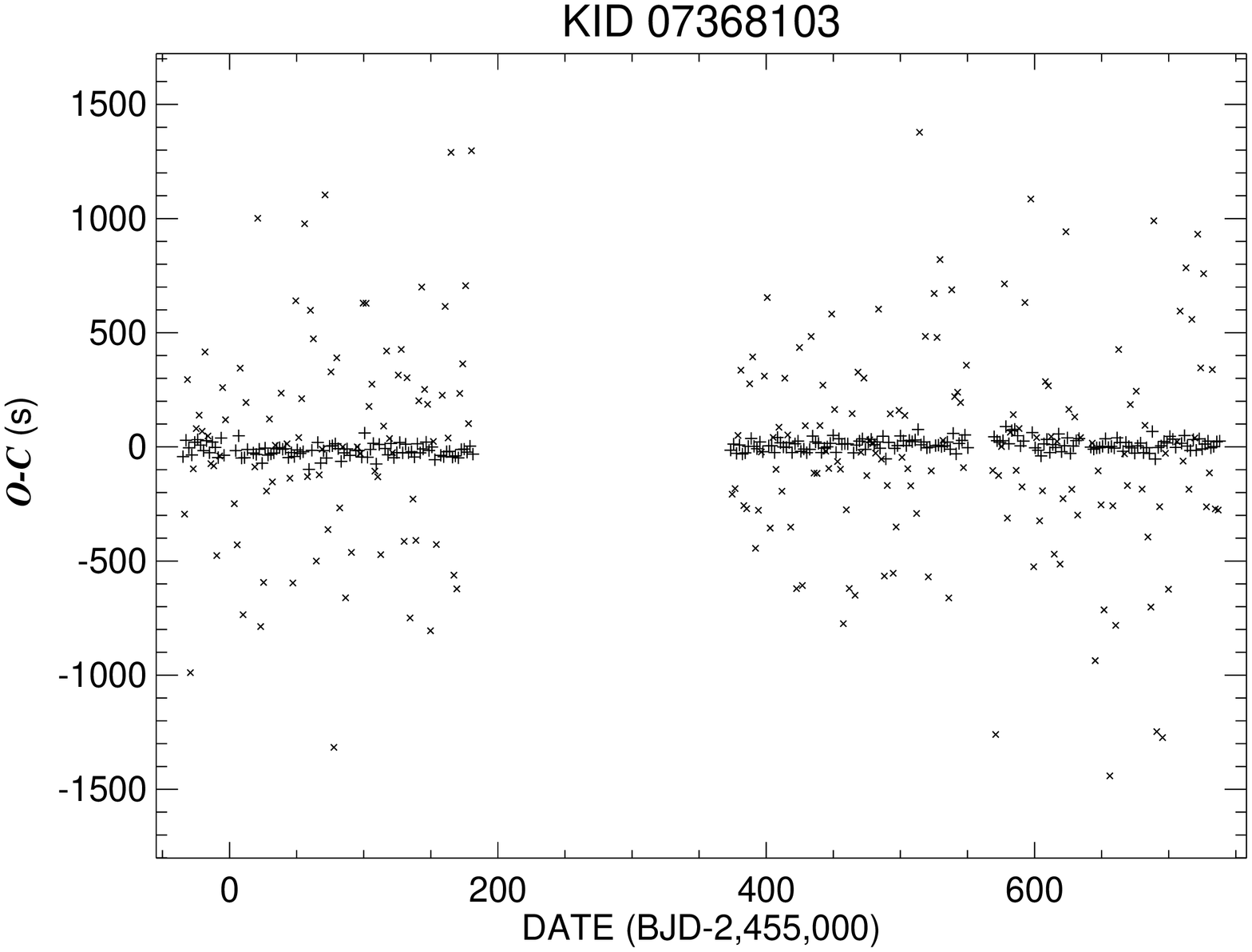}}
\end{center}
\caption{The observed minus calculated eclipse times relative to
a linear ephemeris.  The primary and secondary eclipse
times are indicated by $+$ and $\times$ symbols, 
respectively.}
\end{figure}

\clearpage
\setcounter{figure}{1}
\renewcommand{\thefigure}{\arabic{figure}.19}
\begin{figure}
\begin{center}
{\includegraphics[angle=90,height=12cm]{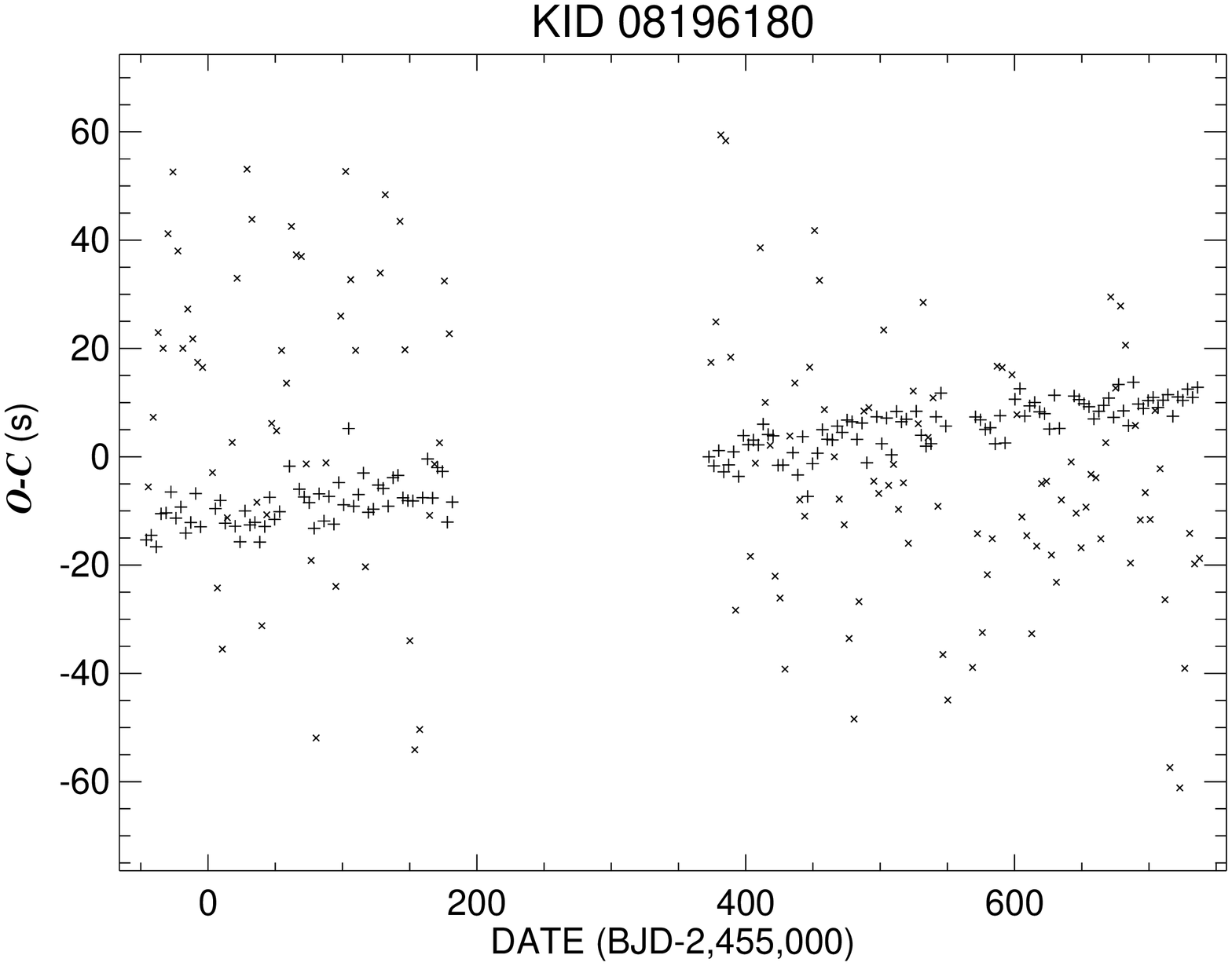}}
\end{center}
\caption{The observed minus calculated eclipse times relative to
a linear ephemeris.  The primary and secondary eclipse
times are indicated by $+$ and $\times$ symbols, 
respectively.}
\end{figure}

\clearpage
\setcounter{figure}{1}
\renewcommand{\thefigure}{\arabic{figure}.20}
\begin{figure}
\begin{center}
{\includegraphics[angle=90,height=12cm]{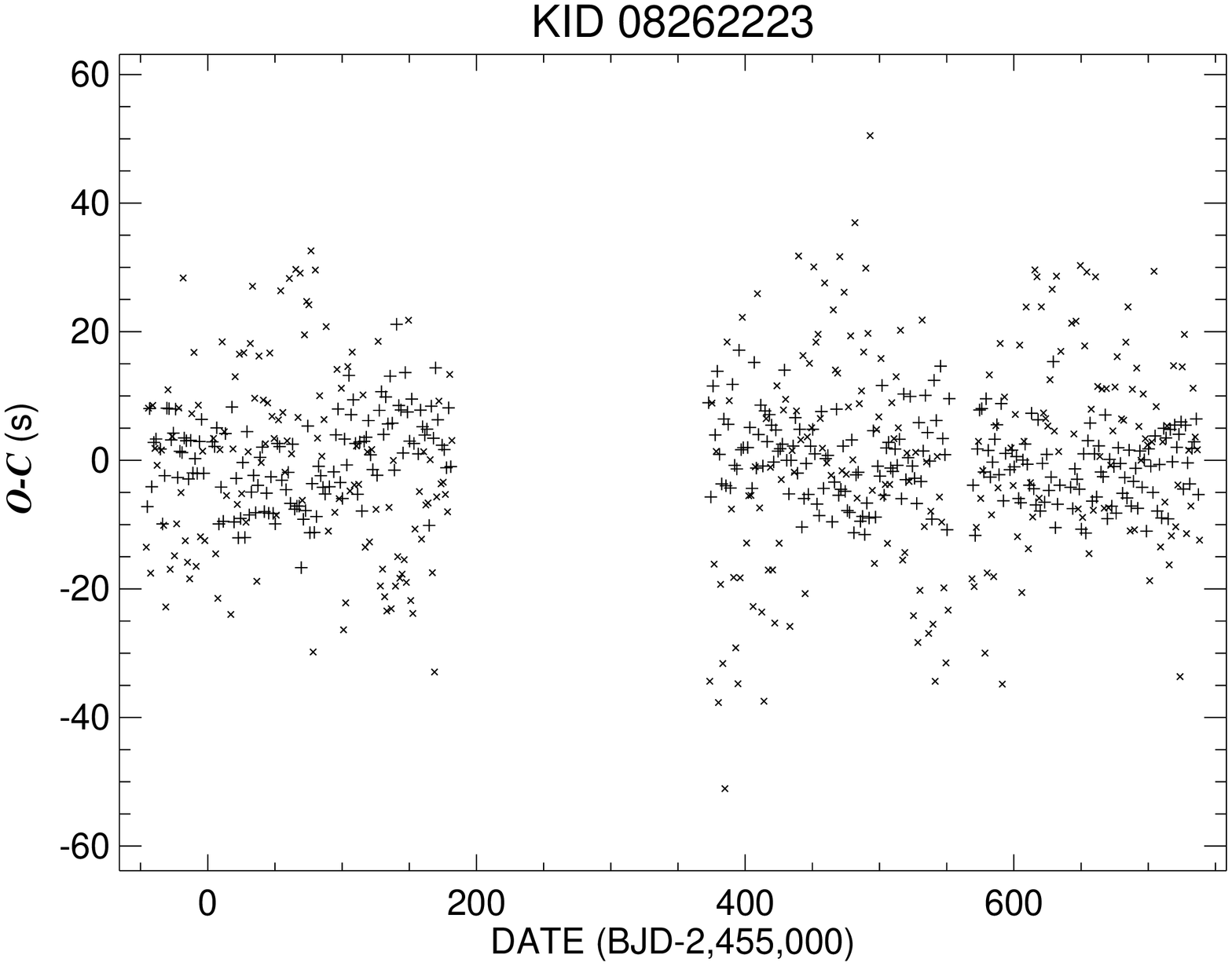}}
\end{center}
\caption{The observed minus calculated eclipse times relative to
a linear ephemeris.  The primary and secondary eclipse
times are indicated by $+$ and $\times$ symbols, 
respectively.}
\end{figure}

\clearpage
\setcounter{figure}{1}
\renewcommand{\thefigure}{\arabic{figure}.21}
\begin{figure}
\begin{center}
{\includegraphics[angle=90,height=12cm]{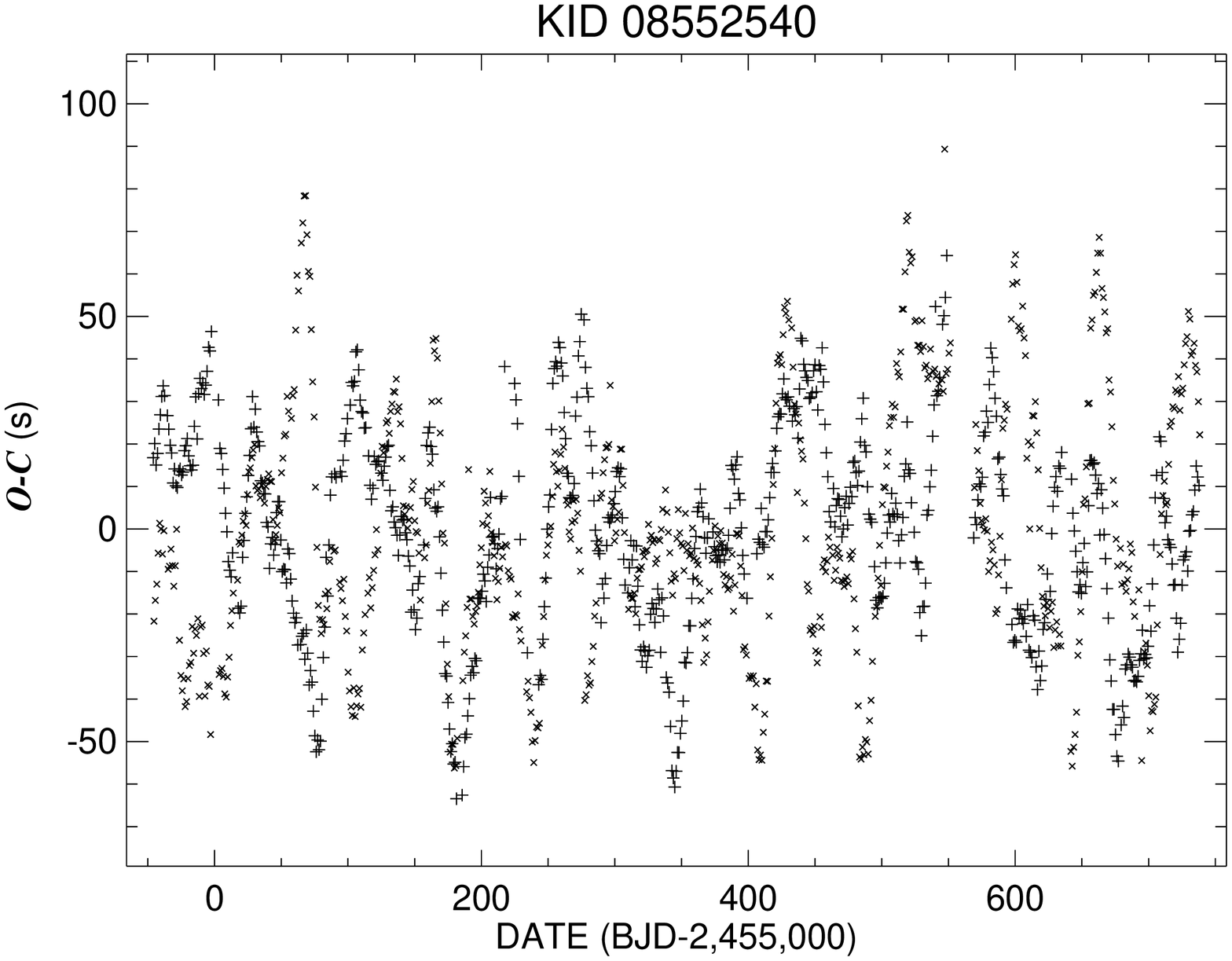}}
\end{center}
\caption{The observed minus calculated eclipse times relative to
a linear ephemeris.  The primary and secondary eclipse
times are indicated by $+$ and $\times$ symbols, 
respectively.}
\end{figure}

\clearpage
\setcounter{figure}{1}
\renewcommand{\thefigure}{\arabic{figure}.22}
\begin{figure}
\begin{center}
{\includegraphics[angle=90,height=12cm]{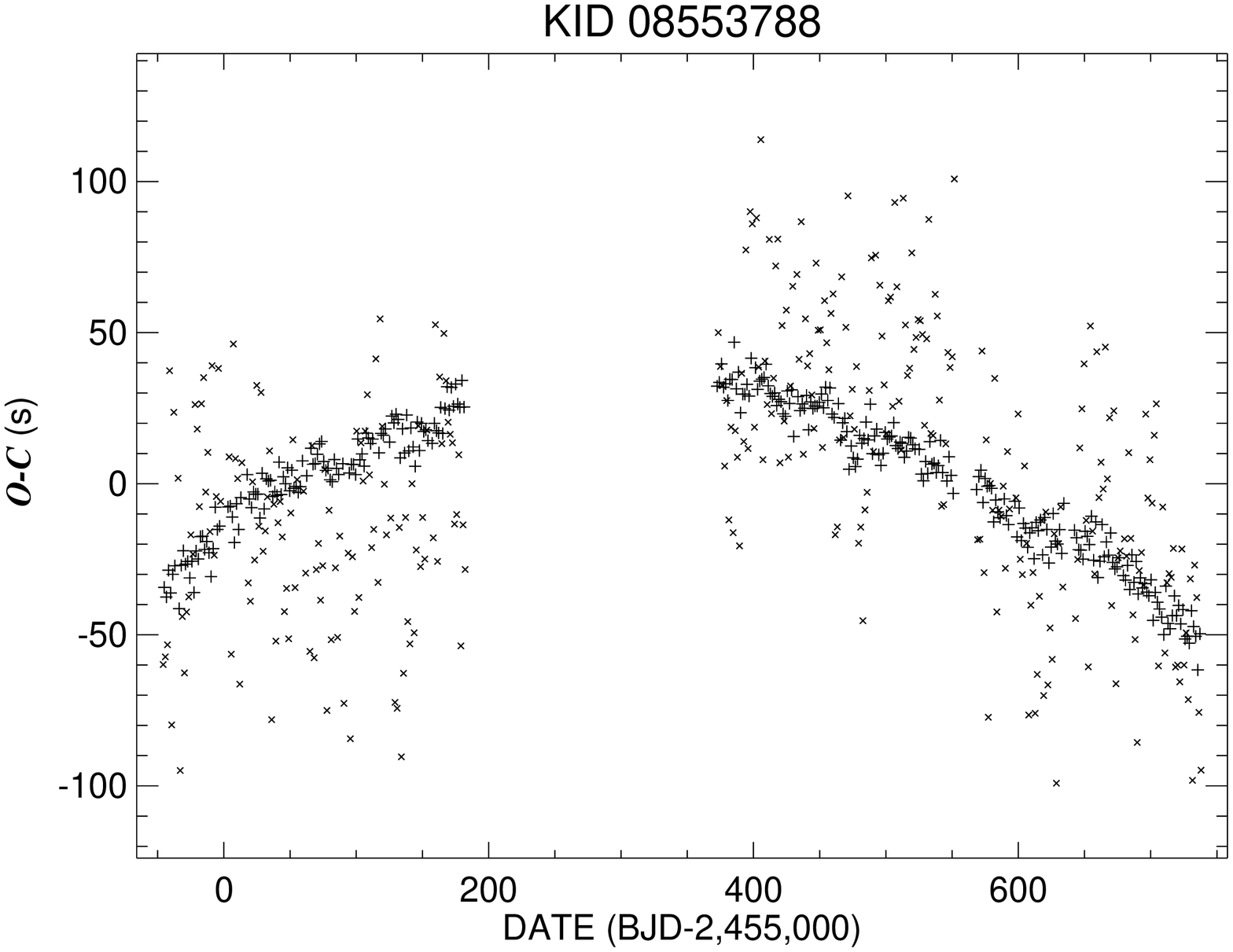}}
\end{center}
\caption{The observed minus calculated eclipse times relative to
a linear ephemeris.  The primary and secondary eclipse
times are indicated by $+$ and $\times$ symbols, 
respectively.}
\end{figure}

\clearpage
\setcounter{figure}{1}
\renewcommand{\thefigure}{\arabic{figure}.23}
\begin{figure}
\begin{center}
{\includegraphics[angle=90,height=12cm]{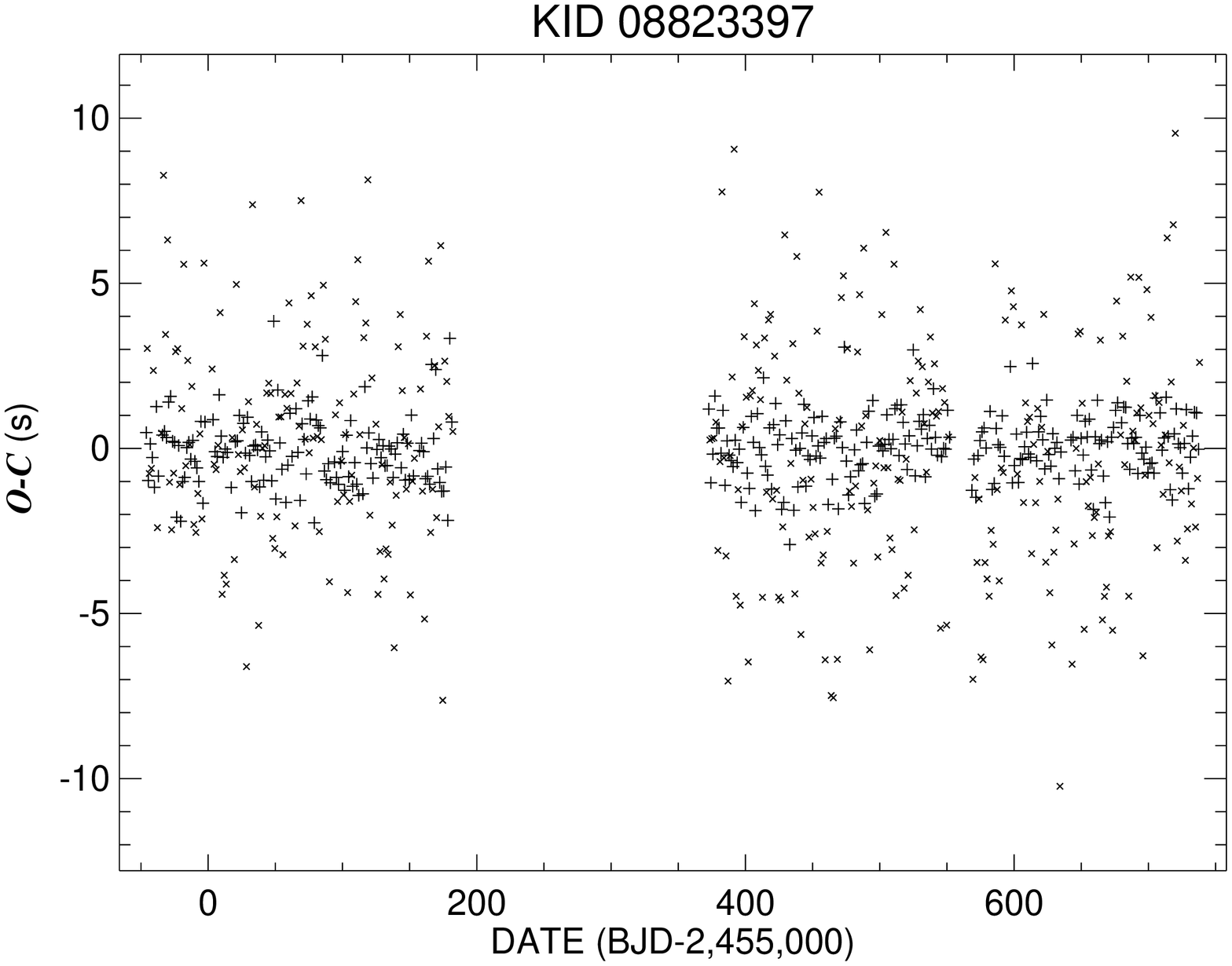}}
\end{center}
\caption{The observed minus calculated eclipse times relative to
a linear ephemeris.  The primary and secondary eclipse
times are indicated by $+$ and $\times$ symbols, 
respectively.}
\end{figure}

\clearpage
\setcounter{figure}{1}
\renewcommand{\thefigure}{\arabic{figure}.24}
\begin{figure}
\begin{center}
{\includegraphics[angle=90,height=12cm]{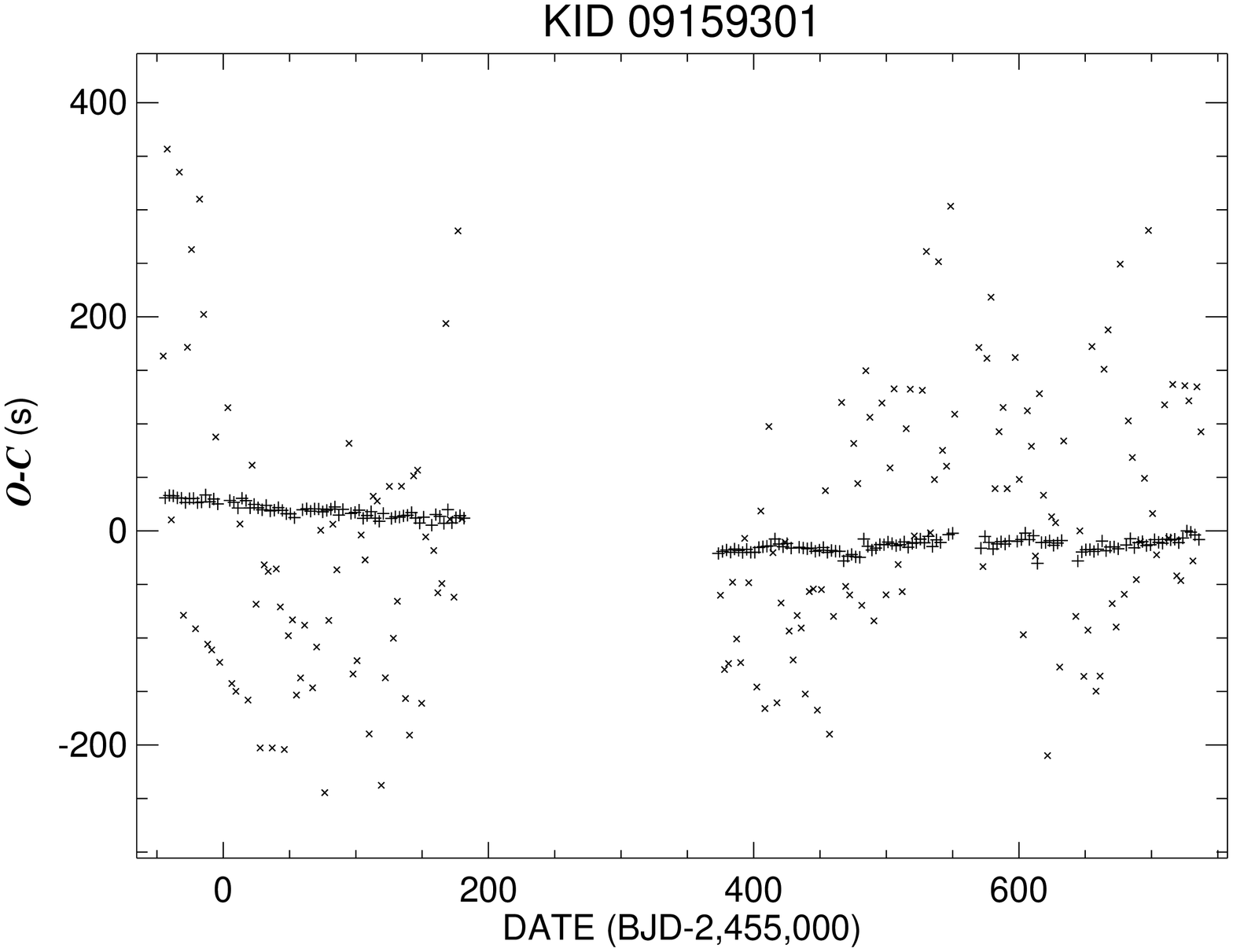}}
\end{center}
\caption{The observed minus calculated eclipse times relative to
a linear ephemeris.  The primary and secondary eclipse
times are indicated by $+$ and $\times$ symbols, 
respectively.}
\end{figure}

\clearpage
\setcounter{figure}{1}
\renewcommand{\thefigure}{\arabic{figure}.25}
\begin{figure}
\begin{center}
{\includegraphics[angle=90,height=12cm]{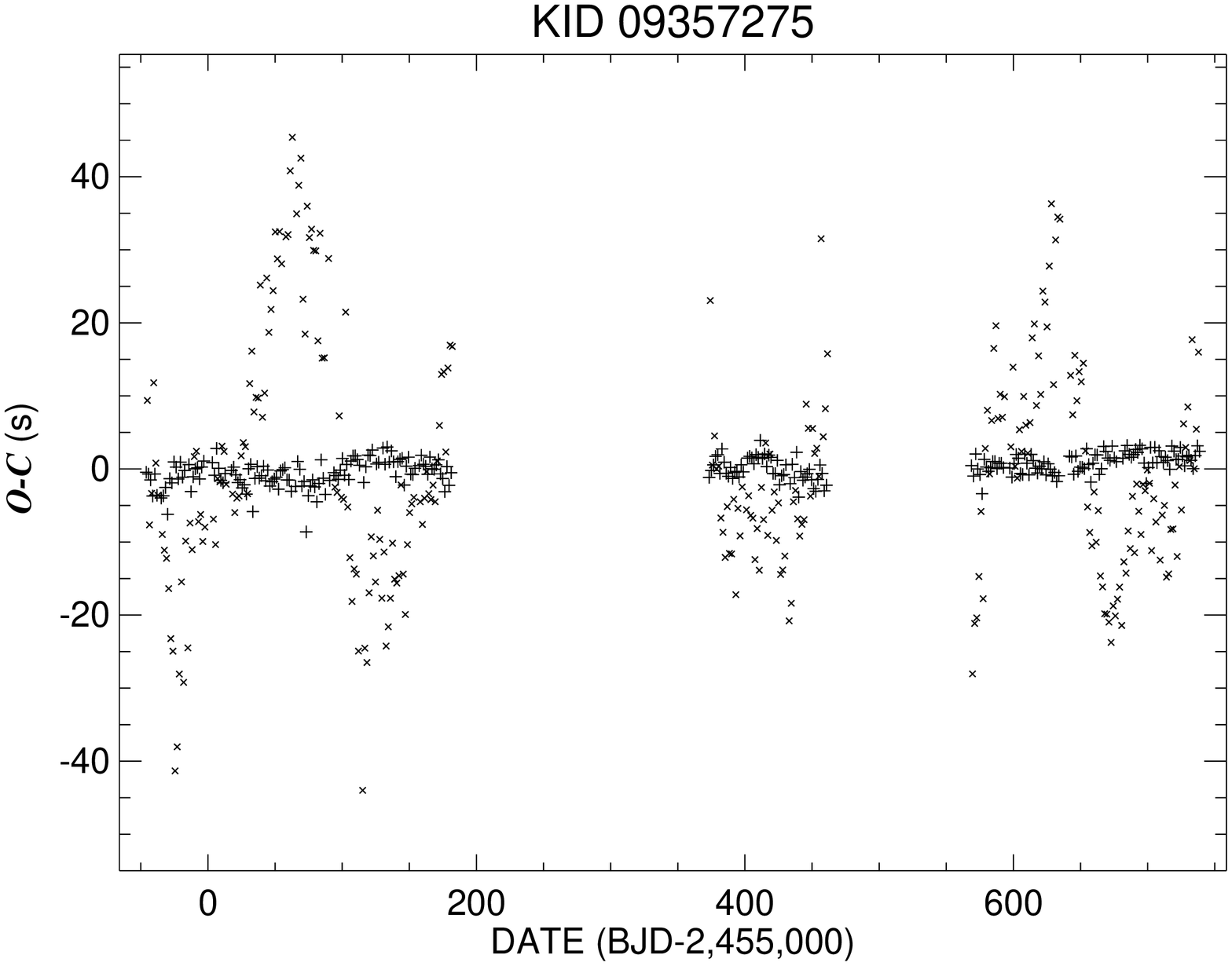}}
\end{center}
\caption{The observed minus calculated eclipse times relative to
a linear ephemeris.  The primary and secondary eclipse
times are indicated by $+$ and $\times$ symbols, 
respectively.}
\end{figure}

\clearpage
\setcounter{figure}{1}
\renewcommand{\thefigure}{\arabic{figure}.26}
\begin{figure}
\begin{center}
{\includegraphics[angle=90,height=12cm]{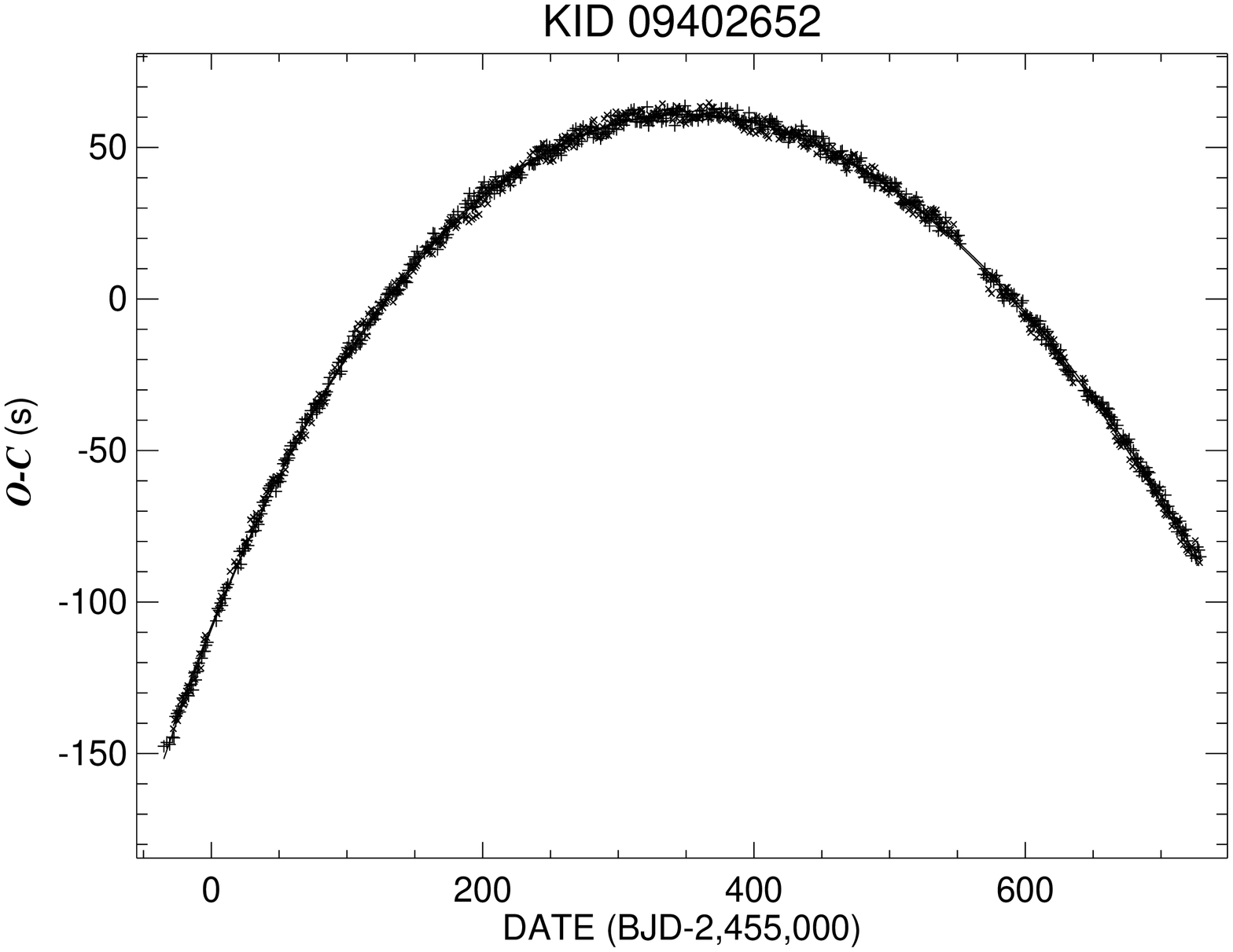}}
\end{center}
\caption{The observed minus calculated eclipse times relative to
a linear ephemeris.  The primary and secondary eclipse
times are indicated by $+$ and $\times$ symbols, 
respectively.}
\end{figure}

\clearpage
\setcounter{figure}{1}
\renewcommand{\thefigure}{\arabic{figure}.27}
\begin{figure}
\begin{center}
{\includegraphics[angle=90,height=12cm]{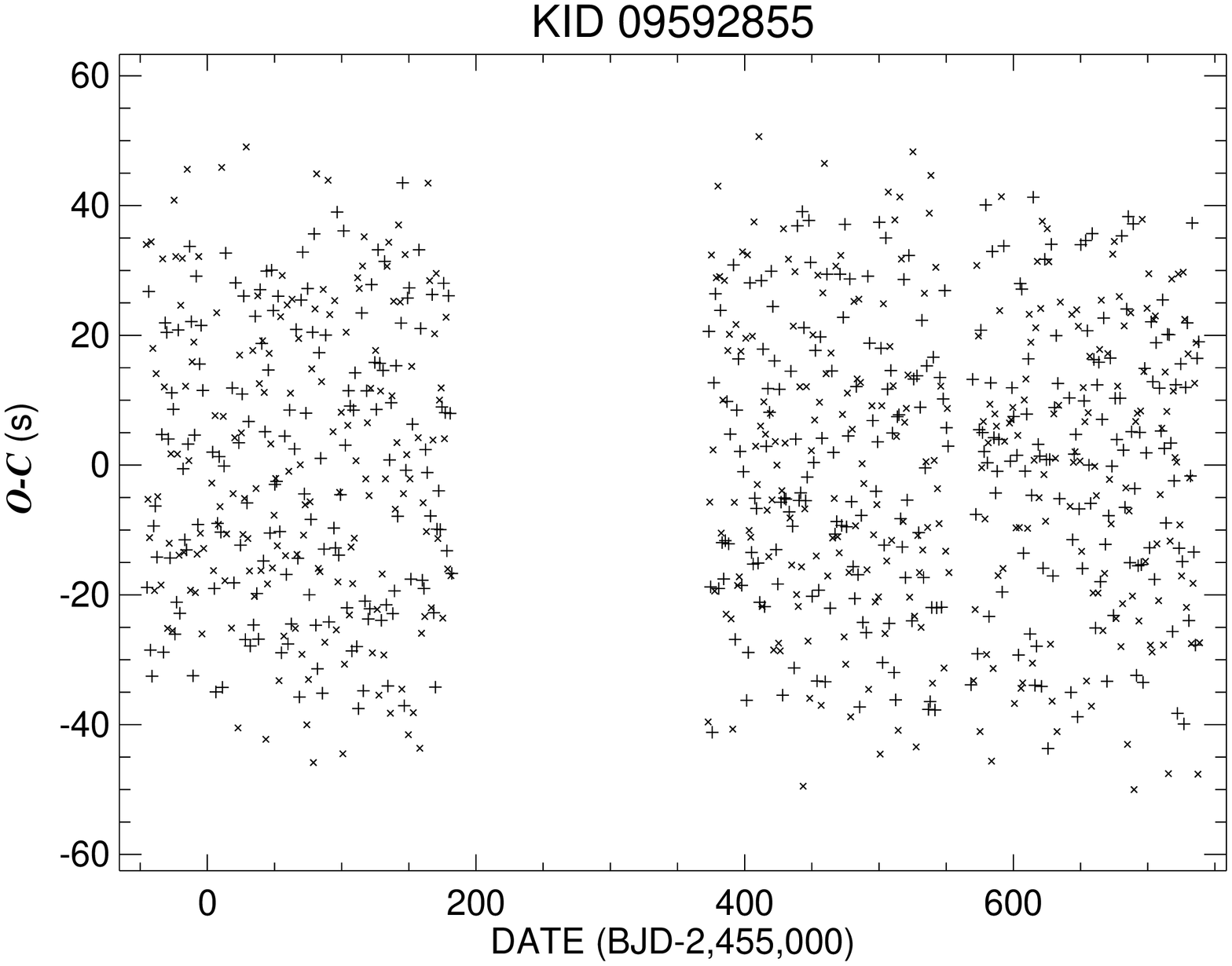}}
\end{center}
\caption{The observed minus calculated eclipse times relative to
a linear ephemeris.  The primary and secondary eclipse
times are indicated by $+$ and $\times$ symbols, 
respectively.}
\end{figure}

\clearpage
\setcounter{figure}{1}
\renewcommand{\thefigure}{\arabic{figure}.28}
\begin{figure}
\begin{center}
{\includegraphics[angle=90,height=12cm]{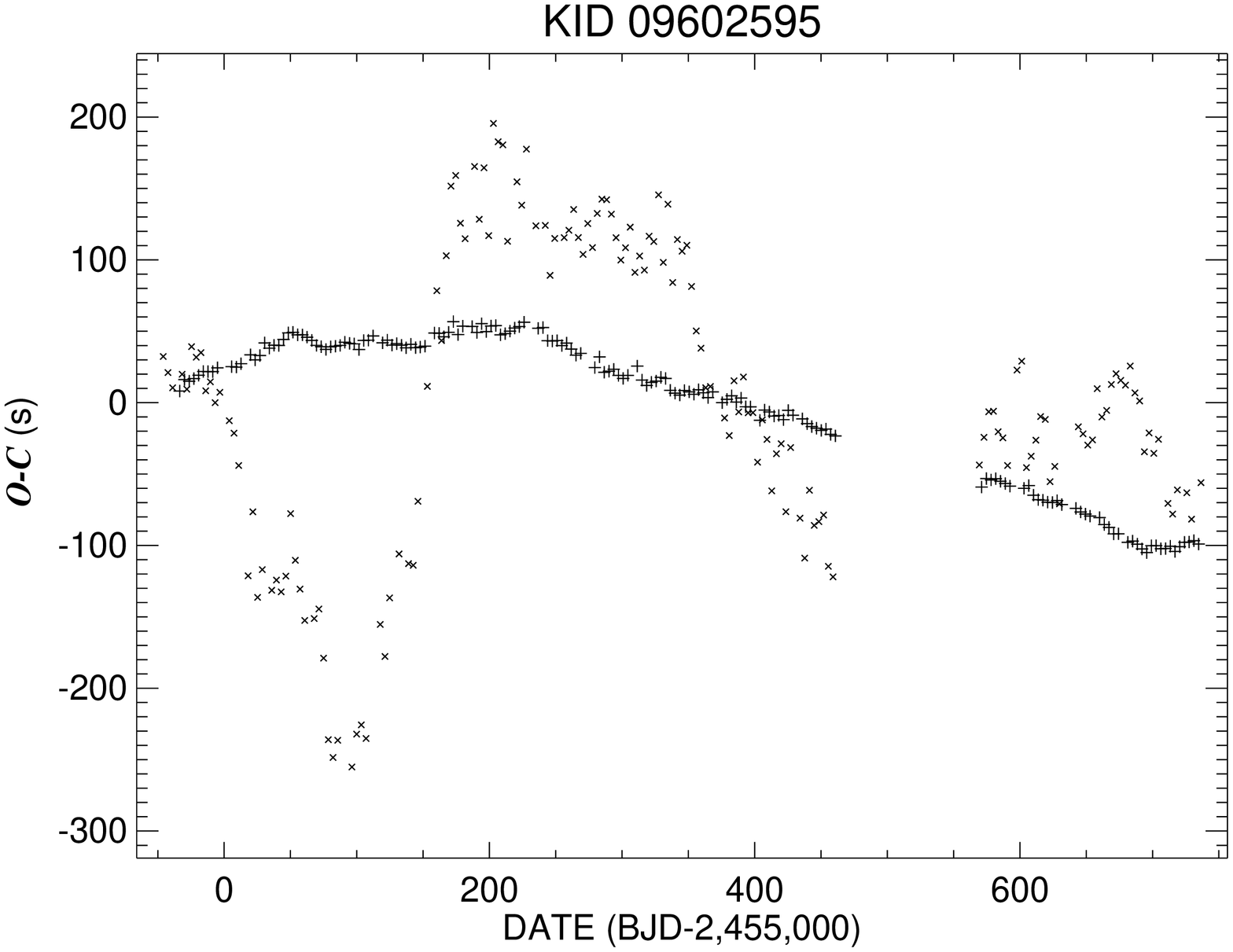}}
\end{center}
\caption{The observed minus calculated eclipse times relative to
a linear ephemeris.  The primary and secondary eclipse
times are indicated by $+$ and $\times$ symbols, 
respectively.}
\end{figure}

\clearpage
\setcounter{figure}{1}
\renewcommand{\thefigure}{\arabic{figure}.29}
\begin{figure}
\begin{center}
{\includegraphics[angle=90,height=12cm]{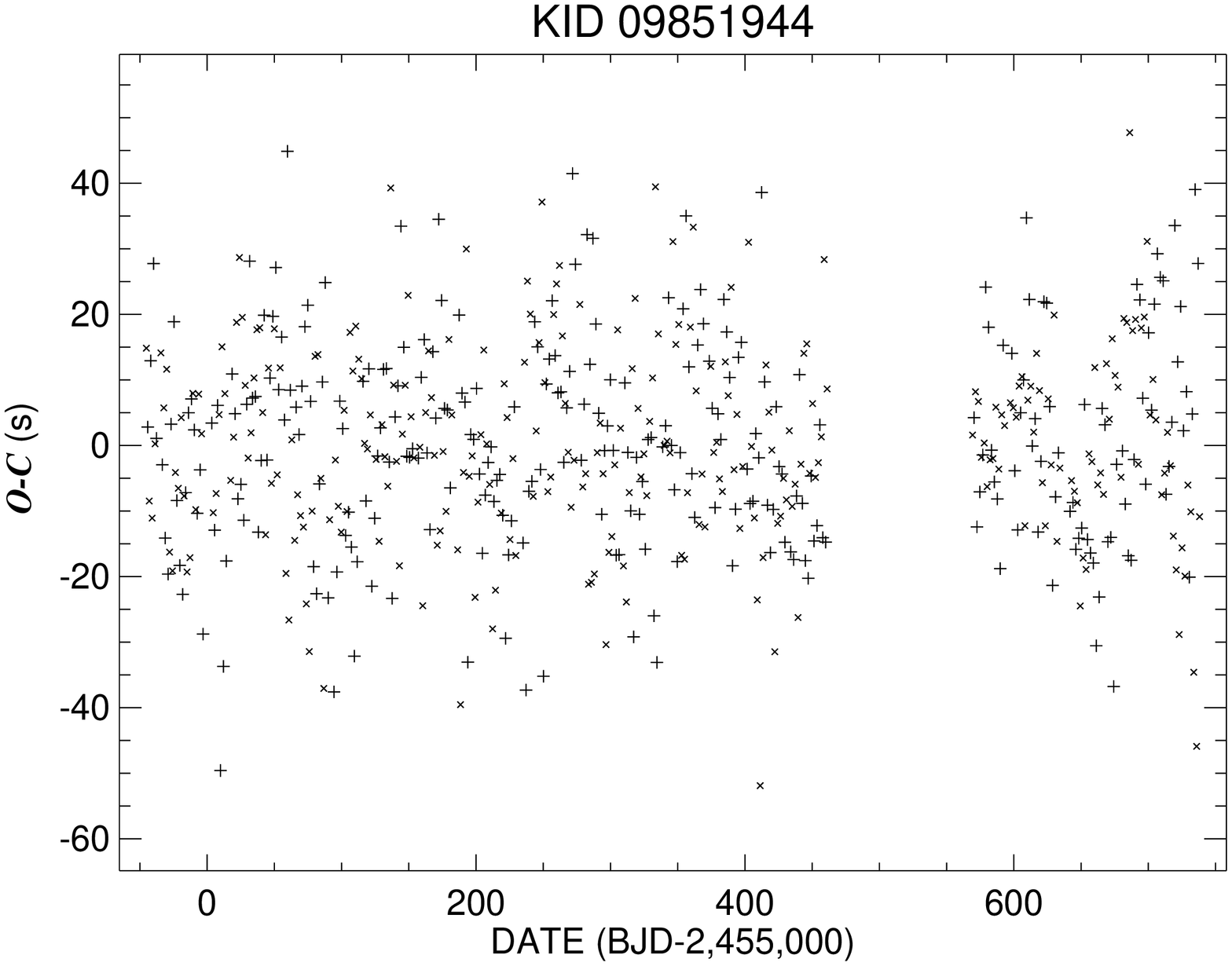}}
\end{center}
\caption{The observed minus calculated eclipse times relative to
a linear ephemeris.  The primary and secondary eclipse
times are indicated by $+$ and $\times$ symbols, 
respectively.}
\end{figure}

\clearpage
\setcounter{figure}{1}
\renewcommand{\thefigure}{\arabic{figure}.30}
\begin{figure}
\begin{center}
{\includegraphics[angle=90,height=12cm]{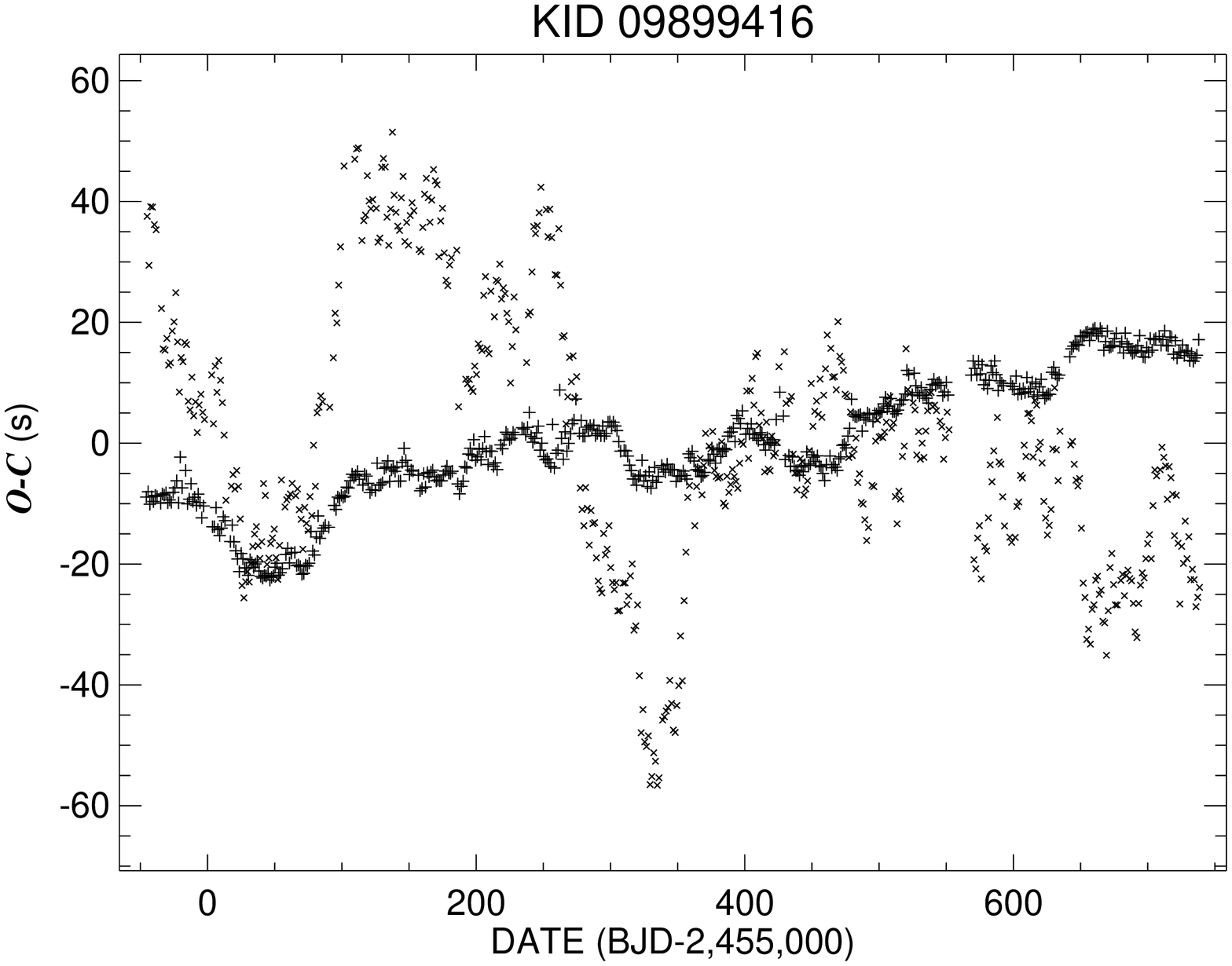}}
\end{center}
\caption{The observed minus calculated eclipse times relative to
a linear ephemeris.  The primary and secondary eclipse
times are indicated by $+$ and $\times$ symbols, 
respectively.}
\end{figure}

\clearpage
\setcounter{figure}{1}
\renewcommand{\thefigure}{\arabic{figure}.31}
\begin{figure}
\begin{center}
{\includegraphics[angle=90,height=12cm]{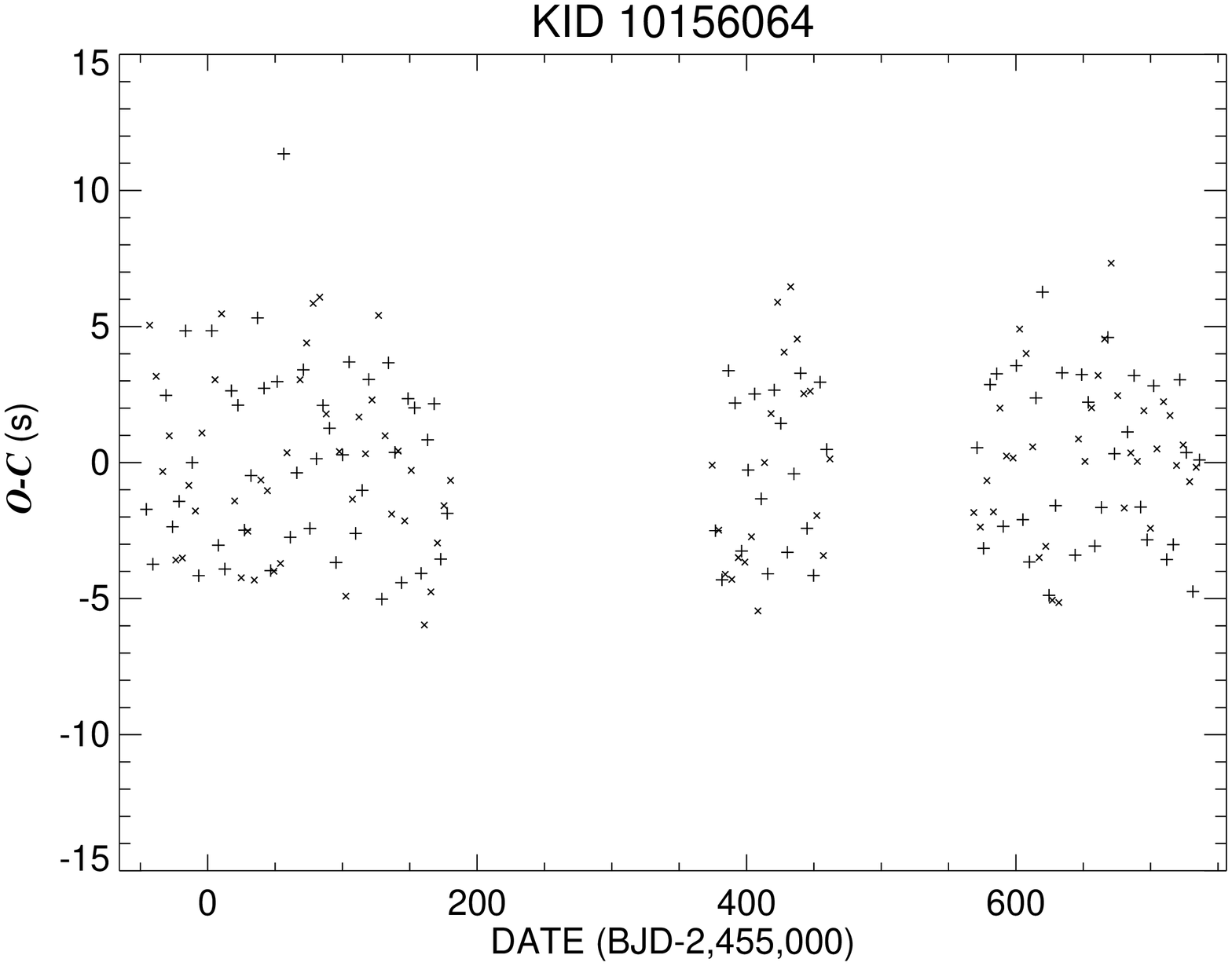}}
\end{center}
\caption{The observed minus calculated eclipse times relative to
a linear ephemeris.  The primary and secondary eclipse
times are indicated by $+$ and $\times$ symbols, 
respectively.}
\end{figure}

\clearpage
\setcounter{figure}{1}
\renewcommand{\thefigure}{\arabic{figure}.32}
\begin{figure}
\begin{center}
{\includegraphics[angle=90,height=12cm]{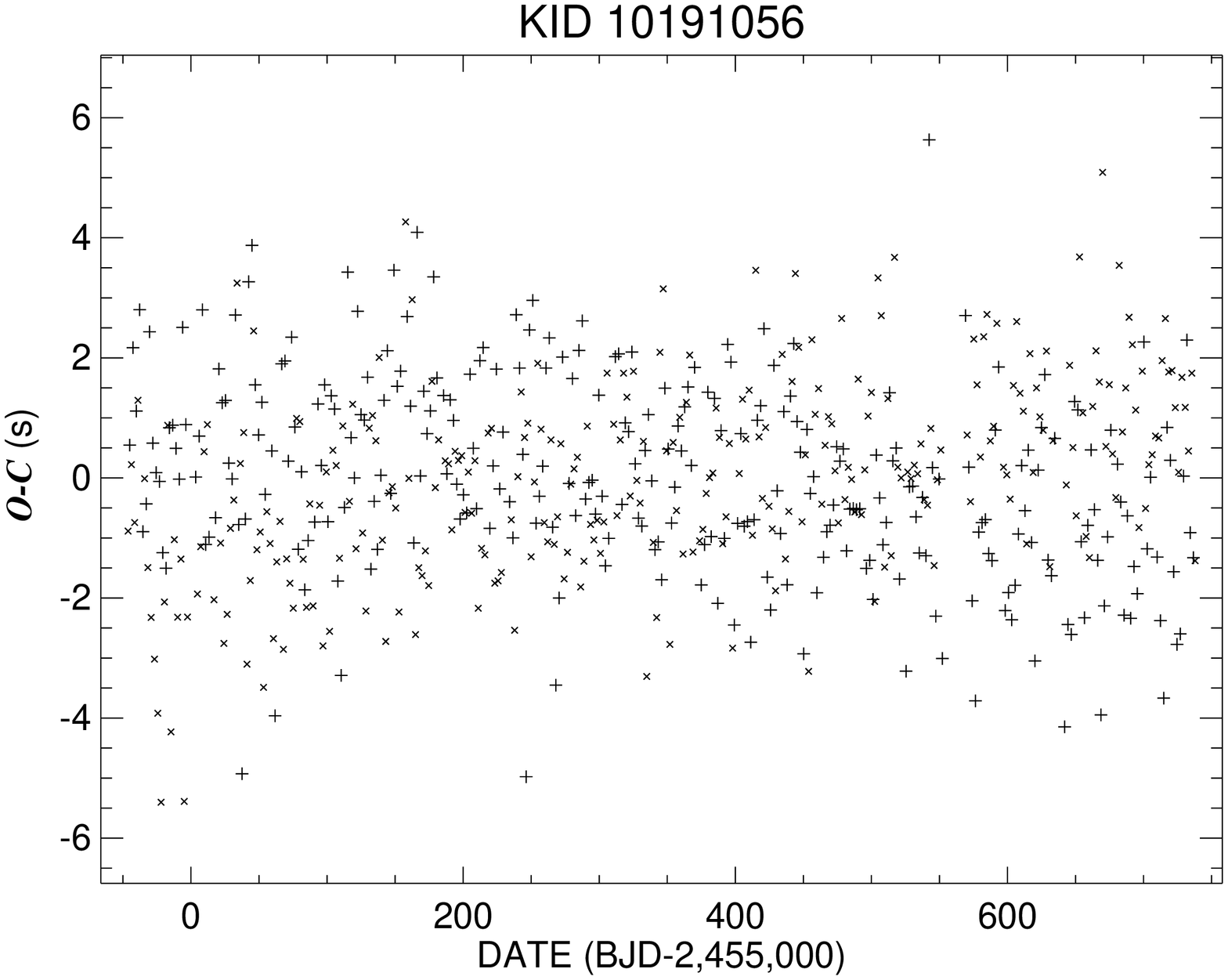}}
\end{center}
\caption{The observed minus calculated eclipse times relative to
a linear ephemeris.  The primary and secondary eclipse
times are indicated by $+$ and $\times$ symbols, 
respectively.}
\end{figure}

\clearpage
\setcounter{figure}{1}
\renewcommand{\thefigure}{\arabic{figure}.33}
\begin{figure}
\begin{center}
{\includegraphics[angle=90,height=12cm]{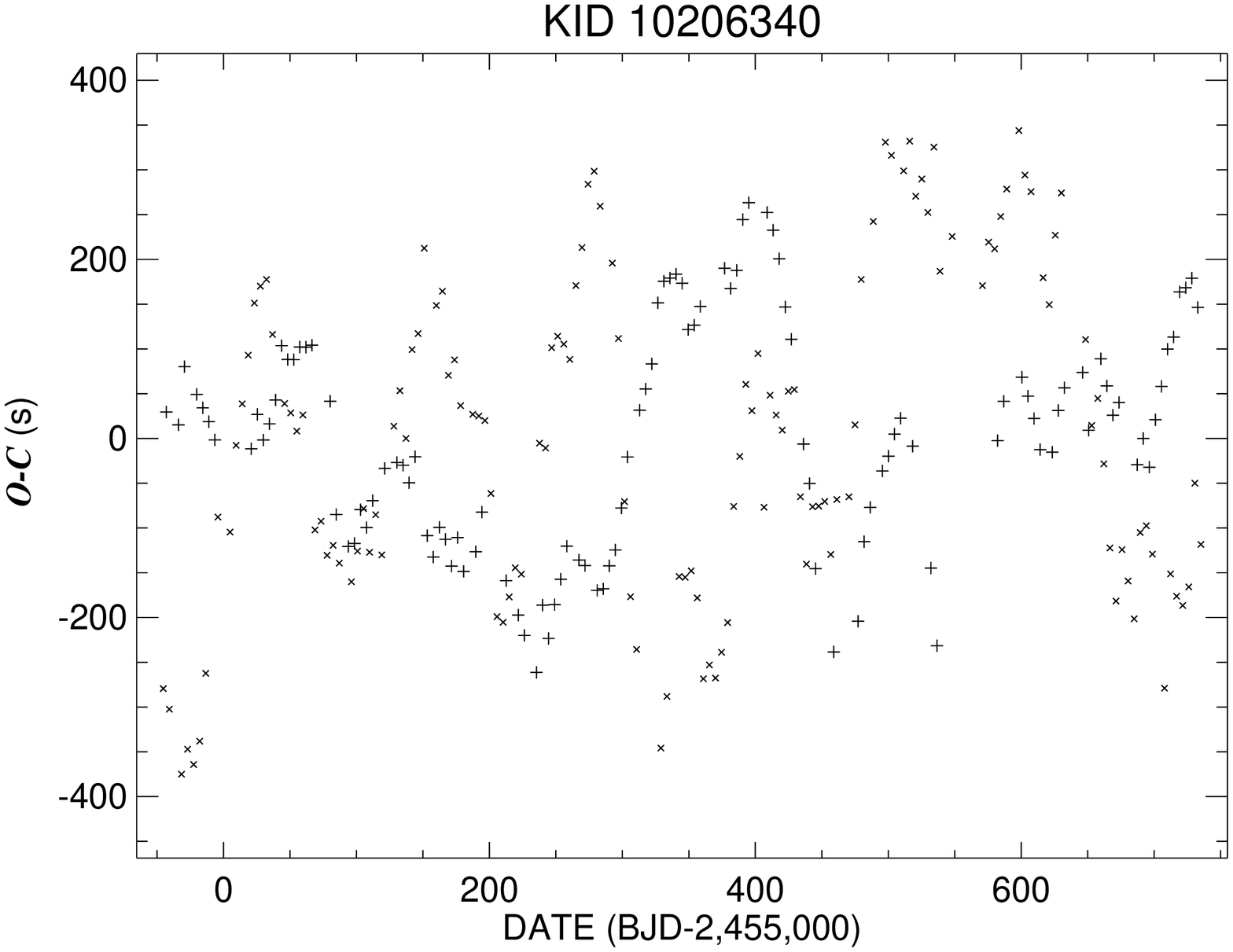}}
\end{center}
\caption{The observed minus calculated eclipse times relative to
a linear ephemeris.  The primary and secondary eclipse
times are indicated by $+$ and $\times$ symbols, 
respectively.}
\end{figure}

\clearpage
\setcounter{figure}{1}
\renewcommand{\thefigure}{\arabic{figure}.34}
\begin{figure}
\begin{center}
{\includegraphics[angle=90,height=12cm]{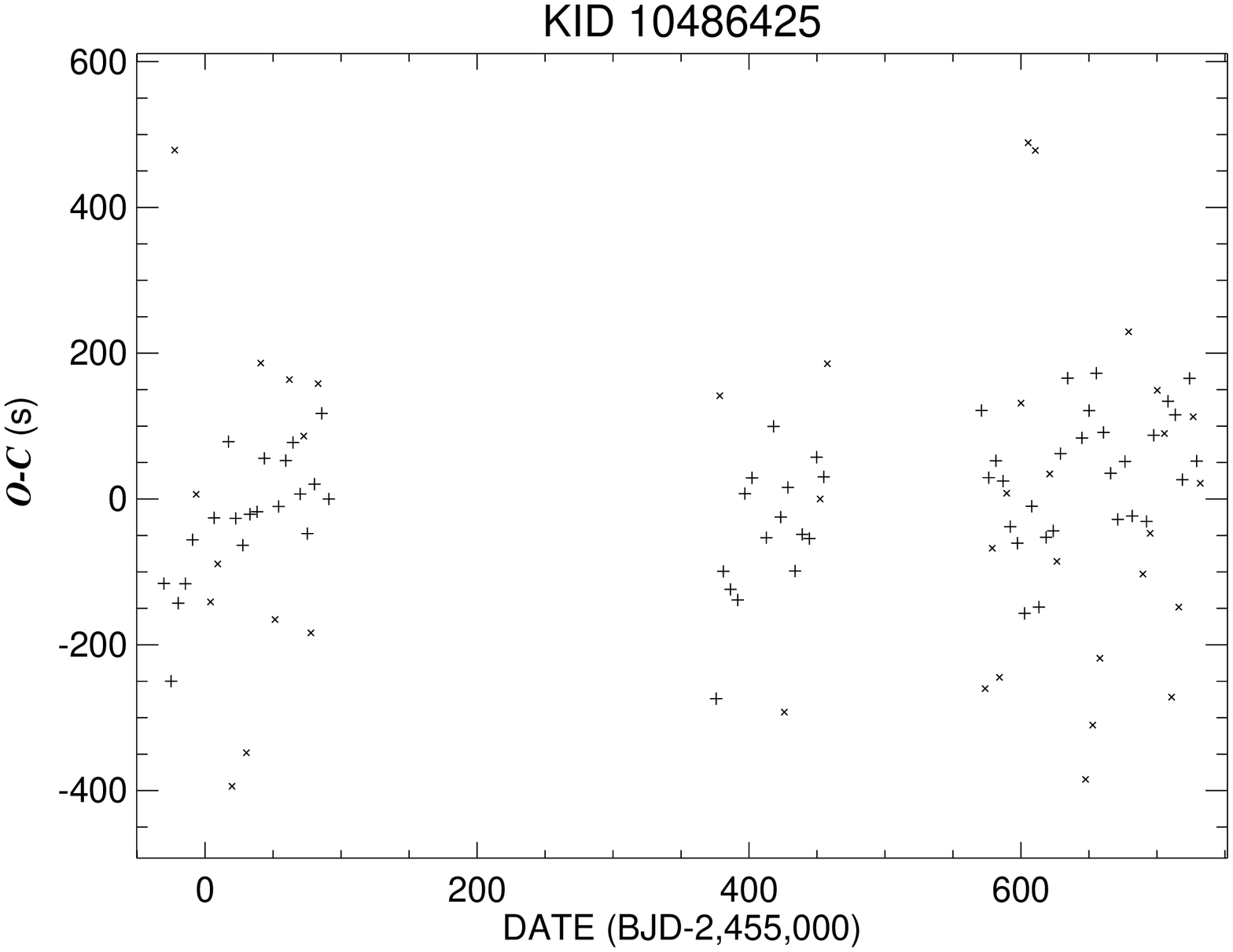}}
\end{center}
\caption{The observed minus calculated eclipse times relative to
a linear ephemeris.  The primary and secondary eclipse
times are indicated by $+$ and $\times$ symbols, 
respectively.}
\end{figure}

\clearpage
\setcounter{figure}{1}
\renewcommand{\thefigure}{\arabic{figure}.35}
\begin{figure}
\begin{center}
{\includegraphics[angle=90,height=12cm]{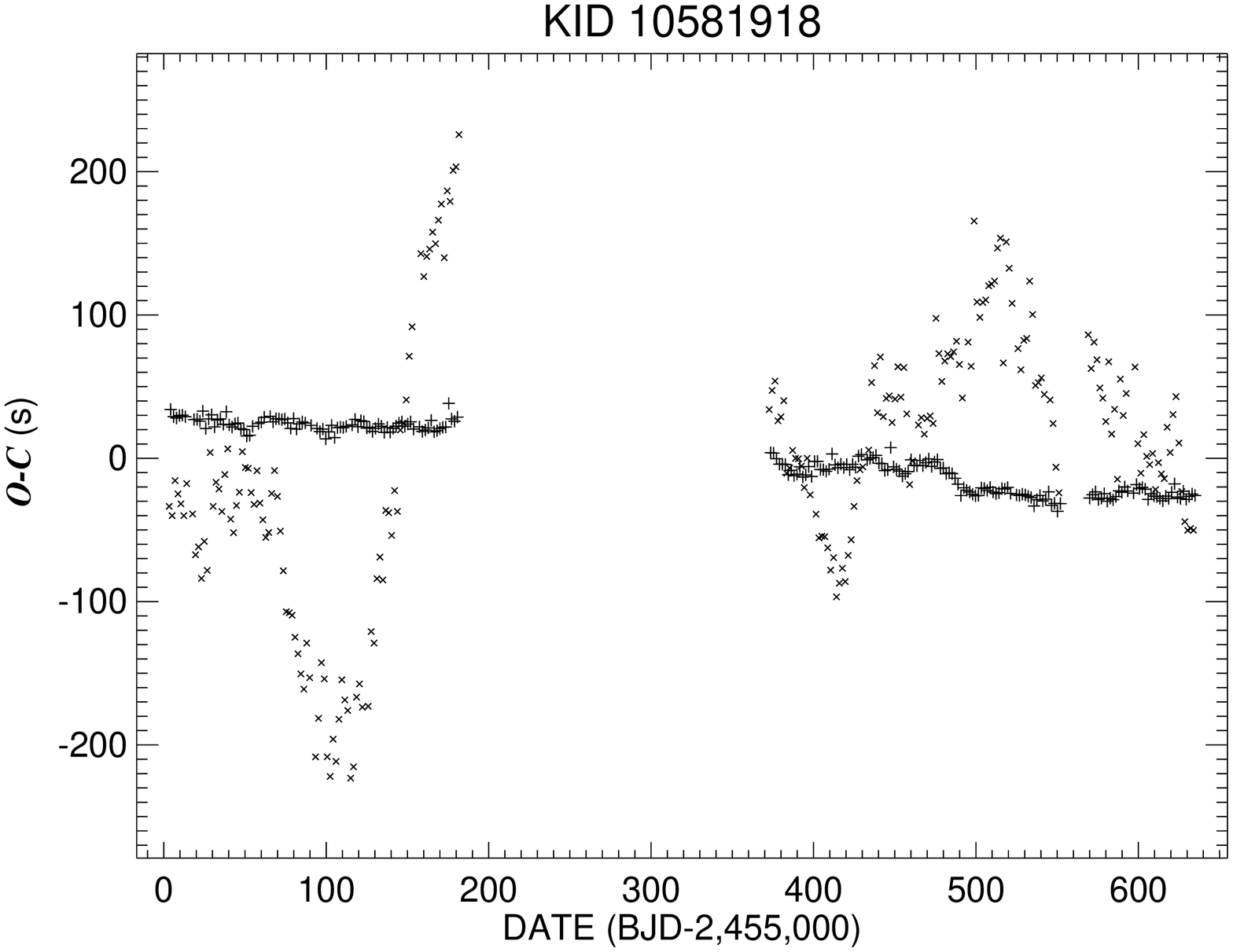}}
\end{center}
\caption{The observed minus calculated eclipse times relative to
a linear ephemeris.  The primary and secondary eclipse
times are indicated by $+$ and $\times$ symbols, 
respectively.}
\end{figure}

\clearpage
\setcounter{figure}{1}
\renewcommand{\thefigure}{\arabic{figure}.36}
\begin{figure}
\begin{center}
{\includegraphics[angle=90,height=12cm]{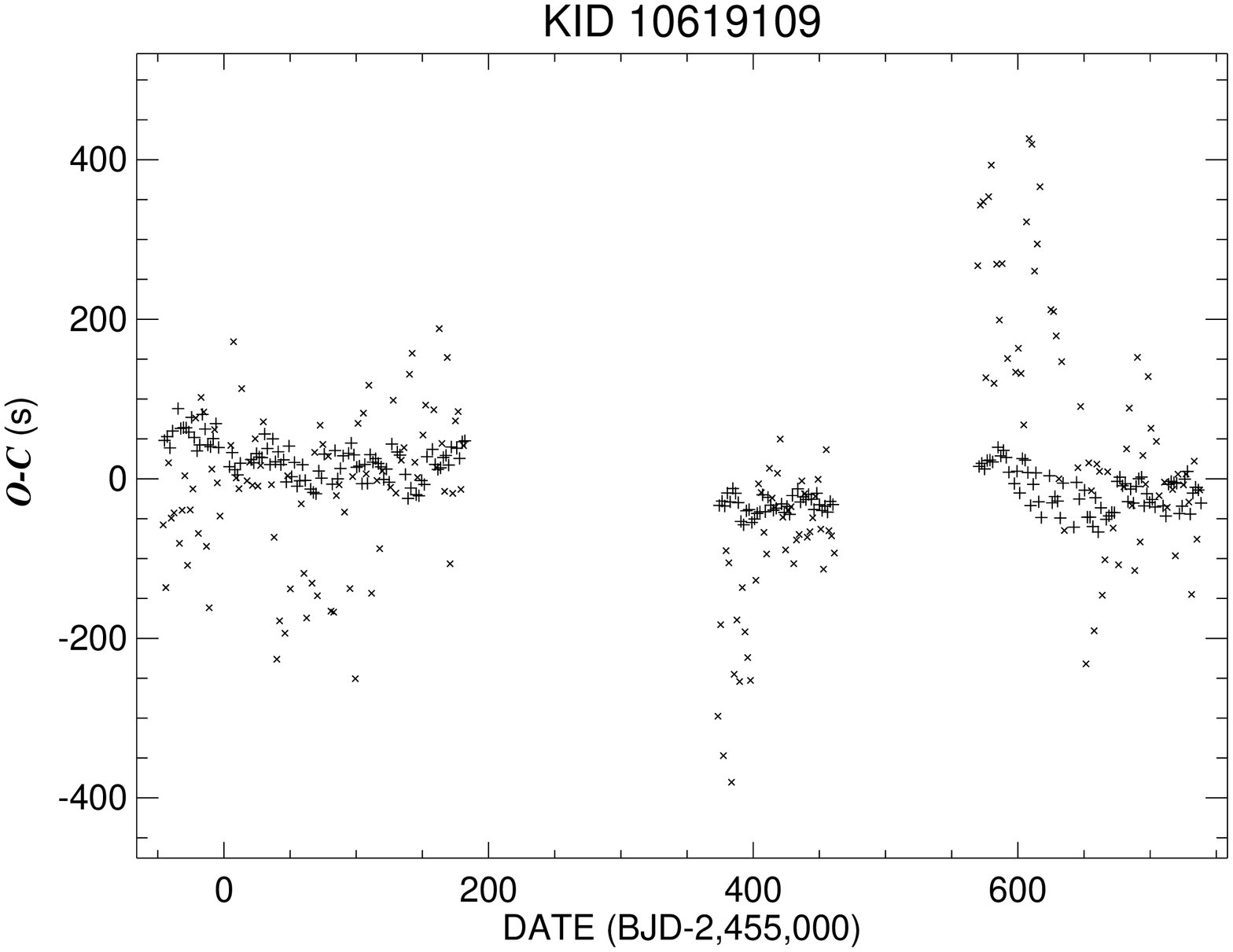}}
\end{center}
\caption{The observed minus calculated eclipse times relative to
a linear ephemeris.  The primary and secondary eclipse
times are indicated by $+$ and $\times$ symbols, 
respectively.}
\end{figure}

\clearpage
\setcounter{figure}{1}
\renewcommand{\thefigure}{\arabic{figure}.37}
\begin{figure}
\begin{center}
{\includegraphics[angle=90,height=12cm]{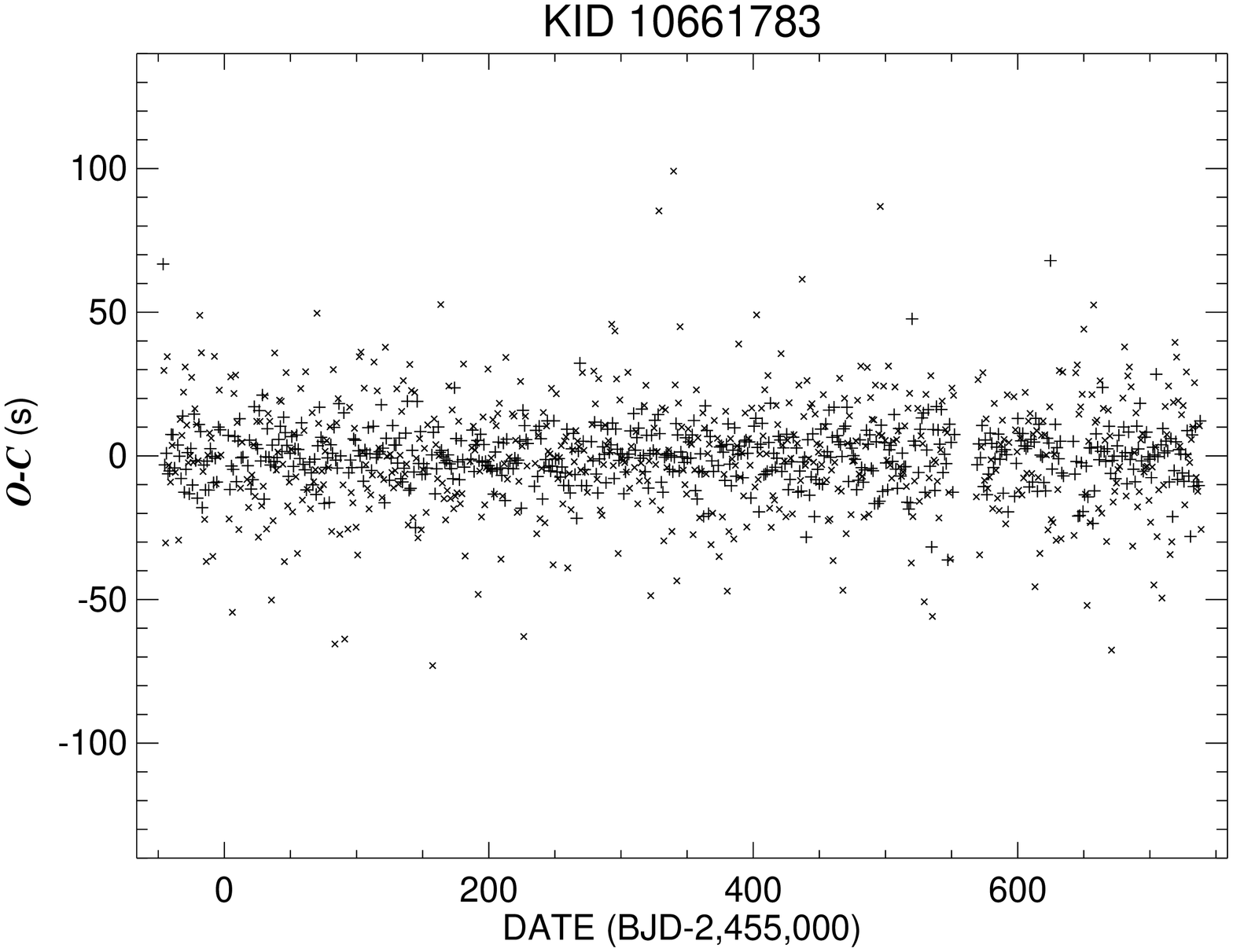}}
\end{center}
\caption{The observed minus calculated eclipse times relative to
a linear ephemeris.  The primary and secondary eclipse
times are indicated by $+$ and $\times$ symbols, 
respectively.}
\end{figure}

\clearpage
\setcounter{figure}{1}
\renewcommand{\thefigure}{\arabic{figure}.38}
\begin{figure}
\begin{center}
{\includegraphics[angle=90,height=12cm]{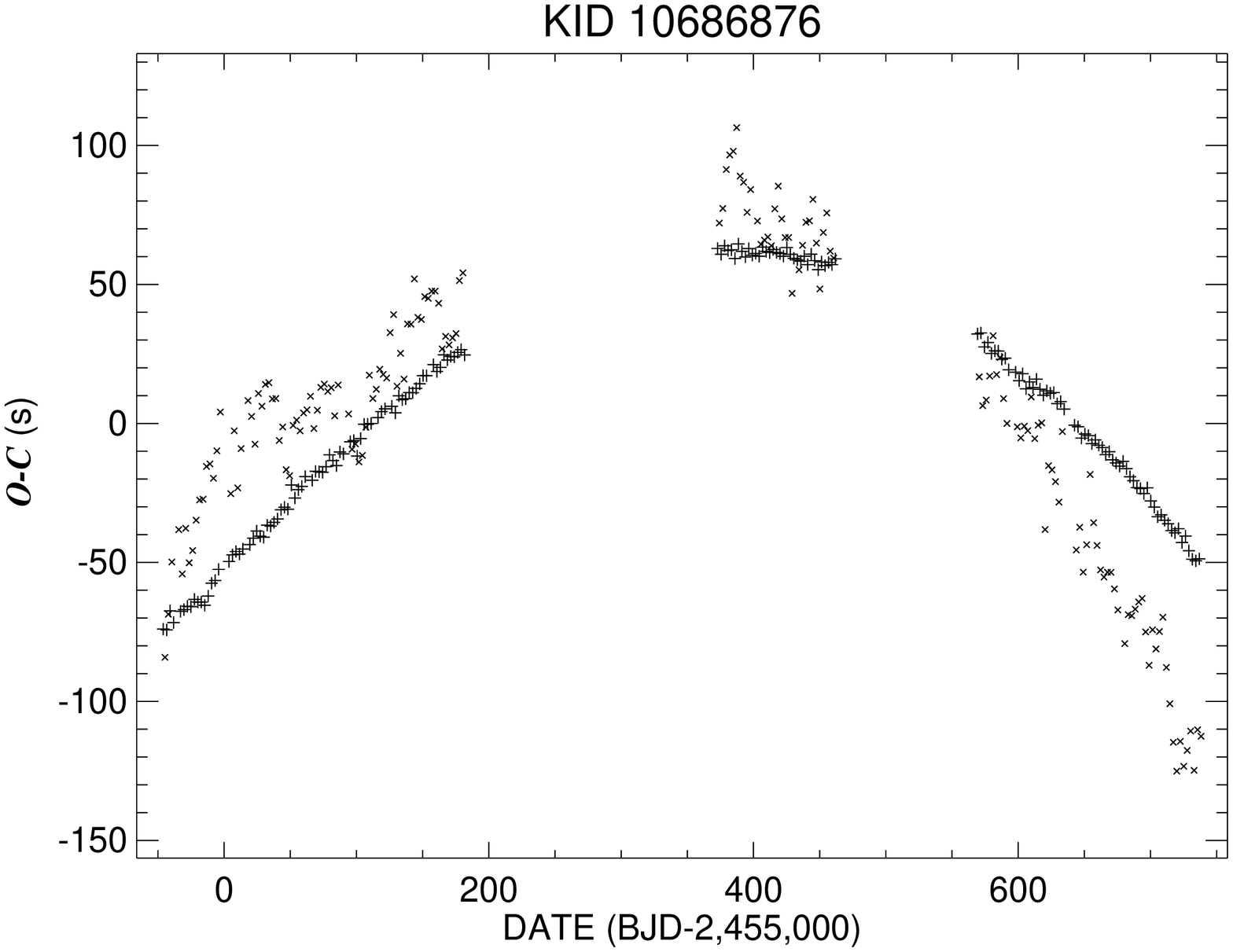}}
\end{center}
\caption{The observed minus calculated eclipse times relative to
a linear ephemeris.  The primary and secondary eclipse
times are indicated by $+$ and $\times$ symbols, 
respectively.}
\end{figure}

\clearpage
\setcounter{figure}{1}
\renewcommand{\thefigure}{\arabic{figure}.39}
\begin{figure}
\begin{center}
{\includegraphics[angle=90,height=12cm]{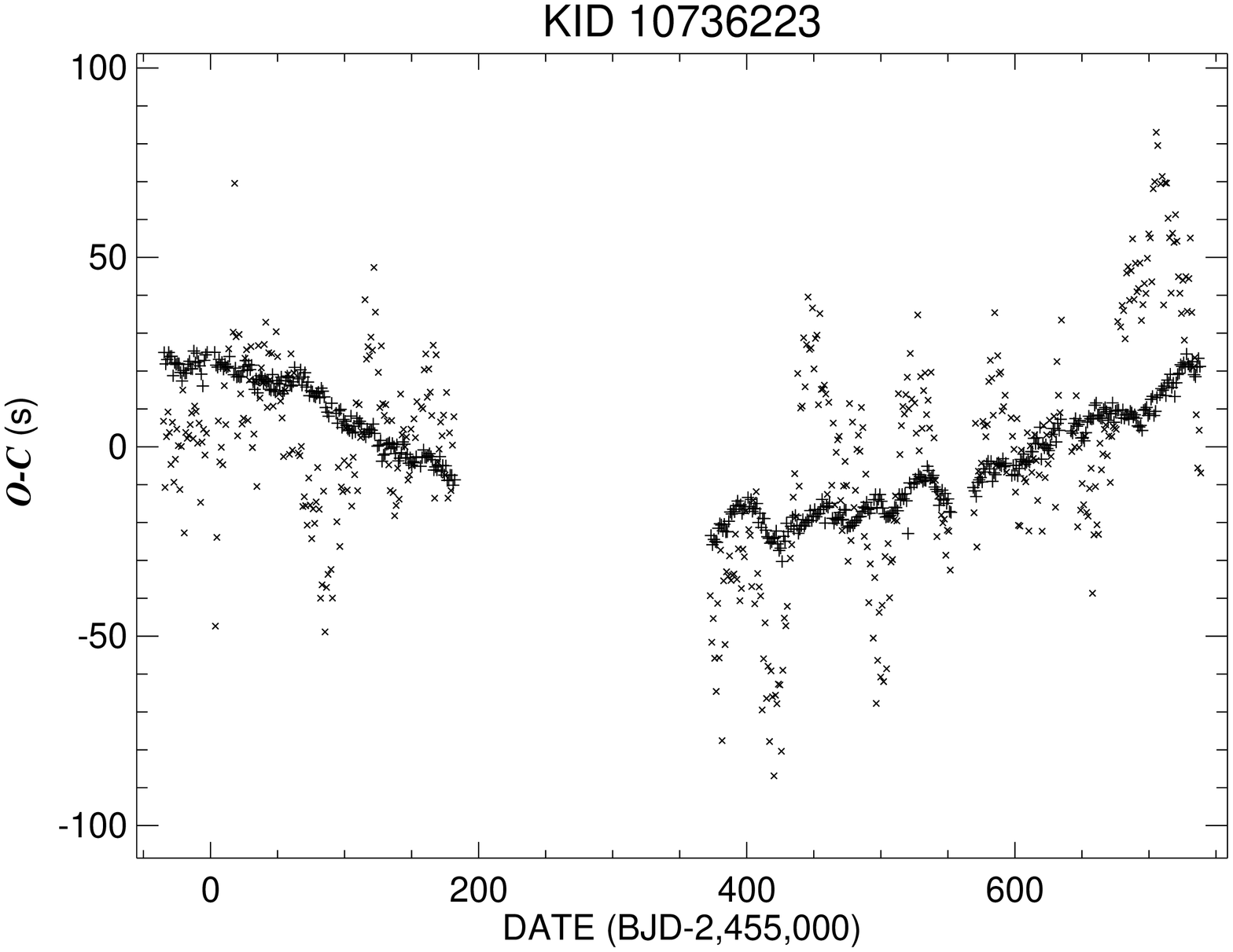}}
\end{center}
\caption{The observed minus calculated eclipse times relative to
a linear ephemeris.  The primary and secondary eclipse
times are indicated by $+$ and $\times$ symbols, 
respectively.}
\end{figure}

\clearpage
\setcounter{figure}{1}
\renewcommand{\thefigure}{\arabic{figure}.40}
\begin{figure}
\begin{center}
{\includegraphics[angle=90,height=12cm]{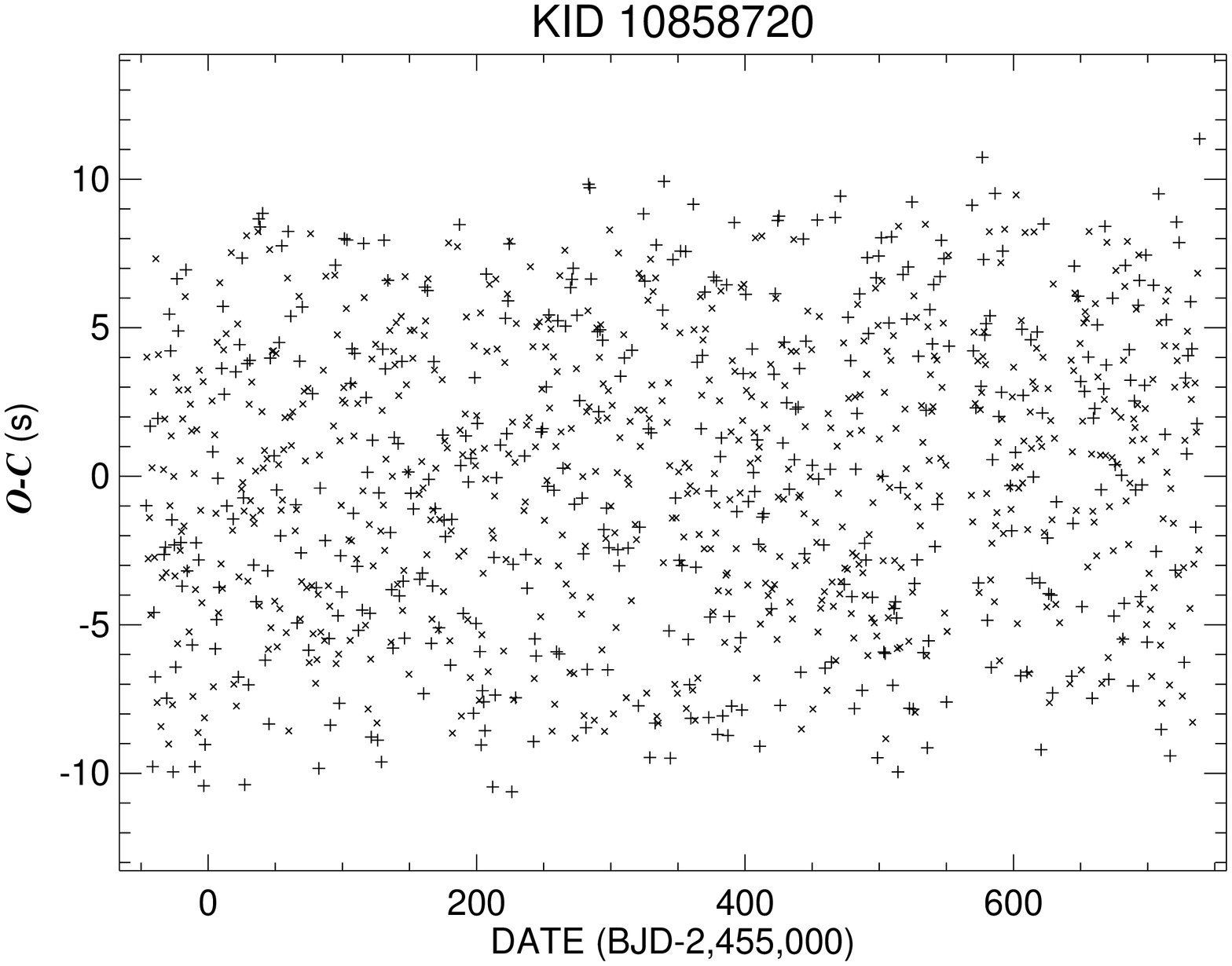}}
\end{center}
\caption{The observed minus calculated eclipse times relative to
a linear ephemeris.  The primary and secondary eclipse
times are indicated by $+$ and $\times$ symbols, 
respectively.}
\end{figure}

\clearpage
\setcounter{figure}{1}
\renewcommand{\thefigure}{\arabic{figure}.41}
\begin{figure}
\begin{center}
{\includegraphics[angle=90,height=12cm]{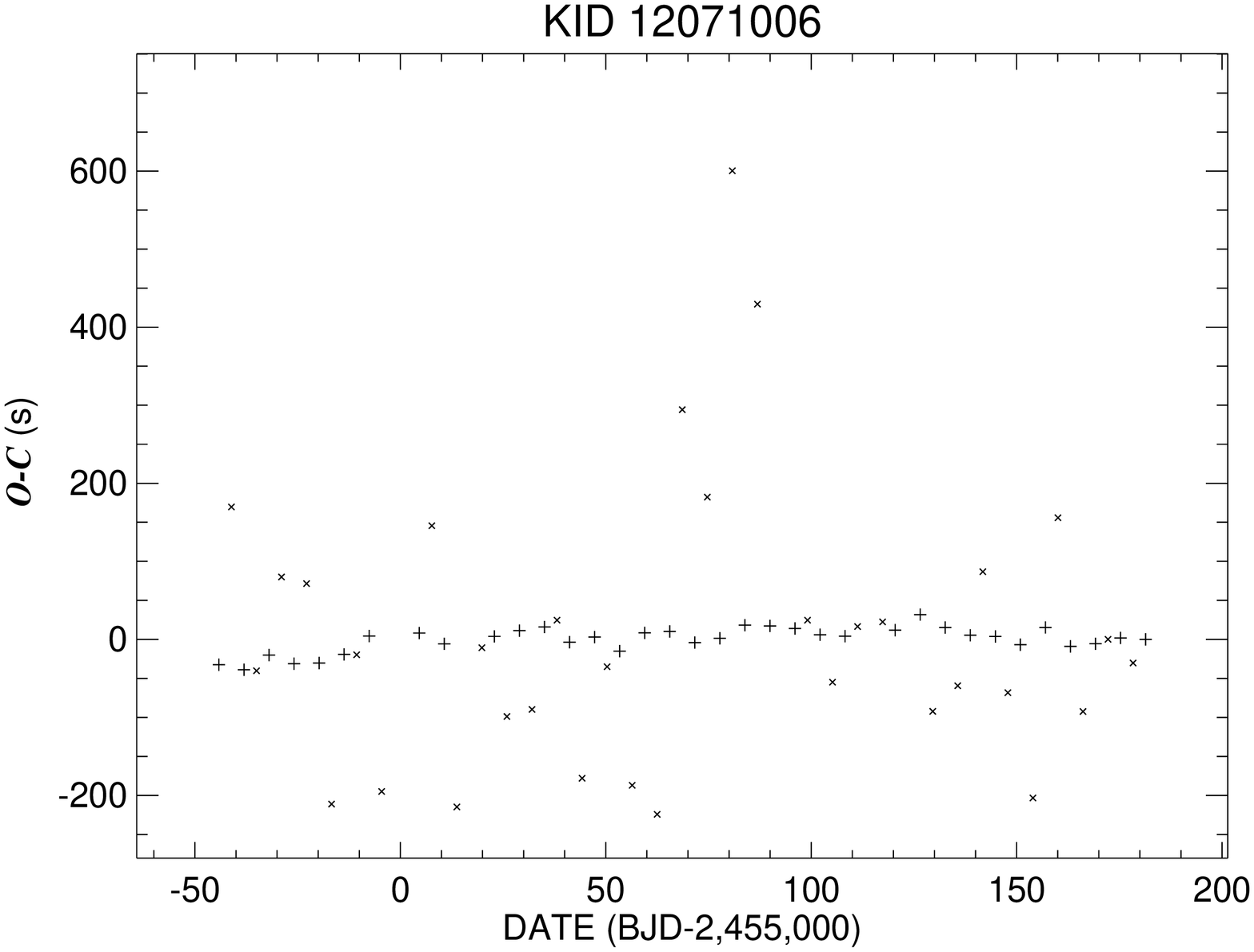}}
\end{center}
\caption{The observed minus calculated eclipse times relative to
a linear ephemeris.  The primary and secondary eclipse
times are indicated by $+$ and $\times$ symbols, 
respectively.}
\end{figure}


\begin{figure}
\figurenum{3}
\begin{center}
{\includegraphics[angle=90,height=12cm]{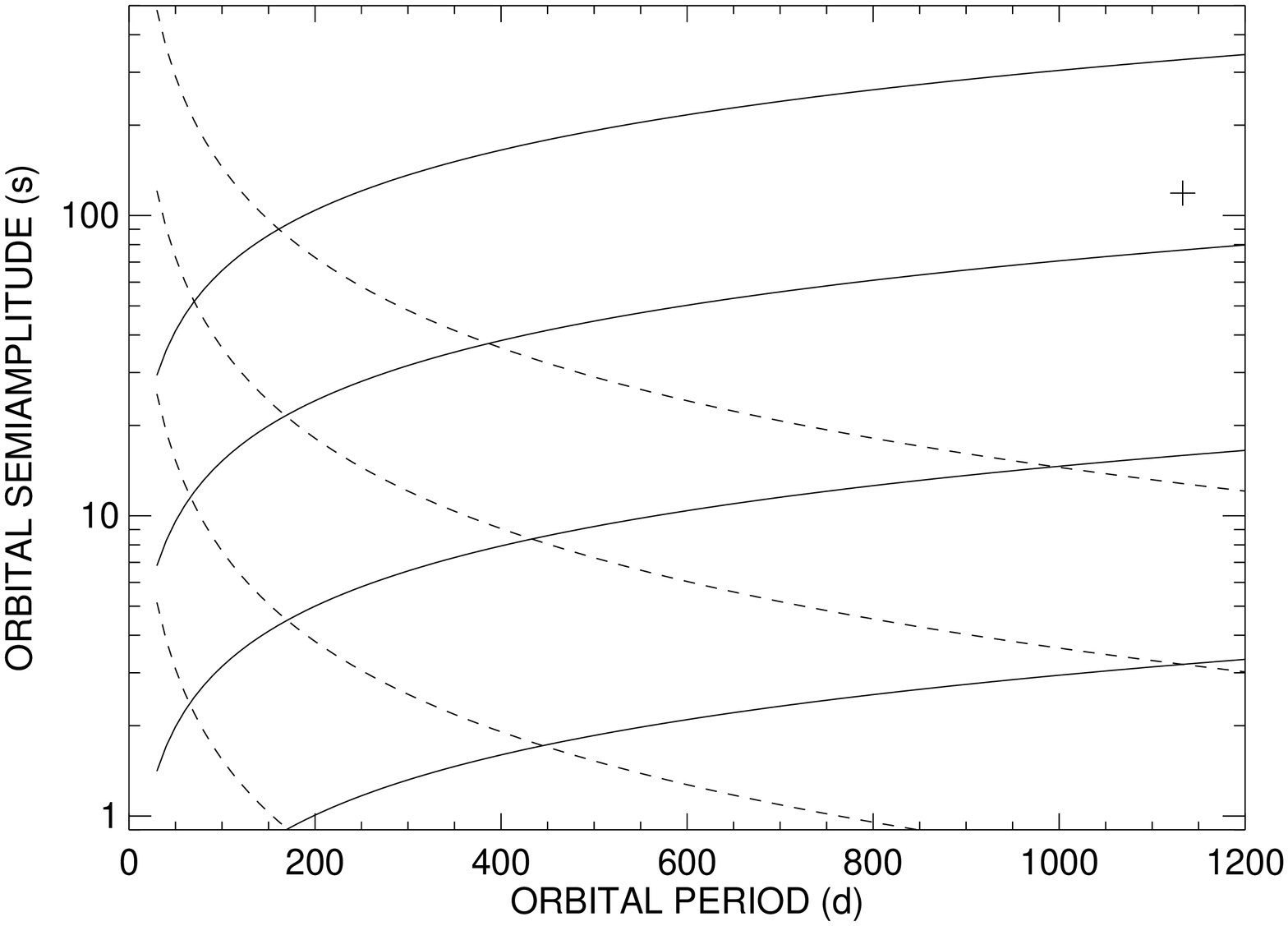}}
\end{center}
\caption{The predicted semiamplitudes for the light travel time effect
({\it solid lines}) and the dynamical effect ({\it dashed lines})
for a third body mass of 0.008, 0.04, 0.2, and $1 M_\odot$, 
from bottom to top, respectively.  The plus sign marks the 
preliminary period and LITE semiamplitude for KID 9402652.}
\label{f3}
\end{figure}


\end{document}